\documentclass[12pt,letterpaper]{article}

\usepackage[
letterpaper,
top=0.45in,
bottom=0.60in,
left=0.65in,
right=0.65in,
includeheadfoot
]{geometry}

\usepackage[T1]{fontenc}
\usepackage[utf8]{inputenc} 
\usepackage{lmodern}
\usepackage{microtype}
\usepackage{setspace}

\usepackage{titlesec}

\titlespacing*{\section}
{0pt}
{2.3ex plus 0.6ex minus 0.25ex}
{1.15ex plus 0.25ex}

\titlespacing*{\subsection}
{0pt}
{1.75ex plus 0.45ex minus 0.20ex}
{0.9ex plus 0.20ex}

\titlespacing*{\subsubsection}
{0pt}
{1.3ex plus 0.35ex minus 0.15ex}
{0.7ex plus 0.15ex}

\titleformat{\section}
{\normalfont\Large\bfseries}
{\thesection}
{0.75em}
{}

\titleformat{\subsection}
{\normalfont\large\bfseries}
{\thesubsection}
{0.75em}
{}

\titleformat{\subsubsection}
{\normalfont\normalsize\bfseries}
{\thesubsubsection}
{0.75em}
{}

\usepackage{
	amsmath,
	amssymb,
	amsfonts,
	amsthm,
	bm,
	mathtools,
	empheq
}

\allowdisplaybreaks[2]

\makeatletter
\@addtoreset{equation}{section}
\@addtoreset{equation}{subsection}
\@addtoreset{equation}{subsubsection}

\renewcommand{\theequation}{%
	\ifnum\value{subsubsection}>0
	\thesubsubsection.\arabic{equation}%
	\else
	\ifnum\value{subsection}>0
	\thesubsection.\arabic{equation}%
	\else
	\thesection.\arabic{equation}%
	\fi
	\fi
}
\makeatother

\usepackage{graphicx}

\graphicspath{{figures/}}

\usepackage{xcolor}
\usepackage{caption}
\usepackage{subcaption}
\usepackage{float}
\usepackage{booktabs}
\usepackage{longtable}
\usepackage{array}
\usepackage{tabularx}
\usepackage{xltabular}
\usepackage{ragged2e}
\usepackage{enumitem}
\usepackage{tikz}

\setlist{
	topsep=3pt,
	itemsep=1.5pt,
	parsep=0pt,
	partopsep=0pt,
	leftmargin=2.0em
}

\renewcommand{\arraystretch}{1.08}

\newcolumntype{L}[1]{>{\RaggedRight\arraybackslash}p{#1}}
\newcolumntype{C}[1]{>{\centering\arraybackslash}p{#1}}
\newcolumntype{R}[1]{>{\raggedleft\arraybackslash}p{#1}}
\newcolumntype{B}[1]{>{\RaggedRight\bfseries\arraybackslash}p{#1}}
\newcolumntype{Y}{>{\RaggedRight\arraybackslash}X}

\newcommand{\EqCell}[1]{%
	\begin{minipage}[t]{\linewidth}
		\centering
		\scriptsize
		\setlength{\abovedisplayskip}{2pt}
		\setlength{\belowdisplayskip}{2pt}
		\[
		#1
		\]
	\end{minipage}
}

\newcommand{\CenteredTableHead}[1]{\hfill\textbf{#1}\hfill\mbox{}}

\usepackage{siunitx}

\usepackage{fancyhdr}

\fancypagestyle{titlepagefooter}{
	\fancyhf{}
	\fancyfoot[C]{Email: \texttt{lillian.kleess@baruchmail.cuny.edu}}
}

\definecolor{linkblue}{RGB}{35,65,120}
\definecolor{citeblue}{RGB}{45,85,140}

\usepackage[
colorlinks=true,
linkcolor=linkblue,
citecolor=citeblue,
urlcolor=linkblue
]{hyperref}

\usepackage[nameinlink,noabbrev]{cleveref}

\newtheoremstyle{paperplain}
{8pt}
{8pt}
{\itshape}
{}
{\bfseries}
{.}
{0.6em}
{\thmname{#1}\thmnumber{ #2}\thmnote{ \textbf{(#3)}}}

\newtheoremstyle{paperdefinition}
{8pt}
{8pt}
{\normalfont}
{}
{\bfseries}
{.}
{0.6em}
{\thmname{#1}\thmnumber{ #2}\thmnote{ \textbf{(#3)}}}

\newtheoremstyle{paperremark}
{7pt}
{7pt}
{\normalfont}
{}
{\bfseries\itshape}
{.}
{0.6em}
{\thmname{#1}\thmnumber{ #2}\thmnote{ \textbf{\textit{(#3)}}}}

\theoremstyle{paperplain}
\newtheorem{theorem}{Theorem}[section]
\newtheorem{proposition}[theorem]{Proposition}
\newtheorem{corollary}[theorem]{Corollary}

\theoremstyle{paperdefinition}
\newtheorem{definition}[theorem]{Definition}

\theoremstyle{paperremark}
\newtheorem{remark}[theorem]{Remark}

\crefname{equation}{Eq.}{Eqs.}
\Crefname{equation}{Equation}{Equations}

\crefname{theorem}{Theorem}{Theorems}
\Crefname{theorem}{Theorem}{Theorems}

\crefname{proposition}{Proposition}{Propositions}
\Crefname{proposition}{Proposition}{Propositions}

\crefname{corollary}{Corollary}{Corollaries}
\Crefname{corollary}{Corollary}{Corollaries}

\crefname{lemma}{Lemma}{Lemmas}
\Crefname{lemma}{Lemma}{Lemmas}

\crefname{definition}{Definition}{Definitions}
\Crefname{definition}{Definition}{Definitions}

\crefname{axiom}{Axiom}{Axioms}
\Crefname{axiom}{Axiom}{Axioms}

\crefname{assumption}{Assumption}{Assumptions}
\Crefname{assumption}{Assumption}{Assumptions}

\crefname{model}{Model}{Models}
\Crefname{model}{Model}{Models}

\crefname{principle}{Principle}{Principles}
\Crefname{principle}{Principle}{Principles}

\crefname{remark}{Remark}{Remarks}
\Crefname{remark}{Remark}{Remarks}

\crefname{interpretation}{Interpretation}{Interpretations}
\Crefname{interpretation}{Interpretation}{Interpretations}

\crefname{example}{Example}{Examples}
\Crefname{example}{Example}{Examples}

\crefname{figure}{Fig.}{Figs.}
\Crefname{figure}{Figure}{Figures}

\crefname{table}{Table}{Tables}
\Crefname{table}{Table}{Tables}

\title{
	\vspace{-1.5em}
	{\LARGE\bfseries Experimental Collapse in Virophysics}
	\\[0.55em]
	{\large\itshape Protocol-Resolved Observation, Inference, and Plaque-Assay Blindness}
}
\author{Lillian St. Kleess}
\date{May 27, 2026}

\begin{document}
\begin{abstract}
	Virological measurements are often interpreted as reporting direct properties
	of virions, such as structure, stiffness, mobility, dielectric response,
	infectivity, or titer. In practice, however, an experiment rarely observes the
	full latent virion--environment ensemble. It reports a
	protocol-conditioned observed ensemble shaped by sample preparation, surfaces,
	fields, structured media, timing, biological selection, amplification,
	detection thresholds, and readout. This paper develops a mathematical
	framework for this process, termed \emph{experimental collapse}, within a
	protocol-resolved virophysics. The central object is a null-inclusive protocol
	observation operator,
	\[
	P_{\mathrm{obs},t}^{\varnothing}(\cdot\mid E)
	=
	\mathcal M_E^{\varnothing}P_{\mathrm{ref},t},
	\]
	which maps a reference latent ensemble to the observed ensemble generated by a
	specified protocol \(E\), including possible null outcomes. This formulation
	separates latent-state transformation, survival or detection weighting,
	readout, and non-observation, allowing protocol effects to be treated as
	explicit components of the measurement rather than as informal sources of
	bias.
	
	The framework defines protocol-conditioned latent ensembles, observed
	ensembles, detection yields, collapse functionals, protocol blindness,
	observation equivalence, Fisher-information observability, inverse inference,
	and multi-protocol consistency. It also provides a mechanism-resolved language
	for common collapse channels, including preparation and interface effects,
	geometric projection, reconstruction, surface immobilization, mechanical
	loading, field steering, medium filtering, time-window selection, biological
	amplification, neutralization, endpoint censoring, and detection thresholds.
	In this view, protocol dependence is not only a limitation of measurement. It
	is also the forward mechanism by which latent virion--environment mechanics
	becomes experimentally inferable.
	
	As a worked example, the plaque assay is developed as a classical instance of
	experimental collapse. A plaque count is shown to estimate an effective
	protocol-conditioned infectious concentration,
	\[
	\Lambda_{\mathrm{PFU}}
	=
	\int_{\Psi}
	\pi_{\mathrm{PFU}}(x;E_{\mathrm{PFU}})
	n_{\mathrm{ref}}(x)\,dx,
	\]
	rather than the total physical particle concentration. In the dilute,
	well-mixed, non-overlapping regime, this quantity recovers the standard
	Poisson plaque-count model and PFU titer formula. Extensions to
	overdispersed counts, zero inflation, plaque merging, endpoint dilution,
	neutralization, and morphology-augmented readouts show how deviations from
	the ideal model can be interpreted as protocol-conditioned information rather
	than as merely statistical noise. Overall, the paper argues that virological
	data are most precisely interpreted as outputs of explicit protocol kernels.
	This perspective clarifies what a measurement reports, what it leaves
	unresolved, and how complementary protocols can be designed to recover
	otherwise hidden latent structure.
\end{abstract}
	\maketitle
	\tableofcontents
	\newpage
	
\section{Introduction \& Conceptual Overview}
\label{sec:introduction}

Virological experiments often appear to answer direct questions. How many
infectious particles are present in a sample? What is the structure of the
virion? How stiff is the capsid or envelope? How does a particle move through
mucus, gel, or extracellular matrix? How does it respond to an applied electric
field? These questions are experimentally well defined, but each contains an
important theoretical distinction. A measurement does not usually report the full
microscopic state of a virion population. It reports the part of that state made
visible, stable, selectable, amplifiable, or readable under a particular
experimental protocol.

This paper develops that distinction into a framework called
\emph{protocol-resolved virophysics}. The central idea is that a virological
measurement should not be treated as a protocol-free window onto a virion
population, but as a protocol-conditioned map from latent
virion--environment states to observed data. In its most compact form,
\begin{empheq}[box=\fbox]{equation}
	\text{Latent virion--environment ensemble}
	\;\xrightarrow{\;\text{Experimental protocol}\;}
	\text{Observed ensemble}.
	\label{eq:intro_protocol_conditioned_map}
\end{empheq}
The observed ensemble is genuine experimental data. The purpose of this
distinction is not to weaken the status of such data, but to identify what kind
of object it is: the output of a specified preparation, conditioning, selection,
transformation, amplification, and readout procedure.

We call this protocol-conditioned reduction
\textbf{\emph{experimental collapse}}. The term is used here in a classical
statistical and biophysical sense, \emph{not} as a claim about quantum
measurement. Experimental collapse refers to the way a high-dimensional latent
virion--environment ensemble is transformed, selected, projected, amplified, or
partially erased by an experimental procedure before it becomes reported data.

\medskip

\noindent
At the level of measures, the organizing equation of the paper is
\begin{empheq}[box=\fbox]{equation}
	P_{\mathrm{obs},t}^{\varnothing}(\cdot\mid E)
	=
	\mathcal M_{E,t}^{\varnothing}
	P_{\mathrm{ref},t}.
	\label{eq:intro_protocol_observation_operator_compact}
\end{empheq}
Here \(P_{\mathrm{ref},t}\) is a reference latent ensemble,
\(\mathcal M_{E,t}^{\varnothing}\) is the null-inclusive observation operator
associated with protocol \(E\), and
\(P_{\mathrm{obs},t}^{\varnothing}(\cdot\mid E)\) is the observed ensemble,
including possible null outcomes. Equation
\eqref{eq:intro_protocol_observation_operator_compact} is deliberately compact.
It plays the same introductory role that \(H=T+V\) often plays in mechanics: it
states the structure of the theory first, before the detailed components are
unfolded.

\medskip

\noindent
To write the same idea more explicitly, let \(\Psi\) denote the latent state
space and let \(\mathcal Y\) denote the non-null observed data space. The
null-enlarged observation space is
\begin{equation}
	\mathcal Y^{\varnothing}
	=
	\mathcal Y \cup \{\varnothing\},
	\label{eq:intro_null_enlarged_observation_space}
\end{equation}
where \(\varnothing\) denotes the event that no accepted datum is reported. For
any measurable set \(A\subseteq \mathcal Y^{\varnothing}\),
\begin{empheq}[box=\fbox]{equation}
	P_{\mathrm{obs},t}^{\varnothing}(A\mid E)
	=
	\int_{\Psi}
	K_E^{\varnothing}(A\mid x)\,
	P_{\mathrm{ref},t}(dx).
	\label{eq:intro_protocol_observation_kernel}
\end{empheq}
The kernel \(K_E^{\varnothing}\) is the protocol's observation kernel. It
collects the transformations between a latent state \(x\in\Psi\) and a reported
datum \(y\in\mathcal Y\), including physical conditioning, biological filtering,
selection, loss, readout, and the possibility that no accepted datum is reported
at all.

\begin{figure}[t]
	\centering
	\includegraphics[width=0.99\linewidth]{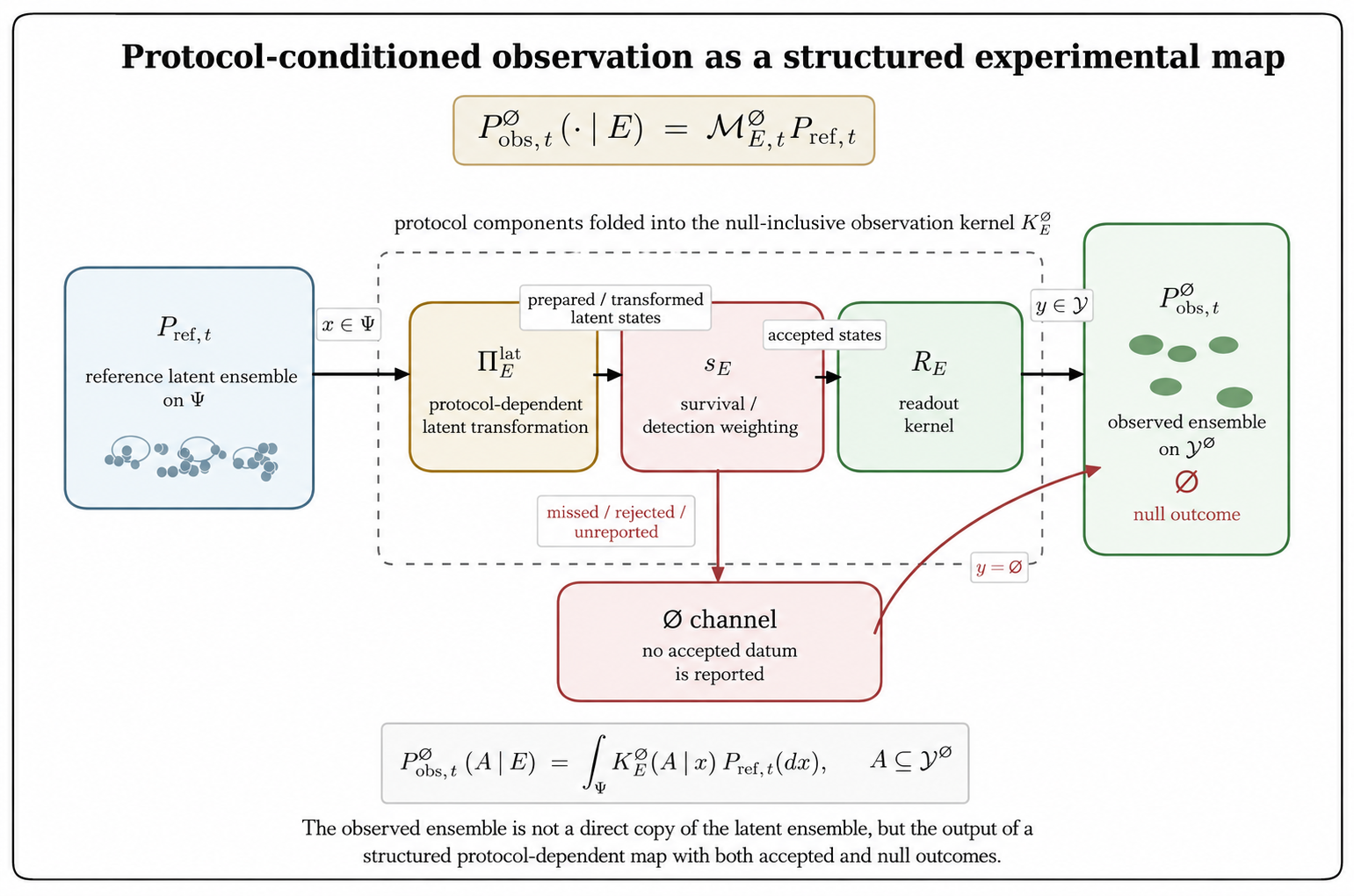}
	\caption{
		Protocol-conditioned observation as a structured experimental map.
		A reference latent ensemble \(P_{\mathrm{ref},t}\) on the latent state
		space \(\Psi\) is acted on by protocol-dependent components, including
		a latent transformation \(\Pi_E^{\mathrm{lat}}\), survival or detection
		weighting \(s_E\), and a readout kernel \(R_E\). These components are
		absorbed into the null-inclusive observation kernel
		\(K_E^{\varnothing}\), which maps latent states either to reported
		observations in \(\mathcal Y\) or to the null outcome \(\varnothing\).
		The observed ensemble \(P_{\mathrm{obs},t}^{\varnothing}\) is therefore
		not a direct copy of the latent ensemble, but the output of a structured,
		protocol-conditioned experimental map.
	}
	\label{fig:intro_protocol_conditioned_observation}
\end{figure}

The need for this framework is visible across virology. In cryo-EM and cryo-ET,
the final density map or tomogram depends not only on virion structure, but also
on grid preparation, air--water interface exposure, ice thickness, vitrification,
particle orientation, particle picking, alignment, classification, and
reconstruction
\cite{Cheng2015,Noble2018,Chen2019,Liu2023,DImprima2019,Hirst2024}. In AFM, a
reported height, rupture force, or indentation stiffness is conditioned by
surface attachment, hydration, tip geometry, loading rate, indentation depth, and
the mechanical model used to interpret the force--deformation curve
\cite{Baclayon2010,Mateu2012,Marchetti2016,Kiss2021,Lyonnais2021}. In
dielectrophoresis and electrorotation, the observed trajectory, trapping
behavior, crossover frequency, or rotation rate depends on the imposed electric
field, medium conductivity, permittivity, hydrodynamic drag, and particle
polarizability \cite{Hughes1998,Hughes2002,Pethig2010,Kim2019}. In mucus or gel
tracking, a reported diffusivity or trajectory class is a joint property of the
virion and the structured medium, including mesh scale, mucin interactions,
adhesion, antibody-mediated immobilization, local rheology, and tracking
thresholds \cite{Boukari2009,Wang2017,Kaler2022,Vahey2019,Abrami2024}.

\medskip

\noindent
Even the plaque assay, one of the most classical and useful measurements in
virology, has this structure. A plaque count is not a direct count of physical
particles. It is a count of visible infectious lesions produced under a specified
cell line, adsorption time, overlay, incubation time, staining method, and
counting rule
\cite{Dulbecco1952,DulbeccoVogt1954,Cooper1961,Baer2014,Mendoza2020}. A
structurally intact particle, a defective particle, a damaged particle, an
aggregate, a neutralized particle, a cell-line-incompatible particle, and an
entry-competent but replication-incompetent particle may all produce the same
observed result: no plaque. Conversely, a single visible plaque reports a
successful biological pathway, not a full microscopic history. The plaque assay
therefore provides a particularly clear example of experimental collapse: a
biological protocol converts a heterogeneous latent particle population into a
countable set of visible lesions.

The purpose of this paper is not to argue that protocol-conditioned measurements
are unreliable. On the contrary, protocol-conditioned measurements are the
foundation of experimental virology. Cryo-EM is powerful because it preserves and
reconstructs structure. AFM is powerful because it applies force. Field-driven
assays are powerful because they couple to electrical and hydrodynamic response.
Tracking is powerful because it records motion through specified environments.
Plaque assays are powerful because they amplify infectious activity into
countable lesions. The point is more precise: each protocol asks a particular
question of the virion--environment system, and the observed ensemble is the
answer to that question.

This distinction matters because protocol-conditioned observations can be
misread in two opposite ways. The first mistake is to treat the observed ensemble
as if it were a direct, protocol-free sample of the latent virion population. The
second is to dismiss protocol-conditioned data as mere artifact. The framework
developed here avoids both mistakes. It treats the observed ensemble as valid,
but valid as the output of a specified experimental map.

\subsection{Latent, Protocol-Conditioned, and Observed Ensembles}
\label{subsec:intro_latent_protocol_observed}

The basic mathematical object of the paper is a chain of ensembles:
\begin{empheq}[box=\fbox]{equation}
	P_{\mathrm{ref},t}
	\;\xrightarrow{\;\Pi_E^{\mathrm{lat}}\;}
	\widetilde P_{E,t}^{\mathrm{lat}}
	\;\xrightarrow{\;s_E,\;R_E,\;\varnothing\;}
	P_{\mathrm{obs},t}^{\varnothing}(\cdot\mid E).
	\label{eq:intro_ensemble_chain}
\end{empheq}
The reference latent ensemble \(P_{\mathrm{ref},t}\) describes the distribution
of virion or virion--environment states before the additional conditioning
imposed by the experimental protocol. The latent transformation kernel
\(\Pi_E^{\mathrm{lat}}\) describes how preparation, surfaces, fields, loading,
media, timing, or biological conditions transform the latent state before
readout. The survival or detection weight \(s_E\) describes which
protocol-conditioned states remain visible to the accepted readout channel. The
readout kernel \(R_E\) describes how surviving states become reported data. The
null outcome \(\varnothing\) records states that are present in the latent
ensemble but produce no accepted observation under protocol \(E\).

A useful expanded form of the observation kernel is
\begin{equation}
	K_E^{\varnothing}(A\mid x)
	=
	\int_{\Psi_E}
	\left[
	s_E(z)\,
	R_E(A\cap\mathcal Y\mid z)
	+
	\bigl(1-s_E(z)\bigr)\,
	\mathbf 1_A(\varnothing)
	\right]
	\Pi_E^{\mathrm{lat}}(dz\mid x),
	\label{eq:intro_unfolded_observation_kernel}
\end{equation}
where \(z\in\Psi_E\) denotes a protocol-conditioned latent state,
\(s_E(z)\in[0,1]\) is the probability or weight that the state survives into the
accepted readout channel, and \(R_E(\cdot\mid z)\) is the non-null readout
kernel. This expression separates three effects that are often conflated in
informal experimental language: latent transformation, survival or detection
weighting, and readout.

\begin{table}[H]
	\centering
	\renewcommand{\arraystretch}{1.18}
	\begin{tabularx}{0.96\linewidth}{@{}>{\centering\arraybackslash}p{0.20\linewidth}X@{}}
		\toprule
		\textbf{Component} & \textbf{Interpretation} \\
		\midrule
		
		\(\Pi_E^{\mathrm{lat}}\)
		&
		Physical or biological transformation of the latent ensemble before
		readout; examples include adsorption, field steering, dilution,
		mechanical loading, incubation, or passage through a structured medium.
		\\[0.35em]
		
		\(s_E\)
		&
		Survival, visibility, detectability, infectivity, localization, or
		acceptance weighting; this determines which conditioned states remain in
		the reported data channel.
		\\[0.35em]
		
		\(R_E\)
		&
		Instrumental, computational, or biological readout; examples include
		image reconstruction, force-curve fitting, trajectory estimation,
		fluorescence thresholding, or plaque counting.
		\\[0.35em]
		
		\(\varnothing\)
		&
		Null channel for states that are present in the latent ensemble but
		produce no accepted observation under protocol \(E\).
		\\
		\bottomrule
	\end{tabularx}
	\caption{
		Interpretation of the main components of the null-inclusive observation
		kernel in Eq.~\eqref{eq:intro_unfolded_observation_kernel}.
	}
	\label{tab:intro_observation_kernel_components}
\end{table}

This separation is important because different parts of a protocol have
different meanings. Surface adsorption may physically transform the state.
Particle picking may select which states are reported. Finite resolution may
project a surviving state into a lower-dimensional representation. A plaque
assay may biologically amplify some states while sending others into the null
channel. These effects are not interchangeable. They are distinct components of
the observation operator.

\subsection{Protocol Blindness}
\label{subsec:intro_protocol_blindness}

Once measurement is written as a map from latent states to observed data, it
becomes possible to ask a precise complementary question: what does a given
protocol fail to distinguish? This failure is not necessarily an experimental
mistake. It is a structural property of the protocol. A protocol can only report
distinctions to which its preparation, selection, transformation, and readout
channels are sensitive.

At the state level, two latent states \(x_1,x_2\in\Psi\) are
\emph{observation-equivalent} under protocol \(E\) if they induce the same
null-inclusive distribution over observed outcomes:
\begin{empheq}[box=\fbox]{equation}
	x_1 \sim_E x_2
	\quad\Longleftrightarrow\quad
	K_E^{\varnothing}(\cdot\mid x_1)
	=
	K_E^{\varnothing}(\cdot\mid x_2).
	\label{eq:intro_observation_equivalence}
\end{empheq}
No amount of repeated observation under protocol \(E\) alone can distinguish
states in the same equivalence class, unless additional assumptions, additional
measurements, or protocol variation are introduced. In this sense, the protocol
defines a quotient of the latent state space:
\begin{empheq}[box=\fbox]{equation}
	\Psi
	\longrightarrow
	\Psi/{\sim_E}.
	\label{eq:intro_protocol_quotient_space}
\end{empheq}
The quotient \(\Psi/{\sim_E}\) is the part of the latent state space that remains
distinguishable after the protocol has acted. Distinctions that survive the map
are observable under \(E\); distinctions identified by the equivalence relation
are collapsed by \(E\).

This is the basic meaning of \emph{protocol blindness}. A protocol is blind to a
latent distinction when different latent states, mechanisms, or parameter values
produce the same observed distribution under that protocol. Protocol blindness
therefore does not mean that the hidden distinction is physically absent. It
means that the distinction is not resolved by the experimental map being used. At the parameter level, protocol blindness can be expressed through Fisher
information. 
\medskip 

\noindent Let
\[
p_E^{\varnothing}(y\mid\theta)
\]
be a protocol-conditioned likelihood, with density or mass taken with respect to
a common dominating measure on \(\mathcal Y^{\varnothing}\), including the atom
at \(\varnothing\). The Fisher-information matrix is
\begin{equation}
	\mathcal I_E(\theta)
	=
	\mathbb E_{Y\sim p_E^{\varnothing}(\cdot\mid\theta)}
	\left[
	\nabla_\theta\log p_E^{\varnothing}(Y\mid\theta)
	\nabla_\theta\log p_E^{\varnothing}(Y\mid\theta)^{\mathsf T}
	\right].
	\label{eq:intro_fisher_information}
\end{equation}
A parameter direction \(v\) is locally blind under protocol \(E\) when the
likelihood does not change to first order in that direction. Equivalently, under
the usual regularity assumptions,
\begin{empheq}[box=\fbox]{equation}
	v\in\mathcal B_E(\theta)
	\quad\Longleftrightarrow\quad
	v^{\mathsf T}\mathcal I_E(\theta)v=0,
	\label{eq:intro_fisher_blind_direction}
\end{empheq}
where \(\mathcal B_E(\theta)\) is the local blind subspace of the protocol at
parameter value \(\theta\). Thus, \(\mathcal I_E(\theta)\) identifies which
parameter combinations are locally visible to the protocol and which parameter
combinations lie in directions of vanishing statistical sensitivity.

This gives a local, information-theoretic form of protocol blindness. If
\(v^{\mathsf T}\mathcal I_E(\theta)v=0\), then infinitesimal changes in
\(\theta\) along \(v\) do not produce a first-order change in the observed
likelihood. Such a direction may correspond to a genuine physical parameter, but
one that is not coupled to the protocol, is selected away, is averaged over, is
below threshold, or is mapped into the same observed category as other states.

\begin{table}[H]
	\centering
	\renewcommand{\arraystretch}{1.18}
	\begin{tabularx}{0.96\linewidth}{@{}p{0.28\linewidth}X@{}}
		\toprule
		\textbf{Mode of blindness} & \textbf{Interpretation} \\
		\midrule
		
		Uncoupled sector
		&
		The latent variable exists, but the protocol does not couple to it. For
		example, an electrical assay may be insensitive to a mechanical parameter
		if the parameter does not affect polarizability, charge distribution, or
		hydrodynamic response.
		\\[0.35em]
		
		Selection blindness
		&
		The protocol preferentially admits some states and excludes others before
		readout. The excluded states enter the null channel or are absent from the
		accepted data set.
		\\[0.35em]
		
		Projection or averaging
		&
		Distinct latent states are mapped to the same lower-dimensional observable,
		such as a scalar intensity, fitted stiffness, apparent diffusivity, or
		plaque count.
		\\[0.35em]
		
		Threshold blindness
		&
		The relevant distinction changes the observed distribution, but only below
		the resolution, detection, counting, or classification threshold of the
		experiment.
		\\[0.35em]
		
		Confounding
		&
		Two or more latent parameters affect the observed likelihood in nearly the
		same direction, making their individual contributions difficult or
		impossible to separate under protocol \(E\) alone.
		\\
		\bottomrule
	\end{tabularx}
	\caption{
		Common mechanisms by which a protocol can become blind to latent
		distinctions. These mechanisms are not mutually exclusive; a single
		experimental procedure can combine selection, projection, thresholding,
		and confounding.
	}
	\label{tab:intro_modes_of_protocol_blindness}
\end{table}

The experimental consequence is important. Absence of a signal under one
protocol is not automatically evidence of absence in the latent mechanics. It may
instead mean that the relevant sector is not coupled to the protocol, is removed
by selection, is averaged over by the readout, falls below threshold, or is
statistically confounded with another parameter. Protocol-resolved interpretation
therefore asks not only what a measurement reports, but also what distinctions
the measurement is structurally unable to resolve.

\subsection{Collapse as an Inverse Problem}
\label{subsec:intro_collapse_inverse_problem}

Experimental collapse is not only a limitation. It is also the forward map that
makes inference possible. If the protocol kernel is known, calibrated,
estimated, or parametrized, then observed data can be used to infer latent virion
parameters, environmental parameters, and protocol parameters. In this sense,
collapse is not merely distortion or loss. It is also signal generation: the
protocol perturbs, selects, amplifies, or reads out the system in a way that
makes certain latent properties experimentally accessible.

Let the full parameter vector be decomposed as
\begin{equation}
	\theta
	=
	\left(
	\theta_{\mathrm{vir}},
	\theta_{\mathrm{env}},
	\theta_E
	\right),
	\label{eq:intro_parameter_decomposition}
\end{equation}
where \(\theta_{\mathrm{vir}}\) denotes virion parameters,
\(\theta_{\mathrm{env}}\) denotes environmental parameters, and \(\theta_E\)
denotes protocol parameters. A protocol-resolved forward model then has the
compact form
\begin{empheq}[box=\fbox]{equation}
	\theta
	\longmapsto
	P_{\mathrm{obs},t}^{\varnothing}(\cdot\mid E,\theta)
	=
	\mathcal M_{E,t}^{\varnothing}(\theta_E)
	P_{\mathrm{ref},t}
	(\cdot\mid\theta_{\mathrm{vir}},\theta_{\mathrm{env}}).
	\label{eq:intro_inverse_forward_model}
\end{empheq}
This is the forward model for a protocol-resolved inverse problem
\cite{Tarantola2005,KaipioSomersalo2005,Stuart2010}. The inverse problem asks
which values of \(\theta\) are compatible with the observed ensemble, given the
experimental map that produced it. The same relation can be summarized schematically as
\begin{empheq}[box=\fbox]{equation}
	\text{Observed protocol-conditioned data}
	\;\longrightarrow\;
	\left\{
	\begin{array}{c}
		\text{Virion parameters} \\
		\text{Environmental parameters} \\
		\text{Protocol parameters}
	\end{array}
	\right\}.
	\label{eq:intro_inverse_environmental_inference_at_a_glance}
\end{empheq}
The arrow in Eq.~\eqref{eq:intro_inverse_environmental_inference_at_a_glance}
should not be read as a direct algebraic inversion. It represents an inference
problem constrained by the forward operator
\(\mathcal M_{E,t}^{\varnothing}\), the latent ensemble model
\(P_{\mathrm{ref},t}\), the data likelihood, and any prior physical,
biological, or experimental information.

This formulation is especially useful for strongly conditioning protocols. AFM
loading is inferentially useful because it makes mechanical compliance visible.
Dielectrophoresis and electrorotation are useful because applied fields make
dielectric response, charge asymmetry, hydrodynamic drag, and polarizability
visible. Mucus and gel tracking are useful because the medium makes transport,
adhesion, immobilization, and local rheological response visible. Plaque assays
are useful because biological amplification makes infectious activity visible.
In each case, the protocol changes the ensemble, but if that change is modeled,
calibrated, or statistically constrained, it can be used to infer properties that
would otherwise remain hidden.

The central practical question is therefore not whether the protocol perturbs the
system. It does. The question is whether the perturbation is characterized well
enough to be used as part of the inference. A protocol-resolved theory treats the
experiment and the virion--environment system together: the observed data are not
a direct copy of the latent population, but neither are they merely artifact.
They are the output of a structured forward map, and that map determines both
what can be inferred and what remains blind.
\subsection{Multi-Protocol Consistency}
\label{subsec:intro_multi_protocol_consistency}

Different experimental protocols generally collapse different latent sectors of
the same virion--environment ensemble. Accordingly, multi-protocol comparison
should not be formulated as a demand for raw observational agreement. A density
map, a force--indentation curve, a field-response spectrum, a trajectory
ensemble, and a plaque count are not expected to be identical objects, because
they are not the same kind of measurement. They live in different observation
spaces, are produced by different physical or biological transformations, and
are filtered through different selection and readout mechanisms.

The relevant question is therefore not whether the raw observables look the
same. The relevant question is whether these different observables can be
explained by a \emph{common latent model} viewed through different
protocol-conditioned observation kernels. Multi-protocol consistency is thus a
statement about shared latent explanation, not raw equality of measured outputs.

At the level of the present framework, suppose that \(M\) protocols
\(E_1,\ldots,E_M\) are applied to comparable virion--environment systems. The
observed ensemble associated with protocol \(E_j\) is written as
\(P_{\mathrm{obs},t}^{(j),\varnothing}\). Multi-protocol consistency asks
whether each observed ensemble can be approximated by applying its own
protocol-conditioned observation operator to the same underlying latent model:
\begin{empheq}[box=\fbox]{equation}
	P_{\mathrm{obs},t}^{(j),\varnothing}(\cdot)
	\approx
	\mathcal M_{E_j,t}^{\varnothing}(\lambda_j)\,
	P_{\mathrm{ref},t}(\cdot\mid\theta),
	\qquad
	j=1,\ldots,M.
	\label{eq:intro_multi_protocol_consistency}
\end{empheq}
Here \(\theta\) denotes the shared latent parameter vector, while
\(\lambda_j\) collects protocol-specific nuisance parameters associated with
protocol \(E_j\). These nuisance parameters may represent calibration constants,
instrument settings, background effects, detection thresholds, preparation
artifacts, analysis choices, or other protocol-local quantities that influence
the observation map without belonging to the common latent virion model itself.

The approximation symbol in
Eq.~\eqref{eq:intro_multi_protocol_consistency} is important. In practice, one
does not expect exact equality between an empirical dataset and an idealized
forward model. Finite-sample variation, measurement noise, model error,
biological heterogeneity, imperfect calibration, and residual protocol
uncertainty all contribute to discrepancies. The question is whether the
discrepancies are compatible with a single latent explanation once each protocol
has been passed through its own observation map.
\begin{figure}[H]
	\centering
	\includegraphics[width=0.98\linewidth]{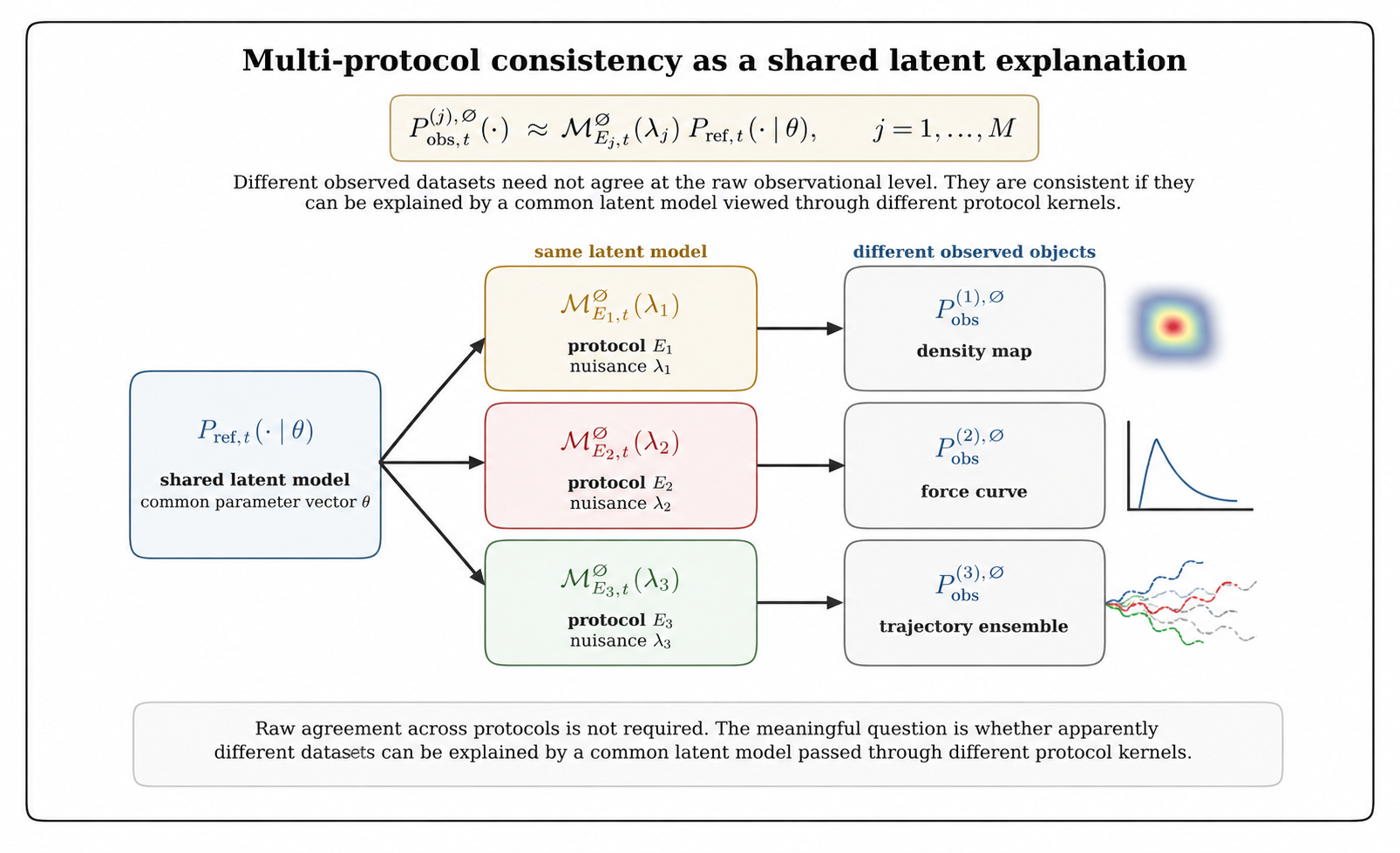}
	\caption{
		\textbf{Multi-protocol consistency as a shared latent explanation.}
		A single reference latent ensemble,
		\(P_{\mathrm{ref},t}(\cdot\mid\theta)\), is mapped through distinct
		protocol-specific observation operators
		\(\mathcal M_{E_j,t}^{\varnothing}(\lambda_j)\), each with its own
		protocol-local nuisance parameters \(\lambda_j\). The resulting observed
		ensembles,
		\(P_{\mathrm{obs},t}^{(j),\varnothing}\), need not agree at the raw
		observational level: one protocol may report a density map, another a
		force curve, and another a trajectory ensemble. In the
		protocol-resolved formulation, the relevant consistency question is not
		whether these observed data objects are identical, but whether they can
		be explained by a common latent virion--environment model after each
		protocol kernel has transformed, selected, projected, or nulled the
		latent ensemble. Apparent disagreement between protocols can therefore
		reflect complementary protocol-conditioned projections rather than
		incompatibility of the underlying latent model.
	}
	\label{fig:multi_protocol_consistency_shared_latent_explanation}
\end{figure}
Equivalently, multi-protocol inference separates two kinds of variation:
\emph{latent variation}, which belongs to the virion--environment system itself,
and \emph{protocol-local variation}, which belongs to the experimental map. The
shared parameter vector \(\theta\) describes the former. The nuisance parameters
\(\lambda_1,\ldots,\lambda_M\) describe the latter. A successful
multi-protocol explanation is one in which the same latent model can generate
the different observed ensembles after the appropriate protocol-specific
conditioning has been applied.

\subsection{Plaque Assays as a Worked Example of Experimental Collapse}
\label{subsec:intro_plaque_assay_worked_example}

The plaque assay provides a concrete worked example of the framework developed
above. It is experimentally familiar, quantitative, and biologically meaningful:
a virus-containing sample is diluted, applied to a susceptible cell monolayer,
allowed to adsorb, constrained by an overlay, incubated, stained or otherwise
visualized, and finally reported as a plaque count
\cite{Dulbecco1952,DulbeccoVogt1954,Cooper1961,Baer2014,Mendoza2020}. Yet the
final observable is far lower-dimensional than the latent process that produces
it. A plaque assay does not directly report total physical particles, total
genomes, antigen abundance, virion orientation, aggregation history, mechanical
state, receptor-binding competence, entry probability, replication competence,
or local spread efficiency. It reports the number of infection-competent units
that successfully pass through a specified cell-line, adsorption, overlay,
incubation, staining, detection, and counting protocol.

This distinction makes the plaque assay a useful anchor for protocol-resolved
virophysics. The assay is not a noisy particle counter. It is a functional
readout: it asks how much of the plated sample can generate visible, countable
infectious lesions under the chosen experimental conditions. Thus, the relevant
object is not a protocol-free property called ``infectiousness,'' but a
protocol-conditioned probability of producing a visible plaque.

The central protocol-resolved quantity is the plaque-forming probability of a
latent state:
\begin{empheq}[box=\fbox]{equation}
	\pi_{\mathrm{PFU}}(x;E_{\mathrm{PFU}})
	=
	\Pr\!\left(
	\begin{array}{c}
		\text{latent state }x
		\text{ generates a visible, countable plaque} \\
		\text{under plaque-assay protocol }E_{\mathrm{PFU}}
	\end{array}
	\right).
	\label{eq:intro_plaque_forming_probability}
\end{empheq}
Here \(E_{\mathrm{PFU}}\) denotes the full plaque-assay protocol, including the
cell line, adsorption time, inoculum volume, overlay composition, incubation
time, staining or reporter method, plaque-detection threshold, and counting
rule. The probability
\(\pi_{\mathrm{PFU}}(x;E_{\mathrm{PFU}})\) is therefore not a context-free
property of the particle alone. It belongs to the state--protocol pair
\((x,E_{\mathrm{PFU}})\). A particle may be physically intact but
non-plaque-forming under one cell line, overlay, incubation time, or detection
threshold, while becoming plaque-forming under a different protocol. Plaque
competence is therefore a protocol-conditioned functional property, not merely a
structural label.

In the language of the observation-kernel formalism, the individual latent state
\(x\) is mapped either into the visible-plaque channel or into the null channel:
\begin{empheq}[box=\fbox]{equation}
	K_{E_{\mathrm{PFU}}}^{\varnothing}
	(\{\mathrm{plaque}\}\mid x)
	=
	\pi_{\mathrm{PFU}}(x;E_{\mathrm{PFU}}),
	\qquad
	K_{E_{\mathrm{PFU}}}^{\varnothing}
	(\{\varnothing\}\mid x)
	=
	1-\pi_{\mathrm{PFU}}(x;E_{\mathrm{PFU}}).
	\label{eq:intro_plaque_kernel_binary_reduction}
\end{empheq}
This binary reduction is deliberately simplified, but it captures the key
collapse: many physically distinct latent states may be mapped to the same
observed outcome, especially the null outcome \(\varnothing\). A structurally
intact but cell-line-incompatible virion, a neutralized virion, a defective
particle, a damaged particle, an aggregate that fails to initiate a countable
lesion, and a particle that enters but fails to spread can all contribute the
same observed result: no plaque.

The corresponding effective plaque-forming concentration is
\begin{empheq}[box=\fbox]{equation}
	\Lambda_{\mathrm{PFU}}(E_{\mathrm{PFU}})
	=
	\int_{\Psi}
	\pi_{\mathrm{PFU}}(x;E_{\mathrm{PFU}})
	n_{\mathrm{ref}}(x)\,dx .
	\label{eq:intro_effective_plaque_forming_concentration}
\end{empheq}
Here \(n_{\mathrm{ref}}(x)\) is the latent number-density distribution over the
state space \(\Psi\). Depending on the modeling resolution, the latent object
\(x\) may represent an individual physical particle, an aggregate, an
infection-competent unit, a damaged or defective particle, or another
assay-relevant object. More generally, if the latent population is represented
by a number measure \(N_{\mathrm{ref}}(dx)\), then
Eq.~\eqref{eq:intro_effective_plaque_forming_concentration} may be written as
\[
\Lambda_{\mathrm{PFU}}(E_{\mathrm{PFU}})
=
\int_{\Psi}
\pi_{\mathrm{PFU}}(x;E_{\mathrm{PFU}})
N_{\mathrm{ref}}(dx).
\]
The density notation is used here for readability.
Equation~\eqref{eq:intro_effective_plaque_forming_concentration} is the central
collapse equation for the plaque assay. It states that PFU is not the total
physical particle concentration,
\begin{equation}
	C_{\mathrm{part}}
	=
	\int_{\Psi} n_{\mathrm{ref}}(x)\,dx,
	\label{eq:intro_total_particle_concentration}
\end{equation}
but a protocol-weighted infectious projection of that concentration. The plaque
assay therefore maps a high-dimensional latent ensemble into one scalar
functional: the expected concentration of visible plaque-forming events under
the specified protocol.
\begin{figure}[H]
	\centering
	\includegraphics[width=1\linewidth]{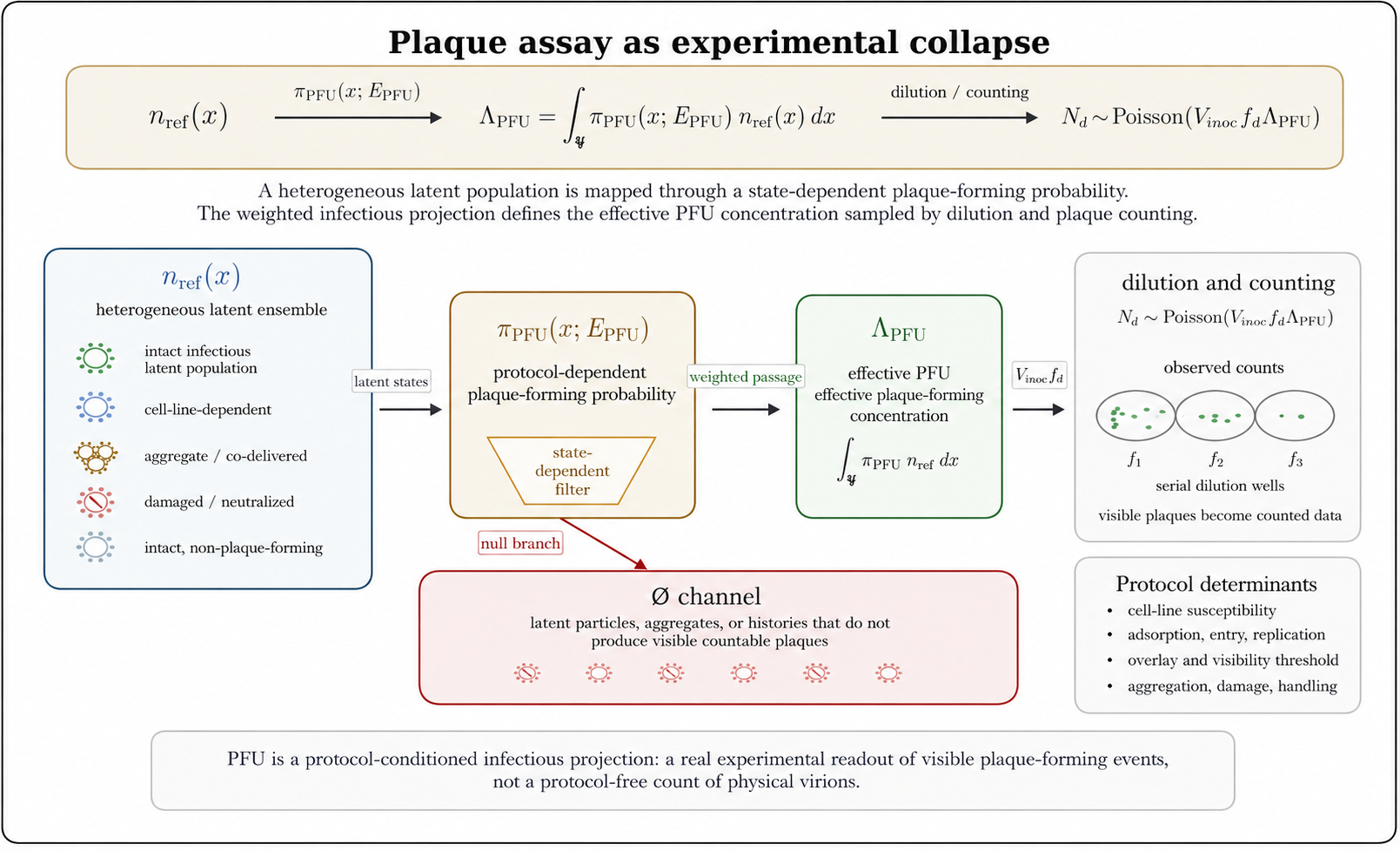}
	\caption{
		\textbf{Plaque assay as experimental collapse.}
		A heterogeneous latent virion or infectious-unit population is mapped
		through the protocol-dependent plaque-forming probability
		\(\pi_{\mathrm{PFU}}(x;E_{\mathrm{PFU}})\). This state-dependent filter
		assigns different plaque-forming weights to intact infectious particles,
		cell-line-dependent particles, aggregates or co-delivered units, damaged
		or neutralized particles, and intact but non-plaque-forming particles.
		The weighted population defines the effective plaque-forming
		concentration \(\Lambda_{\mathrm{PFU}}\), which is then sampled through
		dilution and plaque counting to produce the observed count \(N_d\). The
		null channel contains latent particles, aggregates, or assay histories
		that fail to produce visible countable plaques under the selected
		protocol. Thus PFU is a protocol-conditioned infectious projection of the
		latent ensemble, not a protocol-free count of total physical virions.
	}
	\label{fig:plaque_assay_experimental_collapse1}
\end{figure}
This observation also clarifies the meaning of particle-to-PFU or genome-to-PFU
ratios. Such ratios are not merely measures of experimental inefficiency. They
are coarse summaries of how much latent material is mapped into the visible
plaque-forming channel under a specified protocol. Two samples with similar
particle counts can have different plaque titers if their latent distributions
place different weight on plaque-forming states. Conversely, two samples with
different latent compositions can produce similar plaque counts if their
protocol-weighted plaque-forming concentrations agree. The plaque-forming probability
\(\pi_{\mathrm{PFU}}(x;E_{\mathrm{PFU}})\) can be understood as the probability
that a latent virion, aggregate, or infectious unit passes through a sequence of
protocol-dependent biological and observational stages. Plaque formation is
therefore not a single intrinsic binary property of a particle. It is a terminal
readout of a staged assay pathway.

Schematically, the pathway begins with a latent assay input and proceeds through
delivery, adsorption, attachment and entry, productive replication, local spread,
visibility, and counting. Failure at any required stage sends the latent state
into the null channel for the plaque assay: no visible countable plaque is
reported. This null outcome may reflect handling loss, poor delivery, failed
adsorption, receptor incompatibility, failed entry, nonproductive replication,
spread limitation under the overlay, subthreshold visibility, plaque merging, or
a counting rule that excludes the event. This staged view is useful because different protocol variations act on
different parts of the pathway. Adsorption time primarily perturbs delivery and
cell contact. Cell line or receptor condition perturbs attachment, entry, and
intracellular permissiveness. Neutralization can alter binding, entry, or
aggregation. Overlay composition and incubation time affect local spread and
visibility. Staining threshold and counting criteria act near the final readout.
Thus changing a plaque-assay condition changes the biological observation
kernel, not merely the numerical value of the final titer.

	A low plaque count does not identify a unique failure mechanism by itself. It
	may arise because few competent particles are present, because competent
	particles are poorly coupled to the chosen cell line, because adsorption or
	entry is inefficient, because replication or spread is weak, or because the
	final lesion does not cross the visibility or counting threshold. Treating
	plaque formation as a staged biological pathway turns a scalar PFU readout
	into a structured set of possible bottlenecks that can be probed by controlled
	protocol variation.

\begin{figure}[H]
	\centering
	\includegraphics[width=1\linewidth]{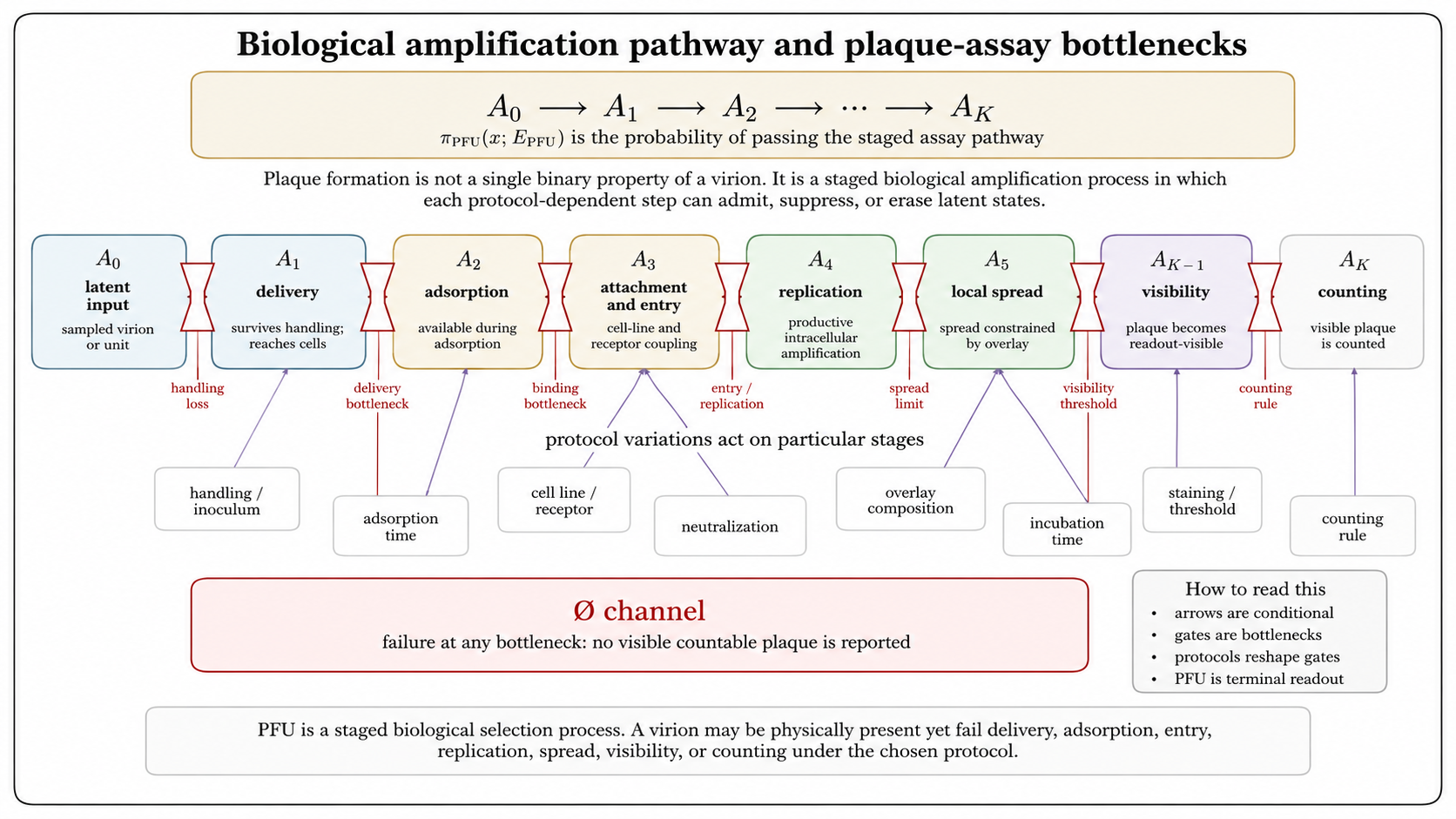}
	\caption{
		\textbf{Biological amplification pathway and plaque-assay bottlenecks.}
		Plaque formation is represented as a staged biological selection process
		rather than as a single intrinsic property of a virion. A latent virion,
		aggregate, or infectious unit must pass through delivery, adsorption,
		attachment and entry, productive replication, local spread, visibility,
		and final counting before it contributes to the PFU readout. Failure at
		any stage sends the event into the plaque-assay null channel, meaning no
		visible countable plaque is reported. The figure also indicates where
		common protocol variations act: inoculum handling, adsorption time, cell
		line or receptor condition, neutralization, overlay composition,
		incubation time, staining threshold, and counting rule. In this view,
		\(\pi_{\mathrm{PFU}}(x;E_{\mathrm{PFU}})\) is the probability that a
		latent state passes the full staged assay pathway under the specified
		protocol.
	}
	\label{fig:biological_amplification_pathway_plaque_bottlenecks}
\end{figure}
\begin{table}[H]
	\centering
	\renewcommand{\arraystretch}{1.18}
	\begin{tabularx}{0.96\linewidth}{@{}p{0.24\linewidth}X@{}}
		\toprule
		\textbf{Protocol component} & \textbf{Interpretation} \\
		\midrule
		
		\(E_{\mathrm{mod}}\)
		&
		Measurement modality, such as cryo-EM, cryo-ET, AFM, DEP,
		electrorotation, particle tracking, plaque assay, focus assay, or
		endpoint dilution assay.
		\\[0.35em]
		
		\(E_{\mathrm{prep}}\)
		&
		Sample preparation pipeline, including dilution, purification,
		vitrification, staining, fixation, labeling, buffer exchange, or other
		pre-observation processing.
		\\[0.35em]
		
		\(E_{\mathrm{field}}\)
		&
		Externally imposed fields or forcing, including electric fields,
		mechanical loading, hydrodynamic flow, optical forcing, or other active
		perturbations.
		\\[0.35em]
		
		\(E_{\mathrm{surf}}\)
		&
		Substrate, confinement, adsorption, grid, interface, or surface-contact
		conditions.
		\\[0.35em]
		
		\(E_{\mathrm{medium}}\)
		&
		Surrounding medium, such as buffer, gel, mucus, extracellular fluid,
		aerosol droplet, overlay, or cellular environment.
		\\[0.35em]
		
		\(E_{\mathrm{time}}\)
		&
		Temporal window, incubation time, sampling cadence, exposure time, or
		duration of forcing.
		\\[0.35em]
		
		\(E_{\mathrm{sel}}\)
		&
		Selection, rejection, survival, detectability, thresholding, particle
		picking, or acceptance criteria.
		\\[0.35em]
		
		\(E_{\mathrm{read}}\)
		&
		Instrumental readout, reconstruction, fitting, counting, classification,
		or post-processing rule.
		\\
		\bottomrule
	\end{tabularx}
	\caption{
		Schematic components of an experimental protocol. The tuple in
		Eq.~\eqref{eq:experimental_protocol_tuple} is not meant to impose a
		unique decomposition on every experiment; rather, it identifies the
		main channels through which a protocol can transform, select, reweight,
		or project a latent virion ensemble.
	}
	\label{tab:experimental_protocol_components}
\end{table}
\noindent The abstract protocol variable \(E\) includes, for example:
	\begin{enumerate}[label=(\roman*),leftmargin=2.2em]
		\item Cryo-electron microscopy or cryo-electron tomography, where sample
		preparation, vitrification, air--water interface exposure, support
		interaction, particle orientation, particle picking, and reconstruction
		determine which configurations are preserved and how they are represented
		\cite{Cheng2015,Thompson2016,Noble2018,Chen2019,Liu2023,DImprima2019};
		
		\item Atomic force microscopy, where virions or virus-like particles are
		imaged, indented, or mechanically interrogated under surface-contact and
		probe-loading conditions
		\cite{Mateu2012,Marchetti2016,Kiss2021,Lyonnais2021};
		
		\item Dielectrophoresis or electrorotation, where nonuniform or rotating
		electric fields actively drive polarization, translation, trapping,
		reorientation, or frequency-dependent response
		\cite{Hughes1998,Hughes2002,Pethig2010,Kim2019};
		
		\item Single-particle tracking in mucus or gel-like media, where observed
		trajectories are shaped jointly by virion size, surface chemistry,
		glycoprotein binding, medium microstructure, adhesion, confinement, and
		the instrumental time window
		\cite{Boukari2009,Wang2017,Kaler2022};
		
		\item Plaque, focus-forming, or endpoint dilution assays, where the
		reported quantity is not the total number of physical virions but a
		protocol-conditioned measure of infectious activity under specified
		host-cell, overlay, incubation, staining, endpoint, and counting
		conditions
		\cite{Dulbecco1952,Cooper1961,Baer2014}.
	\end{enumerate}
The probability \(\pi_{\mathrm{PFU}}\) may be interpreted as the probability of
passing the assay pathway. Schematically, plaque formation requires survival
through handling, delivery to the monolayer, productive adsorption, entry,
replication, local spread under the overlay, and visibility under the final
readout criterion. A useful reduced representation is
\begin{equation}
	\begin{aligned}
		\pi_{\mathrm{PFU}}(x;E_{\mathrm{PFU}})
		&=
		p_{\mathrm{surv}}(x)\,
		p_{\mathrm{deliv}}(x\mid\mathrm{surv})\,
		p_{\mathrm{ads}}(x\mid\mathrm{deliv})\,
		p_{\mathrm{entry}}(x\mid\mathrm{ads})
		\\
		&\quad {}\times
		p_{\mathrm{rep}}(x\mid\mathrm{entry})\,
		p_{\mathrm{spread}}(x\mid\mathrm{rep})\,
		p_{\mathrm{vis}}(x\mid\mathrm{spread}) .
	\end{aligned}
	\label{eq:intro_plaque_pathway_factorization}
\end{equation}
This expression should be read as a schematic conditional factorization of the
assay pathway, not as a claim that the microscopic events are independent. A
particle may be present in the sample and yet fail any one of these filters.
Such a particle contributes to the latent physical population, but not to the
final visible plaque count.

\begin{table}[H]
	\centering
	\renewcommand{\arraystretch}{1.18}
	\begin{tabularx}{0.96\linewidth}{@{}p{0.27\linewidth}X@{}}
		\toprule
		\textbf{Assay stage} & \textbf{Protocol-conditioned interpretation} \\
		\midrule
		
		Handling and dilution
		&
		Particles may be lost, damaged, aggregated, disaggregated, diluted,
		inactivated, or unevenly delivered before contacting the monolayer.
		\\[0.35em]
		
		Adsorption and entry
		&
		The cell line, receptor availability, temperature, adsorption time, and
		medium conditions determine which particles enter the productive
		infection pathway.
		\\[0.35em]
		
		Replication and local spread
		&
		A particle that enters a cell must still generate progeny and spread
		locally through the overlay-restricted environment to form a visible
		lesion.
		\\[0.35em]
		
		Visualization and counting
		&
		Staining, reporter expression, incubation time, plaque morphology,
		thresholding, and counting rules determine which lesions are accepted as
		countable plaques.
		\\[0.35em]
		
		Null channel
		&
		Any state that fails to generate an accepted visible plaque is mapped to
		\(\varnothing\), even if it was physically present in the original
		sample.
		\\
		\bottomrule
	\end{tabularx}
	\caption{
		Protocol-conditioned stages of plaque-assay collapse. Each stage acts as
		a filter or transformation between the latent particle population and the
		final visible plaque count.
	}
	\label{tab:intro_plaque_assay_pathway}
\end{table}
\medskip

\noindent In the dilute, independent, non-overlapping-plaque regime, the effective
plaque-forming concentration enters the usual count model through the Poisson
mean:
\begin{empheq}[box=\fbox]{equation}
	N_d
	\mid
	\Lambda_{\mathrm{PFU}}
	\sim
	\operatorname{Poisson}
	\left(
	V_{\mathrm{inoc}}\,f_d\,
	\Lambda_{\mathrm{PFU}}(E_{\mathrm{PFU}})
	\right),
	\label{eq:intro_poisson_plaque_count_model}
\end{empheq}
where \(N_d\) is the observed plaque count at dilution fraction \(f_d\), and
\(V_{\mathrm{inoc}}\) is the plated inoculum volume. The assumptions behind
Eq.~\eqref{eq:intro_poisson_plaque_count_model} are themselves
protocol-dependent: plaques must be sufficiently rare to avoid strong overlap,
the counted lesions must be approximately independent, and the dilution and
plating process must be well mixed enough for the Poisson approximation to be
appropriate.
\begin{figure}[H]
	\centering
	\includegraphics[width=1\linewidth]{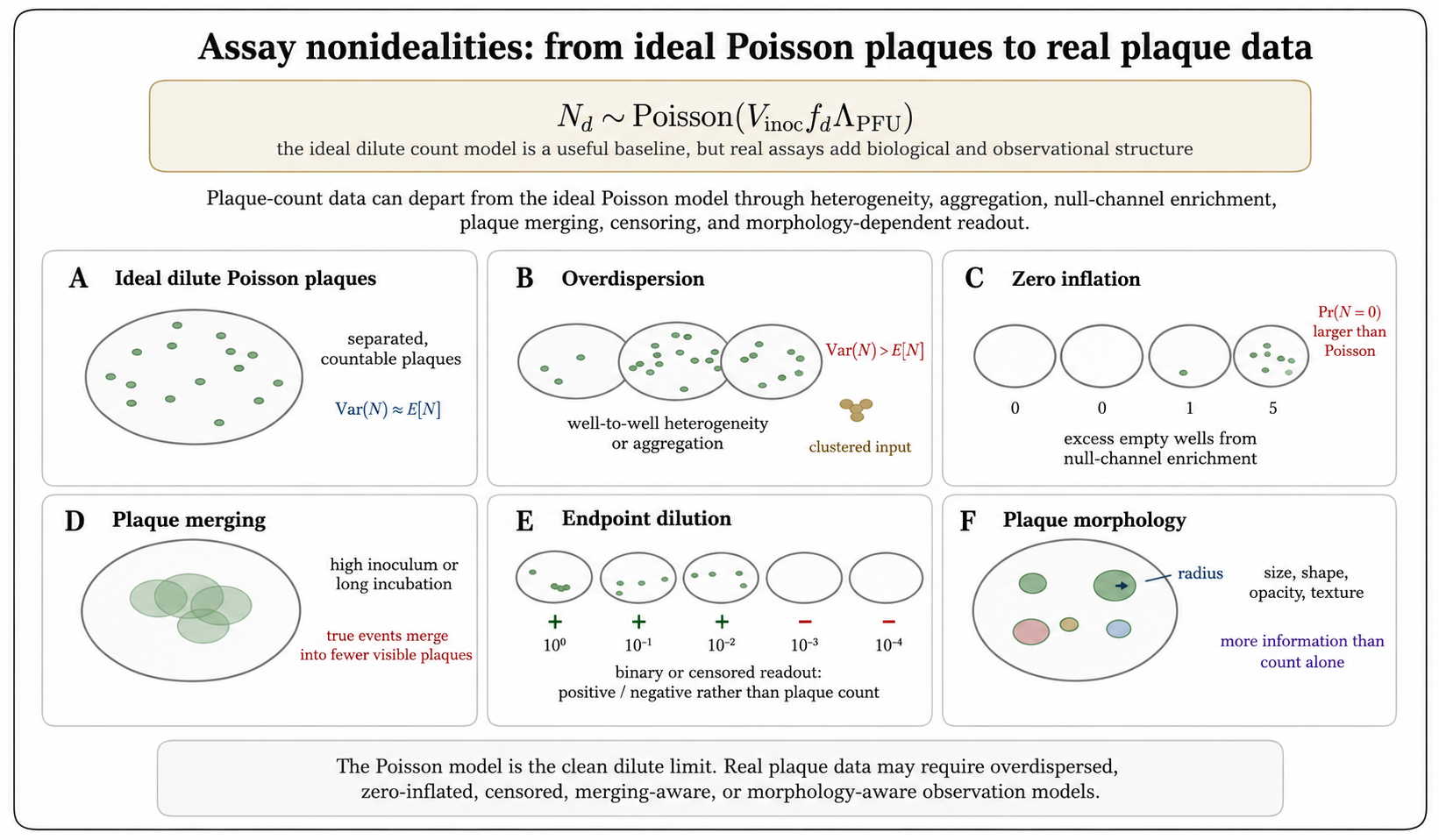}
	\caption{
		\textbf{Assay nonidealities beyond the ideal Poisson plaque-count model.}
		The dilute Poisson model provides a useful baseline for plaque counting
		when infectious events are independent, spatially separated, and directly
		countable. Real plaque-assay data, however, may depart from this ideal
		regime through biological, spatial, and observational structure. The
		panels summarize common deviations: ideal separated plaques with
		approximately Poisson variance; overdispersion from well-to-well
		heterogeneity, aggregation, or clustered input; zero inflation from
		null-channel enrichment and excess empty wells; plaque merging at high
		inoculum or long incubation; endpoint dilution as a binary or censored
		readout rather than a direct plaque count; and plaque morphology as an
		enriched observation carrying information about size, shape, opacity,
		texture, and growth behavior. These nonidealities do not invalidate the
		plaque assay. Rather, they indicate when the readout model should be
		expanded beyond the clean dilute limit to include overdispersed,
		zero-inflated, censored, merging-aware, or morphology-aware observation
		kernels.
	}
	\label{fig:assay_nonidealities_poisson_to_real_plaque_data}
\end{figure}

\noindent Thus, the standard plaque-titer calculation is recovered as an estimator of
\(\Lambda_{\mathrm{PFU}}\), not of \(C_{\mathrm{part}}\):
\begin{empheq}[box=\fbox]{equation}
	\widehat{\Lambda}_{\mathrm{PFU},d}
	=
	\frac{N_d}{V_{\mathrm{inoc}}\,f_d}.
	\label{eq:intro_plaque_titer_estimator}
\end{empheq}
Replicate wells and dilution series improve the statistical precision of this
scalar estimate, but they do not by themselves reveal the latent composition of
the sample. Increasing the number of replicate plaque counts reduces uncertainty
in \(\Lambda_{\mathrm{PFU}}\); it does not automatically identify how that
effective concentration decomposes into particle integrity, receptor
compatibility, replication competence, neutralization state, aggregation state,
or visibility under the counting rule.

The inverse problem becomes especially transparent in a two-subpopulation
reduction. Suppose the latent ensemble is decomposed into a plaque-competent
sector with concentration \(C_{\mathrm C}\) and a plaque-incompetent, weakly
competent, damaged, defective, neutralized, aggregated, or assay-incompatible
sector with concentration \(C_{\mathrm I}\). If the subpopulation-averaged
plaque-forming probabilities are \(\overline p_{\mathrm C}\) and
\(\overline p_{\mathrm I}\), then
\begin{empheq}[box=\fbox]{equation}
	\Lambda_{\mathrm{PFU}}
	=
	\overline p_{\mathrm C}C_{\mathrm C}
	+
	\overline p_{\mathrm I}C_{\mathrm I}.
	\label{eq:intro_two_subpopulation_effective_pfu}
\end{empheq}
This expression states exactly what a plaque count identifies: a weighted sum
over latent sectors. A low plaque titer may reflect a small competent fraction,
a low plaque-forming probability of otherwise competent particles, weak coupling
between the virus and the chosen cell line, suppression by the overlay,
insufficient incubation time, a stringent readout threshold, or loss at
adsorption, entry, replication, local spread, or visibility. PFU alone cannot
distinguish these mechanisms without additional measurements, controlled
protocol variation, or prior biological assumptions.

The same point can be written as an identifiability statement. A single scalar
measurement of \(\Lambda_{\mathrm{PFU}}\) cannot determine the four quantities
\((C_{\mathrm C},C_{\mathrm I},\overline p_{\mathrm C},\overline p_{\mathrm I})\)
without additional constraints:
\begin{equation}
	\Lambda_{\mathrm{PFU}}
	=
	\overline p_{\mathrm C}C_{\mathrm C}
	+
	\overline p_{\mathrm I}C_{\mathrm I}
	\quad
	\text{is one equation for several latent unknowns.}
	\label{eq:intro_pfu_identifiability_limitation}
\end{equation}
This is not a failure of the plaque assay. It is the mathematical statement of
what the plaque assay is designed to report. The assay collapses a heterogeneous
latent population into a countable functional measure of infectious lesion
formation under specified conditions.

\begin{figure}[H]
	\centering
	\includegraphics[width=0.98\linewidth]{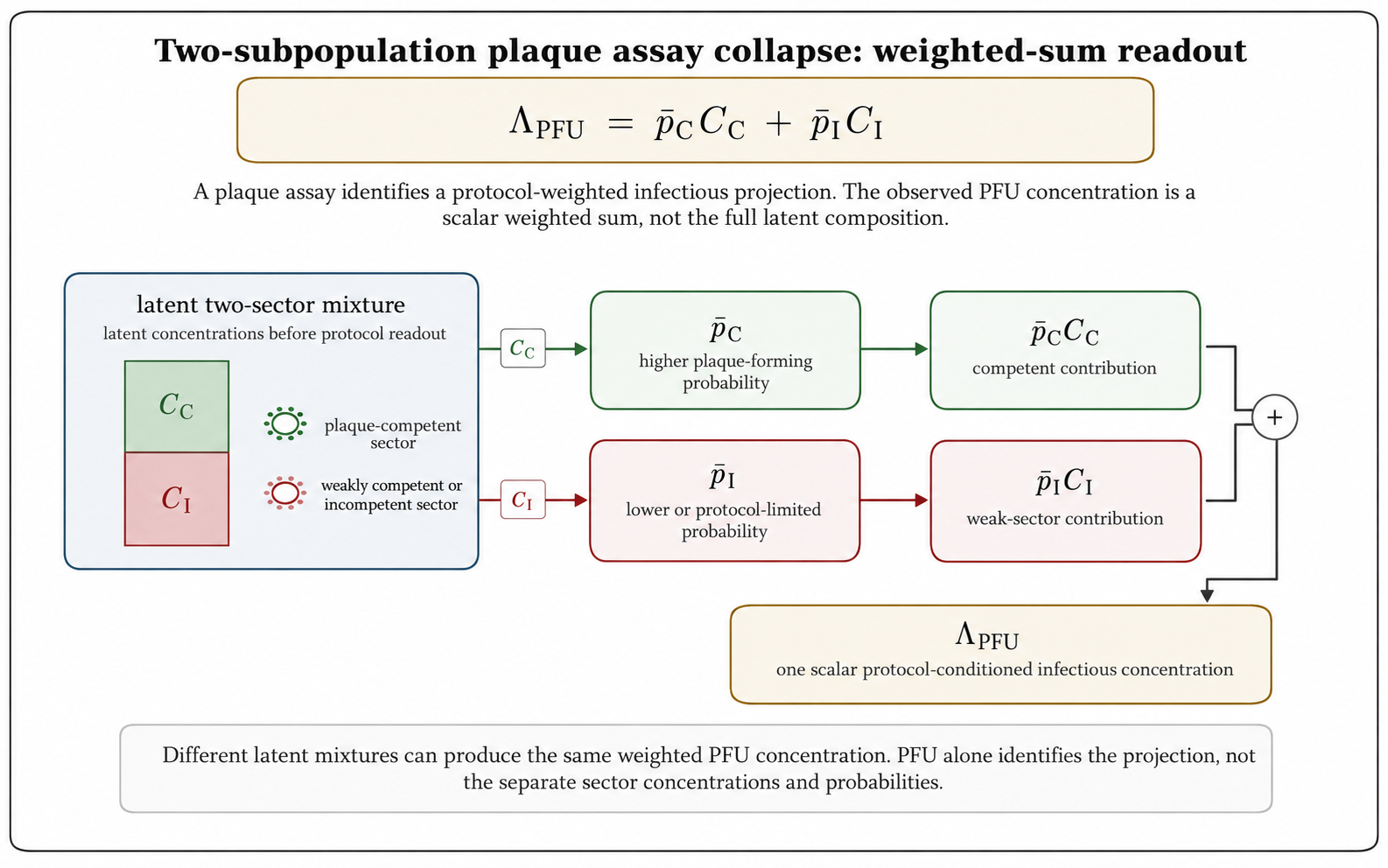}
	\caption{
		\textbf{Plaque-assay collapse as a weighted projection of a latent
			two-subpopulation ensemble.}
		The latent population is represented by two concentration sectors: a
		plaque-competent sector \(C_{\mathrm C}\) and a weakly competent or
		incompetent sector \(C_{\mathrm I}\). Under a fixed plaque-assay
		protocol, these sectors are not observed as independent latent
		concentrations. Instead, each sector is weighted by its
		protocol-conditioned plaque-forming probability,
		\(\overline p_{\mathrm C}\) or \(\overline p_{\mathrm I}\), and the
		assay reports the scalar infectious readout
		The figure emphasizes that PFU concentration is a weighted projection of
		the latent ensemble rather than a direct measurement of the full latent
		composition. Distinct combinations of \(C_{\mathrm C}\),
		\(C_{\mathrm I}\), \(\overline p_{\mathrm C}\), and
		\(\overline p_{\mathrm I}\) can produce the same observed
		\(\Lambda_{\mathrm{PFU}}\). PFU alone therefore does not identify the
		separate sector concentrations or their underlying
		protocol-conditioned plaque-forming probabilities.
	}
	\label{fig:plaque_assay_weighted_sum_collapse}
\end{figure}
\noindent
Equivalently, under the constant-probability approximation, let
\[
C_{\mathrm{part}}
=
C_{\mathrm C}+C_{\mathrm I},
\qquad
\varphi_{\mathrm C}
=
\frac{C_{\mathrm C}}{C_{\mathrm{part}}},
\qquad
\varphi_{\mathrm I}
=
1-\varphi_{\mathrm C}.
\]
Here \(C_{\mathrm{part}}\) is the total latent particle concentration in the
two-sector reduction, and \(\varphi_{\mathrm C}\) is the fraction of that
population assigned to the plaque-competent sector. If the sector-averaged
plaque-forming probabilities are written as \(p_{\mathrm C}\) and
\(p_{\mathrm I}\), then
\begin{empheq}[box=\fbox]{equation}
	\Lambda_{\mathrm{PFU}}
	=
	C_{\mathrm{part}}
	\left[
	p_{\mathrm I}
	+
	\bigl(p_{\mathrm C}-p_{\mathrm I}\bigr)
	\varphi_{\mathrm C}
	\right].
	\label{eq:intro_lambda_pfu_competent_fraction_form}
\end{empheq}
This form separates physical abundance from plaque-forming efficiency. The
factor \(C_{\mathrm{part}}\) describes how much latent material is present,
whereas the bracketed term describes how effectively that latent material is
mapped into the visible plaque-forming channel under the specified protocol.

Equation~\eqref{eq:intro_lambda_pfu_competent_fraction_form} also makes the
non-identifiability explicit. A plaque assay alone identifies the combined
quantity
\[
C_{\mathrm{part}}
\left[
p_{\mathrm I}
+
\bigl(p_{\mathrm C}-p_{\mathrm I}\bigr)
\varphi_{\mathrm C}
\right],
\]
not the separate latent factors \(C_{\mathrm{part}}\), \(\varphi_{\mathrm C}\),
\(p_{\mathrm C}\), and \(p_{\mathrm I}\). Consequently, a large particle-to-PFU
ratio may arise from several distinct mechanisms: a small competent fraction,
poor assay coupling to the competent sector, a weakly competent sector with low
plaque-forming probability, or a protocol that suppresses one or more stages of
plaque development. These possibilities are biologically different, but they can
collapse to the same scalar PFU readout.
\medskip

\noindent
A short numerical example makes the distinction concrete. Suppose two viral
preparations have the same total latent particle concentration,
\[
C_{\mathrm{part}} = 10^{8}\;\mathrm{mL}^{-1},
\]
but differ in the fraction of particles assigned to the plaque-competent
sector. Let the plaque protocol have sector-averaged plaque-forming
probabilities
\[
p_{\mathrm C}=0.80,
\qquad
p_{\mathrm I}=10^{-3},
\]
and consider two preparations with
\[
\varphi_{\mathrm C}^{(A)}=0.10,
\qquad
\varphi_{\mathrm C}^{(B)}=0.01.
\]
Using
\[
\Lambda_{\mathrm{PFU}}
=
C_{\mathrm{part}}
\left[
p_{\mathrm I}
+
\bigl(p_{\mathrm C}-p_{\mathrm I}\bigr)\varphi_{\mathrm C}
\right],
\]
the corresponding effective plaque-forming concentrations are
\begin{align}
	\Lambda_{\mathrm{PFU}}^{(A)}
	&=
	10^{8}
	\left[
	10^{-3}
	+
	(0.80-10^{-3})(0.10)
	\right]
	\approx
	8.09\times 10^{6}\;\mathrm{PFU/mL},
	\label{eq:intro_numerical_lambda_sample_A}
	\\[0.35em]
	\Lambda_{\mathrm{PFU}}^{(B)}
	&=
	10^{8}
	\left[
	10^{-3}
	+
	(0.80-10^{-3})(0.01)
	\right]
	\approx
	8.99\times 10^{5}\;\mathrm{PFU/mL}.
	\label{eq:intro_numerical_lambda_sample_B}
\end{align}
If \(0.1\;\mathrm{mL}\) is plated at dilution fraction \(10^{-4}\), the expected
plaque counts are therefore
\begin{align}
	\mathbb E[N_d^{(A)}]
	&=
	0.1\times 10^{-4}\times 8.09\times 10^{6}
	\approx
	80.9,
	\label{eq:intro_numerical_expected_count_A}
	\\[0.35em]
	\mathbb E[N_d^{(B)}]
	&=
	0.1\times 10^{-4}\times 8.99\times 10^{5}
	\approx
	9.0.
	\label{eq:intro_numerical_expected_count_B}
\end{align}
Thus, two preparations with the same physical particle concentration can fall
into very different plaque-count regimes under the same dilution and plating
conditions. In this example, the difference is not the amount of latent material
present, but the protocol-conditioned projection of that material into the
visible plaque-forming channel. The implied particle-to-PFU ratios also differ:
approximately \(12.4\) for preparation A and \(111.2\) for preparation B. The
particle-to-PFU ratio is therefore not, in this formulation, an intrinsic
constant of the virus alone; it is a compressed readout of latent composition
and protocol-conditioned plaque-forming probabilities.
\medskip

\noindent
The same non-identifiability appears in Fisher-information form. For one plaque
protocol and one dilution, let
\begin{equation}
	\theta
	=
	\begin{pmatrix}
		C_{\mathrm C} \\
		C_{\mathrm I}
	\end{pmatrix},
	\qquad
	\mathbf p
	=
	\begin{pmatrix}
		p_{\mathrm C} \\
		p_{\mathrm I}
	\end{pmatrix}.
	\label{eq:intro_plaque_theta_p_vectors}
\end{equation}
The Poisson mean can then be written compactly as
\begin{equation}
	\mu_d
	=
	V_{\mathrm{inoc}}\,f_d\,
	\mathbf p^{\mathsf T}\theta .
	\label{eq:intro_plaque_poisson_mean_vector_form}
\end{equation}
For a Poisson count with mean \(\mu_d\), the Fisher information for the two
latent concentrations is
\begin{empheq}[box=\fbox]{equation}
	\mathcal I_E(\theta)
	=
	\frac{
		\left(V_{\mathrm{inoc}}f_d\right)^2
	}{
		\mu_d
	}
	\mathbf p\mathbf p^{\mathsf T}.
	\label{eq:intro_rank_one_fisher_two_subpopulation_plaque}
\end{empheq}
This matrix is rank one whenever \(\mathbf p\neq 0\). The protocol is locally
informative only along the sensitivity direction \(\mathbf p\). It is blind to
directions \(v\) satisfying
\begin{equation}
	\mathbf p^{\mathsf T}v=0.
	\label{eq:intro_plaque_blind_direction_condition}
\end{equation}
For example, in the two-sector case, a local blind direction is proportional to
\[
v_{\perp}
=
\begin{pmatrix}
	p_{\mathrm I} \\
	-p_{\mathrm C}
\end{pmatrix},
\qquad
\mathbf p^{\mathsf T}v_{\perp}=0.
\]
Motion in this direction changes the latent sector concentrations while leaving
the scalar combination
\[
p_{\mathrm C}C_{\mathrm C}
+
p_{\mathrm I}C_{\mathrm I}
\]
unchanged to first order. This is the Fisher-information form of
plaque-assay protocol blindness.

Multiple dilutions and replicate wells increase precision along the same
sensitivity direction, but they do not by themselves rotate that direction in
latent subpopulation space. If the same protocol is repeated across dilution
levels \(d=1,\ldots,D\), the information takes the form
\begin{equation}
	\mathcal I_E^{\mathrm{rep}}(\theta)
	=
	\left[
	\sum_{d=1}^{D}
	\frac{
		\left(V_{\mathrm{inoc}}f_d\right)^2
	}{
		\mu_d
	}
	\right]
	\mathbf p\mathbf p^{\mathsf T}.
	\label{eq:intro_replicate_plaque_fisher_same_direction}
\end{equation}
Thus, repeated plaque counts under the same protocol reduce uncertainty in
\(\mathbf p^{\mathsf T}\theta\), but they do not identify the components of
\(\theta\) separately. This is a precise mathematical version of the statement
that replication improves precision but does not automatically resolve latent
composition.

Protocol variation can reduce this blindness when it changes the probability
vector. For independent plaque-count data from protocols
\(E_1,\ldots,E_M\), define
\begin{equation}
	\mathbf p_m
	=
	\begin{pmatrix}
		p_{\mathrm C}^{(m)}\\
		p_{\mathrm I}^{(m)}
	\end{pmatrix},
	\qquad
	\mu_m
	=
	V_m f_m\,\mathbf p_m^{\mathsf T}\theta .
	\label{eq:intro_plaque_protocol_specific_probability_vectors}
\end{equation}
The multi-protocol Fisher information becomes
\begin{empheq}[box=\fbox]{equation}
	\mathcal I_{\mathrm{multi}}(\theta)
	=
	\sum_{m=1}^{M}
	\frac{
		(V_m f_m)^2
	}{
		\mu_m
	}
	\mathbf p_m\mathbf p_m^{\mathsf T}.
	\label{eq:intro_multi_protocol_fisher_two_subpopulation_plaque}
\end{empheq}
Each term is rank one, but the sum can have higher rank if the protocol
probability vectors point in different directions. In the two-sector case, the
rank can increase from one to two only if at least two protocol vectors are not
collinear:
\begin{empheq}[box=\fbox]{equation}
	\operatorname{rank}\mathcal I_{\mathrm{multi}}(\theta)=2
	\quad\Longleftrightarrow\quad
	\exists\,m,n
	\text{ such that }
	\det
	\begin{pmatrix}
		p_{\mathrm C}^{(m)} & p_{\mathrm I}^{(m)} \\
		p_{\mathrm C}^{(n)} & p_{\mathrm I}^{(n)}
	\end{pmatrix}
	\neq 0,
	\label{eq:intro_plaque_noncollinear_protocol_condition}
\end{empheq}
up to the usual requirements that the corresponding means are finite and the
protocols provide nonzero information. This is the plaque-assay version of the
general multi-protocol principle: complementary protocols are valuable not
because their raw observables are identical, but because their observation
kernels probe different latent directions.

Changing the cell line, adsorption time, overlay composition, incubation time,
temperature, neutralization condition, receptor availability, or readout
threshold may rotate the protocol sensitivity direction by changing the
probability vector \(\mathbf p_m\). When this rotation is large enough,
controlled protocol variation can expose latent sectors that are collapsed by
the original assay.

\begin{figure}[H]
	\centering
	\includegraphics[width=0.99\linewidth]{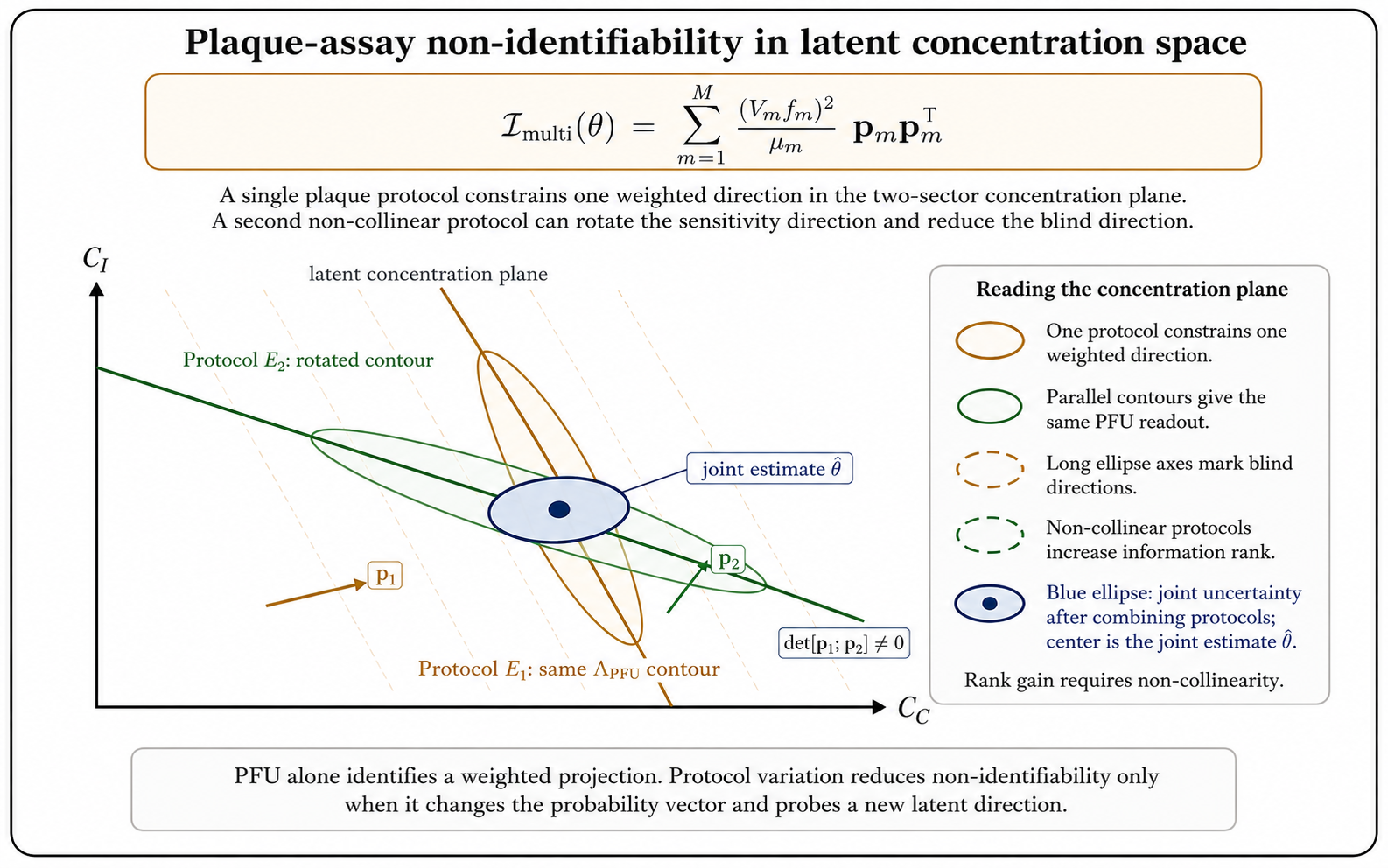}
	\caption{
		\textbf{Plaque-assay non-identifiability in the two-sector latent
			concentration plane.}
		The latent state is reduced to two concentration coordinates:
		\(C_{\mathrm C}\), the plaque-competent sector, and \(C_{\mathrm I}\),
		the weakly competent or incompetent sector. A single plaque protocol
		constrains only one weighted direction in this plane. All points lying
		on the same \(\Lambda_{\mathrm{PFU}}\) contour produce the same scalar
		PFU readout and are therefore indistinguishable under that protocol
		alone. The elongated uncertainty ellipses indicate that each individual
		protocol leaves a blind or weakly constrained direction approximately
		parallel to its equal-readout contour. Introducing a second protocol
		increases the identifiable rank only if the new protocol has a
		non-collinear probability vector \(\mathbf p_m\), thereby rotating the
		sensitivity direction in latent concentration space.}
	\label{fig:plaque_assay_latent_concentration_nonidentifiability}
\end{figure}
The worked example therefore translates experimental collapse into the language
of an everyday virological assay. A plaque count is real experimental data, but
it is not a protocol-free census of virions. It is the visible count generated
after a latent population has been filtered through survival, delivery,
adsorption, entry, replication, local spread, visibility, and counting rules.
This interpretation preserves the practical value of PFU while clarifying its
target: PFU estimates a protocol-conditioned infectious projection.

Additional particle counts, genome counts, antigen measurements, structural
imaging, neutralization assays, receptor-binding assays, tracking assays, or
controlled plaque-protocol variations add complementary observation kernels.
These additional kernels can help separate physical abundance from
plaque-forming efficiency and can constrain latent sectors that the plaque assay
alone cannot resolve. The general mathematical development below formalizes this
logic for arbitrary protocol-conditioned observation maps, of which the plaque
assay is one biologically familiar example.
\medskip 

\noindent This perspective leads to a practical criterion for protocol complementarity. A
new protocol is especially valuable when it contributes information in a
direction that is blind, weakly resolved, or nuisance-confounded under the
existing protocol set. In other words, the goal of multi-protocol inference is
not redundancy for its own sake. The goal is to reduce protocol blindness by
combining experimental maps that are sensitive to different latent sectors.

This can be expressed at the level of likelihoods. If \(D_j\) denotes the data
obtained under protocol \(E_j\), then a protocol-resolved joint inference problem
has the schematic form
\begin{equation}
	p(D_1,\ldots,D_M\mid\theta,\lambda_1,\ldots,\lambda_M)
	=
	\prod_{j=1}^{M}
	p_{E_j}^{\varnothing}(D_j\mid\theta,\lambda_j),
	\label{eq:intro_multi_protocol_likelihood_factorization}
\end{equation}
when the datasets are conditionally independent given the shared latent
parameters and the protocol-specific nuisance parameters. This factorization is
not a claim that the observed datasets are the same kind of object. Rather, it
states that each dataset provides a different view of the same latent model,
conditioned through its own protocol.

Under the same conditional-independence assumption, and when nuisance parameters
are fixed, calibrated, marginalized, or appropriately profiled, the informational
contributions of the separate protocols combine through Fisher information:
\begin{empheq}[box=\fbox]{equation}
	\mathcal I_{\mathrm{multi}}(\theta)
	=
	\sum_{j=1}^{M}
	\mathcal I_{E_j}(\theta).
	\label{eq:intro_multi_protocol_fisher}
\end{empheq}
Equation~\eqref{eq:intro_multi_protocol_fisher} gives a compact expression of
why complementary protocols are powerful. Each protocol constrains some
directions in parameter space more strongly than others. A direction that is
nearly blind under one protocol may be visible under another. Their combination
can therefore shrink the joint uncertainty region even when no single protocol
is individually sufficient.
\begin{figure}[H]
	\centering
	\includegraphics[width=0.98\linewidth]{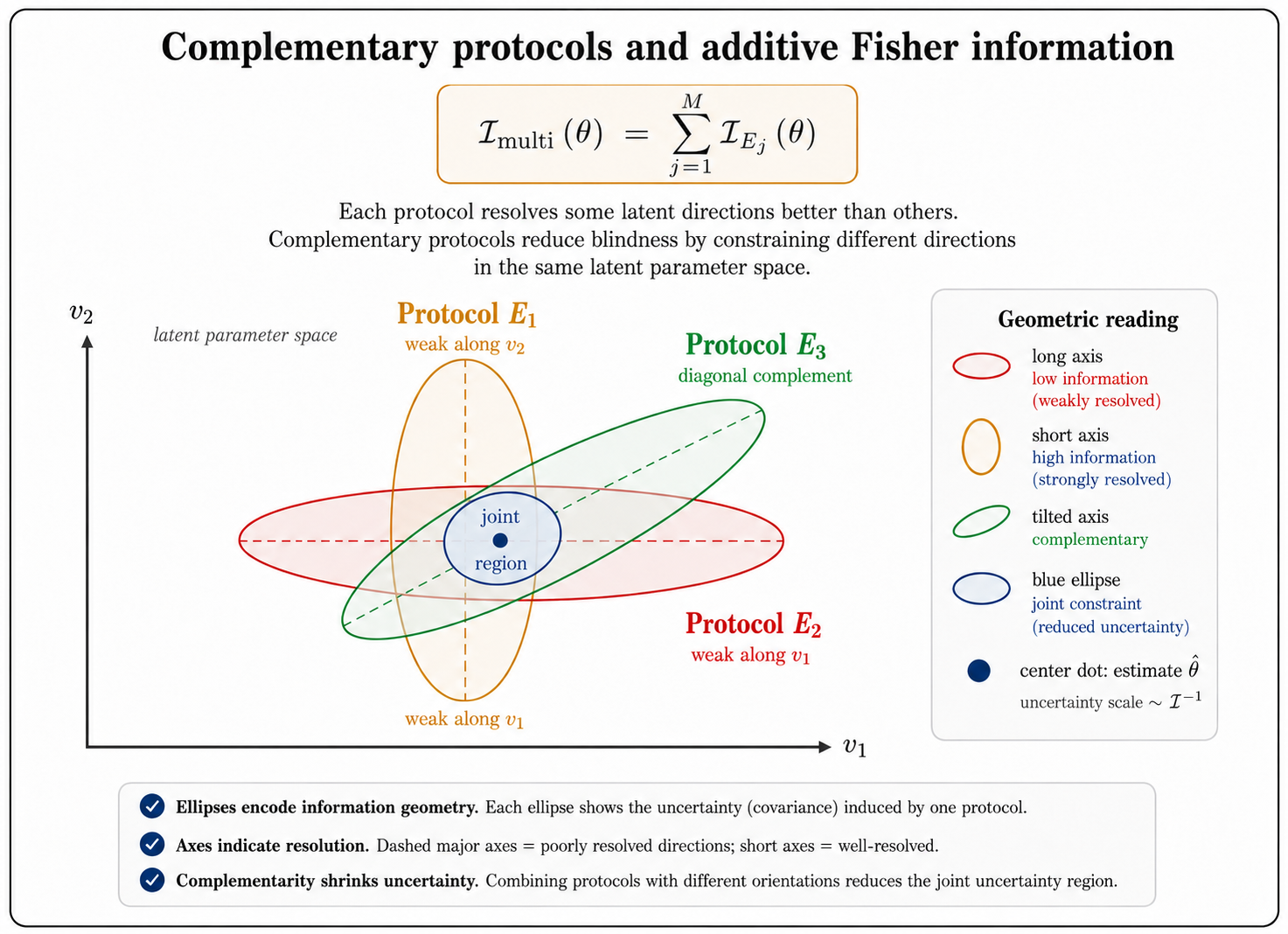}
	\caption{
		\textbf{Complementary protocols and additive Fisher information.}
		Protocol-specific Fisher information constrains different directions in
		the latent parameter space. In this schematic, the axes \(v_1\) and
		\(v_2\) represent local parameter directions in the shared latent
		virion--environment model. Each colored ellipse represents the
		uncertainty region associated with one protocol: long axes indicate
		weakly resolved directions, while short axes indicate strongly resolved
		directions. Protocol \(E_1\), protocol \(E_2\), and protocol \(E_3\)
		provide distinct sensitivity directions, so their combination reduces
		the joint uncertainty region. The blue ellipse represents the combined
		multi-protocol constraint. The figure illustrates why additional data
		from one protocol mainly improves precision along directions that the
		protocol already sees, whereas adding a genuinely complementary protocol
		can reduce protocol blindness by constraining a previously weak or
		unresolved parameter direction.
	}
	\label{fig:complementary_protocols_additive_fisher_information}
\end{figure}
\noindent A corresponding local blind subspace for the combined protocol set may be
written as
\begin{equation}
	\mathcal B_{\mathrm{multi}}(\theta)
	=
	\ker \mathcal I_{\mathrm{multi}}(\theta).
	\label{eq:intro_multi_protocol_blind_subspace}
\end{equation}
When the individual Fisher-information matrices probe different parameter
directions, the kernel of the sum can be smaller than the kernels associated
with the separate protocols. In this sense, multi-protocol inference reduces
blindness not by forcing measurements to look the same, but by combining
different observation kernels that resolve different directions of the same
latent model.

This is the sense in which disagreement between protocols must be interpreted
carefully. Apparent disagreement may indicate a genuine inconsistency in the
latent model, but it may also indicate that the protocols are sensitive to
different latent sectors, that a nuisance parameter has been miscalibrated, or
that one of the observation operators is misspecified. Conversely, apparent
agreement can be misleading if multiple protocols share the same blind direction
or the same selection bias. Multi-protocol consistency is therefore not merely a
comparison of outputs; it is a test of whether a shared latent explanation can
survive multiple, differently conditioned experimental views.

This point motivates the more detailed examples developed below. The plaque
assay is especially useful as an introductory case because its output is simple
and familiar---a count of visible lesions---while its protocol-conditioned
structure is rich. It combines adsorption, entry, replication, spread through a
cell layer, biological amplification, incubation, staining, thresholding, and
counting. For that reason, it provides a clear example of how a classical
virological measurement can be understood as a null-inclusive observation
operator acting on a heterogeneous latent particle population.

\newpage

\section{Experimental Collapse: Concept and Mathematical Formulation}
\label{sec:experimental_collapse_concept_and_math}

The preceding discussion motivates a more explicit mathematical formulation.
The central object is no longer an isolated measurement outcome, but a map
between ensembles: a reference latent ensemble, a protocol-conditioned latent
ensemble, and an observed ensemble. This section defines these objects and the
operations that connect them.

A virological measurement does not generally report the full physical state of a
virion or virion population. The latent state may include position, velocity,
orientation, angular velocity, deformation state, spike or surface
configuration, charge distribution, binding state, local medium variables, and
collective mechanical mode amplitudes. Only some of these degrees of freedom are
made accessible by a given experiment. Which variables become visible depends on
the preparation, forcing, selection, survival, detection, and readout operations
of the protocol.

This is the mathematical content of \emph{experimental collapse}. A protocol may
preserve some latent distinctions, transform others, eliminate some states from
the accepted data channel, reweight the surviving population, or project many
distinct latent states onto the same observed output. The experimentally
reported ensemble is therefore not, in general, a direct sample from the
reference latent ensemble. It is the ensemble produced after the reference
population has passed through a specified experimental map.

\medskip

\noindent
In compact form, the structure is
\begin{empheq}[box=\fbox]{equation}
	\begin{gathered}
		P_{\mathrm{ref},t}
		\;\xrightarrow{\;\mathcal C_{E,t}\;}\;
		P_{E,t}
		\;\xrightarrow{\;\mathcal M_{E,t}^{\varnothing}\;}\;
		P_{\mathrm{obs},t}^{\varnothing}(\cdot \mid E)
		\\[0.5em]
		\scriptstyle
		\text{reference latent ensemble}
		\qquad
		\text{protocol-conditioned latent ensemble}
		\qquad
		\text{observed ensemble}
	\end{gathered}
	\label{eq:experimental_collapse_at_a_glance}
\end{empheq}
The term \emph{reference} is important. The reference latent ensemble is not an
absolute, context-free state of a virion. It is the ensemble associated with a
specified biological or physical environment before the additional conditioning
introduced by the measurement protocol. Thus, the reference ensemble may already
be specific to a buffer, mucus layer, extracellular fluid, aerosol droplet,
surface microenvironment, temperature, ionic condition, or host-tissue context.
Experimental collapse describes the further conditioning imposed by the
measurement procedure itself.

In what follows, a protocol is represented by three mathematical operations: a
latent-state transformation kernel, a survival or detection weight, and a
readout kernel. Together, these operations define how a latent ensemble becomes
an observed ensemble, including the possibility of null observation.

\begin{remark}[Why this is mechanical, not merely epistemic]
	\label{rem:experimental_collapse_mechanical_not_merely_epistemic}
	Experimental collapse is not introduced merely to say that measurements are
	incomplete. Incompleteness alone would be an epistemic limitation. The
	stronger claim is that many virological protocols act as additional physical,
	chemical, mechanical, or biological environments. They can alter forces,
	torques, boundary conditions, accessible orientations, adhesion states,
	deformation pathways, survival probabilities, biological amplification
	pathways, and detectable subpopulations. The discrepancy between the
	reference latent ensemble and the observed ensemble is therefore not only
	informational. It can be dynamical, mechanical, biochemical, or
	assay-dependent.
\end{remark}
	The concept prevents two opposite errors. The first is to treat every observed
	configuration as a faithful sample from an unconstrained or biologically
	natural ensemble. The second is to dismiss protocol-conditioned observations
	as mere artifacts. A better interpretation is that each experiment reports a
	real ensemble, but one produced by a specified mechanical and observational
	map. Once that map is represented explicitly, different experiments can be
	compared as distinct protocol-conditioned projections of a shared latent
	system.

\subsection{Experimental Protocols}
\label{subsec:experimental_protocols}

\begin{definition}[Experimental protocol]
	\label{def:experimental_protocol}
	An \emph{experimental protocol} is the collection of experimental choices,
	physical conditions, biological conditions, selection rules, and readout
	procedures that determine how a latent virion--environment ensemble is
	conditioned and observed. Schematically, we write
	\begin{empheq}[box=\fbox]{equation}
		E
		=
		\left(
		E_{\mathrm{mod}},
		E_{\mathrm{prep}},
		E_{\mathrm{field}},
		E_{\mathrm{surf}},
		E_{\mathrm{medium}},
		E_{\mathrm{time}},
		E_{\mathrm{sel}},
		E_{\mathrm{read}}
		\right).
		\label{eq:experimental_protocol_tuple}
	\end{empheq}
	The entries of this tuple need not be statistically independent or sharply
	separable in a real experiment. The notation is an organizing device: it
	records the major protocol components through which experimental collapse can
	occur.
\end{definition}

\subsection{Latent and Observed State Spaces}
\label{subsec:latent_and_observed_state_spaces}

Let
\[
(\Psi,\Sigma_X)
\]
be the measurable space of latent virion--environment states. A schematic
single-particle latent state may be written as
\begin{equation}
	X
	=
	\left(
	\mathbf x,
	\mathbf v,
	Q,
	\boldsymbol\omega,
	\mathbf s,
	\mathbf c,
	\mathbf m
	\right),
	\label{eq:latent_virion_state}
\end{equation}
where \(\mathbf x\) and \(\mathbf v\) are position and velocity, \(Q\) is an
orientation variable, \(\boldsymbol\omega\) is angular velocity,
\(\mathbf s\) denotes surface, spike, or conformational variables,
\(\mathbf c\) denotes medium-contact, adhesion, receptor-binding, or
environmental-contact variables, and \(\mathbf m\) denotes internal or
collective mechanical mode amplitudes. The specific content of \(X\) depends on
the modeling scale and on the experimental question. For example, \(Q\) may be
irrelevant for an isotropic coarse-grained particle model, but essential for a
spiked, anisotropic, field-responsive, or surface-bound virion model.
\medskip

\noindent The reference latent ensemble is a time-dependent probability measure
\begin{equation}
	P_{\mathrm{ref},t}(dx)
	\quad \text{on} \quad
	(\Psi,\Sigma_X).
	\label{eq:reference_latent_measure}
\end{equation}
This measure represents the distribution of latent states in a specified
biological or physical environment before the additional conditioning imposed by
the measurement protocol. Thus \(P_{\mathrm{ref},t}\) may refer to virions in a
particular extracellular fluid, mucus layer, aerosol droplet, buffer, surface
microenvironment, temperature, ionic condition, or host-tissue context. It is
not a universal free-particle ensemble.

\noindent An experiment generally does not report \(X\) itself. It reports an observed
variable
\[
Y_E \in \mathcal Y_E,
\]
where \((\mathcal Y_E,\Sigma_{Y,E})\) is the protocol-specific non-null
observation space. Depending on \(E\), \(Y_E\) may be a reconstructed density
map, an AFM height image, a force--indentation curve, a particle trajectory, an
electrorotation spectrum, a fluorescence focus count, a plaque count, or an
endpoint dilution readout. The observation space is therefore itself
protocol-dependent.

To include non-detection, rejection, loss, failed infection, failed
classification, or any other accepted absence of a reported datum, we enlarge the
observation space by a null outcome:
\begin{equation}
	\mathcal Y_E^{\varnothing}
	=
	\mathcal Y_E\cup\{\varnothing\}.
	\label{eq:null_enlarged_protocol_observation_space}
\end{equation}
The null outcome \(\varnothing\) does not mean that nothing existed in the
latent ensemble. It means that, under protocol \(E\), the latent state did not
enter the accepted non-null readout channel.

\begin{remark}[Why measurable spaces are used]
	\label{rem:why_measurable_spaces_used}
	Measure notation is used because virological observables are not all
	ordinary Euclidean vectors. Some are continuous trajectories, some are
	discrete counts, some are reconstructed images, some are force curves, some
	are categorical classifications, and some are null outcomes. Writing
	ensembles as probability measures rather than only as densities allows the
	same formalism to cover imaging, tracking, mechanical readout, electrical
	forcing, and infectivity assays.
\end{remark}

\begin{remark}[Protocol dependence of the observation space]
	\label{rem:protocol_dependence_observation_space}
	The notation \(\mathcal Y_E\) emphasizes that different experiments need not
	share the same observed state space. A cryo-EM reconstruction, an AFM
	force--indentation curve, a DEP trajectory, a single-particle tracking path,
	and a plaque count are not raw observations of the same mathematical type.
	They become comparable only after they are interpreted as outputs of
	protocol-specific observation maps acting on a common latent model.
\end{remark}
\subsection{Protocol Transformation Kernels}
\label{subsec:protocol_transformation_kernels}

The first mathematical operation associated with a protocol is the transformation
of the reference latent ensemble into a protocol-conditioned latent ensemble.
This stage represents the part of the experiment that acts on the virion or
virion--environment state before the final accepted readout is formed. We write
this operation as a Markov kernel
\begin{equation}
	\Pi_E^{\mathrm{lat}}(dx_E\mid x),
	\label{eq:protocol_latent_transformation_kernel}
\end{equation}
from the reference latent space \((\Psi,\Sigma_X)\) to a
protocol-conditioned latent space \((\Psi_E,\Sigma_{X,E})\). The variable
\(x\in\Psi\) denotes the reference latent state, while
\(x_E\in\Psi_E\) denotes the state after conditioning by protocol \(E\).

The protocol-conditioned state \(x_E\) may differ from \(x\) because the
experiment can impose constraints, modify boundary conditions, induce
deformation, add field-dependent degrees of freedom, alter medium-contact
variables, remove unresolved coordinates, or define states that exist only under
experimental conditions. For example, an adsorbed AFM state, a vitrified cryo-EM
state, a field-polarized DEP state, a mucus-bound tracking state, and an
infection-pathway state in a plaque assay are not merely different readouts of
the same object. They are different protocol-conditioned latent states.
\medskip 

\noindent The protocol-conditioned latent ensemble before survival, detection, or
acceptance selection is
\begin{empheq}[box=\fbox]{equation}
	\widetilde P_{E,t}^{\mathrm{lat}}(A)
	=
	\int_{\Psi}
	\Pi_E^{\mathrm{lat}}(A\mid x)\,
	P_{\mathrm{ref},t}(dx),
	\qquad
	A\in\Sigma_{X,E}.
	\label{eq:preselection_protocol_conditioned_latent_ensemble}
\end{empheq}
Equivalently,
\[
\widetilde P_{E,t}^{\mathrm{lat}}
=
P_{\mathrm{ref},t}\Pi_E^{\mathrm{lat}},
\]
where the right-hand side denotes the action of the kernel on the reference
measure. Equation~\eqref{eq:preselection_protocol_conditioned_latent_ensemble}
is the first formal stage of experimental collapse: the reference latent
ensemble is carried into a protocol-conditioned latent ensemble by the physical,
biological, and procedural action of the experiment.

\begin{table}[H]
	\centering
	\renewcommand{\arraystretch}{1.18}
	\begin{tabularx}{0.96\linewidth}{@{}p{0.26\linewidth}X@{}}
		\toprule
		\textbf{Protocol action} & \textbf{Role in \(\Pi_E^{\mathrm{lat}}\)} \\
		\midrule
		
		Surface adsorption
		&
		Maps freely suspended or weakly interacting particles into
		surface-associated states with altered orientation, adhesion, deformation,
		or accessible mechanical response.
		\\[0.35em]
		
		Vitrification or fixation
		&
		Transforms a dynamic hydrated ensemble into a preserved structural
		ensemble conditioned by preparation, ice thickness, interface exposure, or
		chemical fixation.
		\\[0.35em]
		
		Electric or mechanical forcing
		&
		Adds field-dependent polarization, trapping, torque, deformation, loading,
		or driven response variables to the conditioned state.
		\\[0.35em]
		
		Structured medium exposure
		&
		Conditions the state by mucus, gel, extracellular matrix, overlay,
		buffer, ionic strength, confinement, adhesion, or local rheology.
		\\[0.35em]
		
		Biological pathway conditioning
		&
		Represents states after adsorption to cells, entry attempts, replication
		competence, spread constraints, neutralization, or other assay-specific
		biological filters.
		\\
		\bottomrule
	\end{tabularx}
	\caption{
		Examples of protocol actions represented by the latent transformation
		kernel \(\Pi_E^{\mathrm{lat}}\). These actions occur before the final
		readout and may physically, chemically, mechanically, or biologically
		condition the latent ensemble.
	}
	\label{tab:examples_protocol_latent_transformation_kernel}
\end{table}

\begin{remark}[Why \(\Pi_E^{\mathrm{lat}}\) is separated from the readout]
	\label{rem:why_latent_kernel_separated_from_readout}
	The latent transformation kernel \(\Pi_E^{\mathrm{lat}}\) describes what the
	protocol does to the virion state before final readout. This should be kept
	separate from the readout kernel, which describes how a surviving
	protocol-conditioned state is represented as data. Surface adsorption,
	vitrification, electric-field forcing, mechanical loading, incubation, and
	medium conditioning belong naturally to the latent transformation stage.
	Finite optical resolution, reconstruction, force-curve fitting, particle
	picking, thresholding, and counting belong to later selection or readout
	stages.
\end{remark}

\begin{remark}[Deterministic maps as a special case]
	\label{rem:deterministic_maps_as_special_case}
	If the protocol maps each reference latent state deterministically to a
	conditioned state \(T_E(x)\), then
	\[
	\Pi_E^{\mathrm{lat}}(dx_E\mid x)
	=
	\delta_{T_E(x)}(dx_E),
	\]
	and the protocol-conditioned latent ensemble is the pushforward
	\[
	\widetilde P_{E,t}^{\mathrm{lat}}
	=
	(T_E)_{\#}P_{\mathrm{ref},t}.
	\]
	The kernel formulation is more general because it also allows stochastic
	preparation, heterogeneous adsorption, variable vitrification outcomes,
	field-dependent fluctuations, stochastic cell entry, heterogeneous survival,
	and protocol-to-protocol variability.
\end{remark}

\subsection{Survival, Detection, and Null Observations}
\label{subsec:survival_detection_and_null_observations}

Not every protocol-conditioned latent state contributes to the final observed
ensemble. Some particles are destroyed, excluded, immobilized outside the field
of view, rejected during reconstruction, too dim to track, unable to infect the
selected cell type, unable to form a visible plaque under the selected overlay
condition, or otherwise absent from the accepted data channel. We therefore
introduce a survival, detection, or acceptance weight
\begin{equation}
	s_E:\Psi_E\rightarrow[0,1],
	\label{eq:survival_detection_weight}
\end{equation}
where \(s_E(x_E)\) is the probability that the protocol-conditioned latent state
\(x_E\) survives the experimental pipeline and contributes to a non-null readout.

The total non-null detection probability is
\begin{equation}
	\eta_E(t)
	=
	\int_{\Psi_E}
	s_E(x_E)\,
	\widetilde P_{E,t}^{\mathrm{lat}}(dx_E).
	\label{eq:total_detection_probability}
\end{equation}
This quantity is the mass of the protocol-conditioned latent ensemble that
passes into the accepted non-null channel. When \(0<\eta_E(t)\leq 1\), the
detected protocol-conditioned latent ensemble is the conditional measure
\begin{empheq}[box=\fbox]{equation}
	P_{E,t}^{\mathrm{lat}}(A\mid \mathrm{det})
	=
	\frac{1}{\eta_E(t)}
	\int_A
	s_E(x_E)\,
	\widetilde P_{E,t}^{\mathrm{lat}}(dx_E),
	\qquad
	A\in\Sigma_{X,E}.
	\label{eq:detected_protocol_conditioned_latent_ensemble}
\end{empheq}
This equation separates transformation from selection. The kernel
\(\Pi_E^{\mathrm{lat}}\) describes how the protocol changes or conditions the
latent state distribution. The weight \(s_E\) describes which conditioned states
remain in the accepted data channel. The detected ensemble
\(P_{E,t}^{\mathrm{lat}}(\cdot\mid\mathrm{det})\) is therefore not generally the
same object as the preselection ensemble
\(\widetilde P_{E,t}^{\mathrm{lat}}\). It is a reweighted and normalized
subensemble.

If \(\eta_E(t)=0\), the protocol produces no accepted non-null observations at
time \(t\). In that case, the conditional detected ensemble is not defined, but
the null-inclusive formulation remains meaningful because the full observed
measure assigns all mass to the null channel.

Null observations are part of the protocol. Let \(\varnothing\) denote the null
readout: no particle, no accepted track, no accepted reconstruction, no plaque,
no focus, no detectable spectrum, no measurable response, or no accepted
classification. Then
\begin{equation}
	\mathbb P_E(\varnothing)
	=
	1-\eta_E(t).
	\label{eq:null_observation_probability}
\end{equation}
The null outcome does not necessarily mean that no virions were present in the
reference ensemble. It means that, under protocol \(E\), the relevant latent
states did not pass into the accepted non-null readout channel.

\begin{remark}[Why the null channel matters]
	\label{rem:why_null_channel_matters}
	The null channel prevents a common interpretive mistake: treating the reported
	ensemble as if it were the entire protocol-conditioned population. A protocol
	may return a clean observed distribution precisely because it has removed,
	rejected, destroyed, failed to amplify, or failed to detect many states. The
	missing probability mass is therefore part of the experiment. It records how
	strongly the protocol selects, filters, thresholds, rejects, or loses latent
	states.
\end{remark}

\begin{remark}[Selection is not the same as absence]
	\label{rem:selection_not_absence}
	A non-detected state is not automatically an absent state. In cryo-EM, a
	particle may be absent from the reconstruction because it was rejected during
	picking or classification. In tracking, a particle may be absent from the
	trajectory ensemble because it was below threshold, immobilized, or outside
	the observation window. In a plaque assay, a particle may be absent from the
	plaque count because it failed adsorption, entry, replication, spread, or
	visibility under the chosen protocol. In each case, the null channel records a
	protocol-conditioned failure to enter the accepted data stream, not
	necessarily the nonexistence of the latent object.
\end{remark}
\subsection{Readout Kernels and Observed Ensembles}
\label{subsec:readout_kernels_and_observed_ensembles}

Selection determines which protocol-conditioned latent states enter the accepted
data channel, but it does not yet specify what is measured. The final non-null
readout is represented by a Markov kernel
\begin{equation}
	R_E(dy\mid x_E),
	\label{eq:readout_kernel}
\end{equation}
from the protocol-conditioned latent space
\((\Psi_E,\Sigma_{X,E})\) to the non-null observation space
\((\mathcal Y_E,\Sigma_{Y,E})\). The readout kernel describes how a surviving
protocol-conditioned state is represented as data. Depending on the protocol,
\(R_E\) may include projection, finite resolution, reconstruction,
segmentation, thresholding, counting, fitting, classification, or measurement
noise.
\medskip

\noindent For a measurable non-null observation set \(B\in\Sigma_{Y,E}\), the detected
observed ensemble is
\begin{empheq}[box=\fbox]{equation}
	P_{\mathrm{obs},t}(B\mid E,\mathrm{det})
	=
	\int_{\Psi_E}
	R_E(B\mid x_E)\,
	P_{E,t}^{\mathrm{lat}}(dx_E\mid \mathrm{det}).
	\label{eq:detected_observed_ensemble}
\end{empheq}
Using the detected latent ensemble from
Eq.~\eqref{eq:detected_protocol_conditioned_latent_ensemble}, this can be
written equivalently as
\begin{equation}
	P_{\mathrm{obs},t}(B\mid E,\mathrm{det})
	=
	\frac{1}{\eta_E(t)}
	\int_{\Psi_E}
	R_E(B\mid x_E)\,
	s_E(x_E)\,
	\widetilde P_{E,t}^{\mathrm{lat}}(dx_E),
	\label{eq:observed_ensemble_expanded}
\end{equation}
provided \(0<\eta_E(t)\leq 1\). Thus the detected observed ensemble is obtained
by three successive operations: latent protocol transformation, survival or
detection weighting, and readout.

To retain non-detection explicitly, we enlarge the observation space by a null
outcome:
\begin{equation}
	\mathcal Y_E^{\varnothing}
	=
	\mathcal Y_E\cup\{\varnothing\},
	\qquad
	\Sigma_{Y,E}^{\varnothing}
	=
	\sigma\!\left(\Sigma_{Y,E}\cup\{\{\varnothing\}\}\right).
	\label{eq:null_enlarged_observation_sigma_space}
\end{equation}
The full observed measure, including the null outcome, is then
\begin{empheq}[box=\fbox]{equation}
	\begin{aligned}
		P_{\mathrm{obs},t}^{\varnothing}(B\mid E)
		&=
		\int_{\Psi_E}
		s_E(x_E)\,
		R_E(B\cap\mathcal Y_E\mid x_E)\,
		\widetilde P_{E,t}^{\mathrm{lat}}(dx_E)
		\\
		&\quad+
		\left(1-\eta_E(t)\right)
		\delta_{\varnothing}(B),
		\qquad
		B\in\Sigma_{Y,E}^{\varnothing}.
	\end{aligned}
	\label{eq:observed_ensemble_with_null}
\end{empheq}
Here \(\delta_{\varnothing}\) is the point mass at the null outcome. This
equation makes the probability accounting explicit. The first term is the
probability mass assigned to accepted non-null observations. The second term is
the probability mass assigned to states that fail to enter the accepted readout
channel.

The null-inclusive observed measure reduces to the detected observed ensemble by
conditioning on the non-null observation space:
\begin{equation}
	P_{\mathrm{obs},t}(B\mid E,\mathrm{det})
	=
	P_{\mathrm{obs},t}^{\varnothing}
	(B\mid E,\,Y\in\mathcal Y_E)
	=
	\frac{
		P_{\mathrm{obs},t}^{\varnothing}(B\mid E)
	}{
		P_{\mathrm{obs},t}^{\varnothing}(\mathcal Y_E\mid E)
	},
	\qquad
	B\in\Sigma_{Y,E},
	\label{eq:detected_observed_as_conditioned_null_inclusive_measure}
\end{equation}
whenever \(P_{\mathrm{obs},t}^{\varnothing}(\mathcal Y_E\mid E)=\eta_E(t)>0\).
This relation is useful because it separates two questions that are often
conflated: what distribution is observed among accepted events, and how much
latent probability mass was excluded, lost, rejected, or mapped to null.

\begin{definition}[Full null-inclusive protocol kernel]
	\label{def:full_null_inclusive_protocol_kernel}
	The full null-inclusive protocol kernel is the Markov kernel
	\[
	K_E^{\varnothing}(dy\mid x)
	\]
	from the reference latent state space \((\Psi,\Sigma_X)\) to the augmented
	observation space
	\((\mathcal Y_E^{\varnothing},\Sigma_{Y,E}^{\varnothing})\). It is obtained
	by composing the latent transformation kernel \(\Pi_E^{\mathrm{lat}}\), the
	survival or detection weight \(s_E\), the readout kernel \(R_E\), and the
	null channel. For \(B\in\Sigma_{Y,E}^{\varnothing}\),
	\begin{empheq}[box=\fbox]{equation}
		\begin{aligned}
			K_E^{\varnothing}(B\mid x)
			&=
			\int_{\Psi_E}
			s_E(x_E)\,
			R_E(B\cap\mathcal Y_E\mid x_E)\,
			\Pi_E^{\mathrm{lat}}(dx_E\mid x)
			\\
			&\quad+
			\left[
			1-
			\int_{\Psi_E}
			s_E(x_E)\,
			\Pi_E^{\mathrm{lat}}(dx_E\mid x)
			\right]
			\delta_{\varnothing}(B).
		\end{aligned}
		\label{eq:full_null_inclusive_protocol_kernel}
	\end{empheq}
\end{definition}

The first term in
Eq.~\eqref{eq:full_null_inclusive_protocol_kernel} describes all ways in which
the original latent state \(x\) can be transformed into a
protocol-conditioned state \(x_E\), survive into the accepted readout channel,
and generate an observed non-null datum. The second term assigns the remaining
probability mass to \(\varnothing\). Since \(R_E(\mathcal Y_E\mid x_E)=1\) and
\(0\leq s_E(x_E)\leq 1\), the kernel is normalized:
\begin{equation}
	K_E^{\varnothing}(\mathcal Y_E^{\varnothing}\mid x)=1.
	\label{eq:null_inclusive_kernel_normalization}
\end{equation}

\begin{definition}[Protocol observation operator]
	\label{def:protocol_observation_operator}
	The \emph{protocol observation operator} associated with \(E\) is the map
	\[
	\mathcal M_E^{\varnothing}:
	P_{\mathrm{ref},t}
	\longmapsto
	P_{\mathrm{obs},t}^{\varnothing}(\cdot\mid E),
	\]
	defined by
	\begin{empheq}[box=\fbox]{equation}
		P_{\mathrm{obs},t}^{\varnothing}(B\mid E)
		=
		\left(
		\mathcal M_E^{\varnothing}P_{\mathrm{ref},t}
		\right)(B)
		=
		\int_{\Psi}
		K_E^{\varnothing}(B\mid x)\,
		P_{\mathrm{ref},t}(dx),
		\qquad
		B\in\Sigma_{Y,E}^{\varnothing}.
		\label{eq:protocol_observation_operator}
	\end{empheq}
	It maps a reference latent ensemble to the full observed ensemble, including
	null observations.
\end{definition}

\begin{remark}[Linearity before conditioning]
	\label{rem:linearity_before_detection_conditioning}
	The null-inclusive operator \(\mathcal M_E^{\varnothing}\) is linear in the
	input probability measure. If
	\(P_{\mathrm{ref},t}=\alpha P_1+(1-\alpha)P_2\), with
	\(0\leq\alpha\leq1\), then
	\[
	\mathcal M_E^{\varnothing}P_{\mathrm{ref},t}
	=
	\alpha\mathcal M_E^{\varnothing}P_1
	+
	(1-\alpha)\mathcal M_E^{\varnothing}P_2.
	\]
	The detected non-null ensemble
	\(P_{\mathrm{obs},t}(\cdot\mid E,\mathrm{det})\) is obtained only after
	conditioning on \(\mathcal Y_E\), and this conditioning introduces the
	normalization by \(\eta_E(t)\). Thus the full probability accounting is
	linear before detection conditioning and generally nonlinear after
	conditioning on non-null observation.
\end{remark}

\begin{remark}[What the readout kernel adds]
	\label{rem:what_readout_kernel_adds}
	The readout kernel is not a duplicate of the latent transformation kernel.
	The latent transformation kernel describes how the protocol changes or
	conditions the state before observation. The readout kernel describes how an
	accepted conditioned state becomes data. For example, deformation under an
	AFM tip belongs to the conditioned latent state, whereas fitting a
	force--indentation curve to an elastic model belongs to readout. Particle
	orientation in vitrified ice belongs to the conditioned latent state, whereas
	classification and reconstruction belong to readout. Local infection and
	spread belong to the plaque-assay pathway, whereas staining, thresholding,
	and counting belong to readout.
\end{remark}

\subsection{Collapse Functionals}
\label{subsec:collapse_functionals}

Experimental collapse is not a single universal number. It must be quantified
relative to a specified observable, distribution, or inference target. A protocol
may preserve one part of the latent ensemble while strongly transforming,
selecting, or projecting another. For this reason, collapse functionals should
state explicitly which feature of the latent system is being compared before and
after protocol conditioning.

Let
\[
g:\Psi\rightarrow\mathbb R,
\qquad
g_E:\Psi_E\rightarrow\mathbb R
\]
be comparable latent and protocol-conditioned quantities, such as radius,
orientation, stiffness, mobility, infectious activity, angular response,
adhesion state, or projected transport coefficient. Assume these quantities are
integrable with respect to the relevant measures. A scalar collapse functional
may be defined as
\begin{empheq}[box=\fbox]{equation}
	\mathcal C_E[g,g_E;t]
	=
	\left|
	\int_{\Psi_E}
	g_E(x_E)\,
	P_{E,t}^{\mathrm{lat}}(dx_E\mid \mathrm{det})
	-
	\int_{\Psi}
	g(x)\,
	P_{\mathrm{ref},t}(dx)
	\right|.
	\label{eq:scalar_collapse_functional}
\end{empheq}
This quantity measures how strongly the detected protocol-conditioned ensemble
displaces a chosen latent quantity from its reference expectation. It is useful
when the target of interest is a scalar feature, such as mean apparent radius,
mean stiffness, mean mobility, mean infectious activity, or mean projected
transport coefficient.
\medskip

\noindent The same idea can be written without conditioning on detection by using the
null-inclusive measure and a readout-level observable \(h\) on
\(\mathcal Y_E^{\varnothing}\):
\begin{equation}
	\mathcal C_E^{\varnothing}[g,h;t]
	=
	\left|
	\int_{\mathcal Y_E^{\varnothing}}
	h(y)\,
	P_{\mathrm{obs},t}^{\varnothing}(dy\mid E)
	-
	\int_{\Psi}
	g(x)\,
	P_{\mathrm{ref},t}(dx)
	\right|.
	\label{eq:null_inclusive_scalar_collapse_functional}
\end{equation}
This form is useful when the null channel itself is part of the quantity being
studied, such as detection probability, infectivity failure, non-plaque-forming
fraction, or rejection probability. The function \(h\) must specify how the null
outcome is scored; for example, one may set \(h(\varnothing)=0\), or define a
separate indicator \(h(y)=\mathbf 1_{\{y=\varnothing\}}\) to measure collapse
into the null channel.

A distributional collapse functional may also be defined. Because
\(\Psi\), \(\Psi_E\), and \(\mathcal Y_E\) need not be the same space, the
distributions being compared must first be mapped to a common comparison space.
Let
\[
L_0:\Psi\rightarrow\mathsf W,
\qquad
L_E:\mathcal Y_E\rightarrow\mathsf W
\]
be measurable maps into a common comparison space
\((\mathsf W,\Sigma_W)\). Then
\begin{empheq}[box=\fbox]{equation}
	\mathfrak C_E^{\mathsf W}(t)
	=
	D_{\mathsf W}
	\left(
	(L_0)_{\#}P_{\mathrm{ref},t},
	(L_E)_{\#}P_{\mathrm{obs},t}(\cdot\mid E,\mathrm{det})
	\right),
	\label{eq:distributional_collapse_functional}
\end{empheq}
where \(D_{\mathsf W}\) is a discrepancy on probability measures over
\(\mathsf W\). Depending on the geometry of the comparison target,
\(D_{\mathsf W}\) may be total variation distance, Wasserstein distance, maximum
mean discrepancy, Jensen--Shannon divergence, Kullback--Leibler divergence when
absolute continuity conditions are satisfied, or another problem-appropriate
measure of distributional discrepancy.

\begin{remark}[Why comparison spaces are needed]
	\label{rem:why_common_comparison_spaces_needed}
	Writing a divergence directly between the latent ensemble and the observed
	ensemble is generally not meaningful unless they live on the same measurable
	space. A plaque count, a force--indentation curve, a cryo-EM reconstruction,
	an electrorotation spectrum, and a trajectory are different mathematical
	objects. The maps \(L_0\) and \(L_E\) state exactly which quantity is being
	compared. They make explicit whether the comparison concerns size,
	orientation, stiffness, transport, infectious activity, detection
	probability, or some other derived feature.
\end{remark}

\begin{remark}[Collapse is observable only through chosen functionals]
	\label{rem:collapse_observable_only_through_functionals}
	There is generally no single number called ``the'' collapse of a protocol. A
	protocol may strongly collapse orientation while preserving size, strongly
	select infectious competence while discarding structural heterogeneity, or
	preserve morphology while eliminating dynamic information. Collapse must
	therefore be reported relative to specified observables, distributions, or
	inference targets. The choice of collapse functional is part of the scientific
	question, not an afterthought.
\end{remark}

\begin{remark}[Collapse functionals versus inference]
	\label{rem:collapse_functionals_versus_inference}
	Collapse functionals quantify how a protocol changes, selects, projects, or
	reweights a chosen feature of the ensemble. They do not by themselves solve
	the inverse problem. Rather, they identify which aspects of the latent system
	are preserved, distorted, or erased by a protocol. This information can then
	be used to design complementary experiments, choose informative protocol
	variations, or determine which latent parameters are identifiable from the
	available observations.
\end{remark}

\subsection{Protocol Blindness and Fisher-Information Observability}
\label{subsec:protocol_blindness_preview}

The same operator formalism also identifies what a protocol cannot see. Once an
observed ensemble is written as the image of a latent ensemble under a
null-inclusive protocol operator, protocol blindness becomes a question about
which changes in the latent model produce distinguishable changes in the
observed law. Suppose the reference latent ensemble is parameterized by
\begin{equation}
	P_{\mathrm{ref},t}
	=
	P_{\theta,t},
	\qquad
	\theta\in\Theta,
	\label{eq:parameterized_reference_latent_ensemble}
\end{equation}
where \(\theta\) may encode stiffness, charge, spike-state heterogeneity,
adhesive binding, rotational mobility, cell-entry competence, infectious
activity, or other latent physical or biological features. Under protocol \(E\),
the null-inclusive observation operator induces an observed law
\begin{equation}
	P_{\mathrm{obs},t}^{\varnothing}(\cdot\mid E,\theta)
	=
	\mathcal M_E^{\varnothing}P_{\theta,t}
	\label{eq:parameterized_observed_law}
\end{equation}
on the augmented observation space
\((\mathcal Y_E^{\varnothing},\Sigma_{Y,E}^{\varnothing})\).
\medskip

\noindent When this observed law admits a density or mass function with respect to a common
dominating measure \(\nu_E^{\varnothing}\), we write
\begin{equation}
	p_E^{\varnothing}(y\mid\theta)
	=
	\frac{
		dP_{\mathrm{obs},t}^{\varnothing}(\cdot\mid E,\theta)
	}{
		d\nu_E^{\varnothing}
	}(y),
	\qquad
	y\in\mathcal Y_E^{\varnothing}.
	\label{eq:protocol_observed_likelihood}
\end{equation}
The Fisher-information matrix for protocol \(E\) is then
\begin{equation}
	\mathcal I_E(\theta)
	=
	\mathbb E_{Y\sim p_E^{\varnothing}(\cdot\mid\theta)}
	\left[
	\nabla_{\theta}\log p_E^{\varnothing}(Y\mid\theta)
	\nabla_{\theta}\log p_E^{\varnothing}(Y\mid\theta)^{\mathsf T}
	\right],
	\label{eq:protocol_fisher_information}
\end{equation}
whenever the usual differentiability and integrability conditions are satisfied.
Equivalently,
\[
\mathcal I_E(\theta)
=
\int_{\mathcal Y_E^{\varnothing}}
\nabla_{\theta}\log p_E^{\varnothing}(y\mid\theta)
\nabla_{\theta}\log p_E^{\varnothing}(y\mid\theta)^{\mathsf T}
p_E^{\varnothing}(y\mid\theta)\,
\nu_E^{\varnothing}(dy).
\]

A parameter direction \(v\in T_{\theta}\Theta\) is locally unobservable under
protocol \(E\) when
\begin{empheq}[box=\fbox]{equation}
	v^{\mathsf T}
	\mathcal I_E(\theta)
	v
	=
	0.
	\label{eq:fisher_blind_direction}
\end{empheq}
The local blind subspace of the protocol is therefore
\begin{equation}
	\mathcal B_E(\theta)
	=
	\ker \mathcal I_E(\theta)
	=
	\left\{
	v\in T_{\theta}\Theta:
	v^{\mathsf T}\mathcal I_E(\theta)v=0
	\right\}.
	\label{eq:local_blind_subspace}
\end{equation}
Under standard regularity assumptions, this condition means that the directional
score in direction \(v\) vanishes almost surely under the observed law. Thus, to
first order, moving the latent model along \(v\) does not produce a statistically
distinguishable change in the data generated by protocol \(E\). Such a direction corresponds to an under-resolved latent sector: a change in
the latent virion--environment ensemble that is not coupled to the protocol, is
removed by selection, is averaged over by the readout, is mapped into the null
channel, or is confounded with another parameter direction. Fisher-information
observability is therefore a local property of the protocol-conditioned forward
map, not a context-free property of the latent variable itself.

\noindent A stronger, global form of protocol blindness occurs when two distinct parameter
values induce the same observed law:
\begin{empheq}[box=\fbox]{equation}
	P_{\mathrm{obs},t}^{\varnothing}(\cdot\mid E,\theta_1)
	=
	P_{\mathrm{obs},t}^{\varnothing}(\cdot\mid E,\theta_2),
	\qquad
	\theta_1\neq\theta_2.
	\label{eq:global_protocol_blindness}
\end{empheq}
Equivalently, when densities exist,
\[
p_E^{\varnothing}(\cdot\mid\theta_1)
=
p_E^{\varnothing}(\cdot\mid\theta_2)
\quad
\nu_E^{\varnothing}\text{-a.e.}
\]
In that case, protocol \(E\) cannot distinguish the two latent hypotheses even
in the infinite-data limit. Local Fisher degeneracy describes infinitesimal
blindness near a parameter value; global observational equivalence describes
exact non-identifiability of distinct latent hypotheses under the protocol.

Multiple protocols can reduce blindness when they probe different latent
directions. If protocols \(E_1,\ldots,E_M\) generate conditionally independent
datasets given \(\theta\), and if nuisance parameters are fixed, calibrated, or
appropriately handled, then their Fisher-information matrices add:
\begin{empheq}[box=\fbox]{equation}
	\mathcal I_{\mathrm{multi}}(\theta)
	=
	\sum_{j=1}^{M}
	\mathcal I_{E_j}(\theta).
	\label{eq:multiprotocol_fisher_information}
\end{empheq}
The corresponding local blind subspace is
\begin{equation}
	\mathcal B_{\mathrm{multi}}(\theta)
	=
	\ker \mathcal I_{\mathrm{multi}}(\theta).
	\label{eq:multiprotocol_blind_subspace}
\end{equation}
A latent sector that is invisible to one protocol may be partially visible to
another. For example, a structural imaging protocol may strongly constrain
morphology while remaining weakly sensitive to mechanical compliance or
infectious activity; a mechanical protocol may constrain stiffness while
discarding free-particle rotational dynamics; and a plaque assay may constrain
protocol-conditioned infectivity while remaining blind to non-plaque-forming
structural heterogeneity. This is the mathematical basis for multi-protocol
consistency and complementarity: protocols are combined not because their raw
observables are the same, but because their observation operators constrain
different directions of a shared latent model.

\begin{table}[H]
	\centering
	\caption{
		Schematic examples of protocol-conditioned observability. The table is
		conceptual: the relevant latent variables, observation kernels, and
		collapse functionals must be specified for each experimental system.
	}
	\label{tab:protocol_observability_examples}
	\renewcommand{\arraystretch}{1.18}
	\begin{tabularx}{0.98\linewidth}{@{}p{0.18\linewidth}p{0.25\linewidth}p{0.27\linewidth}X@{}}
		\toprule
		\textbf{Protocol} &
		\textbf{Reported ensemble} &
		\textbf{Likely collapsed variables} &
		\textbf{Possible blind sectors} \\
		\midrule
		
		Cryo-EM/cryo-ET
		&
		Density maps, particle classes, reconstructed structures
		&
		Orientation distribution, conformation selection, interface survival,
		particle picking, reconstruction-dependent representation
		&
		Dynamics, rare states, preparation-sensitive conformations, fragile or
		transient configurations.
		\\[0.35em]
		
		AFM
		&
		Height maps, force--indentation curves, rupture forces, apparent
		stiffness
		&
		Surface association, loading pathway, deformation state, hydration,
		tip--sample contact geometry
		&
		Free rotational dynamics, non-adhered populations, fragile states,
		unloaded mechanical modes.
		\\[0.35em]
		
		DEP/electrorotation
		&
		Field-driven motion, trapping behavior, crossover frequencies,
		rotation spectra
		&
		Polarization state, medium-dependent dielectric response,
		hydrodynamic drag, field-induced orientation
		&
		Non-dielectric mechanical variables, field-fragile states, latent
		structural heterogeneity weakly coupled to polarization.
		\\[0.35em]
		
		Mucus or gel tracking
		&
		Trajectories, diffusivities, confinement metrics, mobility classes
		&
		Accessible paths, adhesive binding, local rheology, time-windowed
		mobility, tracking thresholds
		&
		Non-trackable particles, short-lived binding states, structural
		variables not expressed through transport, infectivity.
		\\[0.35em]
		
		Plaque/focus assay
		&
		PFU, FFU, endpoint infectious dose, visible lesions or foci
		&
		Cell-entry competence, replication, local spread, overlay survival,
		visibility threshold, counting rule
		&
		Noninfectious particles, structurally intact but non-plaque-forming
		virions, defective particles, neutralized or cell-line-incompatible
		states.
		\\
		\bottomrule
	\end{tabularx}
\end{table}

This completes the operator-level formulation needed for the more detailed
development below. A measurement does not simply reveal a latent virion ensemble;
it maps that ensemble through a protocol-conditioned transformation, selection,
null channel, and readout. The resulting data are therefore interpretable only
relative to the experimental map that produced them. The next sections develop
the three ensemble levels---latent, protocol-conditioned, and observed---in
greater detail and use them to analyze protocol blindness, inverse inference,
multi-protocol consistency, and experimentally collapsed infectivity.
\newpage

\section{Latent, Protocol-Conditioned, and Observed Ensembles}
\label{sec:latent_protocol_conditioned_observed_ensembles}

The preceding section defined the protocol observation operator. We now separate
the ensemble levels on which that operator acts. The basic object in
protocol-resolved virophysics is not a single observed virion, but a sequence of
probability measures: a reference latent ensemble, a protocol-conditioned latent
ensemble, and an observed ensemble. These measures need not live on the same
state space, and they need not preserve the same physical distinctions.

The distinction may be summarized as
\begin{empheq}[box=\fbox]{equation}
	P_{\mathrm{ref},t}
	\;\xrightarrow{\;\Pi_E^{\mathrm{lat}}\;}
	\widetilde P_{E,t}^{\mathrm{lat}}
	\;\xrightarrow{\;s_E,\;R_E,\;\varnothing\;}
	P_{\mathrm{obs},t}^{\varnothing}(\cdot\mid E).
	\label{eq:ensemble_level_chain}
\end{empheq}
The reference latent ensemble describes the virion--environment system before
the additional conditioning imposed by the measurement protocol. The
protocol-conditioned latent ensemble describes the population after preparation,
forcing, medium exposure, surface interaction, biological pathway conditioning,
or other protocol-induced transformations, but before final accepted readout.
The observed ensemble describes the reported data, including the possibility
that some latent states are mapped to the null outcome.

The goal of this distinction is not to make experimental data appear less
reliable. Rather, it is to state precisely what kind of ensemble a protocol
reports. A protocol-conditioned observation may be highly reproducible,
quantitative, and biologically meaningful while still being different from a
direct sample of the reference latent population.

\subsection{Reference Latent Ensembles}
\label{subsec:reference_latent_ensembles}

\begin{definition}[Latent state space and reference latent ensemble]
	\label{def:reference_latent_ensemble}
	Let
	\[
	(\Psi,\Sigma_X)
	\]
	be the measurable space of latent virion--environment states appropriate to
	the level of description under study. The \emph{reference latent ensemble} is
	a time-dependent probability measure
	\begin{equation}
		P_{\mathrm{ref},t}(dx),
		\qquad
		x\in\Psi,
		\label{eq:ensemble_reference_latent_measure}
	\end{equation}
	describing the distribution of virion, virion--environment, or virion-assembly
	states in the biological or physical environment of interest before the
	additional conditioning imposed by the experimental protocol \(E\).
\end{definition}
The word \emph{reference} should be read relative to a specified physical or
biological setting. It does not mean that the ensemble is universal,
equilibrium, unperturbed, or experimentally inaccessible in principle. Rather,
it identifies the baseline ensemble whose transformation by the protocol is to
be analyzed. Thus \(P_{\mathrm{ref},t}\) may describe virions in buffer, mucus,
extracellular fluid, a gel, an aerosol droplet, a surface microenvironment, an
overlay, or a host-tissue context.

When \(P_{\mathrm{ref},t}\) is absolutely continuous with respect to a reference
measure \(\mu_X\), one may write
\begin{equation}
	P_{\mathrm{ref},t}(dx)
	=
	p_{\mathrm{ref}}(x,t)\,\mu_X(dx),
	\label{eq:ensemble_reference_latent_density}
\end{equation}
where \(p_{\mathrm{ref}}(x,t)\) is a density. The measure notation is retained
because many virological observables and latent descriptions are not naturally
represented by ordinary Euclidean densities. They may include continuous
mechanical coordinates, discrete conformational classes, particle counts,
aggregates, damaged states, rejected states, and null outcomes.

\begin{remark}[Interpretation of the reference ensemble]
	\label{rem:interpretation_of_reference_ensemble}
	The reference ensemble is not required to be a free-particle ensemble. It may
	already include hydrodynamic forcing, electrolyte screening, thermal
	fluctuations, mucus confinement, extracellular matrix structure, receptor-rich
	surfaces, biologically relevant adhesion, or host-environment chemistry. These
	features belong to the physical system being modeled. Experimental collapse
	refers to the additional transformation, selection, reweighting, or projection
	introduced by the measurement protocol.
\end{remark}

\begin{remark}[Natural versus protocol-conditioned]
	\label{rem:natural_versus_protocol_conditioned}
	In some contexts it is useful to call \(P_{\mathrm{ref},t}\) a natural latent
	ensemble. This terminology should always be understood relative to the chosen
	environment. A virion ensemble in mucus, in buffer, and in an aerosol droplet
	may all be natural relative to their respective physical settings. None is
	automatically the protocol-conditioned ensemble produced by cryo-EM, AFM,
	dielectrophoresis, tracking, or a plaque assay.
\end{remark}

\subsection{Protocol-Conditioned Latent Ensembles}
\label{subsec:protocol_conditioned_latent_ensembles}

Let \(E\) be a fixed experimental protocol. The protocol may include sample
preparation, vitrification, staining, adsorption, field forcing, surface
attachment, medium exchange, imaging cadence, cell type, overlay medium,
particle-picking rules, thresholding, reconstruction, or any other operation
that changes which virion states become experimentally accessible.

\begin{definition}[Protocol-induced latent-state kernel]
	\label{def:protocol_induced_latent_state_kernel}
	Let
	\[
	(\Psi_E,\Sigma_{X,E})
	\]
	be the protocol-conditioned latent state space. The
	\emph{protocol-induced latent-state kernel} is a Markov kernel
	\begin{equation}
		\Pi_E^{\mathrm{lat}}(dx_E\mid x),
		\qquad
		x\in\Psi,
		\quad
		x_E\in\Psi_E,
		\label{eq:ensemble_protocol_induced_latent_state_kernel}
	\end{equation}
	giving the conditional distribution of protocol-conditioned latent states
	\(x_E\) generated from a reference latent state \(x\) before survival,
	selection, and readout.
\end{definition}
\noindent For each fixed \(x\in\Psi\), the kernel is normalized:
\begin{equation}
	\Pi_E^{\mathrm{lat}}(\Psi_E\mid x)
	=
	1.
	\label{eq:ensemble_latent_kernel_normalization}
\end{equation}
This normalization means that \(\Pi_E^{\mathrm{lat}}\) accounts for how a
reference state is transformed into some protocol-conditioned latent state. It
does not by itself describe whether that state is ultimately detected,
accepted, counted, reconstructed, or reported. Those effects are represented
separately by the survival or detection weight and by the null channel.

The protocol-conditioned latent ensemble before survival or detection selection
is the pushforward of \(P_{\mathrm{ref},t}\) through the kernel
\(\Pi_E^{\mathrm{lat}}\):
\begin{empheq}[box=\fbox]{equation}
	\widetilde P_{E,t}^{\mathrm{lat}}(A)
	=
	\int_{\Psi}
	\Pi_E^{\mathrm{lat}}(A\mid x)\,
	P_{\mathrm{ref},t}(dx),
	\qquad
	A\in\Sigma_{X,E}.
	\label{eq:ensemble_preselection_protocol_conditioned_latent_ensemble}
\end{empheq}
This measure is the latent ensemble after protocol conditioning but before
survival, detection, or acceptance selection. It is therefore distinct both from
the reference ensemble \(P_{\mathrm{ref},t}\) and from the final observed
ensemble \(P_{\mathrm{obs},t}^{\varnothing}(\cdot\mid E)\).

If densities exist, and if \(\Pi_E^{\mathrm{lat}}\) admits a transition density
\(\pi_E^{\mathrm{lat}}(x_E\mid x)\) with respect to a reference measure
\(\mu_{X,E}\) on \(\Psi_E\), then
\begin{equation}
	\widetilde P_{E,t}^{\mathrm{lat}}(dx_E)
	=
	\widetilde p_{E}^{\mathrm{lat}}(x_E,t)\,
	\mu_{X,E}(dx_E),
	\label{eq:ensemble_preselection_protocol_conditioned_density_measure}
\end{equation}
where
\begin{equation}
	\widetilde p_{E}^{\mathrm{lat}}(x_E,t)
	=
	\int_{\Psi}
	\pi_E^{\mathrm{lat}}(x_E\mid x)\,
	p_{\mathrm{ref}}(x,t)\,
	\mu_X(dx).
	\label{eq:ensemble_preselection_protocol_conditioned_latent_density}
\end{equation}
This density form is useful when the state spaces are continuous and dominated
by convenient reference measures. The measure form in
Eq.~\eqref{eq:ensemble_preselection_protocol_conditioned_latent_ensemble}
remains the more general statement.

\begin{table}[t]
	\centering
	\renewcommand{\arraystretch}{1.18}
	\begin{tabularx}{0.96\linewidth}{@{}p{0.27\linewidth}X@{}}
		\toprule
		\textbf{Example} & \textbf{Protocol-conditioned latent interpretation} \\
		\midrule
		
		Cryo-EM or cryo-ET
		&
		A hydrated, dynamic particle ensemble is transformed into a vitrified,
		grid-associated, orientation-conditioned structural ensemble before
		picking, classification, and reconstruction.
		\\[0.35em]
		
		AFM
		&
		A free or weakly interacting particle ensemble is transformed into a
		surface-associated and mechanically loaded ensemble whose deformation
		path depends on tip geometry, loading rate, hydration, and adhesion.
		\\[0.35em]
		
		DEP or electrorotation
		&
		A particle ensemble in medium is transformed into a field-conditioned
		ensemble with induced polarization, hydrodynamic response, torque,
		trapping, or frequency-dependent motion.
		\\[0.35em]
		
		Mucus or gel tracking
		&
		A particle ensemble is conditioned by mesh structure, adhesion, local
		rheology, confinement, and the time window over which motion is sampled.
		\\[0.35em]
		
		Plaque assay
		&
		A particle population is transformed into an assay-pathway ensemble
		conditioned by cell line, adsorption, entry, replication, overlay-limited
		spread, incubation, staining, and counting criteria.
		\\
		\bottomrule
	\end{tabularx}
	\caption{
		Examples of protocol-conditioned latent ensembles. The purpose of
		\(\widetilde P_{E,t}^{\mathrm{lat}}\) is to represent the ensemble after
		protocol-induced physical, mechanical, chemical, or biological
		conditioning, but before the final accepted observed data are formed.
	}
	\label{tab:examples_protocol_conditioned_latent_ensembles}
\end{table}

\begin{remark}[Transformation versus selection]
	\label{rem:transformation_versus_selection_kernel_correction}
	It is important not to force all protocol effects into a normalized
	transformation kernel. Some effects physically, chemically, mechanically, or
	biologically transform a virion state: field forcing, surface adsorption,
	vitrification, dehydration, tip loading, confinement, medium exposure,
	chemical treatment, cell attachment, or entry-pathway conditioning can change
	\(x\) into \(x_E\). Other effects determine whether a state is reported at all:
	particles may be lost during preparation, excluded during particle picking,
	immobilized outside the field of view, fail a signal-to-noise threshold, fail
	to infect the chosen cell type, fail to form a visible plaque, or move too
	quickly to be tracked. These are survival, selection, or detection effects.
	They require an additional weight and an explicit null-observation channel.
\end{remark}

\subsection{Survival, Detection, and Null Observations}
\label{subsec:survival_detection_null_observations}

The preselection ensemble
\(\widetilde P_{E,t}^{\mathrm{lat}}\) describes the population after protocol
conditioning, but before the experiment determines which states enter the
accepted data channel. The next step is therefore not readout itself, but
survival, detection, acceptance, or biological success. These effects are
represented by a protocol-dependent weight.

\begin{definition}[Survival and detection weight]
	\label{def:ensemble_survival_detection_weight}
	The \emph{survival/detection weight} of protocol \(E\) is a measurable
	function
	\begin{equation}
		s_E:\Psi_E\rightarrow[0,1],
		\label{eq:ensemble_protocol_detection_survival_weight}
	\end{equation}
	where \(s_E(x_E)\) is the probability that a protocol-conditioned latent
	state \(x_E\) survives the experimental pipeline and contributes to a
	non-null experimental record.
\end{definition}

The total probability that a state drawn from \(P_{\mathrm{ref},t}\) produces a
non-null record is
\begin{empheq}[box=\fbox]{equation}
	\eta_E(t)
	=
	\int_{\Psi_E}
	s_E(x_E)\,
	\widetilde P_{E,t}^{\mathrm{lat}}(dx_E).
	\label{eq:ensemble_detection_yield_eta}
\end{empheq}
We call \(\eta_E(t)\) the \emph{detection yield}. It measures the probability
mass of the protocol-conditioned latent ensemble that reaches the accepted
non-null data channel. When \(0<\eta_E(t)\leq 1\), the detected
protocol-conditioned latent ensemble is the conditional measure
\begin{empheq}[box=\fbox]{equation}
	P_{E,t}^{\mathrm{lat}}(A\mid\mathrm{det})
	=
	\frac{1}{\eta_E(t)}
	\int_A
	s_E(x_E)\,
	\widetilde P_{E,t}^{\mathrm{lat}}(dx_E),
	\qquad
	A\in\Sigma_{X,E}.
	\label{eq:ensemble_detected_protocol_conditioned_latent_ensemble}
\end{empheq}
Thus \(P_{E,t}^{\mathrm{lat}}(\cdot\mid\mathrm{det})\) is not simply the
protocol-conditioned latent ensemble. It is the protocol-conditioned ensemble
after reweighting by survival, visibility, detectability, acceptance, or
biological success, followed by normalization. If \(\eta_E(t)=0\), no detected
non-null latent ensemble is defined, but the null-inclusive observed measure
remains well defined and assigns all mass to the null channel.

\begin{definition}[Augmented observation space]
	\label{def:ensemble_augmented_observation_space}
	Let \((\mathcal Y_E,\Sigma_{Y,E})\) be the non-null observation space
	associated with protocol \(E\). The augmented observation space is
	\begin{equation}
		\mathcal Y_E^{\varnothing}
		=
		\mathcal Y_E\cup\{\varnothing\},
		\label{eq:ensemble_augmented_observation_space_null}
	\end{equation}
	where \(\varnothing\) denotes non-detection, rejection, loss, failure to pass
	an acceptance threshold, failed amplification, or exclusion from the reported
	dataset. The corresponding augmented sigma-algebra is
	\begin{equation}
		\Sigma_{Y,E}^{\varnothing}
		=
		\sigma\!\left(\Sigma_{Y,E}\cup\{\{\varnothing\}\}\right).
		\label{eq:ensemble_augmented_observation_sigma_algebra}
	\end{equation}
\end{definition}

\begin{remark}[Why a null state is needed]
	\label{rem:why_null_state_needed_experimental_collapse}
	The null state \(\varnothing\) keeps probability accounting explicit. Without
	it, a protocol that reports only a selected subset of states can be mistaken
	for one that faithfully transforms every latent state into a measured state.
	In cryo-EM, for example, air--water interface exposure, ice thickness,
	support interaction, preferred orientation, particle picking, and
	reconstruction criteria may determine which particles contribute to the final
	density map
	\cite{Cheng2015,Thompson2016,Noble2018,Chen2019,Liu2023}. In tracking
	experiments, particles that bind irreversibly, leave the focal volume,
	photobleach, or cannot be localized disappear from the reported trajectory
	ensemble. In plaque assays, particles that fail adsorption, entry,
	replication, local spread, visibility, or counting criteria disappear from
	the plaque count. The null channel represents this missing but
	experimentally meaningful part of the probability flow.
\end{remark}

\subsection{Readout Kernels and Observed Ensembles}
\label{subsec:readout_kernels_observed_ensembles}

Detection determines which protocol-conditioned latent states enter the accepted
data channel. It does not yet specify what datum is recorded. The final readout
step maps a detected protocol-conditioned latent state to an observed outcome.

\begin{definition}[Readout kernel]
	\label{def:ensemble_readout_kernel}
	The \emph{readout kernel} of protocol \(E\) is a Markov kernel
	\begin{equation}
		R_E(dy\mid x_E),
		\qquad
		x_E\in\Psi_E,
		\quad
		y\in\mathcal Y_E,
		\label{eq:ensemble_readout_kernel}
	\end{equation}
	from \((\Psi_E,\Sigma_{X,E})\) to
	\((\mathcal Y_E,\Sigma_{Y,E})\), conditional on non-null detection. Hence
	\begin{equation}
		R_E(\mathcal Y_E\mid x_E)
		=
		1
		\qquad
		\text{for each fixed }x_E\in\Psi_E.
		\label{eq:ensemble_readout_kernel_normalization}
	\end{equation}
	The kernel \(R_E\) includes the final representation of a surviving state as
	data: projection, finite resolution, reconstruction, segmentation,
	thresholding, counting, fitting, classification, and measurement noise.
\end{definition}

For a fixed reference latent state \(x\in\Psi\), define the state-specific
non-null passage probability
\begin{equation}
	q_E(x)
	=
	\int_{\Psi_E}
	s_E(x_E)\,
	\Pi_E^{\mathrm{lat}}(dx_E\mid x).
	\label{eq:ensemble_state_specific_detection_probability}
\end{equation}
This is the probability that the latent state \(x\), after protocol
conditioning, reaches the accepted non-null readout channel. The full
null-inclusive protocol observation kernel is then
\begin{empheq}[box=\fbox]{equation}
	\begin{aligned}
		K_E^{\varnothing}(S\mid x)
		&=
		\int_{\Psi_E}
		s_E(x_E)\,
		R_E(S\cap\mathcal Y_E\mid x_E)\,
		\Pi_E^{\mathrm{lat}}(dx_E\mid x)
		\\
		&\quad
		+
		\bigl[1-q_E(x)\bigr]\,
		\delta_{\varnothing}(S),
		\qquad
		S\in\Sigma_{Y,E}^{\varnothing}.
	\end{aligned}
	\label{eq:ensemble_full_protocol_kernel_with_null}
\end{empheq}
For each fixed \(x\), this is a normalized probability measure on
\(\mathcal Y_E^{\varnothing}\):
\begin{equation}
	K_E^{\varnothing}(\mathcal Y_E^{\varnothing}\mid x)
	=
	1.
	\label{eq:ensemble_full_protocol_kernel_normalization}
\end{equation}
The normalization follows from the fact that \(R_E(\mathcal Y_E\mid x_E)=1\)
and \(0\leq s_E(x_E)\leq 1\).

The full observed ensemble is the image of the reference latent ensemble under
this augmented protocol kernel:
\begin{empheq}[box=\fbox]{equation}
	P_{\mathrm{obs},t}^{\varnothing}(S\mid E)
	=
	\int_{\Psi}
	K_E^{\varnothing}(S\mid x)\,
	P_{\mathrm{ref},t}(dx),
	\qquad
	S\in\Sigma_{Y,E}^{\varnothing}.
	\label{eq:ensemble_full_observed_ensemble_with_null}
\end{empheq}
In particular, the probability of a null observation is
\begin{equation}
	P_{\mathrm{obs},t}^{\varnothing}(\{\varnothing\}\mid E)
	=
	1-\eta_E(t),
	\label{eq:ensemble_null_probability_from_observed_measure}
\end{equation}
because
\[
\eta_E(t)
=
\int_{\Psi}
q_E(x)\,
P_{\mathrm{ref},t}(dx)
=
\int_{\Psi_E}
s_E(x_E)\,
\widetilde P_{E,t}^{\mathrm{lat}}(dx_E).
\]

The reported non-null observed ensemble is obtained by conditioning the
null-inclusive observed measure on the non-null observation space:
\begin{empheq}[box=\fbox]{equation}
	P_{\mathrm{obs},t}(B\mid E,\mathrm{det})
	=
	\frac{
		P_{\mathrm{obs},t}^{\varnothing}(B\mid E)
	}{
		\eta_E(t)
	},
	\qquad
	B\in\Sigma_{Y,E},
	\qquad
	0<\eta_E(t)\leq 1.
	\label{eq:ensemble_observed_ensemble_conditioned_on_detection}
\end{empheq}
Equivalently,
\begin{equation}
	P_{\mathrm{obs},t}(B\mid E,\mathrm{det})
	=
	P_{\mathrm{obs},t}^{\varnothing}
	\left(
	B
	\mid
	Y\in\mathcal Y_E,\,
	E
	\right).
	\label{eq:ensemble_observed_conditional_measure_form}
\end{equation}
This equation separates the reported shape of the non-null data distribution
from the total amount of probability mass that survived into the reported data
channel.

\begin{remark}[Compact interpretation]
	\label{rem:compact_interpretation_observed_ensemble}
	The experimentally reported ensemble is not, in general, the reference latent
	ensemble itself. It is the detected, normalized image of the reference
	ensemble under a protocol-dependent composition of latent-state
	transformation, survival or selection, and readout:
	\[
	P_{\mathrm{ref},t}
	\;\xrightarrow{\;\Pi_E^{\mathrm{lat}}\;}
	\widetilde P_{E,t}^{\mathrm{lat}}
	\;\xrightarrow{\;s_E\;}
	P_{E,t}^{\mathrm{lat}}(\cdot\mid\mathrm{det})
	\;\xrightarrow{\;R_E\;}
	P_{\mathrm{obs},t}(\cdot\mid E,\mathrm{det}).
	\]
	The null-inclusive ensemble
	\(P_{\mathrm{obs},t}^{\varnothing}(\cdot\mid E)\) retains the missing
	probability mass that is lost when one reports only detected observations.
\end{remark}

\begin{definition}[Protocol observation operator]
	\label{def:ensemble_protocol_observation_operator}
	The \emph{protocol observation operator} associated with \(E\) is the map
	\begin{equation}
		\mathcal M_E^{\varnothing}:
		P_{\mathrm{ref},t}
		\longmapsto
		P_{\mathrm{obs},t}^{\varnothing}(\cdot\mid E),
		\label{eq:ensemble_protocol_observation_operator_map}
	\end{equation}
	defined by
	\begin{empheq}[box=\fbox]{equation}
		\left(
		\mathcal M_E^{\varnothing}P_{\mathrm{ref},t}
		\right)(S)
		=
		\int_{\Psi}
		K_E^{\varnothing}(S\mid x)\,
		P_{\mathrm{ref},t}(dx),
		\qquad
		S\in\Sigma_{Y,E}^{\varnothing}.
		\label{eq:ensemble_protocol_observation_operator}
	\end{empheq}
	The non-null reported ensemble is obtained by conditioning
	\(\mathcal M_E^{\varnothing}P_{\mathrm{ref},t}\) on
	\(\mathcal Y_E\).
\end{definition}

\begin{remark}[Why detection yield matters]
	\label{rem:why_detection_yield_matters}
	The same normalized reported distribution
	\(P_{\mathrm{obs},t}(\cdot\mid E,\mathrm{det})\) can arise from very
	different detection yields. A protocol that reports nearly every
	protocol-conditioned state and a protocol that reports only a small selected
	subset may produce superficially similar observed histograms after
	normalization. The detection yield \(\eta_E(t)\) records the missing
	denominator. It is therefore part of experimental collapse, not merely a
	nuisance variable.
\end{remark}

\begin{remark}[Reported data versus probability flow]
	\label{rem:reported_data_versus_probability_flow}
	Reporting only the conditional non-null ensemble can hide important
	protocol effects. For example, two plaque assays may produce similar plaque
	size distributions among counted plaques while having very different
	plaque-forming probabilities. Two tracking experiments may produce similar
	mobility distributions among tracked particles while losing different
	fractions of immobilized or dim particles. Two reconstruction pipelines may
	produce similar particle classes after selection while rejecting different
	fractions of the original images. The null-inclusive formulation keeps both
	the reported distribution and the probability flow into the reported channel
	available for analysis.
\end{remark}
\subsection{Latent versus Observed Virion State}
\label{subsec:latent_versus_observed_virion_state}

The distinction between latent and observed state is the practical center of
experimental collapse. The \emph{latent state} is the state on which the theory
is formulated: it is the state whose evolution, interaction, deformation,
transport, infectivity, or environmental coupling is described by equations of
motion, transition rules, constitutive relations, and biological assumptions.
The \emph{observed state} is the state reported by an instrument, assay,
reconstruction pipeline, tracking algorithm, classification rule, or counting
procedure. In general, these two objects are not identical. They need not even
belong to the same mathematical state space.

This distinction is especially important in virology because different
experimental traditions report fundamentally different kinds of objects. A
cryo-EM reconstruction reports preserved structural density. An AFM experiment
reports surface-conditioned topography or force response. A single-virus
tracking experiment reports localized positions or trajectories over a finite
time window. A dielectrophoretic or electrorotation experiment reports
field-conditioned motion or frequency response. A plaque assay reports visible
infectious lesions. Each is a valid virological observable, but none should be
identified automatically with the complete latent virion state.

\begin{empheq}[box=\fbox]{equation}
	\text{latent state}
	\neq
	\text{observed state},
	\qquad
	X\in\Psi,
	\qquad
	Y_E\in\mathcal Y_E .
	\label{eq:latent_observed_not_equal}
\end{empheq}

The point is not that the latent state is mysterious or metaphysical. A state is
latent only relative to a specified protocol. It is the collection of variables
that the model regards as mechanically, physically, chemically, or biologically
relevant, whether or not a particular experiment records those variables in
full. The observed state is the representation of that latent system after the
protocol has transformed, selected, projected, reconstructed, classified, or
counted it.

\subsubsection{Latent state as the state of the model}
\label{subsubsec:latent_state_as_model_state}

\begin{definition}[Latent virion state]
	\label{def:latent_virion_state}
	A \emph{latent virion state} is the mechanically, physically, chemically, or
	biologically relevant state variable entering the theory before protocol
	reduction. Depending on the scale of description, it may take one of the
	following schematic forms.
	
	\begin{enumerate}[label=(\roman*),leftmargin=2.2em]
		
		\item \textbf{Single-virion level:}
		\begin{equation}
			X^{(1)}
			=
			\left(
			\mathbf r,
			\mathbf v,
			Q,
			\boldsymbol\omega,
			\mathbf s,
			\mathbf c,
			\theta_{\mathrm{env}},
			\theta_{\mathrm{bio}}
			\right),
			\label{eq:single_virion_latent_state}
		\end{equation}
		where \(\mathbf r\) and \(\mathbf v\) are position and velocity,
		\(Q\) is an orientation variable, \(\boldsymbol\omega\) is angular
		velocity, \(\mathbf s\) denotes surface, spike, glycoprotein,
		conformational, or shell variables, \(\mathbf c\) denotes contact,
		adhesion, receptor-binding, or medium-contact state,
		\(\theta_{\mathrm{env}}\) denotes relevant local environmental
		parameters, and \(\theta_{\mathrm{bio}}\) denotes biological competence
		variables such as entry, replication, neutralization, or plaque-forming
		capacity when those variables are part of the model.
		
		\item \textbf{Pairwise interaction level:}
		\begin{equation}
			X^{(2)}
			=
			\left(
			\mathbf r_{ij},
			\mathbf v_{ij},
			Q_i,Q_j,
			\boldsymbol\omega_i,\boldsymbol\omega_j,
			\mathbf c_{ij},
			\theta_{\mathrm{env}}
			\right),
			\label{eq:pairwise_latent_state}
		\end{equation}
		where \(\mathbf r_{ij}\) and \(\mathbf v_{ij}\) are relative position
		and relative velocity, \(Q_i,Q_j\) are orientations,
		\(\boldsymbol\omega_i,\boldsymbol\omega_j\) are angular velocities, and
		\(\mathbf c_{ij}\) denotes contact, near-contact, receptor-mediated,
		electrostatic, hydrodynamic, steric, or orientation-dependent geometric
		data relevant to the interaction.
		
		\item \textbf{Assembly or lattice level:}
		\begin{equation}
			X^{(\mathsf L)}
			=
			\left\{
			\mathbf u(\mathbf R),
			\boldsymbol\eta(\mathbf R),
			\dot{\mathbf u}(\mathbf R),
			\dot{\boldsymbol\eta}(\mathbf R)
			\right\}_{\mathbf R\in\mathsf L},
			\label{eq:lattice_latent_state}
		\end{equation}
		together with shellwise constitutive parameters, boundary conditions,
		interaction coefficients, environmental variables, and any
		protocol-relevant constraints. Here \(\mathbf u(\mathbf R)\) may denote
		translational displacement at lattice site \(\mathbf R\), while
		\(\boldsymbol\eta(\mathbf R)\) denotes rotational, orientational, spike,
		or internal mode variables, depending on the lattice model.
		
		\item \textbf{Joint virion--environment level:}
		\begin{equation}
			X^{(\mathrm{joint})}
			=
			\left(
			X_{\mathrm{vir}},
			X_{\mathrm{env}}
			\right),
			\label{eq:joint_virion_environment_latent_state}
		\end{equation}
		where \(X_{\mathrm{vir}}\) contains virion-level variables and
		\(X_{\mathrm{env}}\) contains local medium or assay variables such as
		viscosity, mesh size, mucin composition, antibody concentration, ionic
		strength, pH, conductivity, surface chemistry, receptor density,
		temperature, cell-line susceptibility, or overlay conditions.
	\end{enumerate}
\end{definition}

These examples are not meant to impose a single universal state vector for all
of virophysics. They show that the latent state is model-dependent. A structural
model may emphasize geometry and conformation; a mechanical model may emphasize
deformation, stiffness, and loading; a transport model may emphasize position,
trajectory, adhesion, and medium structure; an infectivity model may emphasize
entry, replication, spread, and assay compatibility. The latent state should
therefore be chosen to close the theory at the scale of interest.

\begin{remark}[What makes a state latent]
	\label{rem:what_makes_state_latent}
	A state is latent because it is not directly or completely reported by the
	protocol. It is not latent because it is unphysical. For example, a cryo-EM
	density map may preserve structural geometry while eliminating velocity,
	real-time force history, and the preparation pathway that led to the
	preserved state. An AFM image may report morphology and mechanical response
	after surface adsorption, but not the free pre-adsorption trajectory. A
	single-virus tracking experiment may record a projected path while leaving
	orientation, spike presentation, local adhesion state, fusion competence, or
	infectivity unresolved
	\cite{Thompson2016,Noble2018,Chen2019,Liu2023,Kiss2021,Lyonnais2021,
		Liu2020SVT,Wang2018Tracking,NathanDaniel2019,Boukari2009,Kaler2022,Abrami2024}.
\end{remark}

\begin{remark}[Why the latent state matters theoretically]
	\label{rem:why_latent_state_matters_theoretically}
	The latent state is the state on which the theory closes. It is therefore the
	state in which equations of motion, collision rules, shell couplings,
	orientation dynamics, dielectric response, adhesion kinetics, infectivity
	transitions, and collective branch structure are naturally formulated. If a
	protocol compresses or transforms that state before readout, then the
	experimentally reported variables must be interpreted as functionals,
	projections, conditional distributions, or statistical images of the latent
	state rather than as the state itself.
\end{remark}

\subsubsection{Observed state as the state of the protocol}
\label{subsubsec:observed_state_as_protocol_state}

\begin{definition}[Observed virion state]
	\label{def:observed_virion_state}
	For a fixed protocol \(E\), the \emph{observed virion state} is an element
	\begin{equation}
		Y_E\in\mathcal Y_E,
		\label{eq:observed_virion_state_space}
	\end{equation}
	where \((\mathcal Y_E,\Sigma_{Y,E})\) is the protocol-dependent non-null
	observation space. Typical examples include:
	\begin{enumerate}[label=(\roman*),leftmargin=2.2em]
		
		\item reconstructed static geometry, density maps, subtomogram averages,
		particle-class averages, orientation distributions, or selected
		conformational classes in cryo-electron imaging;
		
		\item topography, height profiles, indentation curves, rupture events,
		adhesion estimates, force--distance curves, or stiffness estimates for
		surface-associated particles in AFM;
		
		\item trapping locations, field-driven trajectories, crossover
		frequencies, phase shifts, rotation rates, or response amplitudes in
		dielectrophoresis and electrorotation;
		
		\item tracked positions, mean-square displacements, residence times,
		anomalous-diffusion exponents, confined-transport statistics, pause
		states, directed-motion intervals, jump events, or drift events in
		single-virus tracking;
		
		\item plaque counts, focus counts, endpoint dilution readouts, reporter
		signals, visible lesions, or inferred infectious units in infectivity
		assays.
	\end{enumerate}
\end{definition}

The observed state is therefore not merely a noisy version of \(X\). In many
cases it is a different kind of mathematical object. A latent state may contain
positions, velocities, orientations, spike states, environmental variables, and
biological transition probabilities, while the corresponding observed state may
be an image, curve, count, class label, trajectory, spectrum, or endpoint
readout. This mismatch is precisely why the protocol kernel is needed.

\begin{remark}[The observed state is protocol-specific]
	\label{rem:observed_state_protocol_specific}
	There is no universal observed virion state. A virion observed by cryo-EM is
	represented as a structural object. A virion observed by AFM is represented
	as a surface-associated mechanical object. A virion observed by
	single-particle tracking is represented as a time-indexed trajectory. A
	virion observed by DEP or electrorotation is represented through field
	response. A virion observed through a plaque assay is represented by its
	ability to generate a visible infectious lesion. These are not competing
	definitions of the virion. They are different protocol-conditioned
	representations of the same broader latent system.
\end{remark}

\begin{remark}[Observed does not mean complete]
	\label{rem:observed_does_not_mean_complete}
	An observed state can be precise, reproducible, and experimentally valuable
	without being complete. A high-resolution structure may still omit dynamics,
	mechanical history, or infectivity. A clean trajectory may still omit
	orientation, surface chemistry, or biological competence. A plaque count may
	still omit particle number, defective-particle burden, aggregation state, and
	structural heterogeneity. Protocol-resolved virophysics does not demote these
	observables. It specifies what they are observations of.
\end{remark}

\subsubsection{From latent state to observed state}
\label{subsubsec:from_latent_to_observed_state}

The latent and observed states are connected by the protocol. In the notation of
the preceding sections, a reference latent state \(x\in\Psi\) is first
transformed into a protocol-conditioned latent state \(x_E\in\Psi_E\). It is
then either mapped to a non-null observed datum \(y_E\in\mathcal Y_E\) or sent
to the null channel \(\varnothing\):
\begin{empheq}[box=\fbox]{equation}
	X
	\xrightarrow{\;\Pi_E^{\mathrm{lat}}\;}
	X_E
	\xrightarrow{\;s_E,\;R_E\;}
	Y_E
	\quad
	\text{or}
	\quad
	\varnothing .
	\label{eq:latent_to_observed_state_pipeline}
\end{empheq}
The first arrow represents protocol-induced physical, chemical, biological, or
mechanical conditioning. The second arrow represents survival, detection,
projection, finite resolution, reconstruction, classification, counting, or null
observation. Thus the observed state is not obtained from the latent state by a
single universal measurement map. It is obtained through a protocol-dependent
composition of conditioning, selection, and readout.

\begin{definition}[Observation map with noise]
	\label{def:observation_map_with_noise}
	When the readout can be represented in a vector-valued observation space, a
	useful reduced form is
	\begin{equation}
		Y_E
		=
		\mathcal O_E(X_E)+\nu_E,
		\label{eq:observation_map_with_noise}
	\end{equation}
	where \(X_E\) is the protocol-conditioned latent state,
	\(\mathcal O_E\) is the protocol-dependent observation map, and \(\nu_E\)
	is a measurement-noise term. This representation is less general than the
	kernel formulation, but it is useful when the readout is approximately
	additive or when the observed datum is a finite-dimensional vector. In this
	form, the distinction between latent and observed state is that
	\(\mathcal O_E\) is generally neither the identity map nor invertible.
\end{definition}

The kernel formulation should be regarded as the primary statement. The noisy
observation-map form in Eq.~\eqref{eq:observation_map_with_noise} is a
convenient special case for familiar measurement models. It does not cover every
kind of virological readout, since density maps, force curves, trajectories,
categorical classes, plaque counts, and null outcomes need not be naturally
represented as additive perturbations of a vector-valued latent state.

\begin{definition}[Observation fiber]
	\label{def:observation_fiber}
	If a deterministic observation map
	\[
	\mathcal O_E:\Psi_E\rightarrow\mathcal Y_E
	\]
	is used, the \emph{observation fiber} over an observed datum \(y_E\) is
	\begin{equation}
		\mathcal F_E(y_E)
		=
		\left\{
		x_E\in\Psi_E:
		\mathcal O_E(x_E)=y_E
		\right\}.
		\label{eq:observation_fiber_latent_observed}
	\end{equation}
\end{definition}

\begin{remark}[Interpretation of the observation fiber]
	\label{rem:interpretation_observation_fiber}
	The fiber \(\mathcal F_E(y_E)\) is the set of protocol-conditioned latent
	states that are indistinguishable to the observation map. If this set contains
	mechanically, structurally, dynamically, or biologically distinct states, then
	the protocol has collapsed those distinctions. For example, a plaque count may
	group many structurally different particles into the same infectious-unit
	readout; a trajectory may group different orientation states into the same
	projected path; and a reconstructed density may group different dynamical
	histories into the same preserved structural class.
\end{remark}

\begin{remark}[Fibers and stochastic readout]
	\label{rem:fibers_and_stochastic_readout}
	For stochastic readout kernels, exact fibers are replaced by conditional
	distributions over latent states compatible with an observation. In that case,
	the relevant inverse object is not a set
	\(\mathcal F_E(y_E)\), but a posterior or conditional law over
	\(\Psi_E\) given the observed datum. The deterministic fiber is therefore the
	geometric limit of a more general probabilistic inverse problem.
\end{remark}

\subsubsection{Common forms of latent-to-observed reduction}
\label{subsubsec:common_forms_latent_to_observed_reduction}

The same latent population may generate different observed ensembles under
different protocols. The following reductions are especially common in
virological work.

\begin{enumerate}[label=(\roman*),leftmargin=2.2em]
	
	\item \textbf{State freezing.}
	A dynamic latent process is represented as a preserved structural snapshot.
	Cryo-EM and cryo-ET can provide extraordinary structural information, but the
	observed state is conditioned by preparation, vitrification, air--water
	interface exposure, particle selection, orientation statistics, and
	reconstruction
	\cite{Thompson2016,Noble2018,Chen2019,Liu2023,DImprima2019}.
	
	\item \textbf{Surface conditioning.}
	A freely suspended or weakly interacting particle is represented after
	adsorption, tethering, flattening, hydration change, or probe interaction.
	AFM therefore reports a surface-conditioned mechanical state rather than the
	unconstrained suspension-state dynamics of the same virion
	\cite{Mateu2012,Marchetti2016,Kiss2021,Lyonnais2021}.
	
	\item \textbf{Trajectory projection.}
	A high-dimensional virion--cell or virion--medium process is represented as a
	sequence of localized positions over a finite time window. Single-virus
	tracking can reveal entry, trafficking, confinement, directed motion, and
	transport regimes, but the observed trajectory depends on labeling, imaging
	cadence, localization, segmentation, and the variables chosen for analysis
	\cite{Liu2020SVT,Wang2018Tracking,NathanDaniel2019}.
	
	\item \textbf{Medium filtering.}
	A virion moving through mucus, gel, extracellular matrix, or another
	structured medium is observed through the transport modes that survive the
	medium and the tracking window. In this case, the observed path is a joint
	virion--medium observable, not a property of the virion alone
	\cite{Boukari2009,Wang2017,Kaler2022,Vahey2019,Abrami2024}.
	
	\item \textbf{Field-response projection.}
	A virion in dielectrophoresis or electrorotation is observed through its
	response to an imposed field. The observed state is therefore a
	field-conditioned response object involving virion polarizability, medium
	conductivity and permittivity, hydrodynamic drag, and frequency-dependent
	response
	\cite{Hughes1998,Hughes2002,Pethig2010,Kim2019}.
	
	\item \textbf{Biological amplification.}
	A latent particle population is represented by a functional biological output
	such as PFU, FFU, endpoint dilution, cytopathic effect, or reporter signal.
	These readouts are powerful precisely because they amplify selected
	biological events, but they do not report the total physical particle
	population
	\cite{Dulbecco1952,Cooper1961,Baer2014,Klasse2015,Sanjuan2018}.
\end{enumerate}

\begin{remark}[Examples of latent-to-observed reduction]
	\label{rem:examples_latent_to_observed_reduction}
	In cryo-EM or cryo-ET, the observed state is dominated by preserved geometry
	after preparation, vitrification, particle selection, and reconstruction. In
	AFM, the observed state is conditioned by adsorption, local hydration, tip
	convolution, and mechanical loading. In dielectrophoresis, the observed state
	is a trajectory, trapping response, or spectrum in a field-biased medium. In
	mucus tracking, the observed state is a stochastic path shaped jointly by
	virion properties and the local microstructure, adhesivity, and rheology of
	the mucus network. In plaque assays, the observed state is a visible
	infectious event rather than a direct particle state.
\end{remark}

\subsubsection{Underrepresented dynamical and biological sectors}
\label{subsubsec:underrepresented_dynamical_biological_sectors}

\begin{remark}[Underrepresented dynamical sectors]
	\label{rem:underrepresented_dynamical_sectors}
	A protocol may systematically underrepresent or eliminate entire classes of
	virion behavior. Static imaging may suppress dynamically transient states.
	Surface binding may suppress freely evolving orientations. Field-driven
	measurements may overweight electrically responsive subpopulations.
	Single-virus tracking may collapse orientation, local adhesion, deformation,
	or fusion competence into a projected trajectory. Preparation steps may favor
	configurations that survive the protocol well. Biological assays may collapse
	particle integrity, cell-entry competence, replication, local spread, and
	visibility into a single infectious-unit readout. The result is not that the
	experiment is ``wrong,'' but that the observed state space can be strictly
	smaller than the mechanically and biologically relevant latent state space.
\end{remark}

\begin{remark}[Virions as passive probes of environment]
	\label{rem:virions_as_passive_probes_of_environment}
	In many extracellular contexts, virions behave as passively responsive objects:
	their motion, survival, adhesion, deformation, or detectability can encode
	information about the medium in which they are embedded. The latent state
	\(X\) should therefore often be treated as a joint virion--environment state
	rather than as a particle state alone. Dielectric properties can be inferred
	from virus motion in nonuniform electric fields, while mucus microstructure,
	adhesivity, and barrier properties can be inferred from virion or particle
	tracking data
	\cite{Hughes1998,Hughes2002,Pethig2010,Kim2019,Boukari2009,Wang2017,Kaler2022,
		Abrami2024}. The theoretical significance of experimental collapse is that it
	makes such inverse environmental inference part of the formal structure rather
	than a purely informal interpretation.
\end{remark}

\begin{remark}[Protocol specificity is not a defect]
	\label{rem:protocol_specificity_not_defect}
	The fact that observed states are protocol-specific should not be interpreted
	as a weakness of virological measurement. It is the reason different
	measurements are useful. A plaque assay intentionally discards most
	structural detail in order to report infectious activity. AFM intentionally
	applies force in order to measure mechanical response. Dielectrophoresis
	intentionally imposes a field in order to infer dielectric behavior.
	Single-virus tracking intentionally follows motion in order to identify
	transport regimes. The purpose of the latent--observed distinction is not to
	devalue these measurements, but to state precisely what each one makes
	visible.
\end{remark}

\begin{table}[H]
	\centering
	\caption{
		Examples of latent-to-observed reduction under common virological
		protocols. The table is schematic; a real protocol may combine several
		collapse channels.
	}
	\label{tab:latent_observed_protocol_examples}
	\renewcommand{\arraystretch}{1.18}
	\begin{tabularx}{0.98\linewidth}{@{}p{0.18\linewidth}p{0.27\linewidth}p{0.28\linewidth}X@{}}
		\toprule
		\textbf{Protocol}
		&
		\textbf{Latent variables affected}
		&
		\textbf{Typical observed variables}
		&
		\textbf{Dominant reduction}
		\\
		\midrule
		
		Cryo-EM / cryo-ET
		&
		Geometry, conformation, orientation, survivability in ice, particle-class
		membership
		&
		Density maps, subtomogram averages, class averages, orientation
		distributions
		&
		Preparation, selection, reconstruction
		\\[0.45em]
		
		AFM
		&
		Surface attachment, height, deformation, stiffness, rupture pathway
		&
		Topography, force--distance curves, indentation stiffness
		&
		Surface conditioning, loading, readout
		\\[0.45em]
		
		DEP / electrorotation
		&
		Polarization, charge response, orientation, hydrodynamic drag,
		field-induced motion
		&
		Trajectories, trapping, crossover frequencies, rotation rates
		&
		Field steering, dielectric selection
		\\[0.45em]
		
		Single-virus tracking
		&
		Position, transport regime, confinement, directed motion, time-window
		persistence
		&
		Tracks, MSD, pause states, anomalous exponents, residence times
		&
		Trajectory projection, time-window filtering
		\\[0.45em]
		
		Mucus / gel tracking
		&
		Position, confinement, adhesion, local mesh interaction, viscoelastic
		response
		&
		Tracks, MSD, confinement metrics, mobile/immobile fractions
		&
		Medium filtering, adhesive selection
		\\[0.45em]
		
		Plaque / focus assay
		&
		Cell-entry competence, replication, spread, cytopathic effect, staining
		visibility
		&
		PFU, FFU, endpoint infectious dose
		&
		Biological selection, amplification, counting
		\\
		\bottomrule
	\end{tabularx}
\end{table}

Together, these examples motivate the protocol observation operator introduced
below. Rather than treating observed states as direct copies of latent states,
the operator formalism represents each experiment as a structured map from a
reference latent ensemble to a protocol-specific observed ensemble, including
selection, readout, and null outcomes.

\subsection{Protocol Observation Operators}
\label{subsec:protocol_observation_operators}

The previous sections constructed the full observed ensemble
\(P_{\mathrm{obs},t}^{\varnothing}(\cdot\mid E)\) by composing latent-state
transformation, survival or detection weighting, readout, and the null channel.
It is useful to collect this construction into operator notation. In this form,
an experimental protocol becomes a map from latent probability measures to
observed probability measures.

Let
\[
\operatorname{Prob}(\Psi)
:=
\operatorname{Prob}(\Psi,\Sigma_X)
\]
denote the set of probability measures on the reference latent state space, and
let
\[
\operatorname{Prob}(\mathcal Y_E^{\varnothing})
:=
\operatorname{Prob}
(\mathcal Y_E^{\varnothing},\Sigma_{Y,E}^{\varnothing})
\]
denote the set of probability measures on the augmented observation space
\[
\mathcal Y_E^{\varnothing}
=
\mathcal Y_E\cup\{\varnothing\}.
\]
The full protocol kernel \(K_E^{\varnothing}\) maps each latent state
\(x\in\Psi\) to a probability measure on
\(\mathcal Y_E^{\varnothing}\). Equivalently, by integration against an input
latent ensemble, it defines an operator from latent ensembles to observed
ensembles.

\begin{definition}[Null-inclusive protocol observation operator]
	\label{def:null_inclusive_protocol_observation_operator}
	The \emph{null-inclusive protocol observation operator} associated with
	protocol \(E\) is the map
	\begin{equation}
		\mathcal M_E^{\varnothing}:
		\operatorname{Prob}(\Psi)
		\longrightarrow
		\operatorname{Prob}(\mathcal Y_E^{\varnothing})
		\label{eq:null_inclusive_protocol_observation_operator_mapping}
	\end{equation}
	defined by
	\begin{empheq}[box=\fbox]{equation}
		\left(\mathcal M_E^{\varnothing}P\right)(S)
		=
		\int_{\Psi}
		K_E^{\varnothing}(S\mid x)\,
		P(dx),
		\qquad
		S\in\Sigma_{Y,E}^{\varnothing}.
		\label{eq:null_inclusive_protocol_observation_operator_action}
	\end{empheq}
	For the reference latent ensemble,
	\begin{equation}
		P_{\mathrm{obs},t}^{\varnothing}(\cdot\mid E)
		=
		\mathcal M_E^{\varnothing}P_{\mathrm{ref},t}.
		\label{eq:null_inclusive_observed_ensemble_operator_form}
	\end{equation}
\end{definition}
The detection yield can now be written as a functional of the input latent
measure:
\begin{empheq}[box=\fbox]{equation}
	\eta_E[P]
	=
	\left(\mathcal M_E^{\varnothing}P\right)(\mathcal Y_E)
	=
	\int_{\Psi}
	q_E(x)\,P(dx),
	\label{eq:detection_yield_functional_general}
\end{empheq}
where
\[
q_E(x)
=
K_E^{\varnothing}(\mathcal Y_E\mid x)
\]
is the state-specific probability that latent state \(x\) produces a non-null
observation under protocol \(E\). For the reference ensemble,
\begin{equation}
	\eta_E(t)
	=
	\eta_E[P_{\mathrm{ref},t}]
	=
	P_{\mathrm{obs},t}^{\varnothing}(\mathcal Y_E\mid E).
	\label{eq:detection_yield_reference_ensemble}
\end{equation}

\begin{definition}[Detected protocol observation operator]
	\label{def:detected_protocol_observation_operator}
	When \(\eta_E[P]>0\), the \emph{detected protocol observation operator} is the
	conditional map
	\begin{equation}
		\mathcal M_E^{\mathrm{det}}:
		\operatorname{Prob}(\Psi)
		\longrightarrow
		\operatorname{Prob}(\mathcal Y_E)
		\label{eq:detected_protocol_observation_operator_mapping}
	\end{equation}
	defined by
	\begin{empheq}[box=\fbox]{equation}
		\left(\mathcal M_E^{\mathrm{det}}P\right)(B)
		=
		\frac{
			\left(\mathcal M_E^{\varnothing}P\right)(B)
		}{
			\left(\mathcal M_E^{\varnothing}P\right)(\mathcal Y_E)
		},
		\qquad
		B\in\Sigma_{Y,E}.
		\label{eq:detected_protocol_observation_operator_action}
	\end{empheq}
	Thus
	\begin{equation}
		P_{\mathrm{obs},t}(\cdot\mid E,\mathrm{det})
		=
		\mathcal M_E^{\mathrm{det}}P_{\mathrm{ref},t}.
		\label{eq:detected_observed_ensemble_operator_form}
	\end{equation}
\end{definition}
Equivalently, the detected operator is obtained by conditioning the
null-inclusive output on the non-null observation space:
\begin{equation}
	\mathcal M_E^{\mathrm{det}}P
	=
	\mathcal M_E^{\varnothing}P
	\left(
	\cdot
	\mid
	\mathcal Y_E
	\right),
	\qquad
	\eta_E[P]>0.
	\label{eq:detected_operator_as_conditioned_null_operator}
\end{equation}
This identity makes clear that the null-inclusive operator is the primary
probability-preserving object, while the detected operator is a normalized
conditional object.

\begin{remark}[Linearity before conditioning, nonlinearity after conditioning]
	\label{rem:linearity_before_conditioning_nonlinearity_after}
	The null-inclusive operator \(\mathcal M_E^{\varnothing}\) is linear in the
	input measure \(P\). If
	\[
	P=\alpha P_1+(1-\alpha)P_2,
	\qquad
	0\leq \alpha\leq 1,
	\]
	then
	\[
	\mathcal M_E^{\varnothing}P
	=
	\alpha \mathcal M_E^{\varnothing}P_1
	+
	(1-\alpha)\mathcal M_E^{\varnothing}P_2.
	\]
	The detected operator \(\mathcal M_E^{\mathrm{det}}\) is generally not
	linear, because it conditions on non-null detection by dividing by
	\(\eta_E[P]\). This distinction matters experimentally. A normalized
	reported histogram can hide the fact that the protocol discarded, destroyed,
	failed to detect, or failed to amplify most of the original latent population.
\end{remark}

\begin{remark}[Core meaning of \(K_E^{\varnothing}\)]
	\label{rem:core_meaning_protocol_kernel}
	The kernel \(K_E^{\varnothing}\) is the mathematical object that contains the
	measurement-conditioned passage from latent virion behavior to experimental
	record. In a weakly perturbative experiment, it may be close to a noisy
	projection of the pre-existing latent state. In a strongly conditioning
	experiment, it can include substantial state transformation, survival
	selection, reweighting, biological amplification, and null observation before
	the final record is produced.
\end{remark}

\begin{remark}[Why the null-inclusive operator is primary]
	\label{rem:why_null_inclusive_operator_primary}
	The null-inclusive operator preserves the full probability accounting of the
	experiment. It records both what is reported and what fails to enter the
	accepted data channel. The detected operator is often what appears in
	published histograms, reconstructed classes, tracked-particle summaries, or
	counted events, but it is a conditional distribution. For protocol-resolved
	inference, the missing probability mass can be as informative as the reported
	non-null distribution.
\end{remark}

\subsection{Protocol-Lifted Observables}
\label{subsec:protocol_lifted_observables}

Experimental observables are usually defined on the reported data space. The
kernel formalism allows each such observable to be lifted back to the latent
space, where it becomes an effective protocol-dependent observable. This is the
mathematical reason a reported quantity such as an apparent stiffness,
diffusion coefficient, field-response amplitude, plaque count, or class
membership should not be interpreted as a protocol-free property unless the
protocol dependence has been controlled, calibrated, or modeled.

\begin{definition}[Observable expectations under a protocol]
	\label{def:observable_expectations_under_protocol}
	Let
	\[
	h:\mathcal Y_E\rightarrow\mathbb R
	\]
	be an experimentally reported observable, such as apparent radius, height,
	stiffness, diffusion coefficient, plaque count, focus count, rotation rate,
	capture probability, or class-membership indicator. Its detected
	protocol-conditioned expectation is
	\begin{equation}
		\langle h\rangle_E[P]
		=
		\int_{\mathcal Y_E}
		h(y)\,
		\left(\mathcal M_E^{\mathrm{det}}P\right)(dy),
		\label{eq:observable_expectation_under_protocol}
	\end{equation}
	whenever \(\eta_E[P]>0\) and \(h\) is integrable with respect to
	\(\mathcal M_E^{\mathrm{det}}P\).
\end{definition}
\noindent The expectation in
Eq.~\eqref{eq:observable_expectation_under_protocol} can be written as a ratio
of two latent-space integrals:
\begin{empheq}[box=\fbox]{equation}
	\langle h\rangle_E[P]
	=
	\frac{
		\displaystyle
		\int_{\Psi}
		\widetilde h_E(x)\,
		P(dx)
	}{
		\displaystyle
		\int_{\Psi}
		\widetilde 1_E(x)\,
		P(dx)
	},
	\qquad
	\eta_E[P]>0.
	\label{eq:lifted_observable_expectation}
\end{empheq}
Here
\begin{equation}
	\widetilde h_E(x)
	=
	\int_{\Psi_E}
	s_E(x_E)
	\left[
	\int_{\mathcal Y_E}
	h(y)\,
	R_E(dy\mid x_E)
	\right]
	\Pi_E^{\mathrm{lat}}(dx_E\mid x)
	\label{eq:protocol_lifted_observable}
\end{equation}
is the \emph{protocol-lifted observable} on the reference latent space, and
\begin{equation}
	\widetilde 1_E(x)
	=
	\int_{\Psi_E}
	s_E(x_E)\,
	\Pi_E^{\mathrm{lat}}(dx_E\mid x)
	=
	K_E^{\varnothing}(\mathcal Y_E\mid x)
	=
	q_E(x)
	\label{eq:protocol_lifted_detection_function}
\end{equation}
is the state-specific non-null detection probability. Thus the denominator in
Eq.~\eqref{eq:lifted_observable_expectation} is the detection yield
\(\eta_E[P]\).

\begin{remark}[Interpretive meaning of lifted observables]
	\label{rem:interpretive_meaning_lifted_observables}
	Equation~\eqref{eq:lifted_observable_expectation} states that every
	experimentally measured quantity can be interpreted as an effective
	observable on the latent state space. The protocol determines not only what
	is seen, but also which latent features are weighted, blurred, averaged,
	transformed, amplified, or ignored when a measurement is reported.
\end{remark}

\begin{remark}[Selection bias in lifted observables]
	\label{rem:selection_bias_in_lifted_observables}
	The denominator in Eq.~\eqref{eq:lifted_observable_expectation} is essential.
	The reported expectation is not the latent average of
	\(\widetilde h_E\) alone; it is the latent average among states that enter
	the non-null data channel. Therefore, two protocols can have the same
	readout map but different reported averages if their detection functions
	\(\widetilde 1_E\) weight the latent ensemble differently. This is the
	operator-level expression of selection bias, detectability bias, and
	protocol-conditioned enrichment.
\end{remark}

\begin{remark}[Plaque count as a lifted observable]
	\label{rem:plaque_count_as_lifted_observable}
	In a plaque assay, the reported observable may be the number of plaques or
	the inferred plaque-forming units. In the lifted representation, this
	observable is not simply a count of physical virions. It is an effective
	latent-space observable that weights states by cell-entry competence,
	replication competence, local spread, overlay conditions, incubation time,
	staining visibility, and counting criteria. This makes plaque assays a
	natural worked example of experimental collapse.
\end{remark}

\begin{remark}[Vector-valued observables]
	\label{rem:vector_valued_lifted_observables}
	The same construction applies componentwise to vector-valued observables
	\[
	\mathbf h:\mathcal Y_E\rightarrow\mathbb R^m.
	\]
	This is useful when a protocol reports several coupled summaries, such as
	particle size and orientation class, diffusion coefficient and confinement
	metric, force-curve stiffness and rupture threshold, or plaque count and
	plaque-size distribution.
\end{remark}

\begin{remark}[Null-inclusive observables]
	\label{rem:null_inclusive_observables}
	Some observables are naturally defined on the augmented observation space
	\(\mathcal Y_E^{\varnothing}\), not only on the detected non-null space
	\(\mathcal Y_E\). For example, the indicator
	\[
	h_{\varnothing}(y)=\mathbf 1_{\{y=\varnothing\}}
	\]
	measures collapse into the null channel, while
	\[
	h_{\mathrm{det}}(y)=\mathbf 1_{\{y\in\mathcal Y_E\}}
	\]
	measures detection yield. Such observables should be evaluated using
	\(\mathcal M_E^{\varnothing}P\), not the detected conditional operator
	\(\mathcal M_E^{\mathrm{det}}P\), because conditioning on detection removes
	the null probability by construction.
\end{remark}

\subsection{Mechanism-Resolved Collapse Channels}
\label{subsec:mechanism_resolved_collapse_channels}

The kernel language is intentionally general. For experimental interpretation,
however, it is useful to decompose a protocol into mechanism-specific stages.
Different mechanisms reshape the latent ensemble in different ways: some
physically transform the state, some select which states survive, some amplify
particular biological pathways, and some compress a high-dimensional state into
a lower-dimensional readout. A mechanism-resolved description identifies which
part of the protocol is responsible for a given form of experimental collapse.

\begin{definition}[Mechanism-resolved factorization of the latent-stage kernel]
	\label{def:mechanism_resolved_factorization_latent_kernel}
	Let
	\[
	(\Psi_{E,0},\Sigma_{X,E,0})
	=
	(\Psi,\Sigma_X),
	\qquad
	(\Psi_{E,L},\Sigma_{X,E,L})
	=
	(\Psi_E,\Sigma_{X,E}),
	\]
	and let
	\[
	\Pi_{E,\ell}(dx_\ell\mid x_{\ell-1}),
	\qquad
	\ell=1,\ldots,L,
	\]
	be mechanism-specific Markov kernels between intermediate protocol state
	spaces. With the convention that the rightmost kernel acts first, the
	latent-stage kernel may be factorized as
	\begin{equation}
		\Pi_E^{\mathrm{lat}}
		=
		\Pi_{E,L}
		\circ
		\Pi_{E,L-1}
		\circ
		\cdots
		\circ
		\Pi_{E,1}.
		\label{eq:mechanism_resolved_factorization}
	\end{equation}
	Explicitly, for \(G\in\Sigma_{X,E}\) and \(x_0=x\),
	\begin{empheq}[box=\fbox]{equation}
		\begin{aligned}
			\Pi_E^{\mathrm{lat}}(G\mid x_0)
			&=
			\int_{\Psi_{E,1}}
			\cdots
			\int_{\Psi_{E,L}}
			\mathbf 1_G(x_L)\,
			\Pi_{E,L}(dx_L\mid x_{L-1})
			\cdots
			\Pi_{E,1}(dx_1\mid x_0).
		\end{aligned}
		\label{eq:mechanism_resolved_latent_kernel_integral}
	\end{empheq}
\end{definition}
Typical latent-stage mechanisms may include
\[
\Pi_E^{\mathrm{prep}},
\qquad
\Pi_E^{\mathrm{interface}},
\qquad
\Pi_E^{\mathrm{surf}},
\qquad
\Pi_E^{\mathrm{medium}},
\qquad
\Pi_E^{\mathrm{field}},
\qquad
\Pi_E^{\mathrm{load}},
\qquad
\Pi_E^{\mathrm{time}},
\qquad
\Pi_E^{\mathrm{bio}}.
\]
The ordering is not universal. It must be chosen to match the physical,
chemical, mechanical, biological, and computational sequence of the experiment.

\begin{remark}[Why mechanism order is protocol dependent]
	\label{rem:why_mechanisms_order_protocol_dependent}
	The factorization in Eq.~\eqref{eq:mechanism_resolved_factorization} is a
	modeling choice, not a universal ordering of experimental reality. In one
	protocol, an electric field may steer particles before surface contact. In
	another, surface immobilization may occur before imaging or loading. In
	cryo-EM, preparation, thin-film behavior, air--water interface exposure,
	vitrification, particle orientation, particle picking, and reconstruction are
	tightly coupled
	\cite{Cheng2015,Thompson2016,Noble2018,Chen2019,Liu2023}. In AFM,
	adsorption, hydration state, tip geometry, loading path, and deformation are
	coupled during measurement
	\cite{Mateu2012,Marchetti2016,Kiss2021,Lyonnais2021}. The important point is
	not that every protocol must be decomposed in the same order, but that the
	protocol kernel can be resolved into mechanistically interpretable stages.
\end{remark}

\begin{definition}[Collapse channels]
	\label{def:collapse_channels_experimental_collapse}
	A \emph{collapse channel} is a mechanism by which a protocol transforms,
	selects, reweights, amplifies, or compresses the latent ensemble before it
	becomes reported data. The main collapse channels considered in this paper
	include:
	\begin{enumerate}[label=(\roman*),leftmargin=2.2em]
		
		\item \emph{geometric projection}, in which the protocol reports only a
		lower-dimensional representation, reconstruction, image, trajectory,
		curve, or summary of the latent geometry;
		
		\item \emph{orientation selection}, in which certain orientations are
		preferentially preserved, detected, reconstructed, selected, or
		amplified;
		
		\item \emph{surface immobilization}, in which adsorption or contact changes
		translational freedom, rotational freedom, contact geometry, adhesion
		state, or deformation pathway;
		
		\item \emph{field steering}, in which external fields produce forces,
		torques, polarization, trapping, translation, rotation, sorting, or
		frequency-dependent response;
		
		\item \emph{mechanical loading}, in which the measurement applies force,
		indentation, compression, confinement, stretching, or deformation during
		preparation or readout;
		
		\item \emph{time-window filtering}, in which only states that persist,
		move, bind, relax, amplify, or remain detectable on the instrumental or
		assay timescale contribute to the observed ensemble;
		
		\item \emph{medium filtering}, in which the local medium admits, confines,
		slows, binds, immobilizes, transports, or excludes different virion
		states;
		
		\item \emph{biological amplification and viability selection}, in which the
		protocol reports only states capable of completing a specified biological
		sequence, such as cell attachment, entry, replication, local spread,
		cytopathic effect, focus formation, reporter expression, or plaque
		formation;
		
		\item \emph{survival or detectability selection}, in which only a subset of
		protocol-conditioned states enters the reported dataset at all.
	\end{enumerate}
\end{definition}

These channels are not mutually exclusive. A single protocol may combine several
of them. For example, a plaque assay combines biological amplification, time
windowing, medium or overlay filtering, visibility thresholding, and counting. An
AFM protocol may combine surface immobilization, hydration-dependent
conditioning, mechanical loading, geometric projection, and force-curve readout.
The channel language is therefore not a classification scheme for entire
experiments, but a way to identify which mechanisms participate in the observed
collapse.

\begin{remark}[Latent transformation versus readout selection]
	\label{rem:latent_transformation_vs_readout_selection}
	Some selection effects physically alter the latent state, while others alter
	only the probability that a state is reported. The former belong naturally in
	\(\Pi_E^{\mathrm{lat}}\). The latter belong in \(s_E\), \(R_E\), or the null
	channel \(\varnothing\). The formalism allows either placement when a protocol
	model requires it, provided that the model states clearly where the effect
	enters and what probability mass is being transformed, selected, or nulled.
\end{remark}

\begin{remark}[Why collapse channels are useful]
	\label{rem:why_collapse_channels_are_useful}
	The channel language prevents a protocol from being described only as
	``biased'' or ``artifact-prone.'' It asks a more precise question: which
	mechanism produced the observed conditioning? A preferred-orientation problem
	in cryo-EM, a surface-induced deformation in AFM, a field-induced drift in
	DEP, an adhesive immobilization event in mucus, and a plaque-forming selection
	step in an infectivity assay are different collapse mechanisms. They should
	therefore be modeled differently.
\end{remark}

\begin{table}[H]
	\centering
	\caption{
		Collapse channels and their mechanical interpretation. The listed
		mathematical locations are typical, not exclusive; a real protocol may
		distribute one physical effect across several kernels or weights.
	}
	\label{tab:collapse_channels_mechanical_interpretation}
	\renewcommand{\arraystretch}{1.18}
	\begin{tabularx}{0.98\linewidth}{@{}p{0.23\linewidth}p{0.30\linewidth}X@{}}
		\toprule
		\textbf{Collapse channel}
		&
		\textbf{Typical mathematical location}
		&
		\textbf{Mechanical or experimental interpretation}
		\\
		\midrule
		
		Geometric projection
		&
		\(R_E\) or \(\mathcal O_E:\Psi_E\to\mathcal Y_E\)
		&
		Only selected coordinates, shapes, projections, reconstructions,
		curves, trajectories, or summaries are reported.
		\\[0.45em]
		
		Orientation selection
		&
		\(s_E(x_E)\), \(R_E(dy\mid x_E)\), or
		\(\Pi_E^{\mathrm{prep}}\)
		&
		Some orientations are overrepresented because of preparation,
		interface effects, reconstruction, particle picking, or selection.
		\\[0.45em]
		
		Surface immobilization
		&
		\(\Pi_E^{\mathrm{surf}}\)
		&
		Adsorption or tethering constrains translation, rotation, deformation,
		contact geometry, and accessible mechanical response.
		\\[0.45em]
		
		Field steering
		&
		\(\Pi_E^{\mathrm{field}}\)
		&
		Electric, optical, magnetic, acoustic, or flow fields move, polarize,
		trap, orient, rotate, or sort virions.
		\\[0.45em]
		
		Mechanical loading
		&
		\(\Pi_E^{\mathrm{load}}\) or force-dependent \(R_E\)
		&
		Tip forces, indentation, confinement, compression, or stretching alter
		the measured state or the response inferred from it.
		\\[0.45em]
		
		Time-window filtering
		&
		\(\Pi_E^{\mathrm{time}}\), \(s_E\), or \(R_E\)
		&
		Transient, fast, rare, slow, delayed, or short-lived states are missed
		or underweighted.
		\\[0.45em]
		
		Medium filtering
		&
		\(\Pi_E^{\mathrm{medium}}\) or a joint
		virion--environment latent state
		&
		Mucus, gels, extracellular matrices, droplets, overlays, or buffers
		admit, bind, hinder, confine, immobilize, or exclude different virion
		states.
		\\[0.45em]
		
		Biological amplification
		&
		\(\Pi_E^{\mathrm{bio}}\), \(s_E\), \(R_E\), or an assay-specific
		branching kernel
		&
		Only virions that complete a specified biological sequence generate
		plaques, foci, cytopathic effects, reporter signals, or endpoint
		signals.
		\\[0.45em]
		
		Detection selection
		&
		\(s_E(x_E)\) and \(\varnothing\)
		&
		Some states are absent from the reported ensemble because they are lost,
		destroyed, rejected, noninfectious under the assay, outside the
		observation window, or below threshold.
		\\
		\bottomrule
	\end{tabularx}
\end{table}

\subsection{Quantifying Experimental Collapse}
\label{subsec:quantifying_experimental_collapse}

The word ``collapse'' is useful only if it can be quantified. The protocol
kernel gives a constructive way to do this. Experimental collapse can be
measured as a discrepancy between a reference latent ensemble and a
protocol-conditioned latent ensemble, between preselection and detected
populations, between observed ensembles produced by different protocols, or
between protocol-conditioned values of a chosen experimental observable.

A technical point is essential. The reference latent state space \(\Psi\), the
protocol-conditioned latent state space \(\Psi_E\), and the observation space
\(\mathcal Y_E\) need not be identical. A cryo-EM density map, an AFM
force--indentation curve, a DEP trajectory, a mucus-tracking path, and a plaque
count are different mathematical objects. Collapse functionals should therefore
compare probability measures only after the relevant quantities have been placed
on a common comparison space.

\begin{definition}[Common comparison space]
	\label{def:common_comparison_space}
	Let
	\[
	(\mathsf W,\Sigma_W)
	\]
	be a measurable comparison space. Let
	\[
	L_0:\Psi\rightarrow\mathsf W,
	\qquad
	L_E:\Psi_E\rightarrow\mathsf W
	\]
	be measurable maps that extract comparable quantities from the reference and
	protocol-conditioned latent states. Examples of \(L_0\) and \(L_E\) include
	maps to radius, orientation, mobility, stiffness, infectivity class, adhesion
	state, deformation amplitude, field-response amplitude, normal-mode
	amplitude, branch participation weight, or another experimentally relevant
	summary.
\end{definition}

\begin{remark}[What the comparison space means experimentally]
	\label{rem:comparison_space_experimental_meaning}
	The comparison space \(\mathsf W\) encodes the scientific question. If the
	question concerns apparent size, then \(\mathsf W\) may be a radius or
	diameter space. If the question concerns orientation selection,
	\(\mathsf W\) may be an orientation space. If the question concerns
	infectivity, \(\mathsf W\) may be a binary or multistate competence space. If
	the question concerns transport, \(\mathsf W\) may contain diffusivities,
	residence times, confinement metrics, or anomalous exponents. Collapse is
	therefore always quantified relative to a chosen experimentally meaningful
	comparison.
\end{remark}

\begin{definition}[Latent distributional collapse functional]
	\label{def:latent_distributional_collapse_functional}
	Let \(D_{\mathsf W}(\cdot,\cdot)\) be a divergence, distance, or discrepancy
	between probability measures on \((\mathsf W,\Sigma_W)\). The preselection
	latent collapse magnitude is
	\begin{empheq}[box=\fbox]{equation}
		\mathcal C_{\mathrm{lat}}^{\mathsf W}(E;t)
		=
		D_{\mathsf W}
		\left(
		(L_E)_{\#}\widetilde P_{E,t}^{\mathrm{lat}},
		(L_0)_{\#}P_{\mathrm{ref},t}
		\right).
		\label{eq:latent_distributional_collapse_functional}
	\end{empheq}
	If detection selection is important, the detected latent collapse is
	\begin{empheq}[box=\fbox]{equation}
		\mathcal C_{\mathrm{det}}^{\mathsf W}(E;t)
		=
		D_{\mathsf W}
		\left(
		(L_E)_{\#}P_{E,t}^{\mathrm{lat}}(\cdot\mid\mathrm{det}),
		(L_0)_{\#}P_{\mathrm{ref},t}
		\right),
		\qquad
		0<\eta_E(t)\leq 1.
		\label{eq:detected_latent_collapse_functional}
	\end{empheq}
\end{definition}

\begin{remark}[Pre-detection versus detected collapse]
	\label{rem:pre_detection_vs_detected_collapse}
	The two functionals above answer different questions.
	\(\mathcal C_{\mathrm{lat}}^{\mathsf W}\) measures how much the protocol
	transforms the latent ensemble before survival or detection selection.
	\(\mathcal C_{\mathrm{det}}^{\mathsf W}\) measures the ensemble that actually
	remains visible after selection. Their difference is informative. If
	\(\mathcal C_{\mathrm{lat}}^{\mathsf W}\) is small but
	\(\mathcal C_{\mathrm{det}}^{\mathsf W}\) is large, the dominant collapse
	mechanism is selection rather than mechanical transformation. If both are
	large, the protocol may be both transforming and selecting the latent
	ensemble.
\end{remark}

\begin{remark}[Why pushforwards are needed]
	\label{rem:why_pushforwards_are_needed_for_collapse}
	Writing \(D(P_{E,t}^{\mathrm{lat}},P_{\mathrm{ref},t})\) is only meaningful
	when the two measures live on the same state space and share an appropriate
	reference structure. In many virological settings they do not. A
	protocol-conditioned latent state in AFM, a vitrified cryo-EM state, a
	field-conditioned DEP state, and a plaque-assay pathway state are not
	identical mathematical objects. The maps \(L_0\) and \(L_E\) specify what is
	being compared.
\end{remark}

\begin{remark}[Choice of divergence]
	\label{rem:choice_of_divergence_experimental_collapse}
	The choice of \(D_{\mathsf W}\) depends on the application. A
	Kullback--Leibler divergence is natural when likelihood ratios are
	meaningful and absolute continuity conditions are satisfied. A
	Jensen--Shannon divergence is symmetric and finite in many cases where KL is
	inconvenient. A Wasserstein distance is useful when nearby states in
	\(\mathsf W\) should be treated as similar rather than categorically
	different. Maximum mean discrepancy and other integral-probability metrics
	are useful when distributions are compared through function classes or
	kernels \cite{KullbackLeibler1951,Lin1991,Villani2009,Gretton2012}. The
	theory does not require one universal metric. It requires that the chosen
	metric match the experimental question and the geometry of the data.
\end{remark}

\begin{definition}[Observed protocol-to-protocol collapse]
	\label{def:observed_protocol_to_protocol_collapse}
	Let \(E_1\) and \(E_2\) be two protocols. Let
	\[
	H_{E_1}:\mathcal Y_{E_1}\rightarrow\mathsf W,
	\qquad
	H_{E_2}:\mathcal Y_{E_2}\rightarrow\mathsf W
	\]
	map their non-null observed outputs to a common comparison space. The
	observed protocol-to-protocol collapse is
	\begin{empheq}[box=\fbox]{equation}
		\mathcal C_{\mathrm{obs}}^{\mathsf W}(E_1,E_2;t)
		=
		D_{\mathsf W}
		\left(
		(H_{E_1})_{\#}
		P_{\mathrm{obs},t}(\cdot\mid E_1,\mathrm{det}),
		(H_{E_2})_{\#}
		P_{\mathrm{obs},t}(\cdot\mid E_2,\mathrm{det})
		\right).
		\label{eq:observed_protocol_to_protocol_collapse}
	\end{empheq}
\end{definition}

\begin{remark}[What protocol-to-protocol collapse measures]
	\label{rem:what_protocol_to_protocol_collapse_measures}
	Observed protocol-to-protocol collapse asks whether two protocols produce
	compatible observed summaries after both outputs are mapped to a common
	comparison space. A discrepancy need not mean that one experiment is wrong.
	It may indicate that the protocols observe different projections, select
	different subpopulations, perturb different latent sectors, or amplify
	different functional events. Conversely, a small discrepancy does not prove
	that the protocols see the full latent ensemble; it may mean that they share a
	blind direction or a common selection bias.
\end{remark}

\begin{definition}[Observable-level collapse]
	\label{def:observable_level_collapse}
	Let
	\[
	\mathbf h_E:\mathcal Y_E\rightarrow\mathbb R^m,
	\qquad
	\mathbf h_{E_0}:\mathcal Y_{E_0}\rightarrow\mathbb R^m
	\]
	be comparable reported observables under protocols \(E\) and \(E_0\). Define
	\begin{empheq}[box=\fbox]{equation}
		\mathcal C_{\mathbf h}(E;E_0,t)
		=
		\left\|
		\langle \mathbf h_E\rangle_E(t)
		-
		\langle \mathbf h_{E_0}\rangle_{E_0}(t)
		\right\|,
		\label{eq:observable_level_collapse}
	\end{empheq}
	where the norm is chosen according to the scale, covariance structure, and
	scientific meaning of the observable. A normalized scalar version is
	\begin{equation}
		\widehat{\mathcal C}_{\mathbf h}(E;E_0,t)
		=
		\frac{
			\left\|
			\langle \mathbf h_E\rangle_E(t)
			-
			\langle \mathbf h_{E_0}\rangle_{E_0}(t)
			\right\|
		}{
			\left\|
			\langle \mathbf h_{E_0}\rangle_{E_0}(t)
			\right\|
			+\epsilon_h
		},
		\label{eq:normalized_observable_level_collapse}
	\end{equation}
	with \(\epsilon_h>0\) included to regularize cases where the reference
	expectation is zero or near zero.
\end{definition}

\begin{remark}[What observable-level collapse measures]
	\label{rem:what_observable_level_collapse_measures}
	Observable-level collapse asks a practical question: how much does a reported
	quantity change when the protocol changes? The relevant observable could be
	an apparent radius, height, stiffness, diffusion coefficient, orientation
	distribution, branch speed, branch participation weight, trapping
	probability, rotation rate, plaque count, focus-forming count, endpoint titer,
	or another protocol-defined summary. This makes the abstract kernel framework
	experimentally testable.
\end{remark}

\begin{definition}[Detection-yield collapse]
	\label{def:detection_yield_collapse}
	The detection-yield collapse between protocols \(E\) and \(E_0\) is
	\begin{empheq}[box=\fbox]{equation}
		\mathcal C_{\eta}(E;E_0,t)
		=
		\left|
		\eta_E(t)-\eta_{E_0}(t)
		\right|.
		\label{eq:detection_yield_collapse}
	\end{empheq}
	A logarithmic version,
	\begin{equation}
		\mathcal C_{\log\eta}(E;E_0,t)
		=
		\left|
		\log
		\frac{\eta_E(t)+\epsilon_\eta}
		{\eta_{E_0}(t)+\epsilon_\eta}
		\right|,
		\label{eq:log_detection_yield_collapse}
	\end{equation}
	is useful when detection yields differ by orders of magnitude.
\end{definition}

\begin{remark}[Why detection-yield collapse is not optional]
	\label{rem:why_detection_yield_collapse_not_optional}
	Two protocols may produce similar normalized observed distributions while
	having very different detection yields. A protocol that reports nearly every
	state and a protocol that reports only a small selected subset may appear
	similar after conditioning on detection. The yield \(\eta_E(t)\) records the
	missing probability mass and is therefore part of experimental collapse
	itself. In practical terms, it is the denominator of the experiment.
\end{remark}

\begin{definition}[Selection-strength functional]
	\label{def:selection_strength_functional}
	When the survival/detection weight \(s_E\) is known or estimated, the
	strength of detection selection can be summarized by the variation of
	\(s_E\) over the preselection protocol-conditioned latent ensemble. One simple
	measure is
	\begin{empheq}[box=\fbox]{equation}
		\mathcal S_E(t)
		=
		\operatorname{Var}_{\widetilde P_{E,t}^{\mathrm{lat}}}
		\left[
		s_E(X_E)
		\right].
		\label{eq:selection_strength_functional}
	\end{empheq}
	Large \(\mathcal S_E(t)\) indicates that survival or detection is strongly
	state dependent.
\end{definition}

\begin{remark}[Interpretation of selection strength]
	\label{rem:interpretation_selection_strength}
	A small detection yield \(\eta_E(t)\) and a large selection strength
	\(\mathcal S_E(t)\) describe different aspects of collapse. The yield
	measures how much probability mass reaches the non-null record. The selection
	strength measures how unevenly that probability is distributed across
	protocol-conditioned latent states. A protocol can have low yield because
	every state is weakly detected, or because a small subpopulation is strongly
	detected while most states are nearly invisible. These cases have different
	experimental interpretations.
\end{remark}

\begin{remark}[Quantification is target-dependent]
	\label{rem:quantification_is_target_dependent}
	There is no protocol-independent scalar called ``the collapse.'' A protocol
	may weakly collapse size, strongly collapse orientation, selectively preserve
	one mechanical mode, suppress another, and send a large fraction of the
	population into the null channel. Quantification must therefore specify the
	comparison space, the observable or distribution being compared, the role of
	detection conditioning, and the divergence or norm used. The next subsection
	applies this perspective to modal and dynamical collapse, where the relevant
	comparison targets are time-dependent degrees of freedom, mode amplitudes, and
	dynamical sectors of the latent state.
\end{remark}

\subsection{Modal and Dynamical Collapse}
\label{subsec:modal_and_dynamical_collapse}

When the latent theory includes collective modes, branch amplitudes,
orientation--translation coupling, or time-dependent state variables, a protocol
can collapse not only static features but also dynamical sectors of the theory.
This is especially relevant for assembly-level or lattice-level models in which
displacement, orientation, spike presentation, contact variables, hydrodynamic
response, or electrical response hybridize into collective branches.

The central point is that a protocol may preserve a structural snapshot while
suppressing the dynamical pathway that produced it, or it may make a dynamical
sector visible precisely by forcing, loading, tracking, or amplifying it. Static
and dynamical observability are therefore distinct. A protocol can be highly
informative about morphology while being nearly blind to relaxation, damping,
rotation, adhesion kinetics, or branch participation.

\begin{definition}[Modal collapse]
	\label{def:modal_collapse_experimental_collapse}
	Suppose a latent or protocol-conditioned model defines a branch diagnostic
	\[
	B_{n,E}(\mathbf k),
	\]
	such as a dispersion relation \(\omega_{n,E}(\mathbf k)\), a damping rate
	\(\Gamma_{n,E}(\mathbf k)\), a presentation or tilt weight
	\(P_{\mathrm{tilt},E}^{(n)}(\mathbf k)\), a dielectric-response amplitude,
	an elastic participation factor, or a displacement--orientation participation
	ratio. The modal collapse of this branch diagnostic between protocols \(E\)
	and \(E_0\) is
	\begin{empheq}[box=\fbox]{equation}
		\mathcal C_{B_n}(E;E_0,\mathbf k)
		=
		\left\|
		B_{n,E}(\mathbf k)
		-
		B_{n,E_0}(\mathbf k)
		\right\|.
		\label{eq:modal_collapse_branch_diagnostic}
	\end{empheq}
	The norm is chosen according to the nature of the branch diagnostic. For a
	scalar diagnostic it may be an absolute value; for a vector-valued or
	matrix-valued diagnostic it may be a Euclidean, spectral, Frobenius, or
	problem-specific norm.
\end{definition}

For a comparison over a region \(\Omega_k\) of wave-vector space, one may define
an integrated modal-collapse functional:
\begin{equation}
	\mathcal C_{B_n}^{\Omega_k}(E;E_0)
	=
	\left[
	\int_{\Omega_k}
	w_n(\mathbf k)\,
	\left\|
	B_{n,E}(\mathbf k)
	-
	B_{n,E_0}(\mathbf k)
	\right\|^2
	d\mathbf k
	\right]^{1/2},
	\label{eq:integrated_modal_collapse_functional}
\end{equation}
where \(w_n(\mathbf k)\geq 0\) is a weighting function that encodes the
experimentally relevant region of the branch, such as long-wavelength modes,
high-participation modes, field-coupled modes, or modes within the detectable
frequency band.

\begin{remark}[Why modal collapse matters]
	\label{rem:why_modal_collapse_matters}
	A protocol may suppress, pin, reveal, or reweight different pieces of a latent
	dynamical branch structure. A branch that is hybrid in a weakly constrained
	ensemble may appear translational under a surface-bound protocol, while a
	field-biased protocol may reveal an orientational or dielectric sector that is
	invisible in passive imaging. Modal collapse measures this protocol-dependent
	change directly.
\end{remark}

\begin{remark}[Connection to virological experiments]
	\label{rem:modal_collapse_virological_context}
	In many virological experiments, the relevant dynamical sectors are not
	measured as full normal modes. They appear indirectly through relaxation times
	in tracking data, indentation response in AFM, orientation statistics in
	structural imaging, frequency-dependent response in field-driven assays, or
	time-to-visibility in infectivity assays. The modal language is therefore most
	useful when it is tied to an explicit branch diagnostic that the protocol can
	actually constrain.
\end{remark}

\begin{definition}[Dynamical collapse]
	\label{def:dynamical_collapse_experimental_collapse}
	Let \(X_{0:T}\) denote a latent trajectory over a time interval
	\([0,T]\), and let \(Y_{0:T}^{(E)}\) denote the corresponding observed
	process under protocol \(E\). Let
	\[
	L_0^{\mathrm{dyn}}:\{X_{0:T}\}\rightarrow\mathsf Z,
	\qquad
	L_E^{\mathrm{dyn}}:\{Y_{0:T}^{(E)}\}\rightarrow\mathsf Z
	\]
	map latent and observed dynamical objects into a common dynamical comparison
	space \((\mathsf Z,\Sigma_Z)\), such as relaxation times, residence-time
	distributions, mean-square displacements, mode amplitudes, autocorrelation
	functions, transition rates, or branch participation weights. A dynamical
	collapse functional may be written as
	\begin{empheq}[box=\fbox]{equation}
		\mathcal C_{\mathrm{dyn}}^{\mathsf Z}(E;t,T)
		=
		D_{\mathsf Z}
		\left(
		(L_E^{\mathrm{dyn}})_{\#}
		P_{\mathrm{obs},0:T}(\cdot\mid E,\mathrm{det}),
		(L_0^{\mathrm{dyn}})_{\#}
		P_{\mathrm{ref},0:T}
		\right).
		\label{eq:dynamical_collapse_functional}
	\end{empheq}
\end{definition}

This definition makes explicit that dynamical collapse is not simply the loss of
a coordinate. It can be the loss of a time correlation, relaxation pathway,
transition rate, branch amplitude, dwell-time distribution, or coupling between
degrees of freedom. A protocol that preserves spatial position but destroys
velocity information, or one that preserves average transport but removes
short-lived adhesion states, can therefore be weakly collapsed in one variable
and strongly collapsed in another.

\subsection{Collapse Severity Classes}
\label{subsec:collapse_severity_classes}

The preceding functionals measure collapse quantitatively. It is also useful to
name broad qualitative regimes. These categories are not substitutes for the
functionals above; they are interpretive labels that help explain which kind of
protocol dependence dominates a given experiment.

\begin{definition}[Collapse severity classes]
	\label{def:collapse_severity_classes}
	A protocol may be classified by the size and mechanism of its collapse:
	\begin{enumerate}[label=(\roman*),leftmargin=2.2em]
		
		\item \emph{weak readout collapse}, when
		\(\widetilde P_{E,t}^{\mathrm{lat}}\approx P_{\mathrm{ref},t}\) in the
		relevant comparison space and the dominant effect is noise, finite
		resolution, reconstruction, classification, or projection in \(R_E\);
		
		\item \emph{selection collapse}, when the latent transformation is weak but
		\(P_{E,t}^{\mathrm{lat}}(\cdot\mid\mathrm{det})\) differs strongly from
		the reference ensemble because \(s_E\) is state dependent;
		
		\item \emph{mechanical collapse}, when \(\Pi_E^{\mathrm{lat}}\) physically
		changes the state distribution through fields, surfaces, loading,
		confinement, preparation, or medium interaction;
		
		\item \emph{modal or dynamical collapse}, when the protocol suppresses,
		pins, averages, reweights, or reveals specific time-dependent sectors,
		normal-mode branches, relaxation pathways, or dynamical couplings;
		
		\item \emph{biological amplification collapse}, when a biological assay
		reports only states that pass through a protocol-specific viability,
		attachment, entry, infection, replication, spread, reporter, staining, or
		visibility pathway;
		
		\item \emph{destructive or irreversible collapse}, when the protocol changes
		the accessible state space itself through deformation, rupture,
		aggregation, denaturation, irreversible adsorption, permanent
		inactivation, or loss of infectivity.
	\end{enumerate}
\end{definition}

\begin{remark}[Interpretation of severity classes]
	\label{rem:interpretation_collapse_severity_classes}
	These classes are not judgments about experimental quality. They are
	interpretive labels. A highly perturbative protocol can be extremely valuable
	if its kernel is understood. AFM is valuable precisely because it applies
	force; dielectrophoresis and electrorotation are valuable precisely because
	they impose fields; cryo-EM is valuable precisely because it preserves
	structure; plaque assays are valuable precisely because they amplify
	infectious events into countable lesions. The point is to state what ensemble
	each protocol produces, rather than to treat all protocols as if they sampled
	the same latent distribution.
\end{remark}

\begin{table}[H]
	\centering
	\caption{
		Collapse severity classes and typical examples. The same protocol can
		occupy different regimes depending on the observable, comparison space,
		and latent sector being studied.
	}
	\label{tab:collapse_severity_classes}
	\renewcommand{\arraystretch}{1.18}
	\begin{tabularx}{0.98\linewidth}{@{}p{0.24\linewidth}p{0.35\linewidth}X@{}}
		\toprule
		\textbf{Severity class}
		&
		\textbf{Dominant mathematical signature}
		&
		\textbf{Typical interpretation}
		\\
		\midrule
		
		Weak readout collapse
		&
		\(R_E\) projects, blurs, classifies, or reconstructs, while
		\(\Pi_E^{\mathrm{lat}}\) and \(s_E\) are weakly state dependent
		&
		The protocol mostly observes a pre-existing state with finite resolution
		or readout compression.
		\\[0.45em]
		
		Selection collapse
		&
		\(s_E(x_E)\) is strongly state dependent
		&
		The reported ensemble is a selected, enriched, or thresholded subset of
		the protocol-conditioned population.
		\\[0.45em]
		
		Mechanical collapse
		&
		\(\Pi_E^{\mathrm{lat}}\) substantially changes the latent distribution
		&
		Fields, surfaces, loading, confinement, preparation, or media reshape
		the particle state.
		\\[0.45em]
		
		Modal or dynamical collapse
		&
		Branch diagnostics, relaxation statistics, or dynamical couplings change
		under \(E\)
		&
		The protocol suppresses, pins, averages, reveals, or reweights
		time-dependent sectors of the latent theory.
		\\[0.45em]
		
		Biological amplification collapse
		&
		Observation requires successful passage through an assay-specific
		biological sequence
		&
		Infectious-unit assays report a selected and amplified functional
		subpopulation.
		\\[0.45em]
		
		Destructive or irreversible collapse
		&
		The accessible state space changes, or the protocol cannot be reversed
		&
		Rupture, aggregation, denaturation, irreversible adsorption, or loss of
		infectivity removes states from later observation.
		\\
		\bottomrule
	\end{tabularx}
\end{table}

\subsection{Experimental Collapse and Inverse Inference}
\label{subsec:experimental_collapse_inverse_inference}

The protocol kernel is not only a warning about measurement bias. It is also a
tool for inference. If the protocol is known, estimated, calibrated, or
parametrized, then the observed ensemble can be used to infer latent virion
properties, environmental properties, and protocol-specific coupling
parameters. In this sense, experimental collapse is not merely a loss of
information. It is the forward map through which latent virion--environment
mechanics becomes data.

The central inverse problem is therefore:
\begin{empheq}[box=\fbox]{equation}
	\text{Observed protocol-conditioned data}
	\;\Longrightarrow\;
	\left\{
	\begin{array}{c}
		\text{Latent virion parameters} \\
		\text{Environmental parameters} \\
		\text{Protocol parameters}
	\end{array}
	\right\}.
	\label{eq:experimental_collapse_inverse_problem_at_glance}
\end{empheq}
This arrow should not be read as a direct algebraic inversion. It denotes an
inverse problem constrained by a forward model. A protocol-resolved analysis
must specify how latent virion mechanics, environmental structure, and protocol
operations jointly generate the observed ensemble
\cite{Tarantola2005,KaipioSomersalo2005}.

\begin{definition}[Protocol-resolved parameter vector]
	\label{def:protocol_resolved_parameter_vector}
	Let
	\begin{equation}
		\theta
		=
		\left(
		\theta_{\mathrm{vir}},
		\theta_{\mathrm{env}},
		\theta_E
		\right)
		\in\Theta
		\label{eq:protocol_resolved_parameter_vector}
	\end{equation}
	denote a protocol-resolved parameter vector. Here
	\(\theta_{\mathrm{vir}}\) contains virion mechanical, structural,
	electrical, or biophysical parameters; \(\theta_{\mathrm{env}}\) contains
	environmental parameters; and \(\theta_E\) contains protocol parameters.
\end{definition}

\begin{remark}[Examples of parameter blocks]
	\label{rem:examples_protocol_resolved_parameter_blocks}
	The virion parameter block \(\theta_{\mathrm{vir}}\) may include radius
	distribution, effective charge, spike or surface-state parameters,
	polarizability, stiffness, compliance, adhesion strength, damping
	coefficients, orientation mobility, or infectivity-related latent variables.
	The environmental block \(\theta_{\mathrm{env}}\) may include viscosity,
	ionic strength, dielectric permittivity, conductivity, mucus mesh scale,
	mucin composition, antibody concentration, receptor density, pH,
	temperature, or cell-layer susceptibility. The protocol block \(\theta_E\)
	may include field amplitude, field frequency, surface chemistry, grid type,
	imaging cadence, exposure time, force ramp rate, tip geometry, dilution
	factor, inoculum volume, adsorption time, overlay composition, incubation
	time, staining threshold, particle-picking rule, reconstruction parameters,
	or counting criterion.
\end{remark}

\begin{definition}[Protocol-resolved forward model]
	\label{def:forward_model_experimental_collapse}
	A \emph{protocol-resolved forward model} is the map
	\begin{empheq}[box=\fbox]{equation}
		\theta
		\longmapsto
		P_{\mathrm{obs},t}^{\varnothing}(\cdot\mid E,\theta)
		=
		\mathcal M_E^{\varnothing}(\theta_E)
		P_{\mathrm{ref},t}
		\left(
		\cdot
		\mid
		\theta_{\mathrm{vir}},
		\theta_{\mathrm{env}}
		\right).
		\label{eq:protocol_resolved_forward_model}
	\end{empheq}
	The non-null reported model is obtained by conditioning on detection:
	\begin{equation}
		P_{\mathrm{obs},t}(\cdot\mid E,\theta,\mathrm{det})
		=
		\mathcal M_E^{\mathrm{det}}(\theta_E)
		P_{\mathrm{ref},t}
		\left(
		\cdot
		\mid
		\theta_{\mathrm{vir}},
		\theta_{\mathrm{env}}
		\right),
		\label{eq:protocol_resolved_forward_model_detected}
	\end{equation}
	provided the detection yield is nonzero.
\end{definition}

\begin{remark}[Meaning of the forward model]
	\label{rem:meaning_forward_model_experimental_collapse}
	The forward model states that observed data are produced jointly by virion
	mechanics, environmental mechanics, and protocol mechanics. Thus an apparent
	diffusion coefficient measured in mucus, an indentation stiffness measured by
	AFM, a dielectrophoretic crossover frequency, a cryo-EM orientation
	distribution, or a plaque count should not be interpreted as a property of the
	virion alone unless the environmental and protocol contributions have been
	controlled, marginalized, or modeled.
\end{remark}

\begin{remark}[Why collapse becomes useful for inference]
	\label{rem:why_collapse_becomes_useful_for_inference}
	A strongly conditioning protocol can be inferentially useful precisely because
	it changes the ensemble in a controlled or interpretable way. AFM loading can
	make mechanical compliance visible. Dielectrophoresis and electrorotation can
	make dielectric response visible. Mucus tracking can make adhesive
	confinement and transport heterogeneity visible. Plaque assays can make
	infectious activity visible. In each case, the protocol does not merely obscure
	the latent state; it asks a specific physical or biological question of that
	state.
\end{remark}

This perspective prepares the likelihood-based formulation that follows. Once
the forward model specifies
\(P_{\mathrm{obs},t}^{\varnothing}(\cdot\mid E,\theta)\), one can write
protocol-resolved likelihoods for observed data, separate virion parameters from
environmental and protocol parameters, and analyze which directions in
\(\theta\)-space are identifiable under a given experimental design.

\subsubsection{Protocol-Resolved Likelihoods}
\label{subsubsec:protocol_resolved_likelihoods}

Once a protocol-resolved forward model has been specified, inference proceeds by
writing the likelihood of the observed data under the protocol-conditioned
observed law. The essential point is that the likelihood should be written for
the data that the protocol actually reports, not for an idealized,
protocol-free latent ensemble.

\begin{definition}[Protocol-resolved dataset]
	\label{def:protocol_resolved_dataset}
	Suppose protocol \(E\) produces observed data
	\begin{equation}
		\mathcal D_E
		=
		\{y_1,\ldots,y_N\},
		\qquad
		y_i\in\mathcal Y_E^{\varnothing}.
		\label{eq:protocol_resolved_dataset}
	\end{equation}
	The dataset may include both non-null outcomes \(y_i\in\mathcal Y_E\) and
	null outcomes \(y_i=\varnothing\), depending on whether the experiment records
	rejected, undetected, failed, or otherwise nonaccepted observations.
\end{definition}
For a fixed parameter value \(\theta\), the protocol-resolved forward model
defines the null-inclusive observed law
\[
P_{\mathrm{obs},t}^{\varnothing}(\cdot\mid E,\theta)
\]
on \(\mathcal Y_E^{\varnothing}\). When this law admits a density or mass
function with respect to a dominating measure \(\nu_E^{\varnothing}\), we write
\begin{equation}
	p_E^{\varnothing}(y\mid\theta)
	=
	\frac{
		dP_{\mathrm{obs},t}^{\varnothing}(\cdot\mid E,\theta)
	}{
		d\nu_E^{\varnothing}
	}(y),
	\qquad
	y\in\mathcal Y_E^{\varnothing}.
	\label{eq:protocol_resolved_observed_density}
\end{equation}
The dominating measure \(\nu_E^{\varnothing}\) includes the appropriate
continuous, discrete, or mixed structure of the non-null data, together with the
atom at \(\varnothing\).

\begin{definition}[Null-inclusive protocol-resolved likelihood]
	\label{def:protocol_resolved_likelihood}
	If the observations are conditionally independent given \(\theta\), then the
	null-inclusive protocol-resolved likelihood is
	\begin{empheq}[box=\fbox]{equation}
		\mathcal L_E^{\varnothing}(\theta;\mathcal D_E)
		=
		\prod_{i=1}^{N}
		p_E^{\varnothing}(y_i\mid\theta).
		\label{eq:protocol_resolved_likelihood_iid}
	\end{empheq}
	More generally, \(\mathcal L_E^{\varnothing}\) denotes the probability or
	density of the complete dataset under the protocol-resolved observation
	model, including null outcomes whenever they are recorded.
\end{definition}

\begin{remark}[When the independent-observation likelihood is insufficient]
	\label{rem:when_iid_likelihood_insufficient}
	The product likelihood in Eq.~\eqref{eq:protocol_resolved_likelihood_iid} is a
	clean first representation, but it is not universal. Time-correlated tracking
	data, spatially interacting plaques, shared cell-monolayer variability,
	particle aggregation, batch effects, field-induced correlations, and
	reconstruction pipelines can introduce dependence among observations. In such
	cases, the correct likelihood is not a product of single-observation factors,
	but a joint probability model for the entire dataset under the
	protocol-conditioned observation process.
\end{remark}

\begin{remark}[Why null-inclusive likelihoods matter]
	\label{rem:why_null_inclusive_likelihoods_matter}
	If null observations are discarded before modeling, inference may confuse
	selection with absence. For example, a protocol that produces few accepted
	particles, few trackable trajectories, few reconstructed classes, or few
	plaques may do so because the latent population is rare, because the relevant
	states fail the protocol, because the detection threshold is restrictive, or
	because the protocol maps many states into the null channel. A null-inclusive
	likelihood keeps these possibilities within the same probability model.
\end{remark}

\begin{definition}[Conditional non-null likelihood]
	\label{def:conditional_non_null_likelihood}
	If only non-null observations are retained, then the appropriate likelihood is
	conditional on detection. For a detected dataset
	\[
	\mathcal D_E^{\mathrm{det}}
	=
	\{y_1,\ldots,y_{N_{\mathrm{det}}}\},
	\qquad
	y_i\in\mathcal Y_E,
	\]
	the conditional non-null likelihood is
	\begin{empheq}[box=\fbox]{equation}
		\mathcal L_E^{\mathrm{det}}
		(\theta;\mathcal D_E^{\mathrm{det}})
		=
		\prod_{i=1}^{N_{\mathrm{det}}}
		p_E(y_i\mid\theta,\mathrm{det}),
		\label{eq:conditional_non_null_likelihood}
	\end{empheq}
	where
	\begin{equation}
		p_E(y\mid\theta,\mathrm{det})
		=
		\frac{
			p_E^{\varnothing}(y\mid\theta)
		}{
			\eta_E(\theta)
		},
		\qquad
		y\in\mathcal Y_E,
		\label{eq:conditional_non_null_density}
	\end{equation}
	and
	\[
	\eta_E(\theta)
	=
	P_{\mathrm{obs},t}^{\varnothing}
	(\mathcal Y_E\mid E,\theta)
	\]
	is the detection yield.
\end{definition}

\begin{remark}[Conditional likelihoods lose yield information]
	\label{rem:conditional_likelihoods_lose_yield_information}
	The conditional likelihood is useful when the dataset contains only accepted
	records, but it discards information about the detection yield
	\(\eta_E(\theta)\). Therefore it cannot distinguish a protocol that observes
	nearly all states from one that observes only a small selected subset if the
	normalized non-null distributions are similar. Whenever null outcomes or
	denominators are experimentally available, the null-inclusive likelihood is
	more informative.
\end{remark}

\begin{remark}[Counts, denominators, and missing probability mass]
	\label{rem:counts_denominators_missing_probability_mass}
	In many virological assays, the denominator is scientifically meaningful. The
	number of particles rejected during image processing, the number of particles
	lost from a tracking field, the number of wells with no detectable infection,
	or the number of plaque-negative inoculations can all constrain the null
	channel. A reported non-null distribution describes what was observed after
	selection; the denominator helps determine how much of the latent population
	entered that observed distribution in the first place.
\end{remark}

\subsubsection{Bayesian Inverse Formulation}
\label{subsubsec:bayesian_inverse_formulation}

\begin{definition}[Bayesian inverse problem]
	\label{def:bayesian_inverse_problem_experimental_collapse}
	Given a prior density or probability measure \(q_0(\theta)\) and a
	protocol-resolved likelihood
	\(\mathcal L_E^{\varnothing}(\theta;\mathcal D_E)\), the posterior
	distribution is
	\begin{empheq}[box=\fbox]{equation}
		q(\theta\mid\mathcal D_E,E)
		=
		\frac{
			\mathcal L_E^{\varnothing}(\theta;\mathcal D_E)\,
			q_0(\theta)
		}{
			\displaystyle
			\int_{\Theta}
			\mathcal L_E^{\varnothing}(\vartheta;\mathcal D_E)\,
			q_0(\vartheta)\,d\vartheta
		}.
		\label{eq:bayesian_inverse_problem_experimental_collapse}
	\end{empheq}
	When only detected observations are modeled, the corresponding conditional
	posterior is obtained by replacing
	\(\mathcal L_E^{\varnothing}\) with \(\mathcal L_E^{\mathrm{det}}\).
\end{definition}

\begin{remark}[Why \(q_0\) is used for the prior]
	\label{rem:why_q_prior_not_pi}
	The symbol \(q_0\) is used for the prior to avoid confusing the prior density
	with the latent transformation kernels \(\Pi_E^{\mathrm{lat}}\). This is a
	notation choice only; the posterior has the usual Bayesian form.
\end{remark}

\begin{remark}[Interpretation of the Bayesian form]
	\label{rem:interpretation_bayesian_inverse_problem}
	The Bayesian form is natural for protocol-resolved virophysics because the
	unknowns are separated into virion parameters, environmental parameters, and
	protocol parameters. It also makes uncertainty explicit. Instead of producing
	a single fitted value of ``virion stiffness,'' ``mucus mobility,''
	``dielectric response,'' or ``infectious titer,'' the inference returns a
	posterior distribution whose width reflects experimental noise,
	nuisance-parameter uncertainty, finite sample size, protocol selection, and
	non-identifiability.
\end{remark}

\begin{remark}[Frequentist and Bayesian readings]
	\label{rem:frequentist_bayesian_readings}
	The protocol-resolved framework is not tied to Bayesian inference. The same
	forward model can be used for maximum-likelihood estimation, profile
	likelihoods, confidence intervals, likelihood-ratio tests, posterior
	inference, posterior predictive checks, or model comparison. The central
	requirement is not a particular inferential philosophy. It is that the
	likelihood or objective function be written for the protocol-conditioned
	observed data rather than for an unconditioned latent ensemble.
\end{remark}

\subsection{Protocol Identifiability}
\label{subsec:protocol_identifiability}

Experimental collapse makes identifiability explicitly protocol dependent. A
parameter may be real in the latent mechanics but invisible in a particular
readout. Conversely, a strongly conditioning protocol may reveal a latent sector
that a passive protocol cannot observe. Identifiability is therefore a property
not only of the latent model, but of the latent model composed with the protocol
observation operator.

\begin{definition}[Target and nuisance parameters]
	\label{def:target_and_nuisance_parameters}
	Let the parameter vector be partitioned as
	\begin{equation}
		\theta
		=
		(\theta_a,\lambda),
		\label{eq:target_nuisance_parameter_partition}
	\end{equation}
	where \(\theta_a\) is the target parameter or parameter block, and
	\(\lambda\) contains nuisance parameters. Nuisance parameters may include
	environmental variables, protocol settings, calibration constants, selection
	thresholds, noise parameters, batch effects, reconstruction settings, or
	other quantities that are not the main target of inference but still affect
	the observed law.
\end{definition}

\begin{definition}[Protocol identifiability]
	\label{def:protocol_identifiability_experimental_collapse}
	The parameter block \(\theta_a\) is \emph{identifiable under protocol \(E\)},
	relative to a specified model class and nuisance-parameter treatment, if
	distinct values of \(\theta_a\) produce distinguishable null-inclusive
	observed laws after accounting for nuisance parameters.
	
	In the fixed-nuisance case, this means
	\begin{empheq}[box=\fbox]{equation}
		P_{\mathrm{obs},t}^{\varnothing}
		(\cdot\mid E,\theta_a,\lambda)
		=
		P_{\mathrm{obs},t}^{\varnothing}
		(\cdot\mid E,\theta_a',\lambda)
		\quad
		\Longrightarrow
		\quad
		\theta_a=\theta_a'.
		\label{eq:protocol_identifiability_fixed_nuisance}
	\end{empheq}
	In the profiled-nuisance case, \(\theta_a\) is identifiable only if no
	distinct \(\theta_a'\) can be compensated by an admissible nuisance value
	\(\lambda'\):
	\begin{empheq}[box=\fbox]{equation}
		P_{\mathrm{obs},t}^{\varnothing}
		(\cdot\mid E,\theta_a,\lambda)
		=
		P_{\mathrm{obs},t}^{\varnothing}
		(\cdot\mid E,\theta_a',\lambda')
		\quad
		\Longrightarrow
		\quad
		\theta_a=\theta_a'.
		\label{eq:protocol_identifiability_profiled_nuisance}
	\end{empheq}
\end{definition}

\begin{remark}[Why nuisance parameters matter]
	\label{rem:why_nuisance_parameters_matter_identifiability}
	Nuisance parameters are often where protocol collapse enters inference. An
	AFM stiffness estimate may be confounded by tip geometry, indentation depth,
	hydration, or surface adhesion. A DEP response may be confounded by medium
	conductivity, permittivity, viscosity, and field calibration. A
	mucus-tracking diffusivity may be confounded by local mesh heterogeneity,
	antibody binding, or adhesive trapping. A plaque count may be confounded by
	cell-line susceptibility, adsorption time, overlay composition, incubation
	time, staining threshold, and counting rule. If these quantities are ignored,
	a protocol-conditioned parameter can be mistaken for an intrinsic virion
	property.
\end{remark}

\begin{remark}[Structural versus practical identifiability]
	\label{rem:structural_vs_practical_identifiability}
	It is useful to distinguish two notions. \emph{Structural identifiability}
	asks whether a parameter could be recovered in principle from ideal,
	unlimited data under the assumed protocol model. \emph{Practical
		identifiability} asks whether the parameter can be estimated with useful
	precision from finite, noisy data. A latent sector may be structurally
	identifiable but practically weak if the likelihood is nearly flat, the
	Fisher information is small, the relevant observations are rare, the
	detection yield is low, or nuisance parameters are strongly correlated with
	the target parameter
	\cite{BellmanAstrom1970,WalterPronzato1997,Raue2009}.
\end{remark}

\begin{definition}[Protocol equivalence class]
	\label{def:protocol_equivalence_class}
	For a fixed protocol \(E\), define the observational equivalence class of
	\(\theta\) by
	\begin{equation}
		[\theta]_E
		=
		\left\{
		\theta'\in\Theta:
		P_{\mathrm{obs},t}^{\varnothing}(\cdot\mid E,\theta')
		=
		P_{\mathrm{obs},t}^{\varnothing}(\cdot\mid E,\theta)
		\right\}.
		\label{eq:protocol_equivalence_class}
	\end{equation}
	If \([\theta]_E\) contains more than one parameter point, then protocol \(E\)
	cannot distinguish those latent hypotheses, even in the infinite-data limit.
\end{definition}

\begin{remark}[Why identifiability is protocol dependent]
	\label{rem:why_identifiability_protocol_dependent}
	A variable may be mechanically real but invisible under a given protocol. For
	example, a presentation-tilt sector may be present in the latent mechanical
	theory but unidentifiable in a static geometry-only readout. The same sector
	may become partially identifiable under a field-biased protocol, a
	surface-tilt assay, polarization-sensitive imaging, or another method that
	resolves orientation. Experimental collapse therefore explains why absence of
	evidence in one protocol is not necessarily evidence of absence in the latent
	mechanics.
\end{remark}

\begin{remark}[Experimental design consequence]
	\label{rem:experimental_design_consequence_identifiability}
	The identifiability question suggests a design principle: choose protocol
	variations that change the observation kernel in directions that separate the
	target parameter from nuisance parameters. Varying field frequency may help
	separate dielectric response from hydrodynamic drift. Varying surface
	chemistry or indentation depth may help separate stiffness from adhesion.
	Varying mucus composition or antibody concentration may help separate
	transport from binding. Varying cell line, adsorption time, overlay, or
	incubation time may help separate physical particle abundance from
	plaque-forming efficiency. Protocol variation is therefore not only a
	robustness check; it is a way to improve the rank and conditioning of the
	inverse problem.
\end{remark}

The next subsection develops the local version of this identifiability problem
using Fisher information. The global equivalence class
\([\theta]_E\) describes exact non-identifiability of distinct parameter values;
Fisher-information observability describes which infinitesimal directions in
parameter space are locally visible or blind under a specified protocol.
\subsection{Fisher-Information Observability}
\label{subsec:fisher_information_observability}

Identifiability can also be studied locally through Fisher information. The
preceding definitions asked whether two parameter values can produce the same
observed law. Fisher information asks the differential version of the same
question: if the latent, environmental, or protocol parameters are changed
slightly, does the protocol-conditioned likelihood change in a way that the
experiment can detect?

In this sense, Fisher information measures the local sensitivity of the observed
data distribution to parameter perturbations. A protocol has high Fisher
information in a parameter direction if small changes in that direction produce
distinguishable changes in the observed law. It has low Fisher information if
those changes are weak relative to intrinsic noise, selection, projection,
readout variability, or finite-sample limitations. It has zero Fisher
information in a direction if the protocol is locally blind to that direction.

\begin{empheq}[box=\fbox]{equation}
	\text{Fisher information}
	=
	\text{Local sensitivity of the protocol-conditioned likelihood}.
	\label{eq:fisher_information_at_a_glance}
\end{empheq}
For protocol-resolved virophysics, this is especially important because the
likelihood is not a likelihood for the latent virion state alone. It is the
likelihood of the observed state after latent-state transformation, survival or
detection selection, readout, and possible null observation.

\subsubsection{Score functions and protocol information}
\label{subsubsec:score_functions_protocol_information}

Assume that the null-inclusive observed law admits a density or mass function
\[
p_E^{\varnothing}(y\mid\theta),
\qquad
y\in\mathcal Y_E^{\varnothing},
\]
relative to an appropriate dominating measure
\(\nu_E^{\varnothing}\). Here \(\theta\in\Theta\) may include virion parameters,
environmental parameters, and protocol parameters.

\begin{definition}[Protocol score function]
	\label{def:protocol_score_function}
	The \emph{protocol score function} is
	\begin{empheq}[box=\fbox]{equation}
		\mathbf u_E(y;\theta)
		=
		\nabla_{\theta}
		\log p_E^{\varnothing}(y\mid\theta).
		\label{eq:protocol_score_function}
	\end{empheq}
	It measures how sensitively the log-likelihood of an observed outcome \(y\)
	changes under infinitesimal changes in the parameter vector \(\theta\).
\end{definition}
The word ``observed'' includes the null outcome when the likelihood is
null-inclusive. Therefore, changes in \(\theta\) may affect the score by changing
the distribution of non-null data, the probability of entering the null channel,
or both. In a plaque assay, for example, a parameter may change plaque sizes
among visible plaques, the probability that a plaque forms at all, or both. In a
tracking experiment, a parameter may change the distribution of recorded
trajectories, the probability that a particle is trackable, or both.

\begin{remark}[Interpretation of the score]
	\label{rem:interpretation_protocol_score}
	If a parameter perturbation strongly changes the likelihood of the observed
	data, the score has a large component in that direction. If the perturbation
	barely changes the likelihood, the score has a small component. If the
	perturbation does not change the observed law at all, the score vanishes in
	that direction. Thus the score is the local derivative of the
	protocol-conditioned observation model.
\end{remark}

\begin{definition}[Protocol Fisher information]
	\label{def:protocol_fisher_information}
	Assume the usual differentiability and integrability conditions hold. The
	\emph{Fisher-information matrix} of protocol \(E\) is
	\begin{empheq}[box=\fbox]{equation}
		\mathcal I_E(\theta)
		=
		\mathbb E_{Y\sim p_E^{\varnothing}(\cdot\mid\theta)}
		\left[
		\mathbf u_E(Y;\theta)
		\mathbf u_E(Y;\theta)^{\mathsf T}
		\right].
		\label{eq:protocol_fisher_information_inverse}
	\end{empheq}
	Equivalently,
	\begin{equation}
		\mathcal I_E(\theta)
		=
		\int_{\mathcal Y_E^{\varnothing}}
		\nabla_{\theta}\log p_E^{\varnothing}(y\mid\theta)
		\nabla_{\theta}\log p_E^{\varnothing}(y\mid\theta)^{\mathsf T}
		p_E^{\varnothing}(y\mid\theta)\,
		\nu_E^{\varnothing}(dy).
		\label{eq:protocol_fisher_information_expanded}
	\end{equation}
\end{definition}

\begin{remark}[Fisher information as score covariance]
	\label{rem:fisher_information_as_score_covariance}
	Under standard regularity conditions, the expected score is zero:
	\[
	\mathbb E_{\theta}
	\left[
	\nabla_{\theta}\log p_E^{\varnothing}(Y\mid\theta)
	\right]
	=
	0.
	\]
	Thus \(\mathcal I_E(\theta)\) is the covariance matrix of the score. It
	measures how much the score fluctuates under repeated observations from the
	same protocol-conditioned model. Large score fluctuations indicate that the
	observed data are sensitive to the parameter; small score fluctuations
	indicate weak local sensitivity
	\cite{Fisher1922,Kay1993,VanDerVaart1998}.
\end{remark}

\begin{remark}[Alternative curvature form]
	\label{rem:alternative_curvature_form_fisher_information}
	Under the usual regularity conditions, Fisher information can also be written
	as the negative expected Hessian of the log-likelihood:
	\begin{equation}
		\mathcal I_E(\theta)
		=
		-
		\mathbb E_{Y\sim p_E^{\varnothing}(\cdot\mid\theta)}
		\left[
		\nabla_{\theta}^{2}
		\log p_E^{\varnothing}(Y\mid\theta)
		\right].
		\label{eq:fisher_information_negative_hessian}
	\end{equation}
	This form gives the curvature interpretation. If the likelihood is sharply
	curved near the true parameter, the parameter is locally well constrained. If
	the likelihood is flat in some direction, that direction is weakly
	identifiable or blind under the protocol.
\end{remark}

\subsubsection{Directional information and local blindness}
\label{subsubsec:directional_information_local_blindness}

The full Fisher matrix summarizes information in all parameter directions. To
ask whether a particular latent sector is visible, one evaluates the Fisher
information in a chosen tangent direction.

\begin{definition}[Directional Fisher information]
	\label{def:directional_fisher_information}
	Let \(v\in T_{\theta}\Theta\) be a parameter direction. The
	\emph{directional Fisher information} of protocol \(E\) in direction \(v\) is
	\begin{empheq}[box=\fbox]{equation}
		\mathcal I_E(\theta;v)
		=
		v^{\mathsf T}
		\mathcal I_E(\theta)
		v.
		\label{eq:directional_fisher_information}
	\end{empheq}
	Equivalently,
	\begin{equation}
		\mathcal I_E(\theta;v)
		=
		\mathbb E_{\theta}
		\left[
		\left(
		v^{\mathsf T}
		\nabla_{\theta}\log
		p_E^{\varnothing}(Y\mid\theta)
		\right)^2
		\right].
		\label{eq:directional_fisher_information_score_form}
	\end{equation}
\end{definition}

\begin{remark}[How to read directional Fisher information]
	\label{rem:how_to_read_directional_fisher_information}
	The quantity \(v^{\mathsf T}\mathcal I_E(\theta)v\) measures how strongly the
	observed law changes when the parameter vector is perturbed in direction
	\(v\). For example, \(v\) could represent a change in virion stiffness,
	polarizability, spike-presentation heterogeneity, mucus-binding strength,
	cell-entry competence, adsorption probability, or a hidden orientation-sector
	parameter. If the directional information is large, the protocol is sensitive
	to that perturbation. If it is small, the protocol weakly resolves it.
\end{remark}

\begin{definition}[Locally blind parameter direction]
	\label{def:locally_blind_parameter_direction}
	A tangent direction \(v\in T_{\theta}\Theta\) is \emph{locally blind} under
	protocol \(E\) if
	\begin{empheq}[box=\fbox]{equation}
		v^{\mathsf T}
		\mathcal I_E(\theta)
		v
		=
		0.
		\label{eq:locally_blind_parameter_direction}
	\end{empheq}
	It is \emph{weakly observable} if this quantity is positive but small relative
	to the noise level, sample size, parameter scale, or experimental tolerance.
\end{definition}

\begin{definition}[Local protocol-blind subspace]
	\label{def:local_protocol_blind_subspace}
	The local protocol-blind subspace is
	\begin{empheq}[box=\fbox]{equation}
		\mathcal B_E(\theta)
		=
		\ker \mathcal I_E(\theta)
		=
		\left\{
		v\in T_{\theta}\Theta:
		v^{\mathsf T}\mathcal I_E(\theta)v=0
		\right\}.
		\label{eq:local_protocol_blind_subspace_fisher}
	\end{empheq}
\end{definition}

\begin{remark}[Protocol blindness as a null space]
	\label{rem:protocol_blindness_as_null_space}
	The null space of \(\mathcal I_E(\theta)\) is the local protocol-blind sector.
	Perturbations of \(\theta\) in this sector do not change the observed law to
	first order. In the language of experimental collapse, these are latent
	directions removed, averaged over, nulled, or compressed by the protocol
	observation operator.
\end{remark}

\begin{remark}[Why ``local'' matters]
	\label{rem:why_local_blindness_matters}
	Fisher information is a local object. It describes the behavior of the
	likelihood near a specified parameter value \(\theta\). A direction may be
	blind near one parameter value but visible elsewhere, especially in nonlinear
	models. Thus Fisher-information observability should be interpreted as local
	identifiability, not as a universal statement about the entire parameter
	space.
\end{remark}

\subsubsection{Connection to estimation uncertainty}
\label{subsubsec:fisher_information_estimation_uncertainty}

Fisher information is useful because it connects protocol sensitivity to
estimation uncertainty. In regular parametric models, the Cramér--Rao inequality
states that the covariance of any unbiased estimator is bounded below by the
inverse Fisher information.

\begin{definition}[Cramér--Rao interpretation]
	\label{def:cramer_rao_interpretation}
	For an unbiased estimator \(\widehat\theta\) of \(\theta\), the
	Cramér--Rao bound has the schematic matrix form
	\begin{equation}
		\operatorname{Cov}(\widehat\theta)
		\succeq
		\mathcal I_E(\theta)^{-1},
		\label{eq:cramer_rao_bound_protocol}
	\end{equation}
	when \(\mathcal I_E(\theta)\) is invertible and the regularity conditions for
	the bound hold.
\end{definition}

\begin{remark}[Protocol interpretation of the Cramér--Rao bound]
	\label{rem:protocol_interpretation_cramer_rao}
	Equation~\eqref{eq:cramer_rao_bound_protocol} says that a protocol with more
	Fisher information can, in principle, support more precise parameter
	estimates. If \(\mathcal I_E(\theta)\) has a small eigenvalue, then
	uncertainty is large along the corresponding parameter direction. If
	\(\mathcal I_E(\theta)\) is singular, then at least one direction is locally
	unidentifiable from that protocol alone
	\cite{Rao1945,Cramer1946,Kay1993,VanDerVaart1998}.
\end{remark}
\noindent When the Fisher matrix is singular or nearly singular, the inverse should be
interpreted cautiously. A Moore--Penrose pseudoinverse may describe uncertainty
on the identifiable subspace, but it does not make blind directions identifiable.
Those directions remain unconstrained by the protocol-conditioned likelihood
unless additional structure, regularization, prior information, or protocol
variation is introduced.

For \(N\) independent observations from the same protocol, the Fisher information
adds:
\begin{equation}
	\mathcal I_{E,N}(\theta)
	=
	N\mathcal I_E(\theta).
	\label{eq:fisher_information_sample_size_scaling}
\end{equation}
Thus additional data improve precision in directions that the protocol already
sees. However, more data from the same protocol do not remove a structural blind
direction.

\begin{remark}[More data versus new protocol]
	\label{rem:more_data_versus_new_protocol}
	If \(v^{\mathsf T}\mathcal I_E(\theta)v=0\), then
	\[
	v^{\mathsf T}\mathcal I_{E,N}(\theta)v=0
	\]
	for every sample size \(N\). More observations from the same protocol do not
	make a blind direction visible. To resolve that direction, one must change the
	observation kernel, add a complementary protocol, impose additional prior
	structure, or introduce controlled protocol variation.
\end{remark}

\subsubsection{Nuisance parameters and effective Fisher information}
\label{subsubsec:nuisance_parameters_effective_fisher_information}

In protocol-resolved virophysics, the target parameter is often entangled with
environmental or protocol parameters. This is why Fisher information should be
interpreted after accounting for nuisance directions. Partition the parameter vector as
\[
\theta
=
(\theta_a,\lambda),
\]
where \(\theta_a\) is the target parameter block and \(\lambda\) contains
nuisance parameters. Write the Fisher matrix in block form:
\begin{equation}
	\mathcal I_E(\theta)
	=
	\begin{pmatrix}
		\mathcal I_{aa} & \mathcal I_{a\lambda} \\
		\mathcal I_{\lambda a} & \mathcal I_{\lambda\lambda}
	\end{pmatrix}.
	\label{eq:fisher_block_partition_protocol}
\end{equation}

\begin{definition}[Effective Fisher information after nuisance profiling]
	\label{def:effective_fisher_information_after_nuisance}
	When \(\mathcal I_{\lambda\lambda}\) is invertible, the effective Fisher
	information for \(\theta_a\), after accounting for nuisance parameters, is
	the Schur complement
	\begin{empheq}[box=\fbox]{equation}
		\mathcal I_{a\mid\lambda}
		=
		\mathcal I_{aa}
		-
		\mathcal I_{a\lambda}
		\mathcal I_{\lambda\lambda}^{-1}
		\mathcal I_{\lambda a}.
		\label{eq:effective_fisher_information_after_nuisance}
	\end{empheq}
\end{definition}

\begin{remark}[Interpretation of nuisance-corrected information]
	\label{rem:interpretation_nuisance_corrected_fisher}
	The matrix \(\mathcal I_{a\mid\lambda}\) measures how much information about
	the target parameter remains after nuisance parameters are allowed to adjust.
	If \(\mathcal I_{a\mid\lambda}\) is small, then the protocol may appear
	sensitive to \(\theta_a\) when nuisance parameters are fixed, but weakly
	informative once realistic uncertainty in protocol or environmental
	parameters is included. This is common in virological settings: apparent
	stiffness can be confounded with surface adhesion or indentation depth, field
	response with medium conductivity and viscosity, mucus transport with binding
	heterogeneity, and plaque count with cell-line or overlay conditions.
\end{remark}

\begin{remark}[When the nuisance block is singular]
	\label{rem:singular_nuisance_block_fisher}
	If \(\mathcal I_{\lambda\lambda}\) is singular, then some nuisance directions
	are themselves locally unidentifiable under the protocol. In that case, the
	Schur complement must be replaced by an appropriate generalized inverse,
	reparameterization, regularized information matrix, or explicit prior model.
	The singularity is not merely a technical inconvenience; it indicates that the
	experiment may not separate target and nuisance effects without additional
	protocol variation or external calibration.
\end{remark}

This completes the local observability analysis for a single protocol. Fisher
information identifies which infinitesimal parameter directions are visible,
weakly resolved, nuisance-confounded, or locally blind under the
protocol-conditioned likelihood. The next step is to ask how this structure
changes when additional protocols are introduced.

\subsubsection{Information gain from additional protocols}
\label{subsubsec:information_gain_additional_protocols}

Fisher-information observability gives a direct experimental-design principle:
a new protocol is most useful when it adds information in parameter directions
that are blind, weakly resolved, or nuisance-confounded under the existing
protocol set. In protocol-resolved virophysics, the goal is therefore not merely
to repeat an observation, but to change the observation kernel in a way that
probes the latent virion--environment system from a new direction.

\begin{definition}[Information gain from an additional protocol]
	\label{def:information_gain_additional_protocol}
	Let \(E_1\) be an existing protocol and \(E_2\) a proposed additional
	protocol. For a parameter direction \(v\in T_{\theta}\Theta\), the
	directional information gain supplied by \(E_2\), relative to \(E_1\), is
	\begin{equation}
		\Delta\mathcal I_{E_2\mid E_1}(\theta;v)
		=
		v^{\mathsf T}
		\mathcal I_{E_2}(\theta)
		v.
		\label{eq:directional_information_gain_additional_protocol}
	\end{equation}
	If the datasets generated by \(E_1\) and \(E_2\) are conditionally independent
	given \(\theta\), then the combined information is
	\begin{empheq}[box=\fbox]{equation}
		\mathcal I_{E_1,E_2}(\theta)
		=
		\mathcal I_{E_1}(\theta)
		+
		\mathcal I_{E_2}(\theta).
		\label{eq:combined_two_protocol_information}
	\end{empheq}
	More generally, for conditionally independent protocols
	\(E_1,\ldots,E_M\),
	\begin{equation}
		\mathcal I_{\mathrm{multi}}(\theta)
		=
		\sum_{j=1}^{M}
		\mathcal I_{E_j}(\theta).
		\label{eq:multi_protocol_fisher_information_expanded}
	\end{equation}
\end{definition}
Equation~\eqref{eq:directional_information_gain_additional_protocol} is most
important when \(v\) lies in, or near, the blind subspace of the existing
protocol. If
\[
v\in\mathcal B_{E_1}(\theta)
=
\ker\mathcal I_{E_1}(\theta),
\]
then \(E_1\) supplies no local information in direction \(v\). The proposed
protocol \(E_2\) reduces that blindness precisely when
\[
v^{\mathsf T}\mathcal I_{E_2}(\theta)v>0.
\]
Thus the value of an additional protocol is not determined only by how much total
information it adds, but by whether it adds information in the directions that
are missing from the current experimental design.

\begin{definition}[Blind-subspace information gain]
	\label{def:blind_subspace_information_gain}
	Let \(\mathcal B_{E_1}(\theta)=\ker\mathcal I_{E_1}(\theta)\) be the local
	blind subspace of protocol \(E_1\). The blind-subspace information gain of an
	additional protocol \(E_2\) may be summarized by the restriction
	\begin{equation}
		\mathcal I_{E_2}^{\mathcal B}(\theta)
		=
		\left.
		\mathcal I_{E_2}(\theta)
		\right|_{\mathcal B_{E_1}(\theta)}.
		\label{eq:blind_subspace_information_gain}
	\end{equation}
	If this restricted form is nonzero, then \(E_2\) probes at least one direction
	that was locally blind under \(E_1\).
\end{definition}

\begin{remark}[Experimental design interpretation]
	\label{rem:experimental_design_interpretation}
	Equation~\eqref{eq:multi_protocol_fisher_information_expanded} gives an
	experimental design principle. A new protocol is most valuable when it adds
	information in directions that are blind or weakly constrained under the
	existing protocol. Thus the best complementary experiment is not necessarily
	the one that most closely reproduces the first experiment. It is often the one
	whose observation kernel projects, perturbs, selects, or amplifies the latent
	mechanics differently.
\end{remark}

\begin{remark}[Examples of complementary Fisher information]
	\label{rem:examples_complementary_fisher_information}
	A cryo-EM or cryo-ET protocol may provide strong information about geometry,
	symmetry, morphology, and selected conformational classes, but weak
	information about free dynamics, compliance, or infectious activity. AFM may
	add information about surface-conditioned stiffness, deformation, rupture, or
	adhesion. Dielectrophoresis and electrorotation may add information about
	dielectric response, polarizability, charge asymmetry, hydrodynamic drag, and
	frequency-dependent field coupling. Mucus or gel tracking may add information
	about adhesion, residence time, confinement, mobile and immobile fractions, and
	transport heterogeneity. A plaque or focus assay may add information about
	assay-conditioned infectious competence. Each protocol adds Fisher information
	in the directions to which its observation kernel is sensitive.
\end{remark}

\begin{remark}[When Fisher information does not simply add]
	\label{rem:when_fisher_information_does_not_simply_add}
	The additive formula in
	Eq.~\eqref{eq:multi_protocol_fisher_information_expanded} assumes conditional
	independence of the datasets given the parameters. If protocols share batch
	effects, preparation artifacts, calibration errors, cell-culture variability,
	environmental drift, or correlated sample history, then the joint likelihood
	must include those dependencies explicitly. In that case, the multi-protocol
	Fisher information is obtained from the joint likelihood, not by naively
	summing single-protocol matrices.
\end{remark}

\subsection{Multi-Protocol Consistency}
\label{subsec:multi_protocol_consistency}

Multi-protocol comparison is often described as checking whether different
experiments agree. In protocol-resolved virophysics, the more precise question
is whether different observed ensembles can be explained by a common latent
model passed through different protocol kernels. Agreement is therefore not
required at the level of raw observables. A density map, an indentation curve, a
field-response spectrum, a trajectory ensemble, and a plaque count need not look
alike, because they are not the same kind of object. The relevant question is
whether their differences are compatible with a shared latent
virion--environment model plus protocol-specific transformations, selections,
readouts, and null channels.

\begin{definition}[Multi-protocol consistency]
	\label{def:multi_protocol_consistency_experimental_collapse}
	Let \(E_1,\ldots,E_M\) be experimental protocols with datasets
	\[
	\mathcal D_1,\ldots,\mathcal D_M.
	\]
	A latent model \(P_{\mathrm{ref},t}(\cdot\mid\theta)\) is
	\emph{multi-protocol consistent} if there exists a common latent parameter
	set \(\theta\), together with admissible protocol-specific nuisance parameters
	\(\lambda_1,\ldots,\lambda_M\), such that
	\begin{equation}
		P_{\mathrm{obs},t}^{(j),\varnothing}
		(\cdot\mid E_j,\theta,\lambda_j)
		=
		\mathcal M_{E_j}^{\varnothing}(\lambda_j)
		P_{\mathrm{ref},t}(\cdot\mid\theta),
		\qquad
		j=1,\ldots,M,
		\label{eq:multi_protocol_consistency_model}
	\end{equation}
	adequately explains the observed data for all protocols.
\end{definition}
Here \(\theta\) represents the shared latent virion--environment parameters, while
\(\lambda_j\) represents protocol-local quantities such as calibration constants,
surface chemistry, imaging thresholds, cell-line conditions, field settings,
overlay composition, reconstruction choices, or counting rules. The distinction
between \(\theta\) and \(\lambda_j\) is important: multi-protocol consistency
does not require all protocols to have the same nuisance parameters. It requires
that their protocol-specific differences be compatible with a shared latent
explanation.

Equivalently, when the datasets are conditionally independent given the shared
latent parameters and protocol-specific nuisance parameters, the joint
multi-protocol likelihood is
\begin{empheq}[box=\fbox]{equation}
	\mathcal L_{\mathrm{multi}}
	(\theta,\lambda_1,\ldots,\lambda_M)
	=
	\prod_{j=1}^{M}
	\mathcal L_{E_j}^{\varnothing}
	(\theta,\lambda_j;\mathcal D_j).
	\label{eq:multi_protocol_likelihood}
\end{empheq}
When only detected non-null observations are available for some protocols, the
corresponding conditional likelihoods
\(\mathcal L_{E_j}^{\mathrm{det}}\) may be used, with the understanding that
yield information has then been discarded or must be modeled separately.

\begin{remark}[Why multi-protocol comparison is powerful]
	\label{rem:why_multi_protocol_comparison_powerful}
	Experimental collapse makes multi-protocol comparison more valuable, not less.
	If cryo-ET, AFM, dielectrophoresis, plaque assays, and mucus tracking are
	interpreted as different kernels acting on related latent mechanics, then
	disagreement between their observations can become informative. It may
	indicate that one protocol pins orientation, another selects geometry, another
	probes dielectric response, another amplifies infectivity, and another reports
	medium-filtered transport. The goal is not to force all protocols to agree at
	the observation level. The goal is to ask whether they can be explained by a
	coherent latent model plus distinct protocol kernels.
\end{remark}

\begin{definition}[Protocol inconsistency]
	\label{def:protocol_inconsistency}
	A set of protocols \(E_1,\ldots,E_M\) is
	\emph{protocol-inconsistent} relative to a latent model class \(\mathcal P\)
	if no admissible
	\(P_{\mathrm{ref},t}(\cdot\mid\theta)\in\mathcal P\) and no admissible
	protocol-specific nuisance parameters
	\(\lambda_1,\ldots,\lambda_M\) can jointly account for the observed datasets
	under the specified protocol kernels.
\end{definition}

\begin{remark}[Interpretation of inconsistency]
	\label{rem:interpretation_of_protocol_inconsistency}
	Protocol inconsistency is not automatically experimental failure. It may
	indicate that the latent model is missing a relevant state variable, that a
	protocol kernel is misspecified, that an environmental variable was
	uncontrolled, that the virion population changed between experiments, or that
	a biological assay is selecting a functional subpopulation not represented in
	the mechanical model. The inconsistency is therefore diagnostic. It tells the
	modeler where the assumed latent state, environmental description, or protocol
	kernel may be incomplete.
\end{remark}

\begin{remark}[Agreement can also be misleading]
	\label{rem:agreement_can_be_misleading}
	Just as disagreement is not automatically failure, agreement is not
	automatically full validation. Two protocols can agree because they share the
	same blind direction, select the same subpopulation, or collapse different
	latent states into the same reported summary. Multi-protocol consistency is
	strongest when the protocols are complementary, not merely redundant.
\end{remark}

\begin{remark}[Section-level interpretation]
	\label{rem:fisher_information_section_interpretation}
	Fisher-information observability gives a quantitative language for
	experimental collapse. It asks not only whether a protocol produces data, but
	which latent directions those data can constrain. In this framework, a protocol
	is not simply ``informative'' or ``uninformative.'' It is informative in
	particular parameter directions, weak in others, and blind in others.
	Multi-protocol design is the process of choosing complementary kernels whose
	information directions jointly resolve the latent virion--environment model.
\end{remark}

\subsection{Inverse Environmental Inference}
\label{subsec:inverse_environmental_inference}

In many virological settings, the environment is not merely background. Mucus,
extracellular matrix, aerosols, droplets, buffers, ionic conditions, surfaces,
overlays, and cellular monolayers can determine which virion states are
accessible, mobile, adhesive, mechanically responsive, infectious, detectable,
or countable. The protocol-resolved formalism therefore supports not only
inference about virions, but also inverse inference about the environments
through which virions move, bind, deform, polarize, or amplify.

This point is important because many virological observables are naturally joint
virion--environment observables. A trajectory in mucus reflects both the virion
and the mucus network. A dielectrophoretic response reflects both the particle
and the electrical properties of the suspending medium. An AFM indentation curve
reflects both particle mechanics and surface-contact conditions. A plaque assay
reflects both virion infectious competence and the susceptibility, spatial
restriction, and amplification properties of the cellular assay environment.
Thus the inverse problem is often not
\[
\text{observed data}
\longrightarrow
\text{virion parameters alone},
\]
but rather
\begin{empheq}[box=\fbox]{equation}
	\text{observed protocol-conditioned data}
	\;\longrightarrow\;
	\left\{
	\begin{array}{c}
		\text{virion parameters} \\
		\text{environmental parameters} \\
		\text{protocol parameters}
	\end{array}
	\right\}.
	\label{eq:inverse_environmental_inference_parameter_blocks}
\end{empheq}

\begin{definition}[Joint virion--environment latent state]
	\label{def:inverse_joint_virion_environment_latent_state}
	A joint virion--environment latent state has the schematic form
	\begin{equation}
		X
		=
		\left(
		X_{\mathrm{vir}},
		X_{\mathrm{env}}
		\right),
		\label{eq:inverse_joint_virion_environment_latent_state}
	\end{equation}
	where \(X_{\mathrm{vir}}\) contains virion-level variables and
	\(X_{\mathrm{env}}\) contains local environmental variables. Examples include
	viscosity, viscoelastic relaxation time, mesh size, mucin composition,
	binding-site density, antibody concentration, ionic strength, dielectric
	permittivity, conductivity, flow field, surface chemistry, receptor density,
	cell-layer susceptibility, pH, and temperature.
\end{definition}

\begin{remark}[Environment as a latent variable]
	\label{rem:environment_as_latent_variable}
	The environmental component \(X_{\mathrm{env}}\) may vary from particle to
	particle and from location to location. In mucus or extracellular matrix, two
	virions in the same nominal sample may experience different local mesh sizes,
	adhesive states, or viscoelastic environments. In a plaque assay, different
	regions of the monolayer may differ in cell density, susceptibility, or local
	spread conditions. In a field-driven experiment, the effective medium
	parameters may depend on ionic strength, conductivity, temperature, and field
	frequency. Thus the environment should often be modeled as part of the latent
	state rather than as a fixed background constant.
\end{remark}

\begin{definition}[Environmental parameter block]
	\label{def:environmental_parameter_block}
	Let
	\[
	\theta
	=
	\left(
	\theta_{\mathrm{vir}},
	\theta_{\mathrm{env}},
	\theta_E
	\right)
	\]
	be the protocol-resolved parameter vector. The environmental parameter block
	\(\theta_{\mathrm{env}}\) contains parameters governing the distribution or
	dynamics of \(X_{\mathrm{env}}\). Examples include mucus mesh scale, adhesive
	binding rate, effective viscosity, viscoelastic relaxation time, dielectric
	permittivity, conductivity, surface-binding strength, receptor density,
	cell-layer susceptibility, local spread efficiency, antibody concentration,
	ionic strength, pH, and temperature.
\end{definition}

\begin{remark}[Virions as passive probes]
	\label{rem:virions_as_passive_probes_inverse}
	Because extracellular virions are passively responsive physical objects, their
	observed motion, survival, adhesion, deformation, or field response can encode
	information about the environment. A particle trajectory in mucus may contain
	information about confinement, adhesion, local viscoelastic structure, or
	mobile and immobile subpopulations
	\cite{Boukari2009,Wang2017,Kaler2022,Abrami2024}. A dielectrophoretic or
	electrorotational response may contain information about dielectric contrast,
	medium conductivity, hydrodynamic drag, and field-frequency-dependent response
	\cite{Hughes1998,Hughes2002,Pethig2010,Kim2019}. A plaque assay may contain
	information not only about virion infectivity but also about cell
	susceptibility, overlay restrictions, local spread, incubation time, and
	visibility threshold
	\cite{Dulbecco1952,Cooper1961,Baer2014}.
\end{remark}

\begin{definition}[Environmental inverse problem]
	\label{def:environmental_inverse_problem}
	Given protocol-conditioned data \(\mathcal D_E\), the environmental inverse
	problem is to infer \(\theta_{\mathrm{env}}\), possibly jointly with
	\(\theta_{\mathrm{vir}}\) and \(\theta_E\), from the protocol-resolved
	likelihood
	\begin{equation}
		\mathcal L_E^{\varnothing}
		\left(
		\theta_{\mathrm{vir}},
		\theta_{\mathrm{env}},
		\theta_E;
		\mathcal D_E
		\right).
		\label{eq:environmental_inverse_likelihood}
	\end{equation}
	In Bayesian form, the marginal posterior for the environmental parameters is
	\begin{empheq}[box=\fbox]{equation}
		q(\theta_{\mathrm{env}}\mid \mathcal D_E,E)
		=
		\int
		q(
		\theta_{\mathrm{vir}},
		\theta_{\mathrm{env}},
		\theta_E
		\mid
		\mathcal D_E,E
		)
		\,
		d\theta_{\mathrm{vir}}\,
		d\theta_E.
		\label{eq:environmental_marginal_posterior}
	\end{empheq}
\end{definition}

\begin{remark}[Virion--environment confounding]
	\label{rem:virion_environment_confounding}
	Environmental inference is useful but can be confounded. A slow trajectory in
	mucus may reflect a strongly adhesive virion, a dense local mucus mesh,
	antibody cross-linking, low local water content, or a combination of these
	effects. A weak plaque count may reflect low infectious competence, poor cell
	susceptibility, restrictive overlay conditions, insufficient incubation time,
	or a stringent staining threshold. A weak field response may reflect particle
	polarizability, medium conductivity, viscosity, or hydrodynamic drag.
	Environmental inference is therefore strongest when protocol variation or
	complementary measurements separate virion parameters from environmental
	parameters.
\end{remark}

\begin{definition}[Environmental identifiability under a protocol]
	\label{def:environmental_identifiability_under_protocol}
	An environmental parameter block \(\theta_{\mathrm{env}}\) is identifiable
	under protocol \(E\), relative to a chosen model class and nuisance-parameter
	treatment, if distinct values of \(\theta_{\mathrm{env}}\) produce
	distinguishable null-inclusive observed laws after accounting for virion and
	protocol nuisance parameters. In the profiled-nuisance case, this requires
	\begin{empheq}[box=\fbox]{equation}
		P_{\mathrm{obs},t}^{\varnothing}
		(
		\cdot
		\mid
		E,
		\theta_{\mathrm{env}},
		\lambda
		)
		=
		P_{\mathrm{obs},t}^{\varnothing}
		(
		\cdot
		\mid
		E,
		\theta_{\mathrm{env}}',
		\lambda'
		)
		\quad
		\Longrightarrow
		\quad
		\theta_{\mathrm{env}}=\theta_{\mathrm{env}}',
		\label{eq:environmental_identifiability_condition}
	\end{empheq}
	where \(\lambda,\lambda'\) contain virion and protocol nuisance parameters.
\end{definition}

\begin{remark}[Why inverse environmental inference belongs in virophysics]
	\label{rem:why_inverse_environmental_inference_belongs}
	Inverse environmental inference is not separate from virophysics. Virions do
	not move, bind, polarize, deform, or infect in an abstract vacuum. They do so
	in media, on surfaces, in droplets, in gels, in mucus, on cell layers, and
	under specific experimental constraints. A protocol-resolved theory should
	therefore allow the experiment to teach us about the environment as well as
	about the virion.
\end{remark}

\begin{remark}[Design implication for environmental inference]
	\label{rem:design_implication_environmental_inference}
	Environmental parameters are most identifiable when the experimental design
	changes the environment in controlled ways while also monitoring virion
	response. Examples include varying mucus composition or antibody concentration
	in tracking assays, varying ionic strength or conductivity in DEP and
	electrorotation, varying surface chemistry in AFM, or varying cell line,
	overlay, and incubation time in plaque assays. Such variations change the
	protocol-conditioned observation kernel and can separate environmental effects
	from intrinsic virion properties.
\end{remark}

The worked-example template below translates these ideas into a practical
comparison procedure. Rather than asking whether two protocols produce identical
raw outputs, it asks whether their protocol-conditioned observations can be
mapped to a common comparison target and interpreted as complementary views of a
shared latent virion--environment system.
\newpage
\subsection{Worked Example Template: Comparing Protocols}
\label{subsec:worked_example_template_comparing_protocols}

The formalism can be applied to a pair or family of protocols by following a
standard sequence. The purpose of the template is to prevent protocol comparison
from becoming a direct comparison of raw observables that may not live in the
same state space. Instead, each observed ensemble is interpreted through its own
protocol kernel and then compared through a shared latent model, a common
comparison space, or a specified inference target.

\medskip
\noindent The guiding question is therefore not simply
\[
\text{Do the experiments give the same result?}
\]
but rather
\begin{empheq}[box=\fbox]{equation}
	\text{Can one latent virion--environment model explain all
		protocol-conditioned observations?}
	\label{eq:worked_template_guiding_question}
\end{empheq}
This framing is essential in virophysics because different protocols often
probe different physical, mechanical, electrical, transport, or biological
sectors of the same underlying system.

\begin{enumerate}[label=(\roman*),leftmargin=2.2em]
	
	\item \textbf{Choose the latent model.}
	Specify the latent state \(X\), the latent state space
	\((\Psi,\Sigma_X)\), the reference ensemble
	\(P_{\mathrm{ref},t}(dx\mid\theta)\), and the parameter blocks of interest:
	\[
	\theta
	=
	\left(
	\theta_{\mathrm{vir}},
	\theta_{\mathrm{env}},
	\theta_E
	\right).
	\]
	Here \(\theta_{\mathrm{vir}}\) may describe virion mechanics, structure,
	charge, spike presentation, polarizability, adhesion, or infectivity;
	\(\theta_{\mathrm{env}}\) may describe medium, surface, field, mucus,
	overlay, or cell-layer properties; and \(\theta_E\) describes
	protocol-specific settings and nuisance parameters.
	
	\item \textbf{Specify each protocol.}
	For every protocol \(E_j\), identify the protocol-conditioned latent space
	\((\Psi_{E_j},\Sigma_{X,E_j})\), the latent transformation kernel
	\(\Pi_{E_j}^{\mathrm{lat}}\), the survival or detection weight
	\(s_{E_j}\), the readout kernel \(R_{E_j}\), and the null channel
	\(\varnothing\). This step states how the protocol transforms, selects,
	projects, amplifies, or rejects latent states before they become reported
	data.
	
	\item \textbf{Define observed quantities.}
	Specify the non-null observation spaces
	\[
	(\mathcal Y_{E_j},\Sigma_{Y,E_j})
	\]
	and the reported observables
	\[
	\mathbf h_{E_j}:\mathcal Y_{E_j}\rightarrow\mathbb R^{m_j}.
	\]
	Examples include density-map summaries, force-curve parameters, trajectory
	statistics, field-response amplitudes, plaque counts, focus counts, endpoint
	titers, orientation classes, or inferred stiffness and mobility parameters.
	
	\item \textbf{Map to a common comparison space when needed.}
	If the protocols report different kinds of data, introduce measurable maps
	\[
	H_{E_j}:\mathcal Y_{E_j}\rightarrow\mathsf W
	\]
	into a common comparison space \((\mathsf W,\Sigma_W)\). This step states
	exactly what is being compared: size, stiffness, mobility, infectivity,
	orientation, dielectric response, environmental transport parameters,
	mechanical compliance, plaque-forming efficiency, or another shared summary.
	
	\item \textbf{Compute collapse functionals.}
	Evaluate distributional, observable-level, yield-level, modal, dynamical, or
	Fisher-information collapse measures. Examples include
	\[
	\mathcal C_{\mathrm{lat}}^{\mathsf W}(E_j;t),
	\qquad
	\mathcal C_{\mathrm{obs}}^{\mathsf W}(E_i,E_j;t),
	\qquad
	\mathcal C_{\eta}(E_i;E_j,t),
	\qquad
	\mathcal I_{E_j}(\theta).
	\]
	These quantities determine whether the relevant difference is mainly latent
	transformation, detection selection, readout compression, dynamical-sector
	loss, nuisance confounding, or protocol-to-protocol discrepancy.
	
	\item \textbf{Test multi-protocol consistency.}
	Ask whether a common latent parameter set \(\theta\), together with
	protocol-specific nuisance parameters \(\lambda_1,\ldots,\lambda_M\), can
	explain all observed datasets:
	\begin{equation}
		P_{\mathrm{obs},t}^{(j),\varnothing}
		(\cdot\mid E_j,\theta,\lambda_j)
		=
		\mathcal M_{E_j}^{\varnothing}(\lambda_j)
		P_{\mathrm{ref},t}(\cdot\mid\theta),
		\qquad
		j=1,\ldots,M.
		\label{eq:worked_template_multi_protocol_consistency}
	\end{equation}
	This step asks whether apparent differences between protocols can be explained
	as distinct protocol-conditioned views of the same latent
	virion--environment system.
	
	\item \textbf{Identify blind or weakly resolved sectors.}
	Use sensitivity matrices, Fisher information, yield differences,
	protocol-to-protocol discrepancies, or posterior uncertainty to identify
	latent directions that are invisible, weakly constrained, or confounded with
	nuisance parameters. In Fisher-information language, this means examining
	\[
	\mathcal B_{E_j}(\theta)
	=
	\ker \mathcal I_{E_j}(\theta),
	\qquad
	\mathcal B_{\mathrm{multi}}(\theta)
	=
	\ker \mathcal I_{\mathrm{multi}}(\theta).
	\]
	A useful complementary protocol is one that reduces the blind subspace or
	improves the conditioning of the inverse problem.
	
	\item \textbf{Revise the model or protocol kernels if consistency fails.}
	Inconsistency may signal a missing latent variable, an incorrect protocol
	kernel, an unmodeled environmental effect, uncontrolled nuisance variation,
	population heterogeneity, batch dependence, or genuine biological differences
	between experimental conditions. The inconsistency is therefore diagnostic,
	not merely negative.
\end{enumerate}

\begin{remark}[How to read protocol disagreement]
	\label{rem:how_to_read_protocol_disagreement}
	If two protocols disagree at the level of raw observed data, the disagreement
	is not automatically an error. It may indicate that the protocols project
	different latent sectors, apply different selection weights, impose different
	mechanical or biological conditions, or amplify different functional
	subpopulations. The template above asks whether such disagreement can be
	explained by a shared latent model plus distinct protocol kernels.
\end{remark}

\newpage 
\subsection{Section Reference: Core Objects and Equations}
\label{subsec:section_reference_core_objects_equations}
\addcontentsline{toc}{subsection}{Section Reference: Core Objects and Equations}

\begingroup
\footnotesize
\setlength{\tabcolsep}{4pt}
\renewcommand{\arraystretch}{1.18}

\begin{xltabular}{0.98\linewidth}{
		@{}
		B{0.20\linewidth}
		L{0.43\linewidth}
		Y
		@{}
	}
	\caption{
		Core objects, equations, and interpretations introduced in the
		protocol-resolved ensemble and inference framework. This table is a
		compact reference for the notation used in later sections.
	}
	\label{tab:section_reference_core_objects_equations}
	\\
	\toprule
	\textbf{Object or equation}
	&
	\multicolumn{1}{c}{\textbf{Mathematical form}}
	&
	\textbf{Interpretation}
	\\
	\midrule
	\endfirsthead
	
	\caption[]{
		Core objects, equations, and interpretations introduced in the
		protocol-resolved ensemble and inference framework.
		\emph{Continued from previous page.}
	}
	\\
	\toprule
	\textbf{Object or equation}
	&
	\multicolumn{1}{c}{\textbf{Mathematical form}}
	&
	\textbf{Interpretation}
	\\
	\midrule
	\endhead
	
	\midrule
	\multicolumn{3}{r}{\emph{Continued on next page.}}
	\\
	\endfoot
	
	\bottomrule
	\endlastfoot
	
	Reference latent ensemble
	&
	\EqCell{
		P_{\mathrm{ref},t}(dx)
		\quad\text{on}\quad
		(\Psi,\Sigma_X)
	}
	&
	Baseline distribution of latent virion--environment states before
	measurement-protocol conditioning.
	\\[0.45em]
	
	Protocol parameter vector
	&
	\EqCell{
		\theta
		=
		(\theta_{\mathrm{vir}},\theta_{\mathrm{env}},\theta_E)
	}
	&
	Parameter vector separating virion properties, environmental properties, and
	protocol-specific quantities.
	\\[0.45em]
	
	Protocol-conditioned latent ensemble
	&
	\EqCell{
		\widetilde P_{E,t}^{\mathrm{lat}}
		=
		P_{\mathrm{ref},t}\Pi_E^{\mathrm{lat}}
	}
	&
	Latent ensemble after preparation, forcing, medium exposure, surface contact,
	biological pathway conditioning, or another protocol-induced transformation.
	\\[0.45em]
	
	Latent transformation kernel
	&
	\EqCell{
		\Pi_E^{\mathrm{lat}}(dx_E\mid x)
	}
	&
	Maps a reference latent state \(x\in\Psi\) to a
	protocol-conditioned latent state \(x_E\in\Psi_E\).
	\\[0.45em]
	
	Survival/detection weight
	&
	\EqCell{
		s_E:\Psi_E\rightarrow[0,1]
	}
	&
	Probability that a protocol-conditioned latent state enters the accepted
	non-null data channel.
	\\[0.45em]
	
	Detection yield
	&
	\EqCell{
		\eta_E(t)
		=
		\int_{\Psi_E}
		s_E(x_E)\,
		\widetilde P_{E,t}^{\mathrm{lat}}(dx_E)
	}
	&
	Total probability mass that survives, is detected, is accepted, or
	successfully amplifies into a non-null record.
	\\[0.45em]
	
	Detected latent ensemble
	&
	\EqCell{
		P_{E,t}^{\mathrm{lat}}(A\mid\mathrm{det})
		=
		\frac{
			\int_A s_E(x_E)\,
			\widetilde P_{E,t}^{\mathrm{lat}}(dx_E)
		}{
			\eta_E(t)
		}
	}
	&
	Protocol-conditioned latent ensemble after selection, detection, visibility,
	or biological success, normalized over non-null outcomes.
	\\[0.45em]
	
	Non-null observation space
	&
	\EqCell{
		(\mathcal Y_E,\Sigma_{Y,E})
	}
	&
	Protocol-specific space of reported non-null data: images, curves,
	trajectories, spectra, counts, classes, or endpoint readouts.
	\\[0.45em]
	
	Null-augmented observation space
	&
	\EqCell{
		\mathcal Y_E^{\varnothing}
		=
		\mathcal Y_E\cup\{\varnothing\}
	}
	&
	Observation space enlarged by the null outcome representing non-detection,
	rejection, failed amplification, loss, or failure to enter the accepted
	record.
	\\[0.45em]
	
	Null channel
	&
	\EqCell{
		\varnothing
	}
	&
	Outcome assigned to latent states that fail to produce accepted, visible,
	countable, trackable, reconstructable, or otherwise recorded data under the
	protocol.
	\\[0.45em]
	
	Readout kernel
	&
	\EqCell{
		R_E(dy\mid x_E)
	}
	&
	Maps a detected protocol-conditioned latent state to reported data, including
	finite resolution, fitting, reconstruction, thresholding, classification,
	counting, and measurement noise.
	\\[0.45em]
	
	State-specific detection probability
	&
	\EqCell{
		\begin{aligned}
			q_E(x)
			&=
			K_E^{\varnothing}(\mathcal Y_E\mid x)
			\\
			&=
			\int_{\Psi_E}
			s_E(x_E)\,
			\Pi_E^{\mathrm{lat}}(dx_E\mid x)
		\end{aligned}
	}
	&
	Probability that latent state \(x\) generates a non-null observation under
	protocol \(E\).
	\\[0.45em]
	
	Null-inclusive protocol kernel
	&
	\EqCell{
		\begin{aligned}
			K_E^{\varnothing}(S\mid x)
			&=
			\int_{\Psi_E}
			s_E(x_E)\,
			R_E(S\cap\mathcal Y_E\mid x_E)
			\Pi_E^{\mathrm{lat}}(dx_E\mid x)
			\\
			&\quad+
			[1-q_E(x)]\delta_{\varnothing}(S)
		\end{aligned}
	}
	&
	Full state-to-observation kernel including latent transformation, selection,
	readout, and probability flow into the null channel.
	\\[0.45em]
	
	Null-inclusive observation operator
	&
	\EqCell{
		(\mathcal M_E^{\varnothing}P)(S)
		=
		\int_{\Psi}
		K_E^{\varnothing}(S\mid x)\,P(dx)
	}
	&
	Linear operator mapping a latent ensemble to the full observed ensemble,
	including null outcomes.
	\\[0.45em]
	
	Null-inclusive observed ensemble
	&
	\EqCell{
		P_{\mathrm{obs},t}^{\varnothing}(\cdot\mid E)
		=
		\mathcal M_E^{\varnothing}P_{\mathrm{ref},t}
	}
	&
	Observed ensemble generated by applying the protocol observation operator to
	the reference latent ensemble.
	\\[0.45em]
	
	Detected observation operator
	&
	\EqCell{
		(\mathcal M_E^{\mathrm{det}}P)(B)
		=
		\frac{
			(\mathcal M_E^{\varnothing}P)(B)
		}{
			(\mathcal M_E^{\varnothing}P)(\mathcal Y_E)
		}
	}
	&
	Conditional operator giving the reported non-null observed ensemble after
	normalization by detection yield.
	\\[0.45em]
	
	Protocol-resolved observed density
	&
	\EqCell{
		p_E^{\varnothing}(y\mid\theta)
		=
		\frac{
			dP_{\mathrm{obs},t}^{\varnothing}
			(\cdot\mid E,\theta)
		}{
			d\nu_E^{\varnothing}
		}(y)
	}
	&
	Density or mass function of the null-inclusive observed law with respect to
	an appropriate dominating measure.
	\\[0.45em]
	
	Protocol-lifted observable
	&
	\EqCell{
		\begin{aligned}
			\widetilde h_E(x)
			&=
			\int_{\Psi_E}
			s_E(x_E)
			\left[
			\int_{\mathcal Y_E}
			h(y)R_E(dy\mid x_E)
			\right]
			\\
			&\qquad\qquad\times
			\Pi_E^{\mathrm{lat}}(dx_E\mid x)
		\end{aligned}
	}
	&
	Turns a reported observable \(h\) into an effective protocol-dependent
	observable on the latent state space.
	\\[0.45em]
	
	Observed expectation under a protocol
	&
	\EqCell{
		\langle h\rangle_E[P]
		=
		\frac{
			\int_{\Psi}\widetilde h_E(x)P(dx)
		}{
			\int_{\Psi}q_E(x)P(dx)
		}
	}
	&
	Expected reported value among detected observations; the denominator
	explicitly accounts for protocol-dependent selection.
	\\[0.45em]
	
	Common comparison space
	&
	\EqCell{
		L_0:\Psi\to\mathsf W,
		\qquad
		L_E:\Psi_E\to\mathsf W
	}
	&
	Places latent and protocol-conditioned objects into a shared space before
	computing collapse magnitudes or protocol-to-protocol discrepancies.
	\\[0.45em]
	
	Latent distributional collapse
	&
	\EqCell{
		\mathcal C_{\mathrm{lat}}^{\mathsf W}(E;t)
		=
		D_{\mathsf W}
		\!\left(
		(L_E)_{\#}\widetilde P_{E,t}^{\mathrm{lat}},
		(L_0)_{\#}P_{\mathrm{ref},t}
		\right)
	}
	&
	Measures how strongly the protocol transforms the latent ensemble before
	detection selection.
	\\[0.45em]
	
	Detected latent collapse
	&
	\EqCell{
		\mathcal C_{\mathrm{det}}^{\mathsf W}(E;t)
		=
		D_{\mathsf W}
		\!\left(
		(L_E)_{\#}P_{E,t}^{\mathrm{lat}}(\cdot\mid\mathrm{det}),
		(L_0)_{\#}P_{\mathrm{ref},t}
		\right)
	}
	&
	Measures the collapse of the latent subpopulation that actually reaches the
	detected channel.
	\\[0.45em]
	
	Detection-yield collapse
	&
	\EqCell{
		\mathcal C_{\eta}(E;E_0,t)
		=
		|\eta_E(t)-\eta_{E_0}(t)|
	}
	&
	Compares how much probability mass reaches the non-null data channel under
	two protocols.
	\\[0.45em]
	
	Null-inclusive likelihood
	&
	\EqCell{
		\mathcal L_E^{\varnothing}(\theta;\mathcal D_E)
		=
		\prod_i p_E^{\varnothing}(y_i\mid\theta)
	}
	&
	Likelihood of the complete null-inclusive dataset under the
	protocol-conditioned observed law.
	\\[0.45em]
	
	Conditional detected likelihood
	&
	\EqCell{
		\mathcal L_E^{\mathrm{det}}
		(\theta;\mathcal D_E^{\mathrm{det}})
		=
		\prod_i p_E(y_i\mid\theta,\mathrm{det})
	}
	&
	Likelihood for datasets in which only non-null accepted observations are
	retained.
	\\[0.45em]
	
	Bayesian inverse formulation
	&
	\EqCell{
		q(\theta\mid\mathcal D_E,E)
		=
		\frac{
			\mathcal L_E^{\varnothing}(\theta;\mathcal D_E)q_0(\theta)
		}{
			\int_{\Theta}
			\mathcal L_E^{\varnothing}(\vartheta;\mathcal D_E)
			q_0(\vartheta)\,d\vartheta
		}
	}
	&
	Posterior distribution over latent, environmental, and protocol parameters
	after observing protocol-conditioned data.
	\\[0.45em]
	
	Target/nuisance partition
	&
	\EqCell{
		\theta=(\theta_a,\lambda)
	}
	&
	Separates the target parameter block \(\theta_a\) from nuisance parameters
	\(\lambda\), such as environmental variables, protocol settings, calibration
	constants, thresholds, or noise parameters.
	\\[0.45em]
	
	Protocol identifiability
	&
	\EqCell{
		\begin{aligned}
			&P_{\mathrm{obs},t}^{\varnothing}
			(\cdot\mid E,\theta_a,\lambda)
			\\
			&\qquad =
			P_{\mathrm{obs},t}^{\varnothing}
			(\cdot\mid E,\theta_a',\lambda')
			\Rightarrow
			\theta_a=\theta_a'
		\end{aligned}
	}
	&
	Target parameter block remains identifiable after accounting for admissible
	nuisance-parameter variation.
	\\[0.45em]
	
	Protocol equivalence class
	&
	\EqCell{
		\begin{aligned}
			[\theta]_E
			=
			\{\theta'\in\Theta:
			&\,P_{\mathrm{obs},t}^{\varnothing}
			(\cdot\mid E,\theta')
			\\
			&=
			P_{\mathrm{obs},t}^{\varnothing}
			(\cdot\mid E,\theta)\}
		\end{aligned}
	}
	&
	Set of parameter values that a protocol cannot distinguish, even in the
	ideal infinite-data limit.
	\\[0.45em]
	
	Protocol score function
	&
	\EqCell{
		\mathbf u_E(y;\theta)
		=
		\nabla_{\theta}\log p_E^{\varnothing}(y\mid\theta)
	}
	&
	Local derivative of the protocol-conditioned log-likelihood with respect to
	the parameter vector.
	\\[0.45em]
	
	Fisher information
	&
	\EqCell{
		\mathcal I_E(\theta)
		=
		\mathbb E_{\theta}
		[
		\mathbf u_E(Y;\theta)
		\mathbf u_E(Y;\theta)^{\mathsf T}
		]
	}
	&
	Local sensitivity of the protocol-conditioned likelihood to parameter
	perturbations.
	\\[0.45em]
	
	Directional Fisher information
	&
	\EqCell{
		\mathcal I_E(\theta;v)
		=
		v^{\mathsf T}\mathcal I_E(\theta)v
	}
	&
	Amount of local information supplied by protocol \(E\) in parameter direction
	\(v\).
	\\[0.45em]
	
	Local blind subspace
	&
	\EqCell{
		\mathcal B_E(\theta)
		=
		\ker\mathcal I_E(\theta)
	}
	&
	Infinitesimal parameter directions that the protocol cannot locally resolve.
	\\[0.45em]
	
	Effective Fisher information
	&
	\EqCell{
		\mathcal I_{a\mid\lambda}
		=
		\mathcal I_{aa}
		-
		\mathcal I_{a\lambda}
		\mathcal I_{\lambda\lambda}^{-1}
		\mathcal I_{\lambda a}
	}
	&
	Information about the target block \(\theta_a\) after nuisance parameters
	are allowed to adjust.
	\\[0.45em]
	
	Multi-protocol information
	&
	\EqCell{
		\mathcal I_{\mathrm{multi}}(\theta)
		=
		\sum_{j=1}^{M}\mathcal I_{E_j}(\theta)
	}
	&
	Combined Fisher information for conditionally independent protocols.
	Complementary protocols can reduce blind directions.
	\\[0.45em]
	
	Multi-protocol consistency
	&
	\EqCell{
		P_{\mathrm{obs},t}^{(j),\varnothing}
		=
		\mathcal M_{E_j}^{\varnothing}(\lambda_j)
		P_{\mathrm{ref},t}(\cdot\mid\theta)
	}
	&
	Asks whether different protocol-conditioned datasets can be explained by a
	shared latent virion--environment model with protocol-specific nuisance
	parameters.
	\\
\end{xltabular}

\endgroup
\newpage 

\section{Mechanisms of Experimental Collapse}
\label{sec:mechanisms_of_experimental_collapse}

The preceding sections introduced experimental collapse as a
protocol-conditioned map from a reference latent ensemble to an observed
ensemble. We now resolve that map into mechanism-specific collapse channels.
This decomposition is not merely taxonomic. It identifies which part of an
experiment transforms the latent state, which part selects or rejects states,
which part conditions survival or detectability, and which part projects
surviving states into the reported data space.
\medskip]

\noindent The central laboratory question is:
\begin{empheq}[box=\fbox]{equation}
	\text{What happened to the latent virion--environment ensemble before it
		became reported data?}
	\label{eq:mechanisms_guiding_question}
\end{empheq}
This question is practical rather than philosophical. A cryo-EM reconstruction,
an AFM force curve, a dielectrophoretic trajectory, a mucus-tracking path, and a
plaque count are all real observations. The point is that each is produced by a
specific sequence of preparation, conditioning, selection, survival, detection,
and readout operations.

Furthermore, a protocol is rarely one operation. It is usually a sequence of physical,
chemical, biological, computational, and instrumental steps. Each step may
preserve some latent variables while transforming, filtering, amplifying,
compressing, or erasing others. Before writing the full kernel composition, it
is useful to state the pipeline in its most compact form:
\begin{empheq}[box=\fbox]{equation}
	\text{Latent State}
	\;\longrightarrow\;
	\text{Conditioned State}
	\;\longrightarrow\;
	\text{Selected State}
	\;\longrightarrow\;
	\text{Reported Datum}.
	\label{eq:mechanism_pipeline_minimal}
\end{empheq}
Here ``conditioned'' means that the virion has been placed into the physical,
chemical, biological, and procedural environment of the experiment. A cryo-EM
specimen is conditioned by grid preparation, thin-film formation, air--water
interface exposure, vitrification, particle selection, and reconstruction. An
AFM specimen is conditioned by surface attachment, hydration, probe geometry,
indentation, and loading rate. A DEP or electrorotation specimen is conditioned
by the imposed field, suspending medium, conductivity, permittivity, and
hydrodynamic drag. A mucus-tracking experiment is conditioned by the structured
medium, adhesion landscape, imaging cadence, and tracking window. A plaque assay
is conditioned by the cell line, adsorption time, overlay, incubation time,
staining method, visibility threshold, and counting rule.

A slightly more detailed schematic is
\begin{empheq}[box=\fbox]{equation}
	X
	\;\longrightarrow\;
	X_{\mathrm{prep}}
	\;\longrightarrow\;
	X_{\mathrm{constr}}
	\;\longrightarrow\;
	X_{\mathrm{det}}
	\;\longrightarrow\;
	Y_E
	\;\text{or}\;
	\varnothing .
	\label{eq:mechanism_pipeline_at_glance}
\end{empheq}
The first arrow represents preparation or handling. The second represents
surface, field, medium, loading, confinement, or biological constraints. The
third represents survival, selection, detectability, or acceptance. The final
arrow represents readout or null observation. The variables in
Eq.~\eqref{eq:mechanism_pipeline_at_glance} are schematic rather than universal.
Their purpose is to help identify where each mechanism acts in a real
experiment.

\begin{remark}[How to use this section]
	\label{rem:how_to_use_mechanisms_section}
	For any protocol, the mechanism-resolved analysis proceeds in stages. First,
	list the experimental steps. Second, decide whether each step transforms the
	latent state, changes the probability of detection, or changes the final
	readout. Third, assign that step to a kernel, weight, readout map, or null
	channel. 
\end{remark}

\subsection{Mechanism-Resolved Factorization}
\label{subsec:mechanism_resolved_factorization}

Let \(E\) be a fixed protocol. Instead of treating the latent-stage kernel
\(\Pi_E^{\mathrm{lat}}\) as an indivisible object, we may factor it into
mechanism-specific kernels. The simplest symbolic statement is
\begin{empheq}[box=\fbox]{equation}
	\Pi_E^{\mathrm{lat}}
	=
	\Pi_E^{\mathrm{last}}
	\circ
	\cdots
	\circ
	\Pi_E^{\mathrm{first}}.
	\label{eq:mechanism_factorization_simple}
\end{empheq}
This equation states that the protocol-conditioned latent state is produced by
applying several experimentally meaningful stages in sequence. The rightmost
kernel acts first.
\medskip

\noindent More explicitly, let
\[
(\Psi_{E,0},\Sigma_{X,E,0})
=
(\Psi,\Sigma_X),
\qquad
(\Psi_{E,L},\Sigma_{X,E,L})
=
(\Psi_E,\Sigma_{X,E}),
\]
and let
\[
\Pi_{E,\ell}(dx_\ell\mid x_{\ell-1}),
\qquad
\ell=1,\ldots,L,
\]
be Markov kernels between intermediate protocol-conditioned state spaces.

\begin{definition}[Mechanism-resolved protocol factorization]
	\label{def:mechanism_resolved_protocol_factorization}
	A \emph{mechanism-resolved factorization} of the latent-stage protocol kernel
	is a composition
	\begin{equation}
		\Pi_E^{\mathrm{lat}}
		=
		\Pi_{E,L}
		\circ
		\Pi_{E,L-1}
		\circ
		\cdots
		\circ
		\Pi_{E,1},
		\label{eq:mechanism_resolved_kernel_composition}
	\end{equation}
	where each factor represents a physically, chemically, biologically, or
	procedurally interpretable protocol stage. For \(G\in\Sigma_{X,E}\),
	\begin{empheq}[box=\fbox]{equation}
		\begin{aligned}
			\Pi_E^{\mathrm{lat}}(G\mid x_0)
			&=
			\int_{\Psi_{E,1}}
			\cdots
			\int_{\Psi_{E,L}}
			\mathbf 1_G(x_L)\,
			\Pi_{E,L}(dx_L\mid x_{L-1})
			\cdots
			\Pi_{E,1}(dx_1\mid x_0).
		\end{aligned}
		\label{eq:mechanism_resolved_kernel_integral}
	\end{empheq}
\end{definition}
This expression is the unfolded version of
Eq.~\eqref{eq:mechanism_factorization_simple}. A state begins as
\(x_0\), passes through intermediate protocol states
\[
x_1,\ldots,x_L,
\]
and ends in the final protocol-conditioned latent state
\(x_L\in\Psi_E\). The indicator \(\mathbf 1_G(x_L)\) asks whether the final
conditioned state lies in the set \(G\). The multiple integral sums over all
possible intermediate protocol histories.

\noindent Typical mechanism-specific factors include
\begin{equation}
	\Pi_E^{\mathrm{prep}},
	\qquad
	\Pi_E^{\mathrm{interface}},
	\qquad
	\Pi_E^{\mathrm{surf}},
	\qquad
	\Pi_E^{\mathrm{field}},
	\qquad
	\Pi_E^{\mathrm{load}},
	\qquad
	\Pi_E^{\mathrm{medium}},
	\qquad
	\Pi_E^{\mathrm{time}},
	\qquad
	\Pi_E^{\mathrm{bio}}.
	\label{eq:typical_mechanism_kernels}
\end{equation}
The order of these factors is protocol dependent. A cryo-EM pipeline, an AFM
indentation experiment, a dielectrophoretic trapping experiment, a mucus
tracking experiment, and a plaque assay do not apply the same sequence of
operations.

\begin{remark}[Mechanism order is part of the model]
	\label{rem:mechanism_order_is_part_of_model}
	The factorization in Eq.~\eqref{eq:mechanism_resolved_kernel_composition} is
	not a universal ordering of experimental reality. It is a model of a
	particular protocol. In cryo-EM, sample application, thin-film formation,
	air--water interface exposure, vitrification, particle orientation, particle
	picking, and reconstruction are coupled stages of one
	preparation-and-readout pipeline
	\cite{Noble2018,Chen2019,Liu2023,DImprima2019,Levitz2022,Hirst2024}. In AFM,
	adsorption, hydration, probe geometry, indentation depth, loading rate, and
	deformation pathway are coupled during measurement
	\cite{Mateu2012,Marchetti2016,Kiss2021,Lyonnais2021}. The correct
	factorization is therefore the one that matches the experimental sequence
	being modeled.
\end{remark}

\begin{remark}[Transformation, selection, and readout should not be conflated]
	\label{rem:transformation_selection_readout_not_conflated}
	A mechanism may enter the model in three different ways. It may transform the
	latent state, which belongs in \(\Pi_E^{\mathrm{lat}}\). It may alter whether
	a state produces a usable record, which belongs in \(s_E\) or the null channel
	\(\varnothing\). Or it may alter how a surviving state is represented, which
	belongs in \(R_E\). For example, surface contact may physically deform a
	virion, making it a latent transformation, while particle picking may reject
	low-contrast particles without changing them, making it a selection effect.
\end{remark}

\begin{table}[H]
	\centering
	\caption{
		Practical placement of common collapse mechanisms in the protocol
		formalism. The entries are schematic; a real protocol may distribute one
		mechanism across several mathematical objects.
	}
	\label{tab:mechanism_placement_manual}
	\renewcommand{\arraystretch}{1.18}
	\begin{tabularx}{0.98\linewidth}{@{}p{0.25\linewidth}p{0.26\linewidth}X@{}}
		\toprule
		\textbf{Experimental question}
		&
		\textbf{Mathematical location}
		&
		\textbf{Interpretation}
		\\
		\midrule
		
		Did the protocol physically, chemically, mechanically, or biologically
		change the state?
		&
		\(\Pi_E^{\mathrm{lat}}\)
		&
		Preparation, surface binding, field forcing, loading, confinement,
		medium interaction, cell contact, or assay-pathway conditioning changes
		the latent ensemble before readout.
		\\[0.45em]
		
		Did the state survive or become detectable?
		&
		\(s_E\), \(\eta_E\), \(\varnothing\)
		&
		Only some protocol-conditioned states produce usable, visible, countable,
		trackable, reconstructable, or amplified records.
		\\[0.45em]
		
		How is a surviving state represented as data?
		&
		\(R_E\), \(\mathcal O_E\)
		&
		The readout projects, reconstructs, thresholds, localizes, classifies,
		counts, fits, or otherwise summarizes the surviving state.
		\\[0.45em]
		
		What is lost by reporting only the data object?
		&
		Observation fibers or quotient classes
		&
		Many mechanically or biologically distinct states may map to the same
		observed image, track, spectrum, class, plaque count, or endpoint
		readout.
		\\
		\bottomrule
	\end{tabularx}
\end{table}

\subsection{Preparation and Interface Collapse}
\label{subsec:preparation_interface_collapse}

Preparation collapse occurs when the sample-handling pipeline changes the
population before the instrument or assay reports it. In structural virology,
especially cryo-EM and cryo-ET, this may include dilution, buffer exchange,
blotting, thin-film formation, air--water interface exposure, support-film
interaction, vitrification, ice-thickness variation, beam sensitivity, particle
picking, classification, and reconstruction. The preparation stage is therefore
not a neutral prelude to measurement. It is part of the measurement map.

\noindent The simplest representation is
\begin{empheq}[box=\fbox]{equation}
	X
	\xrightarrow{\;\Pi_E^{\mathrm{prep}}\;}
	X_{\mathrm{prep}}.
	\label{eq:preparation_collapse_simple}
\end{empheq}
The prepared state \(X_{\mathrm{prep}}\) need not have the same distribution as
the reference latent state \(X\). It may be enriched for certain orientations,
depleted of fragile states, altered by surfaces or interfaces, restricted to
states that survive preparation, or reweighted by the criteria used for particle
selection and reconstruction.

\begin{definition}[Preparation/interface collapse]
	\label{def:preparation_interface_collapse}
	\emph{Preparation/interface collapse} is the protocol-induced transformation
	or selection of latent states during sample preparation, surface or interface
	exposure, preservation, fixation, vitrification, staining, buffer exchange, or
	other pre-instrumental conditioning. It may be represented by a combination of
	\[
	\Pi_E^{\mathrm{prep}},
	\qquad
	\Pi_E^{\mathrm{interface}},
	\qquad
	s_E,
	\qquad
	R_E .
	\]
	The precise placement depends on whether the mechanism transforms the state,
	selects which states survive, or changes how surviving states are represented.
\end{definition}
\noindent In cryo-EM, a useful schematic factorization is
\begin{equation}
	\Pi_E^{\mathrm{prep}}
	=
	\Pi_E^{\mathrm{vit}}
	\circ
	\Pi_E^{\mathrm{AWI}}
	\circ
	\Pi_E^{\mathrm{thinfilm}}
	\circ
	\Pi_E^{\mathrm{grid}},
	\label{eq:cryo_prep_factorization}
\end{equation}
where \(\Pi_E^{\mathrm{grid}}\) represents interaction with the grid or support,
\(\Pi_E^{\mathrm{thinfilm}}\) represents formation of the thin aqueous layer,
\(\Pi_E^{\mathrm{AWI}}\) represents air--water interface exposure, and
\(\Pi_E^{\mathrm{vit}}\) represents vitrification. The order is schematic; in a
real preparation, grid chemistry, thin-film formation, interface exposure, and
vitrification are coupled.

\begin{remark}[Cryo-EM as a concrete example]
	\label{rem:cryo_em_concrete_preparation_collapse}
	Cryo-EM is often discussed as a structure-preserving method, and in many cases
	it is extraordinarily successful. However, the ensemble that enters a
	reconstruction can be strongly conditioned by preparation. Air--water
	interface interactions can produce particle adsorption, denaturation,
	dissociation, preferred orientation, or uneven particle distribution; ice
	thickness can determine whether particles are visible; grid and support
	interactions can alter where particles localize; and particle picking can
	transmit only the subset satisfying contrast, shape, and reconstruction
	criteria
	\cite{Cheng2015,Noble2018,Chen2019,Liu2023,DImprima2019,Levitz2022,Hirst2024}.
	In the language of experimental collapse, cryo-EM does not merely project a
	pre-existing ensemble into images. It can also create a
	preparation-conditioned structural ensemble before readout.
\end{remark}

\begin{remark}[Preparation collapse as an experimental variable]
	\label{rem:preparation_collapse_as_experimental_variable}
	Preparation collapse is not only a nuisance. It can become an experimental
	variable. Changing grid type, support chemistry, blotting conditions,
	dispense-to-plunge time, surfactant, buffer composition, vitrification
	strategy, or particle-picking criteria changes the preparation kernel. If the
	observed ensemble changes under these variations, the change is evidence that
	the preparation stage is selecting or transforming latent states. This is why
	grid-preparation metadata should be treated as part of the protocol, not as
	peripheral documentation.
\end{remark}

A simple orientation-selection model illustrates the point. Let
\(Q\in SO(3)\) denote particle orientation, and let \(w_E(Q)\geq 0\) be the
preparation-and-selection weight for orientation \(Q\). The compact idea is
\[
\text{Detected orientation density}
\;\propto\;
\text{Selection weight}
\times
\text{Reference orientation density}.
\]
In normalized form,
\begin{empheq}[box=\fbox]{equation}
	p_{E,\mathrm{det}}(Q)
	=
	\frac{
		w_E(Q)\,p_{\mathrm{ref}}(Q)
	}{
		\displaystyle
		\int_{SO(3)}
		w_E(Q')p_{\mathrm{ref}}(Q')\,dQ'
	}.
	\label{eq:orientation_selection_weight_model}
\end{empheq}
If \(w_E(Q)\) is nearly constant, the protocol approximately preserves the
reference orientation distribution. If \(w_E(Q)\) is sharply peaked, the
protocol creates an orientation-collapsed ensemble. Preferred-orientation
problems in single-particle cryo-EM are a concrete instance of this logic, and
tilted data collection is one strategy for increasing the range of sampled
views \cite{Tan2017,Hirst2024}.

\begin{remark}[How to read the orientation-selection model]
	\label{rem:how_to_read_orientation_selection_model}
	Equation~\eqref{eq:orientation_selection_weight_model} is not meant to
	capture all of cryo-EM preparation. It isolates one mechanism:
	state-dependent survival or representation of orientations. The same
	mathematical structure applies to other selection processes. A tracking
	protocol may overweight mobile particles. An AFM protocol may overweight
	particles that adhere stably to a surface. A plaque assay may overweight
	states that amplify efficiently in the chosen cell line. In each case, the
	observed distribution is a normalized, protocol-weighted version of a larger
	latent distribution.
\end{remark}

\begin{remark}[Laboratory interpretation]
	\label{rem:preparation_collapse_laboratory_interpretation}
	The practical implication is that preparation variables are model variables.
	If changing the grid, support film, plunge conditions, buffer, incubation time,
	surface chemistry, or selection threshold changes the reported ensemble, then
	the protocol has revealed a preparation-sensitive latent sector. The correct
	response is not simply to discard the observation as an artifact, but to ask
	which mechanism in the preparation kernel changed and what latent feature it
	was selecting, suppressing, or transforming.
\end{remark}
\subsection{Geometric Projection and Reconstruction Collapse}
\label{subsec:geometric_projection_reconstruction_collapse}

Geometric projection occurs when the reported object is a lower-dimensional,
processed, reconstructed, or summarized representation of the latent state. This
collapse channel is present whenever a three-dimensional, time-dependent,
deformable, orientation-dependent, or internally heterogeneous state is reported
as an image, density map, class average, trajectory, scalar coefficient,
spectrum, endpoint readout, or count.

The simplest deterministic representation is
\begin{empheq}[box=\fbox]{equation}
	Y_E
	=
	\mathcal O_E(X_E),
	\label{eq:geometric_projection_simple}
\end{empheq}
where \(X_E\in\Psi_E\) is the protocol-conditioned latent state,
\(Y_E\in\mathcal Y_E\) is the reported non-null object, and
\(\mathcal O_E:\Psi_E\rightarrow\mathcal Y_E\) is the protocol-dependent
observation map. The important point is that \(\mathcal O_E\) is usually not
invertible. Many different protocol-conditioned latent states can produce the
same observed datum.

\begin{definition}[Geometric projection collapse]
	\label{def:geometric_projection_collapse}
	\emph{Geometric projection collapse} occurs when the readout map
	\[
	\mathcal O_E:\Psi_E\rightarrow\mathcal Y_E
	\]
	is non-injective, so that distinct protocol-conditioned latent states
	\(x_E,x_E'\in\Psi_E\) satisfy
	\[
	\mathcal O_E(x_E)=\mathcal O_E(x_E')
	\]
	and are therefore indistinguishable at the level of the reported datum. In
	kernel form, this non-invertible readout is represented by
	\(R_E(dy\mid x_E)\), which may include noise, reconstruction, projection,
	classification, thresholding, or summarization.
\end{definition}

The latent information compressed by a deterministic readout can be described by
the observation fiber
\begin{equation}
	\mathcal F_E(y_E)
	=
	\left\{
	x_E\in\Psi_E:
	\mathcal O_E(x_E)=y_E
	\right\}.
	\label{eq:observation_fiber}
\end{equation}
If \(\mathcal F_E(y_E)\) contains mechanically, dynamically, structurally, or
biologically distinct states, then the observation is under-resolved with
respect to those latent directions.

\begin{remark}[How to read the observation fiber]
	\label{rem:how_to_read_observation_fiber}
	The fiber \(\mathcal F_E(y_E)\) is the set of protocol-conditioned latent
	states that become indistinguishable after readout. If \(y_E\) is a
	two-dimensional projected track, the fiber may contain many unobserved
	three-dimensional paths, orientations, and binding histories. If \(y_E\) is a
	density map, the fiber may contain different preparation histories,
	conformational pathways, or dynamical trajectories that led to the same
	preserved structure. If \(y_E\) is a plaque count, the fiber contains many
	possible infection, replication, spread, and staining histories that produce
	the same counted outcome. The fiber is therefore the geometric version of
	protocol blindness.
\end{remark}

\begin{remark}[Stochastic reconstruction and posterior fibers]
	\label{rem:stochastic_reconstruction_posterior_fibers}
	When the readout is stochastic, the exact fiber
	\(\mathcal F_E(y_E)\) should be replaced by a conditional distribution over
	protocol-conditioned latent states given the observation. In that case, the
	relevant inverse object is
	\[
	P_E(dx_E\mid y_E,\mathrm{det}),
	\]
	or an approximation to it. This posterior fiber describes which latent states
	remain compatible with the reported datum after accounting for the protocol
	kernel, readout noise, reconstruction algorithm, and selection rules.
\end{remark}

\begin{remark}[Reconstruction as a protocol-dependent inverse problem]
	\label{rem:reconstruction_as_protocol_dependent_inverse_problem}
	Reconstruction does not simply return the latent state. It estimates a
	structured object from noisy, incomplete, selected, and processed data. In
	single-particle cryo-EM, the observed micrographs are noisy projections of
	particles with unknown orientations, and reconstruction depends on alignment,
	classification, orientation coverage, particle selection, and post-processing.
	In cryo-ET, a limited tilt range can produce missing-wedge effects and
	anisotropic resolution, so the reconstructed tomogram is not a complete
	sampling of all Fourier directions \cite{Tan2017,WiedemannHeckel2024}. In
	protocol-resolved language, reconstruction belongs to the readout kernel
	\(R_E\), and its limitations are part of the experimental collapse rather than
	a separate afterthought.
\end{remark}

\begin{remark}[Laboratory use: diagnosing reconstruction-limited conclusions]
	\label{rem:laboratory_use_reconstruction_limited_conclusions}
	The practical question is not only whether a reconstruction is high resolution,
	but which latent distinctions the reconstruction can and cannot support. A
	density map may strongly constrain morphology while remaining weakly sensitive
	to transient dynamics, preparation history, rare conformational states, or
	orientation-dependent mechanical response. A protocol-resolved analysis asks
	whether an apparent structural absence reflects true absence in the latent
	ensemble, insufficient sampling of the relevant views, rejection during
	classification, or compression by the reconstruction pipeline.
\end{remark}

\begin{remark}[Examples of geometric projection]
	\label{rem:examples_geometric_projection}
	A cryo-EM class average may preserve characteristic views while suppressing
	particle-to-particle variability. A reconstructed density may preserve
	geometry while removing velocity, force history, preparation pathway, and
	short-lived conformational transitions. A single-particle tracking experiment
	may report two-dimensional projected positions while discarding body
	orientation, spike configuration, local binding state, deformation, and
	unresolved pauses. A plaque assay may compress an entire infection,
	replication, spread, and staining history into a single counted lesion. These
	are not failures of the methods. They are examples of protocol-specific
	projection.
\end{remark}

\begin{table}[H]
	\centering
	\caption{
		Examples of geometric projection and reconstruction collapse. The table
		identifies the observed object, the latent content it often preserves,
		and the latent content it commonly compresses or erases.
	}
	\label{tab:geometric_projection_examples}
	\renewcommand{\arraystretch}{1.18}
	\begin{tabularx}{0.98\linewidth}{@{}p{0.20\linewidth}p{0.25\linewidth}p{0.25\linewidth}X@{}}
		\toprule
		\textbf{Protocol}
		&
		\textbf{Observed object}
		&
		\textbf{Often preserved}
		&
		\textbf{Often compressed}
		\\
		\midrule
		
		Cryo-EM / cryo-ET
		&
		Density map, class average, subtomogram average
		&
		Structural geometry, selected conformations, particle classes
		&
		Dynamics, force history, preparation pathway, rare states, missing
		orientations, anisotropic resolution
		\\[0.45em]
		
		Single-virus tracking
		&
		Projected trajectory, MSD, transport-state classification
		&
		Position history, transport regimes, residence times
		&
		Orientation, deformation, spike state, local binding state, unresolved
		short events
		\\[0.45em]
		
		AFM
		&
		Topography or force--indentation curve
		&
		Height, deformation response, rupture or stiffness signature
		&
		Free-suspension state, pre-contact trajectory, full unloaded 3D shape,
		orientation before adsorption
		\\[0.45em]
		
		DEP / electrorotation
		&
		Field-driven trajectory, crossover frequency, rotation rate
		&
		Polarization response, trapping behavior, frequency dependence
		&
		Field-free Brownian dynamics, non-electrical latent variables,
		uncoupled mechanical sectors
		\\[0.45em]
		
		Plaque assay
		&
		Plaque count or plaque-size distribution
		&
		Visible infectious amplification
		&
		Physical particle count, morphology, noninfectious states, failed
		infection histories, cell-line-incompatible particles
		\\
		\bottomrule
	\end{tabularx}
\end{table}

\begin{remark}[Open-problem perspective]
	\label{rem:geometric_projection_open_problem_perspective}
	Several recurring experimental difficulties can be phrased as geometric
	projection problems. Preferred orientation in cryo-EM asks which orientations
	are absent or downweighted before reconstruction. Missing-wedge artifacts in
	cryo-ET ask which Fourier directions are under-sampled. Trajectory projection
	in tracking asks which three-dimensional, orientational, or binding-state
	variables are compressed into a two-dimensional path. Plaque counting asks
	which infection histories have been compressed into the same lesion count. In
	each case, the protocol-resolved remedy is similar: identify the fiber,
	estimate which latent variables vary within it, and add protocol variations
	that split the fiber into experimentally distinguishable subsets.
\end{remark}

\subsection{Medium Filtering and Adhesive Collapse}
\label{subsec:medium_filtering_adhesive_collapse}

Many virological measurements occur in structured media rather than simple
dilute buffers. Mucus, gels, droplets, extracellular matrix, tissue fluids,
airway surface liquid, overlays, and cell-surface glycocalyces can filter
virions by size, shape, surface chemistry, adhesion, charge, receptor mimicry,
enzymatic activity, antibody binding, and local rheology. In such cases, the
medium is not merely background. It is part of the latent state and part of the
protocol-conditioned observation map.

The simplest schematic is
\begin{empheq}[box=\fbox]{equation}
	X
	\;\longrightarrow\;
	X_{\mathrm{medium}}(\theta_{\mathrm{env}}),
	\label{eq:medium_filtering_simple}
\end{empheq}
where \(X\) is the reference latent state and
\(X_{\mathrm{medium}}\) is the state after interaction with a structured medium.
The environmental parameter block \(\theta_{\mathrm{env}}\) may include mesh
size, viscosity, viscoelastic relaxation time, mucin composition, binding-site
density, antibody concentration, ionic strength, pH, local flow, hydration, or
local biochemical composition.

\begin{definition}[Medium-filtering collapse]
	\label{def:medium_filtering_collapse}
	\emph{Medium-filtering collapse} is the state-dependent slowing, trapping,
	binding, exclusion, channeling, immobilization, or mobility-state switching of
	virions by a structured medium. It may be represented by a medium-dependent
	kernel
	\[
	\Pi_E^{\mathrm{medium}}
	(dx_{\mathrm{medium}}\mid x;\theta_{\mathrm{env}}),
	\]
	where \(x\in\Psi\) is the pre-medium latent state,
	\(x_{\mathrm{medium}}\in\Psi_{E,\mathrm{medium}}\) is the
	medium-conditioned state, and \(\theta_{\mathrm{env}}\) parameterizes the
	local medium.
\end{definition}

\begin{remark}[The medium as a selector]
	\label{rem:medium_as_selector}
	A structured medium can act as a mechanical filter, an adhesive filter, and a
	biochemical filter at the same time. A particle may be mobile because it is
	small relative to the mesh, because its surface chemistry avoids adhesive
	binding, because enzymatic activity clears receptor-like obstacles, or because
	local rheology permits motion on the observation timescale. Conversely, a
	particle may be immobilized because it is sterically hindered, adhesively
	bound, cross-linked, trapped in a dense region, or too slow to be tracked
	within the chosen time window.
\end{remark}

For a trajectory-level model, one may write the unfolded medium-conditioned
motion schematically as
\begin{equation}
	d\mathbf r_t
	=
	\mathbf b_E(\mathbf r_t,Q_t,M_t,\theta_{\mathrm{env}})\,dt
	+
	\sqrt{2}\,
	\mathbf B_E(\mathbf r_t,Q_t,M_t,\theta_{\mathrm{env}})\,d\mathbf W_t
	+
	d\mathbf J_t,
	\label{eq:medium_filtered_trajectory_schematic}
\end{equation}
with
\begin{equation}
	D_E(\mathbf r_t,Q_t,M_t,\theta_{\mathrm{env}})
	=
	\mathbf B_E
	\mathbf B_E^{\mathsf T}.
	\label{eq:medium_diffusion_tensor_factorization}
\end{equation}
Here \(\mathbf b_E\) is a drift field, \(D_E\) is an effective diffusion tensor,
\(Q_t\) is orientation, \(M_t\) is a hidden mobility or binding state, and
\(\mathbf J_t\) represents jumps, sticking events, release events, or
transitions between mobility regimes.

\begin{remark}[How to read the medium-filtered trajectory model]
	\label{rem:how_to_read_medium_filtered_trajectory_model}
	Equation~\eqref{eq:medium_filtered_trajectory_schematic} is not meant to
	impose one universal model of mucus or gel transport. It identifies where
	different mechanisms enter. Drift may represent flow, chemophysical bias, or
	directed transport. The tensor \(D_E\) represents local anisotropic or
	heterogeneous mobility. The hidden state \(M_t\) represents mobile, confined,
	adhesive, bound, or immobile modes. The jump term \(\mathbf J_t\) allows
	intermittent release, sticking, hopping, or sudden changes in local
	environment. Thus a measured trajectory is a joint virion--medium process,
	not merely free diffusion with noise added.
\end{remark}

\begin{definition}[Mobility-state decomposition]
	\label{def:mobility_state_decomposition}
	A medium-filtered trajectory may be decomposed into latent mobility states
	\[
	M_t\in
	\{
	\mathrm{mobile},
	\mathrm{confined},
	\mathrm{adhesive},
	\mathrm{immobile},
	\mathrm{released}
	\}.
	\]
	The observed trajectory is then a projection of the joint process
	\begin{equation}
		X_t
		=
		(\mathbf r_t,Q_t,M_t,X_{\mathrm{env},t})
		\label{eq:joint_medium_filtered_latent_process}
	\end{equation}
	onto reported positions, tracks, or trajectory summaries.
\end{definition}

\begin{remark}[Why hidden mobility states matter]
	\label{rem:hidden_mobility_states_matter}
	A single reported diffusivity can hide multiple latent mobility states. A
	trajectory with intermittent pauses may be fit by an effective diffusion
	coefficient, but the pauses may correspond to adhesive binding, steric
	confinement, transient receptor-like interactions, antibody cross-linking, or
	local mesh heterogeneity. The medium-filtering formalism encourages reporting
	not only an effective diffusivity, but also the protocol assumptions that
	determine which mobility states can be distinguished.
\end{remark}

\begin{definition}[Medium-filtered observables]
	\label{def:medium_filtered_observables}
	Typical medium-filtered observables include
	\[
	h_{\mathrm{med}}(Y_E)
	\in
	\left\{
	D_{\mathrm{eff}},
	\alpha_{\mathrm{MSD}},
	\tau_{\mathrm{res}},
	f_{\mathrm{mobile}},
	f_{\mathrm{immobile}},
	\ell_{\mathrm{conf}},
	k_{\mathrm{bind}},
	k_{\mathrm{release}}
	\right\},
	\]
	where \(D_{\mathrm{eff}}\) is an effective diffusivity,
	\(\alpha_{\mathrm{MSD}}\) is an anomalous-transport exponent,
	\(\tau_{\mathrm{res}}\) is a residence time,
	\(f_{\mathrm{mobile}}\) and \(f_{\mathrm{immobile}}\) are mobile and immobile
	fractions, \(\ell_{\mathrm{conf}}\) is a confinement scale, and
	\(k_{\mathrm{bind}},k_{\mathrm{release}}\) are effective binding and release
	rates.
\end{definition}

\begin{remark}[Mucus tracking as a concrete example]
	\label{rem:mucus_tracking_concrete_example}
	Single-particle tracking in mucus reports a joint virion--medium process.
	Observed trajectories depend on virion size and surface chemistry, but also on
	mucus mesh structure, mucin binding, antibodies, adhesivity, hydration, and
	local rheology. HIV virion motion in cervical mucus and influenza motion in
	airway mucus illustrate this point: the trajectory ensemble is not simply a
	free-diffusion ensemble with noise added, but a medium-filtered ensemble whose
	immobilized, confined, and mobile subpopulations may carry biological meaning
	\cite{Boukari2009,Wang2017,Kaler2022,Vahey2019,Abrami2024}.
\end{remark}

\begin{remark}[Medium filtering as inverse environmental inference]
	\label{rem:medium_filtering_as_inverse_environmental_inference}
	Medium-filtered collapse can be useful precisely because the medium changes
	the trajectory. If the protocol kernel is modeled, trajectory data can be used
	to infer environmental quantities such as effective mesh scale, adhesive
	binding, local viscoelasticity, antibody-mediated immobilization, hydration,
	or flow. The risk is confounding: a slow trajectory may reflect a sticky
	virion, a dense medium, a local antibody interaction, low frame rate, poor
	localization, or a tracking threshold. Therefore medium inference is strongest
	when particle properties, medium properties, and protocol settings are varied
	or independently constrained.
\end{remark}

\begin{remark}[Laboratory use: separating virion and medium effects]
	\label{rem:laboratory_use_separating_virion_medium_effects}
	Medium-filtering collapse gives a practical way to design more informative
	tracking experiments. One can vary virion surface chemistry, antibody
	concentration, mucin composition, ionic strength, hydration, or imaging cadence
	and ask which fitted quantities change: \(D_{\mathrm{eff}}\),
	\(\alpha_{\mathrm{MSD}}\), \(f_{\mathrm{immobile}}\), residence times, or
	binding/release rates. If the same virion preparation changes behavior across
	media, the protocol has identified an environmental sector. If different virion
	preparations behave differently in the same medium, the protocol has identified
	a virion-surface or binding sector. If both change together, a joint
	virion--environment model is required.
\end{remark}

\begin{table}[H]
	\centering
	\caption{
		Medium-filtering mechanisms and their interpretation. A structured medium
		may transform the latent state, select the observed population, and alter
		the reported trajectory statistics simultaneously.
	}
	\label{tab:medium_filtering_mechanisms}
	\renewcommand{\arraystretch}{1.18}
	\begin{tabularx}{0.98\linewidth}{@{}p{0.24\linewidth}p{0.30\linewidth}X@{}}
		\toprule
		\textbf{Medium mechanism}
		&
		\textbf{Possible mathematical location}
		&
		\textbf{Experimental interpretation}
		\\
		\midrule
		
		Steric hindrance
		&
		\(\Pi_E^{\mathrm{medium}}\), \(D_E(\cdot)\)
		&
		Mesh geometry slows or excludes particles depending on size, shape,
		orientation, and local density.
		\\[0.45em]
		
		Adhesive binding
		&
		\(\Pi_E^{\mathrm{medium}}\), hidden state \(M_t\), \(s_E\)
		&
		Particles switch between mobile, bound, confined, immobile, and released
		states.
		\\[0.45em]
		
		Antibody or mucin immobilization
		&
		\(\Pi_E^{\mathrm{medium}}\), \(s_E\)
		&
		Binding or cross-linking can shift particles into confined or immobile
		subpopulations.
		\\[0.45em]
		
		Local rheology
		&
		\(D_E(\cdot)\), \(\mathbf b_E(\cdot)\), memory terms
		&
		Viscosity and viscoelasticity determine effective diffusivity, residence
		time, confinement, and anomalous transport.
		\\[0.45em]
		
		Flow or mucociliary clearance
		&
		\(\mathbf b_E(\cdot)\), \(s_E\)
		&
		Advection can bias trajectories and alter the probability that particles
		remain in the observation window.
		\\[0.45em]
		
		Tracking threshold
		&
		\(R_E\), \(s_E\), \(\varnothing\)
		&
		Particles that are too slow, too dim, too fast, too intermittent, or too
		poorly localized may be excluded from the reported trajectory ensemble.
		\\
		\bottomrule
	\end{tabularx}
\end{table}

\begin{remark}[Transition to time-window collapse]
	\label{rem:transition_to_time_window_collapse}
	Medium filtering is closely linked to time-window collapse. A particle may be
	mobile on long timescales but appear immobile during a short observation, or it
	may undergo rare release events that are missed by the tracking window. Thus,
	the next collapse mechanism concerns persistence: which states last long
	enough, recur often enough, or amplify quickly enough to enter the reported
	dataset.
\end{remark}

\subsection{Time-Window and Persistence Collapse}
\label{subsec:time_window_persistence_collapse}

Every protocol observes over a finite time window. Short-lived states may vanish
before measurement begins. Rare transitions may not occur during acquisition.
Slow relaxation may be mistaken for a static state. Fast motion may blur, evade
localization, or be excluded during analysis. Long incubation assays may amplify
small initial differences into countable outcomes. Thus time is not merely the
axis along which data are collected. It is a filtering mechanism.

The simplest schematic is
\begin{empheq}[box=\fbox]{equation}
	\text{Latent time process}
	\;\longrightarrow\;
	\text{Sampled, averaged, or time-integrated record}.
	\label{eq:time_window_collapse_simple}
\end{empheq}
A time-dependent latent process may therefore appear static, averaged, missing,
or amplified depending on the observation window, sampling cadence, exposure
time, preparation delay, incubation time, and analysis rule.

\begin{definition}[Time-window collapse]
	\label{def:time_window_collapse}
	\emph{Time-window collapse} is the state-dependent filtering produced by the
	observation interval, sampling cadence, exposure time, preparation delay,
	incubation time, or analysis window of the protocol. It is represented by a
	time-window kernel
	\[
	\Pi_E^{\mathrm{time}}
	(dx_{\mathrm{time}}\mid x;t_0,t_1,\Delta t_{\mathrm{samp}},\tau_{\mathrm{exp}}),
	\]
	where \(t_0\) and \(t_1\) define the observation window,
	\(\Delta t_{\mathrm{samp}}\) is the sampling interval when the record is
	discrete in time, and \(\tau_{\mathrm{exp}}\) is the exposure or integration
	time when individual measurements average over finite temporal intervals.
\end{definition}
Let \(X_t\) be a latent stochastic process. A discrete-time observation model may
be written as
\begin{equation}
	Y_m
	=
	\mathcal O_E(X_{t_m})+\nu_m,
	\qquad
	t_m=t_0+m\Delta t_{\mathrm{samp}},
	\qquad
	m=1,\ldots,M.
	\label{eq:discrete_time_observation}
\end{equation}
When exposure or integration over a finite frame matters, a more realistic form
is
\begin{equation}
	Y_m
	=
	\frac{1}{\tau_{\mathrm{exp}}}
	\int_{t_m}^{t_m+\tau_{\mathrm{exp}}}
	\mathcal O_E(X_t)\,dt
	+
	\nu_m.
	\label{eq:finite_exposure_time_observation}
\end{equation}
Equation~\eqref{eq:finite_exposure_time_observation} makes explicit that fast
motion, short pauses, transient binding, and rapid conformational changes may be
blurred or averaged within a single recorded datum.

If the relevant state persists for a lifetime \(\tau_{\mathrm{state}}\), the
degree of time-window visibility depends on dimensionless ratios such as
\begin{empheq}[box=\fbox]{equation}
	\Pi_{\mathrm{samp}}
	=
	\frac{\tau_{\mathrm{state}}}{\Delta t_{\mathrm{samp}}},
	\qquad
	\Pi_{\mathrm{win}}
	=
	\frac{\tau_{\mathrm{state}}}{t_1-t_0},
	\qquad
	\Pi_{\mathrm{exp}}
	=
	\frac{\tau_{\mathrm{state}}}{\tau_{\mathrm{exp}}}.
	\label{eq:time_window_dimensionless_ratios}
\end{empheq}
When \(\Pi_{\mathrm{samp}}\ll1\), the state is typically shorter-lived than the
sampling interval and is likely to be missed or averaged over. When
\(\Pi_{\mathrm{samp}}\gg1\), the state persists across many samples and is more
directly visible. When \(\Pi_{\mathrm{win}}\ll1\), the state occupies only a
small part of the observation window. When \(\Pi_{\mathrm{win}}\gg1\), it may
appear static over the experiment. When \(\Pi_{\mathrm{exp}}\ll1\), the state is
short compared with the exposure time and may be motion-blurred or integrated
away.

\begin{remark}[Sampling does not merely discretize time]
	\label{rem:sampling_does_not_merely_discretize_time}
	Sampling changes what is inferable. A fast binding--release process may appear
	as a reduced effective diffusion coefficient. A short-lived conformational
	state may disappear before vitrification. A rare transition may not occur
	during the recorded movie. A slow relaxation may be mistaken for a stable
	state. Thus sampling cadence, exposure time, and observation duration are
	protocol parameters, not only technical metadata.
\end{remark}

\begin{definition}[Persistence-biased detection]
	\label{def:persistence_biased_detection}
	Let \(A\subseteq\Psi\) be a latent state class, such as a bound state, mobile
	state, deformation state, conformational state, fusion intermediate, or
	infection stage. Define the persistence time
	\[
	\tau_A
	=
	\operatorname{Leb}
	\left\{
	t\in[t_0,t_1]:X_t\in A
	\right\},
	\]
	the total time spent in \(A\) during the observation window. A protocol with
	minimum resolvable persistence \(\tau_{\min}\) detects this state class with a
	weight of the schematic form
	\begin{equation}
		s_E^{\mathrm{time}}(A)
		=
		\Pr(\tau_A\geq \tau_{\min}\mid E).
		\label{eq:persistence_biased_detection_weight}
	\end{equation}
\end{definition}

\begin{remark}[Interpretation of persistence-biased detection]
	\label{rem:interpretation_persistence_biased_detection}
	Persistence-biased detection means that long-lived states are easier to
	observe than short-lived states, even if the short-lived states are
	mechanistically important. This is relevant for transient binding, fusion
	intermediates, short pauses in tracking data, fragile structural states during
	preparation, rare orientation states, and early infection events that do not
	persist long enough to become visible under the chosen assay.
\end{remark}

\begin{definition}[Time-integrated readout]
	\label{def:time_integrated_readout}
	Some protocols report a time-integrated functional rather than a sequence of
	instantaneous observations. A general form is
	\begin{equation}
		Y_E
		=
		\int_{t_0}^{t_1}
		h_E(X_t)\,dt
		+
		\nu_E,
		\label{eq:time_integrated_readout}
	\end{equation}
	where \(h_E\) is a protocol-dependent contribution function and \(\nu_E\) is
	measurement noise. Plaque counts, fluorescence accumulation, reporter signals,
	cytopathic-effect readouts, and endpoint assays often have this
	time-integrated character.
\end{definition}

\begin{remark}[Incubation time as a collapse mechanism]
	\label{rem:incubation_time_as_collapse_mechanism}
	In biological amplification assays, incubation time determines which infection
	events become visible. A short incubation may miss slow-growing or weakly
	amplifying events. A long incubation may merge plaques, saturate a reporter
	signal, obscure early kinetic differences, or amplify initially small
	differences into large observed differences. Thus incubation time is part of
	the protocol kernel, not merely a scheduling choice.
\end{remark}

\begin{remark}[Cryo-EM preparation time as a concrete example]
	\label{rem:cryo_em_preparation_time_example}
	In cryo-EM preparation, the time between sample application and vitrification
	sets the interval during which particles can interact with the grid, support
	film, thin aqueous layer, and air--water interface. Studies varying grid-making
	or dispense-to-plunge times show that macromolecules can respond to the
	thin-film environment on experimentally relevant timescales, and that reducing
	this time can alter particle behavior but is not a universal solution for all
	specimens \cite{Levitz2022,Klebl2020,Hirst2024}. In the present language,
	preparation time changes the time-window kernel.
\end{remark}

\begin{remark}[Laboratory use: choosing the observation window]
	\label{rem:laboratory_use_choosing_observation_window}
	Time-window collapse suggests a practical design rule: choose the sampling
	cadence and observation duration to match the latent event being tested. If
	the target is transient binding in mucus, the frame interval must be short
	enough to resolve pauses and release events. If the target is viscoelastic
	relaxation under AFM loading, the loading rate and dwell time must be treated
	as part of the mechanical protocol. If the target is slow plaque-forming
	competence, incubation time and readout threshold must be varied rather than
	assumed fixed. In each case, timing parameters can be used deliberately to
	separate fast, slow, persistent, and rare latent sectors.
\end{remark}

\begin{remark}[Examples of time-window collapse]
	\label{rem:examples_time_window_collapse}
	In tracking experiments, fast binding--unbinding events may be averaged into
	an effective diffusion coefficient. In cryo-EM preparation, the interval
	between sample deposition and vitrification determines how long particles can
	interact with the air--water interface or support. In AFM, loading rate and
	dwell time determine whether a response appears elastic, viscoelastic,
	plastic, or destructive. In DEP and electrorotation, pulse duration, field
	frequency, and observation time determine which polarization and trapping
	responses are visible. In plaque assays, adsorption time, incubation time, and
	readout time determine which infectious events become visible plaques. Thus
	time is not only an acquisition coordinate. It is a selection mechanism.
\end{remark}

\begin{table}[H]
	\centering
	\caption{
		Time-window and persistence collapse in common virological protocols.
		Timing parameters determine which latent events are visible, averaged,
		amplified, saturated, or missed.
	}
	\label{tab:time_window_collapse_examples}
	\renewcommand{\arraystretch}{1.18}
	\begin{tabularx}{0.98\linewidth}{@{}p{0.22\linewidth}p{0.27\linewidth}p{0.28\linewidth}X@{}}
		\toprule
		\textbf{Protocol}
		&
		\textbf{Timing parameter}
		&
		\textbf{Latent events affected}
		&
		\textbf{Risk if ignored}
		\\
		\midrule
		
		Single-virus tracking
		&
		Frame interval, exposure time, trajectory length
		&
		Short pauses, binding--release events, directed intervals, confinement
		states
		&
		Fast states are collapsed into effective diffusion or missed entirely.
		\\[0.45em]
		
		Cryo-EM / cryo-ET
		&
		Application-to-vitrification time, blotting time, plunge speed
		&
		Interface contact, orientation selection, dissociation, denaturation,
		thin-film response
		&
		The prepared ensemble is mistaken for the pre-preparation ensemble.
		\\[0.45em]
		
		AFM
		&
		Loading rate, dwell time, recovery time
		&
		Viscoelastic relaxation, rupture, plastic deformation, recovery
		&
		Rate-dependent response is treated as static stiffness.
		\\[0.45em]
		
		DEP / electrorotation
		&
		Field frequency, pulse duration, observation time
		&
		Polarization relaxation, trapping, aggregation, field-induced loss
		&
		Driven response is mistaken for equilibrium or field-free behavior.
		\\[0.45em]
		
		Plaque / focus assay
		&
		Adsorption time, incubation time, staining/readout time
		&
		Entry, replication, local spread, plaque visibility, plaque merging
		&
		Visible count is treated as protocol-independent infectivity.
		\\
		\bottomrule
	\end{tabularx}
\end{table}

\begin{remark}[Transition to biological amplification]
	\label{rem:transition_time_window_to_biological_amplification}
	Time-window collapse leads naturally to biological amplification collapse. In
	plaque, focus-forming, endpoint dilution, cytopathic-effect, and reporter
	assays, the observed datum is not a direct microscopic state. It is the outcome
	of a time-extended biological pathway. The next subsection treats that pathway
	as a collapse mechanism in its own right.
\end{remark}

\subsubsection{Plaque, Focus, and Endpoint Assays as Biological Kernels}
\label{subsubsec:plaque_focus_endpoint_biological_kernels}

Plaque, focus-forming, endpoint dilution, reporter-cell, cytopathic-effect, and
neutralization assays are biological observation kernels. They do not report the
latent virion population directly. Instead, they place that population into a
specified biological environment and report the subset of states that can pass
through an assay-dependent sequence of delivery, attachment, entry,
replication, spread, signal generation, visibility, and counting.

For a plaque assay, a schematic latent sequence is
\begin{empheq}[box=\fbox]{equation}
	x
	\longrightarrow
	\text{Cell attachment}
	\longrightarrow
	\text{Entry}
	\longrightarrow
	\text{Replication}
	\longrightarrow
	\text{Local spread}
	\longrightarrow
	\text{Visible plaque}
	\longrightarrow
	\text{Count}.
	\label{eq:plaque_assay_sequence}
\end{empheq}
A physical virion that fails any required step contributes to the null channel
for that assay, even if it is structurally intact. Thus the assay readout is not
a direct particle census; it is a protocol-conditioned biological success
measure.

\begin{definition}[Plaque-forming probability]
	\label{def:plaque_forming_probability_mechanisms}
	For a latent state \(x\), define the plaque-forming probability
	\begin{empheq}[box=\fbox]{equation}
		\pi_{\mathrm{PFU}}(x;E_{\mathrm{PFU}})
		=
		\Pr\!\left(
		\begin{array}{c}
			x \text{ generates a visible, countable plaque} \\
			\text{under plaque-assay protocol } E_{\mathrm{PFU}}
		\end{array}
		\right).
		\label{eq:plaque_forming_probability_mechanisms}
	\end{empheq}
	This probability is a protocol-conditioned biological weight. It includes
	delivery, adsorption, entry, replication, local spread, plaque visibility,
	and counting under the specified plaque-assay protocol.
\end{definition}

\begin{remark}[Plaque assay as a concrete example]
	\label{rem:plaque_assay_concrete_example}
	Plaque assays are often reported in plaque-forming units, but a
	plaque-forming unit is not the same object as a physical virion. It is a
	protocol-conditioned infectious event made visible by a cell monolayer,
	inoculation conditions, adsorption time, overlay medium, incubation time,
	staining procedure, and counting rule
	\cite{Dulbecco1952,Cooper1961,Baer2014}. This makes the plaque assay one of
	the clearest examples of experimental collapse: the observed count is a
	selected and amplified functional projection of a larger latent population.
\end{remark}

In the null-inclusive kernel language, the plaque assay may be written as a
binary biological observation kernel:
\begin{equation}
	K_{E_{\mathrm{PFU}}}^{\varnothing}
	(\{\mathrm{plaque}\}\mid x)
	=
	\pi_{\mathrm{PFU}}(x;E_{\mathrm{PFU}}),
	\qquad
	K_{E_{\mathrm{PFU}}}^{\varnothing}
	(\{\varnothing\}\mid x)
	=
	1-\pi_{\mathrm{PFU}}(x;E_{\mathrm{PFU}}).
	\label{eq:plaque_binary_biological_kernel}
\end{equation}
The null outcome includes physical particles that fail delivery, fail
attachment, fail entry, fail replication, fail local spread, fail visibility, or
fail the counting rule. These null states are biologically different, but they
are observationally identical unless additional protocol variation or auxiliary
measurements separate them.

Focus-forming assays have a closely related structure, but the final visible
event is typically an immunostained or reporter-detected focus rather than a
macroscopic plaque. Endpoint dilution assays collapse the readout even further:
instead of counting localized plaques or foci, they report whether infection,
cytopathic effect, antigen expression, or reporter signal is detected at each
dilution. An endpoint quantity such as TCID\(_{50}\) is then estimated from a
binary infection/noninfection pattern across dilutions and replicate wells
\cite{Cresta2021,Lei2021}.

\begin{definition}[Biological readout spaces]
	\label{def:biological_readout_spaces}
	Different biological amplification assays correspond to different observation
	spaces:
	\begin{align}
		\mathcal Y_{\mathrm{PFU}}
		&=
		\{0,1,2,\ldots\},
		&&\text{Plaque count},
		\label{eq:pfu_readout_space}
		\\
		\mathcal Y_{\mathrm{FFU}}
		&=
		\{0,1,2,\ldots\},
		&&\text{Focus count},
		\label{eq:ffu_readout_space}
		\\
		\mathcal Y_{\mathrm{ED}}
		&=
		\{0,1\}^{J\times R},
		&&\text{Endpoint-dilution infection/noninfection pattern},
		\label{eq:endpoint_dilution_readout_space}
		\\
		\mathcal Y_{\mathrm{rep}}
		&\subseteq\mathbb R,
		&&\text{Reporter intensity or accumulated signal}.
		\label{eq:reporter_readout_space}
	\end{align}
	Here \(J\) denotes the number of dilutions and \(R\) the number of replicate
	wells per dilution.
\end{definition}

\begin{remark}[Different biological assays report different projections]
	\label{rem:different_biological_assays_report_different_projections}
	PFU, FFU, TCID\(_{50}\), cytopathic-effect endpoints, and reporter signals are
	not interchangeable direct counts of physical virions. They are different
	protocol-conditioned projections of infectious activity. A virus that forms
	clear plaques under one overlay and cell line may produce foci, reporter
	signals, or endpoint responses under another. The appropriate comparison is not
	whether the raw assay numbers are identical, but whether they can be explained
	by a shared latent infectivity model passed through different biological
	kernels.
\end{remark}
A compact way to compare biological kernels is to write an assay-specific success
probability:
\begin{equation}
	\pi_{\mathrm{bio}}^{(a)}(x;E_a)
	=
	\Pr(
	\text{latent state }x
	\text{ produces assay signal }a
	\mid E_a
	),
	\label{eq:assay_specific_biological_success_probability}
\end{equation}
where \(a\in\{\mathrm{PFU},\mathrm{FFU},\mathrm{ED},\mathrm{rep},\mathrm{CPE}\}\)
labels the assay type. The same latent state \(x\) may have different success
probabilities under different biological kernels:
\[
\pi_{\mathrm{PFU}}(x;E_{\mathrm{PFU}})
\neq
\pi_{\mathrm{FFU}}(x;E_{\mathrm{FFU}})
\neq
\pi_{\mathrm{ED}}(x;E_{\mathrm{ED}}).
\]
This is not inconsistency. It means that each assay asks a different biological
question of the latent population.

\begin{remark}[Laboratory interpretation]
	\label{rem:biological_kernel_laboratory_interpretation}
	A discrepancy between PFU, FFU, reporter signal, and endpoint dilution need not
	mean that one assay is wrong. It may indicate that the assays are sensitive to
	different biological stages. A focus assay may detect infected cells before
	macroscopic plaques form. A reporter assay may detect intracellular expression
	without requiring the same spatial spread. An endpoint dilution assay may
	report whether infection occurred in a well without resolving how many local
	infection events occurred. Treating each assay as a biological kernel makes the
	disagreement interpretable rather than merely inconvenient.
\end{remark}

\subsubsection{Amplification, Thresholds, and Saturation}
\label{subsubsec:amplification_thresholds_saturation}

Biological amplification assays often include thresholds. A local infection may
exist but remain below the staining threshold. A reporter signal may be produced
but remain below detection. A plaque may form but be too small, too diffuse, or
merged with another plaque to be counted. Conversely, strong amplification can
drive a signal into saturation, making different high-infectivity histories
appear similar. Thus biological amplification is coupled to detection selection.

\begin{definition}[Biological visibility threshold]
	\label{def:biological_visibility_threshold}
	Let
	\[
	A_E(x_{\mathrm{bio}})
	\]
	be a biological amplification signal, such as infected-cell number, local
	antigen amount, reporter intensity, cytopathic-effect score, plaque area, or
	focus area. A simple visibility rule is
	\begin{equation}
		s_E^{\mathrm{bio}}(x_{\mathrm{bio}})
		=
		\mathbf 1\{A_E(x_{\mathrm{bio}})\geq a_E^{\ast}\},
		\label{eq:biological_visibility_threshold}
	\end{equation}
	where \(a_E^{\ast}\) is the assay-specific visibility or counting threshold.
	A soft threshold may instead use a smooth detection function
	\begin{equation}
		s_E^{\mathrm{bio}}(x_{\mathrm{bio}})
		=
		\rho_E(A_E(x_{\mathrm{bio}})),
		\qquad
		0\leq \rho_E\leq 1.
		\label{eq:soft_biological_visibility_threshold}
	\end{equation}
\end{definition}

\begin{remark}[Thresholds are part of the protocol]
	\label{rem:thresholds_are_part_of_protocol}
	Visibility thresholds are not minor analysis details. They determine which
	biological events enter the reported dataset. A focus-forming assay may detect
	infection earlier than a plaque assay because it does not require the same
	macroscopic lesion. A cytopathic-effect endpoint may ignore subvisible
	infection that does not cross the scoring threshold. A reporter assay may
	saturate at high signal. These threshold effects belong in \(s_E\), \(R_E\),
	or both.
\end{remark}

\begin{definition}[Amplification gain]
	\label{def:biological_amplification_gain}
	Let \(A_E(x_{\mathrm{bio}},t)\) be a biological signal generated by a latent
	state under protocol \(E\). A schematic amplification gain over an incubation
	window \([t_0,t_1]\) is
	\begin{empheq}[box=\fbox]{equation}
		G_E(x)
		=
		\frac{
			A_E(x_{\mathrm{bio}},t_1)+\epsilon_A
		}{
			A_E(x_{\mathrm{bio}},t_0)+\epsilon_A
		},
		\qquad
		\epsilon_A>0.
		\label{eq:biological_amplification_gain}
	\end{empheq}
	Large \(G_E\) indicates that a small initial latent difference has been
	expanded into a large observable difference by replication, spread, reporter
	expression, antigen accumulation, or cytopathic amplification.
\end{definition}

\begin{remark}[Amplification can both reveal and hide]
	\label{rem:amplification_reveals_and_hides}
	Amplification reveals rare biologically competent events by making them
	countable. At the same time, it hides many microscopic distinctions among
	events that cross the same threshold. Two infection histories may differ
	substantially in entry time, replication speed, local spread, plaque
	morphology, or reporter kinetics and still produce the same final count.
	Conversely, two initially similar states may produce very different
	observations after nonlinear amplification.
\end{remark}
A useful schematic model is to treat the biological signal as a growing,
saturating quantity:
\begin{equation}
	\frac{dA_E}{dt}
	=
	g_E(A_E,x_{\mathrm{bio}},\theta_{\mathrm{env}},\theta_E),
	\qquad
	Y_E
	=
	\rho_E(A_E(t_1))+\nu_E.
	\label{eq:biological_signal_growth_readout}
\end{equation}
Here \(g_E\) represents replication, spread, reporter production, antigen
accumulation, or cytopathic progression, while \(\rho_E\) represents visibility,
thresholding, or saturation. The same final signal can arise from different
growth histories, and the same growth history can be reported differently under
different thresholds.

\begin{remark}[Laboratory use: separating growth from readout]
	\label{rem:laboratory_use_separating_growth_from_readout}
	When possible, measuring multiple readout times can help separate growth from
	thresholding. A single endpoint signal may confound slow replication, delayed
	entry, low initial infectious competence, reporter threshold, and saturation.
	Time-course reporter data, multiple staining times, or plaque-size
	distributions can partially separate these effects by adding temporal or
	morphological structure to the biological kernel.
\end{remark}

\subsection{Detection, Rejection, and Null-Channel Collapse}
\label{subsec:detection_rejection_null_channel_collapse}

Detection collapse occurs whenever the reported dataset includes only accepted,
visible, reconstructable, localizable, countable, or biologically amplified
states. It is represented by the survival/detection weight \(s_E\), the
detection yield \(\eta_E(t)\), and the null observation \(\varnothing\).

The simplest schematic is
\begin{empheq}[box=\fbox]{equation}
	X_E
	\;\longrightarrow\;
	Y_E
	\quad \text{with probability } s_E(X_E),
	\qquad
	X_E
	\;\longrightarrow\;
	\varnothing
	\quad \text{with probability } 1-s_E(X_E).
	\label{eq:detection_selection_simple}
\end{empheq}
The null channel is the mathematical place where missing probability mass is
kept.

\begin{definition}[Detection-selection collapse]
	\label{def:detection_selection_collapse}
	\emph{Detection-selection collapse} is the state-dependent loss of probability
	mass from the non-null observed ensemble. It is quantified by
	\begin{empheq}[box=\fbox]{equation}
		s_E(x_E)\in[0,1],
		\qquad
		\eta_E(t)
		=
		\int_{\Psi_E}
		s_E(x_E)\,
		\widetilde P_{E,t}^{\mathrm{lat}}(dx_E).
		\label{eq:detection_selection_collapse_weight_yield}
	\end{empheq}
	The missing probability \(1-\eta_E(t)\) is assigned to the null channel
	\(\varnothing\).
\end{definition}

\begin{remark}[Why the null channel is mechanistic]
	\label{rem:null_channel_mechanistic_not_bookkeeping}
	The null channel is not merely a bookkeeping device. It is mechanistic. A
	particle may be null because it was destroyed during preparation, adsorbed in
	an unusable orientation, too dim to localize, too fast to track, too deformed
	to reconstruct, unable to infect the chosen cell type, unable to activate a
	reporter, unable to form a visible plaque, or rejected by a computational
	criterion. These are different physical and biological mechanisms, even if
	they produce the same observed outcome: no reported datum.
\end{remark}

\begin{definition}[Null-channel decomposition]
	\label{def:null_channel_decomposition}
	When mechanisms can be distinguished, the null probability may be decomposed
	schematically as
	\begin{equation}
		1-\eta_E(t)
		=
		p_{\mathrm{loss}}
		+
		p_{\mathrm{damage}}
		+
		p_{\mathrm{reject}}
		+
		p_{\mathrm{below}}
		+
		p_{\mathrm{biofail}}
		+
		p_{\mathrm{other}},
		\label{eq:null_channel_decomposition}
	\end{equation}
	where the terms represent loss during preparation, damage or inactivation,
	computational or instrumental rejection, below-threshold signal, biological
	failure, and other null mechanisms. In a rigorous model these terms should be
	defined as probabilities of disjoint null categories or replaced by a
	categorical null variable.
\end{definition}

A more explicit version introduces a categorical null label
\[
Z_{\varnothing}
\in
\{
\mathrm{loss},
\mathrm{damage},
\mathrm{reject},
\mathrm{below},
\mathrm{biofail},
\mathrm{other}
\}.
\]
Then
\begin{equation}
	P_{\mathrm{obs},t}^{\varnothing}
	(\{\varnothing,Z_{\varnothing}=z\}\mid E)
	=
	\int_{\Psi_E}
	p_E(z\mid x_E,\varnothing)\,
	\bigl[1-s_E(x_E)\bigr]\,
	\widetilde P_{E,t}^{\mathrm{lat}}(dx_E),
	\label{eq:categorical_null_channel_model}
\end{equation}
whenever the experiment or model can justify assigning null outcomes to
mechanistic categories.

\begin{remark}[Null categories should not be invented without evidence]
	\label{rem:null_categories_require_evidence}
	The decomposition in Eq.~\eqref{eq:null_channel_decomposition} is useful only
	when the experiment can distinguish null mechanisms or when a justified model
	is being used. Otherwise, the safe statement is that probability mass entered
	the null channel. Assigning that mass to damage, rejection, failed infection,
	or below-threshold signal requires additional evidence.
\end{remark}

\begin{remark}[Why normalized datasets can mislead]
	\label{rem:why_normalized_datasets_can_mislead}
	A dataset conditioned on detection can look clean even when most of the latent
	population was lost. For example, a cryo-EM reconstruction may be high quality
	after extensive particle rejection; a tracking analysis may report smooth
	trajectories after excluding short or dim tracks; a plaque assay may produce
	countable plaques while ignoring non-plaque-forming particles. The normalized
	observed distribution is therefore incomplete without the detection yield and,
	when possible, the null-channel structure.
\end{remark}

\begin{remark}[Laboratory use: recording denominators]
	\label{rem:laboratory_use_recording_denominators}
	The practical recommendation is to record denominators whenever possible:
	particles imaged versus particles picked, picked particles versus particles
	retained after classification, attempted tracks versus accepted tracks, wells
	inoculated versus wells positive, plaques counted versus wells excluded, and
	reporter-positive versus reporter-negative events. These denominators estimate
	parts of the null channel. Without them, the experiment reports only the
	conditional distribution of successful observations.
\end{remark}
\subsubsection{Stage-Resolved Biological Amplification as a Protocol Design Tool}
\label{subsubsec:stage_resolved_biological_amplification_design}

The biological-kernel formulation suggests a practical way to use amplification
assays more diagnostically. Instead of treating a plaque count, focus count,
endpoint readout, or reporter signal as a single scalar measure of
``infectivity,'' one can vary controlled components of the assay and ask which
stage of the biological pathway is most responsible for the observed change.
In this sense, biological amplification assays can be used not only to quantify
infectious activity, but also to probe where the latent population is being
filtered by the assay.

Let the biological pathway be written as
\[
A_0,A_1,\ldots,A_K,
\]
where the stages may represent delivery to the assay environment, adsorption,
cell attachment, entry, genome release, replication, local spread, reporter
activation, staining, visibility, and counting. The assay-success probability
has the conditional factorization
\begin{empheq}[box=\fbox]{equation}
	\pi_{\mathrm{bio}}(x;E)
	=
	\Pr(A_0\mid x,E)
	\prod_{k=1}^{K}
	\Pr(A_k\mid A_0,\ldots,A_{k-1},x,E).
	\label{eq:stage_resolved_biological_success_probability}
\end{empheq}
The protocol \(E\) determines every factor in this product. Changing the cell
line, adsorption time, inoculum volume, overlay, incubation time, staining
method, reporter threshold, or counting rule changes the biological kernel and
therefore changes which latent states enter the visible assay channel.

\begin{definition}[Stage-resolved biological protocol family]
	\label{def:stage_resolved_biological_protocol_family}
	A \emph{stage-resolved biological protocol family} is a collection of related
	assay protocols
	\[
	\mathcal E_{\mathrm{bio}}
	=
	\{E^{(1)},E^{(2)},\ldots,E^{(M)}\},
	\]
	constructed by varying one or more biological-assay parameters while holding
	the remaining conditions fixed when possible. Examples include variation of
	cell line, receptor availability, adsorption time, neutralization condition,
	overlay composition, incubation time, staining threshold, reporter readout, or
	counting criterion.
\end{definition}

For each protocol \(E^{(m)}\), the biological kernel induces a success
probability
\[
\pi_{\mathrm{bio}}(x;E^{(m)}).
\]
The corresponding effective biological signal is
\begin{equation}
	\Lambda_{\mathrm{bio}}(E^{(m)})
	=
	\int_{\Psi}
	\pi_{\mathrm{bio}}(x;E^{(m)})
	n_{\mathrm{ref}}(x)\,dx,
	\label{eq:effective_biological_signal_protocol_family}
\end{equation}
where \(n_{\mathrm{ref}}(x)\) is the latent number-density distribution over
virion or virion--assay-relevant states. For plaque assays,
\(\Lambda_{\mathrm{bio}}\) reduces to an effective plaque-forming concentration;
for focus assays it becomes an effective focus-forming concentration; for
endpoint assays it controls the dilution-dependent probability of detecting
infection.

\begin{remark}[Assay variation as kernel perturbation]
	\label{rem:assay_variation_as_kernel_perturbation}
	Controlled assay variation should be interpreted as a perturbation of the
	biological observation kernel. Varying adsorption time primarily perturbs
	delivery, attachment, and entry factors. Varying cell line perturbs receptor
	availability, entry compatibility, intracellular permissiveness, and local
	spread. Varying overlay composition perturbs diffusion, cell-to-cell spread,
	lesion morphology, and plaque coalescence. Varying incubation time perturbs
	replication, spread, visibility, and saturation. Varying staining or reporter
	threshold perturbs the final readout. Thus, each assay variation asks a
	different question of the same latent virion population.
\end{remark}

\begin{definition}[Stage-sensitivity index]
	\label{def:stage_sensitivity_index}
	Let \(\xi_r\) be a controllable assay parameter, such as adsorption time,
	overlay viscosity, antibody concentration, incubation time, reporter
	threshold, or cell-line condition. The local stage-sensitivity of the
	effective biological signal to \(\xi_r\) may be defined as
	\begin{empheq}[box=\fbox]{equation}
		S_r(E)
		=
		\frac{\partial}{\partial \xi_r}
		\log
		\left[
		\Lambda_{\mathrm{bio}}(E;\xi_r)
		+
		\epsilon_{\Lambda}
		\right],
		\qquad
		\epsilon_{\Lambda}>0.
		\label{eq:stage_sensitivity_index}
	\end{empheq}
	Large \(|S_r(E)|\) indicates that the reported biological signal is sensitive
	to the assay stage controlled by \(\xi_r\).
\end{definition}

\begin{remark}[Interpretation of stage sensitivity]
	\label{rem:interpretation_stage_sensitivity}
	The stage-sensitivity index is not meant to replace detailed mechanistic
	modeling. It is a compact diagnostic. If the signal changes strongly with
	adsorption time, the assay may be limited by delivery, attachment, entry, or
	early cell interaction. If it changes strongly with overlay conditions, local
	spread or plaque morphology may be limiting. If it changes strongly with
	incubation time, growth rate, delayed replication, visibility, or saturation
	may dominate. If it changes strongly with staining threshold, the final readout
	rather than the biological infection process may be the limiting factor.
\end{remark}

\begin{definition}[Biological bottleneck decomposition]
	\label{def:biological_bottleneck_decomposition}
	A biological amplification assay is \emph{bottlenecked} at stage \(A_k\) if
	the conditional factor
	\[
	\Pr(A_k\mid A_0,\ldots,A_{k-1},x,E)
	\]
	is small, highly variable across latent states, or strongly sensitive to a
	protocol parameter. A schematic bottleneck score for stage \(A_k\) is
	\begin{equation}
		\mathcal B_k(E)
		=
		1-
		\int_{\Psi}
		\Pr(A_k\mid A_0,\ldots,A_{k-1},x,E)
		\,P_{\mathrm{ref},t}(dx),
		\label{eq:biological_bottleneck_score}
	\end{equation}
	with larger values indicating stronger average loss at that stage.
\end{definition}

\begin{remark}[Bottlenecks and latent heterogeneity]
	\label{rem:bottlenecks_and_latent_heterogeneity}
	A high bottleneck score may reflect a genuine biological limitation, a
	protocol limitation, or both. For example, weak plaque formation may reflect
	low infectious competence in the latent virion population, poor compatibility
	with the chosen cell line, overly restrictive overlay conditions, insufficient
	incubation time, or a stringent visibility threshold. The bottleneck
	decomposition therefore helps convert a low PFU, FFU, or endpoint readout into
	a structured hypothesis about where the assay is filtering the latent
	population.
\end{remark}

\begin{table}[H]
	\centering
	\caption{
		Stage-resolved perturbations of biological amplification assays. Each
		protocol variation changes a different component of the biological kernel
		and can therefore help identify which assay stage limits the observed
		signal.
	}
	\label{tab:stage_resolved_biological_kernel_perturbations}
	\renewcommand{\arraystretch}{1.18}
	\begin{tabularx}{0.98\linewidth}{@{}p{0.24\linewidth}p{0.32\linewidth}X@{}}
		\toprule
		\textbf{Protocol variation}
		&
		\textbf{Biological stage primarily probed}
		&
		\textbf{Interpretation in the collapse framework}
		\\
		\midrule
		
		Adsorption time
		&
		Delivery, attachment, entry initiation
		&
		Tests whether the assay is limited by early contact with the cell layer
		or by insufficient time for productive adsorption.
		\\[0.45em]
		
		Cell line or receptor condition
		&
		Receptor binding, entry compatibility, intracellular permissiveness
		&
		Separates intrinsic virion competence from host-cell compatibility and
		cell-line-specific amplification.
		\\[0.45em]
		
		Neutralization or antibody condition
		&
		Attachment, entry, aggregation, immobilization, immune blocking
		&
		Determines whether the latent population is being shifted into a
		non-amplifying or null biological channel.
		\\[0.45em]
		
		Overlay composition
		&
		Local spread, lesion morphology, plaque separation
		&
		Tests whether visible plaque formation is limited by diffusion,
		cell-to-cell spread, medium restriction, or plaque coalescence.
		\\[0.45em]
		
		Incubation time
		&
		Replication, spread, signal growth, visibility
		&
		Distinguishes early failure from delayed amplification, slow growth,
		plaque merging, reporter saturation, or late visibility.
		\\[0.45em]
		
		Staining or reporter threshold
		&
		Visibility, classification, counting
		&
		Tests whether events occur biologically but fail to cross the accepted
		readout threshold.
		\\[0.45em]
		
		Dilution series and replicate wells
		&
		Count statistics, endpoint probability, dynamic range
		&
		Improves precision and reveals saturation, overlap, zero-inflation, or
		dilution-dependent deviations from the assumed count model.
		\\
		\bottomrule
	\end{tabularx}
\end{table}

\begin{remark}[Application to difficult assay interpretation]
	\label{rem:application_difficult_assay_interpretation}
	The stage-resolved view is useful in common ambiguous cases. A high particle
	count with low PFU may indicate many defective particles, but it may also
	indicate poor cell-line compatibility, neutralization, aggregation, loss
	during adsorption, restrictive overlay conditions, or an overly stringent
	readout. A low focus count may reflect weak entry, weak antigen expression, or
	thresholding. A positive endpoint dilution pattern with few visible plaques
	may indicate infection without efficient local spread. These distinctions are
	lost if the assay is summarized only as a single infectious titer.
\end{remark}

\begin{definition}[Cross-assay biological consistency]
	\label{def:cross_assay_biological_consistency}
	Let \(E_{\mathrm{PFU}}\), \(E_{\mathrm{FFU}}\), and \(E_{\mathrm{ED}}\) denote
	plaque, focus-forming, and endpoint-dilution protocols. A latent infectivity
	model is \emph{cross-assay biologically consistent} if there exists a common
	latent distribution \(P_{\mathrm{ref},t}(\cdot\mid\theta)\) and
	assay-specific biological kernels such that
	\begin{equation}
		P_{\mathrm{obs}}^{\varnothing}
		(\cdot\mid E_j,\theta,\lambda_j)
		=
		\mathcal M_{E_j}^{\mathrm{bio},\varnothing}(\lambda_j)
		P_{\mathrm{ref},t}(\cdot\mid\theta),
		\qquad
		E_j\in
		\{E_{\mathrm{PFU}},E_{\mathrm{FFU}},E_{\mathrm{ED}}\}.
		\label{eq:cross_assay_biological_consistency}
	\end{equation}
\end{definition}

\begin{remark}[Why cross-assay consistency matters]
	\label{rem:why_cross_assay_consistency_matters}
	PFU, FFU, endpoint dilution, reporter intensity, and cytopathic-effect
	readouts may disagree numerically while still being compatible with a shared
	latent infectivity model. Conversely, apparent agreement can be misleading if
	the assays share the same blind sector or threshold. Cross-assay consistency
	therefore asks a stronger and more useful question than whether titers match:
	it asks whether distinct biological kernels can explain the observed assay
	outputs from the same latent population.
\end{remark}
\subsubsection{Count Statistics, Spatial Interaction, and Assay-Level Nonidealities}
\label{subsubsec:count_statistics_spatial_interaction_assay_nonidealities}

The biological-kernel framework also clarifies why infectious-unit assays often
deviate from ideal count models. In the dilute, well-mixed, independent-event
limit, plaque and focus counts are often modeled as Poisson random variables.
However, this limit is itself a protocol regime. Aggregation, local cell-layer
heterogeneity, clustered infection, variable adsorption efficiency, uneven
overlay conditions, plaque merging, and well-to-well variation can all move the
assay away from Poisson behavior.

The simplest ideal count model is
\begin{empheq}[box=\fbox]{equation}
	N_d
	\mid
	\Lambda_{\mathrm{bio}}
	\sim
	\operatorname{Poisson}
	\left(
	V_{\mathrm{inoc}}\,f_d\,
	\Lambda_{\mathrm{bio}}(E)
	\right),
	\label{eq:ideal_poisson_biological_count_model}
\end{empheq}
where \(N_d\) is the observed plaque or focus count at dilution fraction \(f_d\),
\(V_{\mathrm{inoc}}\) is the inoculum volume, and
\(\Lambda_{\mathrm{bio}}(E)\) is the effective protocol-conditioned biological
signal concentration. For a plaque assay, \(\Lambda_{\mathrm{bio}}\) may be
identified with an effective plaque-forming concentration; for a focus assay, it
may be identified with an effective focus-forming concentration.

\begin{remark}[The Poisson model is a protocol regime]
	\label{rem:poisson_model_protocol_regime}
	The Poisson model assumes that visible biological events arise independently,
	that the inoculum is well mixed, that plaques or foci do not strongly overlap,
	that local cell susceptibility is approximately homogeneous, and that each
	counted event is generated by an independent assay-success pathway. These are
	not merely statistical assumptions. They are statements about the biological
	and spatial regime of the assay.
\end{remark}

\paragraph{Overdispersion.}

Observed plaque or focus counts may have variance larger than the Poisson mean.
This overdispersion can arise from virion aggregation, spatial clustering,
heterogeneous cell susceptibility, variation in local overlay thickness,
well-to-well differences, batch effects, or stochastic differences in adsorption
and spread. A common phenomenological replacement is the negative-binomial
model:
\begin{empheq}[box=\fbox]{equation}
	N_d
	\sim
	\operatorname{NegBin}
	\left(
	\mu_d,\kappa
	\right),
	\qquad
	\mu_d
	=
	V_{\mathrm{inoc}}\,f_d\,\Lambda_{\mathrm{bio}}(E),
	\label{eq:negative_binomial_biological_count_model}
\end{empheq}
with variance
\begin{equation}
	\operatorname{Var}(N_d)
	=
	\mu_d
	+
	\frac{\mu_d^2}{\kappa}.
	\label{eq:negative_binomial_variance}
\end{equation}
Here \(\kappa>0\) is an overdispersion parameter. In the limit
\(\kappa\rightarrow\infty\), the model approaches the Poisson variance
\(\operatorname{Var}(N_d)=\mu_d\).

\begin{remark}[Interpretation of overdispersion]
	\label{rem:interpretation_overdispersion_biological_counts}
	Overdispersion is not merely a statistical inconvenience. It is evidence that
	the assay may contain unmodeled heterogeneity. For example, a high variance
	across wells may indicate variable delivery to the monolayer, local
	differences in cell susceptibility, clustered infection due to aggregates, or
	heterogeneous amplification after entry. In the protocol-kernel language,
	overdispersion suggests that \(\Lambda_{\mathrm{bio}}(E)\) is not fixed across
	replicates, but varies with latent well-level, local environmental, or
	protocol-specific factors.
\end{remark}

A hierarchical way to express this is
\begin{equation}
	N_{dr}\mid \Lambda_{r}
	\sim
	\operatorname{Poisson}
	\left(
	V_{\mathrm{inoc}}f_d\Lambda_r
	\right),
	\qquad
	\Lambda_r
	\sim
	G_E,
	\label{eq:hierarchical_count_model}
\end{equation}
where \(r\) indexes replicate wells and \(G_E\) describes replicate-to-replicate
variation in the effective biological signal. A gamma choice for \(G_E\) yields
a negative-binomial marginal distribution. Other choices may be appropriate when
the dominant heterogeneity is aggregation, spatial clustering, or cell-layer
structure.

\paragraph{Zero inflation and null-enriched assays.}

Some datasets contain more zero-count wells than expected under a Poisson or
negative-binomial model. This may occur when a fraction of wells receive no
effective infectious units, when delivery fails, when adsorption is inefficient,
when neutralization is strong, when the cell layer is locally nonpermissive, or
when the biological signal remains below threshold. A zero-inflated model writes
\begin{equation}
	N_d
	\sim
	\begin{cases}
		0,
		&
		\text{with probability } \zeta_E,
		\\[0.35em]
		\operatorname{Poisson}(\mu_d),
		&
		\text{with probability } 1-\zeta_E,
	\end{cases}
	\label{eq:zero_inflated_poisson_biological_count_model}
\end{equation}
or, more explicitly,
\begin{equation}
	\Pr(N_d=0)
	=
	\zeta_E
	+
	(1-\zeta_E)e^{-\mu_d},
	\qquad
	\Pr(N_d=n)
	=
	(1-\zeta_E)e^{-\mu_d}
	\frac{\mu_d^n}{n!},
	\quad n\geq1.
	\label{eq:zero_inflated_poisson_probabilities}
\end{equation}
The parameter \(\zeta_E\) represents a protocol-conditioned structural-zero
probability: the probability that a well, field, or assay unit is effectively
unable to produce a visible biological signal.

\begin{remark}[Zero inflation as null-channel enrichment]
	\label{rem:zero_inflation_null_channel_enrichment}
	Zero inflation is the count-level expression of null-channel enrichment. A
	zero plaque count may mean that no infectious units were present, but it may
	also mean that the relevant latent states failed delivery, attachment, entry,
	replication, spread, visibility, or counting. The zero-inflated parameter
	\(\zeta_E\) should therefore be interpreted as a protocol-conditioned null
	probability, not automatically as absence of virions.
\end{remark}

\paragraph{Spatial interaction and plaque merging.}

Plaque and focus assays also have a spatial component. At low density and short
enough incubation, visible lesions may be approximately independent. At higher
inoculum, longer incubation, or weaker overlay restriction, lesions may overlap,
merge, compete for susceptible cells, or become difficult to segment. The
``countable range'' is therefore a spatial protocol regime, not only a
convenient laboratory rule.

Let
\[
\Phi_E
=
\{z_1,\ldots,z_{N}\}
\subset \Omega_{\mathrm{well}}
\]
denote the spatial point pattern of visible plaques or foci in the well or
imaging region \(\Omega_{\mathrm{well}}\). A first idealization is an
inhomogeneous Poisson point process with intensity
\[
\lambda_E(z)
=
\lambda_E(z\mid\theta_{\mathrm{vir}},\theta_{\mathrm{env}},\theta_E),
\]
so that
\begin{equation}
	\mathbb E[N]
	=
	\int_{\Omega_{\mathrm{well}}}
	\lambda_E(z)\,dz.
	\label{eq:spatial_point_process_expected_count}
\end{equation}
Spatial heterogeneity in \(\lambda_E(z)\) can represent nonuniform cell
susceptibility, uneven inoculum distribution, local overlay variation, edge
effects, or spatial variation in staining or imaging.

Plaque merging can be represented by a detection or segmentation map
\[
\mathcal S_E:
\Phi_E^{\mathrm{true}}
\longrightarrow
\Phi_E^{\mathrm{obs}},
\]
where \(\Phi_E^{\mathrm{true}}\) is the latent pattern of biological lesions and
\(\Phi_E^{\mathrm{obs}}\) is the observed pattern after overlap, merging,
thresholding, and segmentation. In count form,
\begin{equation}
	N_{\mathrm{obs}}
	=
	\#\,\mathcal S_E(\Phi_E^{\mathrm{true}})
	\leq
	\#\,\Phi_E^{\mathrm{true}},
	\label{eq:plaque_merging_segmentation_map}
\end{equation}
when merging or segmentation failure combines multiple biological events into a
single visible lesion.

\begin{remark}[Countable range as a protocol condition]
	\label{rem:countable_range_as_protocol_condition}
	The usual countable range for plaques or foci can be interpreted as the regime
	in which \(\mathcal S_E\) is close to one-to-one and the spatial process is
	approximately noninteracting. Outside this regime, the observed count is not
	only a noisy version of the latent lesion number. It is a spatially collapsed
	readout affected by lesion growth, overlap, local depletion of susceptible
	cells, staining threshold, and segmentation.
\end{remark}

\paragraph{Endpoint dilution as censored biological observation.}

Endpoint dilution assays, including TCID\(_{50}\)-type readouts, do not directly
count individual infectious lesions. They report binary or categorical outcomes
across dilutions and replicate wells. The observation at dilution \(j\) and
replicate \(r\) may be written as
\[
Z_{jr}\in\{0,1\},
\]
where \(Z_{jr}=1\) indicates detectable infection, cytopathic effect, reporter
signal, or another endpoint criterion.

A simple endpoint model is
\begin{empheq}[box=\fbox]{equation}
	\Pr(Z_{jr}=1\mid\theta,E)
	=
	1-
	\exp
	\left[
	-
	V_{\mathrm{inoc}} f_j
	\Lambda_{\mathrm{ED}}(E,\theta)
	\right],
	\label{eq:endpoint_dilution_binary_model}
\end{empheq}
where \(\Lambda_{\mathrm{ED}}\) is the effective concentration of units capable
of producing a positive endpoint under the endpoint-dilution protocol. This is
not a plaque count. It is a thresholded detection probability.

More generally, if \(A_{jr}\) is a latent biological signal in well \(jr\), then
\begin{equation}
	Z_{jr}
	=
	\mathbf 1
	\{A_{jr}\geq a_E^{\ast}\},
	\label{eq:endpoint_dilution_threshold_model}
\end{equation}
where \(a_E^{\ast}\) is the endpoint scoring threshold. The observed dilution
pattern
\[
\mathcal D_{\mathrm{ED}}
=
\{Z_{jr}:j=1,\ldots,J,\;r=1,\ldots,R\}
\]
is therefore a censored or thresholded representation of a latent biological
amplification process.

\begin{remark}[Endpoint assays as censored kernels]
	\label{rem:endpoint_assays_as_censored_kernels}
	Endpoint dilution assays collapse quantitative biological histories into
	binary or categorical outcomes. A well scored positive may contain very
	different amounts of replication, cytopathic effect, or reporter signal; a
	well scored negative may contain no infection, failed infection, or
	subthreshold infection. Endpoint assays are therefore naturally represented as
	censored biological kernels rather than as direct counts of infectious
	particles.
\end{remark}

\paragraph{Neutralization as biological-kernel deformation.}

Neutralization assays can be described as deformations of the biological kernel.
Antibodies, inhibitors, serum components, or other neutralizing agents may reduce
the probability of attachment, entry, fusion, replication, local spread, or
readout visibility. If \(c\) denotes neutralizing-agent concentration, then the
stage-resolved success probability becomes
\begin{equation}
	\pi_{\mathrm{bio}}(x;E,c)
	=
	\Pr(A_0\mid x,E,c)
	\prod_{k=1}^{K}
	\Pr(A_k\mid A_0,\ldots,A_{k-1},x,E,c).
	\label{eq:neutralization_kernel_deformation}
\end{equation}
The neutralizing condition shifts latent states toward the null biological
channel by reducing one or more stage probabilities.

A compact neutralization curve can be written as
\begin{equation}
	F_{\mathrm{neutral}}(c)
	=
	1
	-
	\frac{
		\Lambda_{\mathrm{bio}}(E,c)
	}{
		\Lambda_{\mathrm{bio}}(E,0)+\epsilon_{\Lambda}
	},
	\qquad
	\epsilon_{\Lambda}>0,
	\label{eq:neutralization_fraction_signal_loss}
\end{equation}
where
\[
\Lambda_{\mathrm{bio}}(E,c)
=
\int_{\Psi}
\pi_{\mathrm{bio}}(x;E,c)\,
n_{\mathrm{ref}}(x)\,dx.
\]
This form emphasizes that neutralization is not merely a change in the observed
count. It is a protocol-conditioned deformation of the biological pathway.

\begin{remark}[Interpreting neutralization mechanisms]
	\label{rem:interpreting_neutralization_mechanisms}
	A reduction in PFU, FFU, endpoint positivity, or reporter signal can arise
	from different neutralization mechanisms. The agent may prevent attachment,
	block fusion or entry, aggregate particles, promote immobilization, reduce
	replication, alter spread, or change readout visibility. These mechanisms can
	produce similar dose--response curves at the level of final signal. A
	stage-resolved kernel helps identify which biological step is being deformed.
\end{remark}

\paragraph{Plaque morphology as additional information.}

A plaque assay need not be reduced to a scalar count. Plaque size, shape,
opacity, boundary sharpness, time-to-visibility, and growth rate can carry
information about replication kinetics, local spread, cell-layer susceptibility,
overlay restriction, cytopathic effect, and immune or inhibitor-mediated
restriction. Let
\[
M_i
=
(A_i,R_i,S_i,T_i,\chi_i,\ldots)
\]
denote a morphology vector for plaque \(i\), where \(A_i\) is area,
\(R_i\) is an effective radius, \(S_i\) is a shape or circularity descriptor,
\(T_i\) is time-to-visibility, and \(\chi_i\) represents intensity, opacity, or
edge structure. The enriched plaque readout is then
\begin{equation}
	Y_{\mathrm{PFU}}^{\mathrm{morph}}
	=
	\left(
	N,
	M_1,\ldots,M_N
	\right),
	\label{eq:plaque_morphology_enriched_readout}
\end{equation}
rather than \(Y_{\mathrm{PFU}}=N\) alone.

\begin{remark}[Why plaque morphology matters]
	\label{rem:why_plaque_morphology_matters}
	Plaque morphology can partially split the fiber associated with a scalar PFU
	count. Two wells with the same plaque count may have different plaque-size
	distributions, growth rates, edge structures, or time-to-visibility. These
	differences can contain information about local spread, replication kinetics,
	cell susceptibility, overlay restriction, or inhibitory effects. Thus plaque
	morphology can turn a scalar biological amplification readout into a richer
	protocol-conditioned observation.
\end{remark}

\begin{definition}[Morphology-augmented biological kernel]
	\label{def:morphology_augmented_biological_kernel}
	A morphology-augmented biological kernel maps a latent state \(x\) not only to
	a binary success or null outcome, but to a distribution over lesion
	morphologies:
	\begin{equation}
		K_E^{\mathrm{morph},\varnothing}
		(dN,dM_1,\ldots,dM_N\mid x).
		\label{eq:morphology_augmented_biological_kernel}
	\end{equation}
	This kernel contains more information than the count-only projection whenever
	morphology depends on latent replication, spread, cell-layer, overlay, or
	inhibitor parameters.
\end{definition}

\begin{remark}[Summary of statistical nonidealities]
	\label{rem:summary_statistical_nonidealities_biological_assays}
	The ideal PFU or FFU model is useful because it provides a clean baseline.
	The protocol-resolved framework does not discard that baseline; it states its
	assumptions. Overdispersion, zero inflation, spatial interaction, endpoint
	censoring, neutralization, and morphology are not merely complications. They
	are additional protocol-conditioned signals that can reveal aggregation,
	heterogeneous cell susceptibility, environmental restriction, biological
	bottlenecks, or latent subpopulations when modeled explicitly.
\end{remark}
\subsection{Mechanism-Level Identifiability and Fisher-Information Observability}
\label{subsec:mechanism_level_identifiability_fisher_observability}

The collapse mechanisms described above are not only qualitative descriptions of
experimental bias. They determine which latent parameters are identifiable under
a protocol. A parameter may be real in the latent virion--environment model and
still be invisible under a particular readout. Conversely, a parameter may become
identifiable only when the protocol intentionally drives, pins, selects,
amplifies, perturbs, or otherwise couples to the relevant latent sector.

Thus identifiability is not a property of a latent parameter alone. It is a
property of the latent model after composition with the protocol observation
operator:
\begin{empheq}[box=\fbox]{equation}
	\text{Identifiability under }E
	\quad
	\text{is a property of}
	\quad
	\mathcal M_E^{\varnothing}P_{\mathrm{ref},t}(\cdot\mid\theta).
	\label{eq:mechanism_identifiability_composed_map_principle}
\end{empheq}
Equivalently, the relevant object is not the latent ensemble by itself, but the
protocol-conditioned observed law
\[
P_{\mathrm{obs},t}^{\varnothing}(\cdot\mid E,\theta)
=
\mathcal M_E^{\varnothing}
P_{\mathrm{ref},t}(\cdot\mid\theta).
\]
This is the law that determines which parameter directions can be estimated,
which are weakly constrained, and which are collapsed into protocol-blind
directions.

This distinction matters because the same latent parameter can be strongly
visible in one protocol and nearly invisible in another. A static geometry-only
readout may be insensitive to a hidden orientational, dynamical, or
presentation-state sector. A surface-bound AFM protocol may be sensitive to
surface-conditioned stiffness but weakly sensitive to free rotational diffusion.
A dielectrophoretic or electrorotation protocol may be sensitive to
polarizability, charge asymmetry, and hydrodynamic drag, but not to the same
morphological statistics that dominate cryo-EM. A mucus-tracking protocol may be
sensitive to adhesive transport, confinement, residence time, and mobile versus
immobile fractions, while remaining blind to full body orientation or spike
presentation. A plaque assay may be sensitive to assay-conditioned infectious
activity while remaining largely blind to noninfectious physical particles
\cite{Noble2018,Chen2019,Kiss2021,Lyonnais2021,Hughes2002,Pethig2010,
	Boukari2009,Kaler2022,Baer2014}.

\subsubsection{Identifiability after mechanism-resolved collapse}
\label{subsubsec:identifiability_after_mechanism_resolved_collapse}

Let
\[
\theta
=
\left(
\theta_{\mathrm{vir}},
\theta_{\mathrm{env}},
\theta_E
\right)
\in\Theta
\]
be the protocol-resolved parameter vector. Here
\(\theta_{\mathrm{vir}}\) contains virion parameters,
\(\theta_{\mathrm{env}}\) contains environmental parameters, and
\(\theta_E\) contains protocol parameters. A mechanism-resolved protocol may
make one block more visible while confounding or suppressing another. For this
reason, identifiability should always be stated relative to the protocol kernel,
the nuisance-parameter treatment, and the experimental regime.

Let
\[
\theta
=
(\theta_a,\lambda),
\label{eq:mechanism_target_nuisance_parameter_partition}
\]
where \(\theta_a\) is the target parameter block and \(\lambda\) contains
nuisance parameters. The nuisance block may include environmental variables,
protocol settings, calibration constants, selection thresholds, noise
parameters, batch effects, reconstruction settings, staining thresholds,
cell-line conditions, or other quantities that affect the observed law but are
not the main target of inference.

A compact statement of protocol identifiability is:
\begin{empheq}[box=\fbox]{equation}
	P_{\mathrm{obs},t}^{\varnothing}
	(\cdot\mid E,\theta_a,\lambda)
	=
	P_{\mathrm{obs},t}^{\varnothing}
	(\cdot\mid E,\theta_a',\lambda')
	\quad
	\Longrightarrow
	\quad
	\theta_a=\theta_a',
	\label{eq:mechanism_protocol_identifiability_profiled}
\end{empheq}
for all admissible nuisance values \(\lambda,\lambda'\). This is the profiled
nuisance version: it asks whether a change in the target parameter can be
mimicked by changing environmental, protocol, calibration, or readout
parameters.

\begin{remark}[Why the nuisance treatment must be stated]
	\label{rem:mechanism_nuisance_treatment_must_be_stated}
	A parameter can appear identifiable when nuisance variables are artificially
	fixed, but become non-identifiable when those nuisance variables are allowed
	to vary. An AFM stiffness estimate may be confounded by tip geometry,
	indentation depth, surface adhesion, or hydration state. A DEP response may be
	confounded by medium conductivity, permittivity, viscosity, and field
	calibration. A mucus-transport parameter may be confounded by local mesh
	heterogeneity, adhesive binding, antibody cross-linking, or tracking
	threshold. A plaque count may be confounded by cell-line susceptibility,
	adsorption time, overlay composition, incubation time, staining threshold, and
	counting rule. Protocol identifiability is therefore always a statement about
	a target parameter together with a nuisance-parameter model.
\end{remark}

\begin{remark}[Structural versus practical identifiability]
	\label{rem:mechanism_structural_practical_identifiability}
	Structural identifiability asks whether a parameter could be recovered in
	principle from ideal, unlimited data generated by the specified protocol
	model. Practical identifiability asks whether it can be estimated with useful
	precision from finite, noisy data under realistic conditions
	\cite{BellmanAstrom1970,WalterPronzato1997,Raue2009}. This distinction is
	especially important in virophysics because rare conformations, weakly
	populated orientations, transient binding events, low-yield infectious events,
	and small mechanically active subpopulations may be structurally present while
	remaining practically difficult to estimate.
\end{remark}

\begin{definition}[Mechanism-level protocol equivalence class]
	\label{def:mechanism_protocol_equivalence_class}
	For a fixed protocol \(E\), define the observational equivalence class of
	\(\theta\) by
	\begin{equation}
		[\theta]_E
		=
		\left\{
		\theta'\in\Theta:
		P_{\mathrm{obs},t}^{\varnothing}
		(\cdot\mid E,\theta')
		=
		P_{\mathrm{obs},t}^{\varnothing}
		(\cdot\mid E,\theta)
		\right\}.
		\label{eq:mechanism_protocol_parameter_equivalence_class}
	\end{equation}
	If \([\theta]_E\) contains more than one parameter point, then protocol \(E\)
	cannot distinguish those latent hypotheses, even in the infinite-data limit.
\end{definition}

\begin{remark}[Protocol blindness as equivalence]
	\label{rem:mechanism_protocol_blindness_as_equivalence}
	The equivalence class \([\theta]_E\) is the parameter-space version of
	protocol blindness. Distinct latent explanations may be observationally
	equivalent because the protocol projects away the relevant distinction,
	because selection weights erase it, because a nuisance parameter compensates
	for it, or because the readout does not couple to that sector. Failure to
	observe a latent feature under one protocol should therefore be interpreted as
	protocol-limited visibility, not automatically as mechanical or biological
	absence.
\end{remark}

\subsubsection{Observable sensitivity as a mechanism diagnostic}
\label{subsubsec:observable_sensitivity_mechanism_diagnostic}

Before using the full likelihood, it is often useful to inspect the sensitivity
of reported observables to model parameters. This is a lower-dimensional,
observable-level diagnostic. It is not a substitute for the likelihood, but it
helps identify which experimental summaries respond to which latent sectors.

Let
\[
\mathbf h_E:\mathcal Y_E\rightarrow\mathbb R^M
\]
be a vector of reported observables, such as apparent radius, height,
orientation-class frequency, indentation stiffness, rupture force, effective
diffusivity, residence time, trapping probability, rotation rate, plaque count,
focus count, plaque area, or time-to-visibility. Define the
protocol-conditioned expectation vector
\begin{equation}
	\mathbf H_E(\theta)
	=
	\mathbb E_{\theta,E}
	\left[
	\mathbf h_E(Y_E)
	\mid \mathrm{det}
	\right].
	\label{eq:mechanism_protocol_conditioned_observable_expectation_vector}
\end{equation}

\begin{definition}[Protocol sensitivity matrix]
	\label{def:mechanism_protocol_sensitivity_matrix}
	The \emph{protocol sensitivity matrix} is
	\begin{empheq}[box=\fbox]{equation}
		\mathbf S_E(\theta)
		=
		\frac{\partial \mathbf H_E}{\partial\theta}
		=
		\left[
		\frac{\partial H_{E,\mu}}
		{\partial\theta_a}
		\right]_{\mu a}.
		\label{eq:mechanism_protocol_sensitivity_matrix}
	\end{empheq}
\end{definition}

\begin{remark}[Reading the sensitivity matrix]
	\label{rem:mechanism_reading_protocol_sensitivity_matrix}
	The column indexed by \(\theta_a\) indicates which reported observables
	respond to that parameter. A nearly zero column means that the parameter is
	weakly visible under the chosen protocol and observable set. Strongly
	correlated columns indicate degeneracy: two different latent mechanisms
	produce similar changes in the reported summaries. For example, a decrease in
	observed mucus mobility may be produced by stronger virion adhesion, denser
	local mucus mesh, antibody cross-linking, or a tracking threshold. A
	sensitivity matrix helps decide whether these mechanisms can be separated by
	the available observables.
\end{remark}

\begin{remark}[Why sensitivity is not the whole story]
	\label{rem:mechanism_sensitivity_not_whole_story}
	Sensitivity matrices measure how chosen summaries change. They may miss
	information contained in distributional shape, higher moments, null frequency,
	time correlations, rare events, spatial structure, or morphology. Fisher
	information extends the same principle to the full protocol-conditioned
	likelihood.
\end{remark}

\subsubsection{Fisher-information observability}
\label{subsubsec:mechanism_fisher_information_observability}

Fisher information gives the likelihood-level version of protocol observability.
It asks how sharply the protocol-conditioned likelihood changes when the
parameter vector is perturbed. Equivalently, it measures local
distinguishability of nearby parameter values under the composed map
\[
\theta
\longmapsto
P_{\mathrm{obs},t}^{\varnothing}(\cdot\mid E,\theta).
\]

\begin{empheq}[box=\fbox]{equation}
	\text{Large Fisher information}
	\;\Longleftrightarrow\;
	\text{Nearby parameter values are easier to distinguish under protocol }E.
	\label{eq:mechanism_fisher_information_plain_language}
\end{empheq}
Small Fisher information indicates weak observability, while zero Fisher
information indicates local protocol blindness.

Assume that the null-inclusive observed law admits a density or probability mass
function
\[
p_E^{\varnothing}(y\mid\theta),
\qquad
y\in\mathcal Y_E^{\varnothing},
\]
and that the usual differentiability and integrability conditions hold. The
protocol score function is
\begin{equation}
	\mathbf u_E(y;\theta)
	=
	\nabla_\theta
	\log p_E^{\varnothing}(y\mid\theta),
	\label{eq:mechanism_protocol_score_function}
\end{equation}
and the Fisher-information matrix is
\begin{empheq}[box=\fbox]{equation}
	\mathcal I_E(\theta)
	=
	\mathbb E_{Y\sim p_E^{\varnothing}(\cdot\mid\theta)}
	\left[
	\mathbf u_E(Y;\theta)
	\mathbf u_E(Y;\theta)^{\mathsf T}
	\right].
	\label{eq:mechanism_fisher_information_protocol}
\end{empheq}
Under standard regularity conditions, this is also the negative expected Hessian
of the log-likelihood,
\begin{equation}
	\mathcal I_E(\theta)
	=
	-
	\mathbb E_{Y\sim p_E^{\varnothing}(\cdot\mid\theta)}
	\left[
	\nabla_\theta^2
	\log p_E^{\varnothing}(Y\mid\theta)
	\right].
	\label{eq:mechanism_fisher_information_negative_expected_hessian}
\end{equation}
This gives the curvature interpretation: sharply curved likelihoods constrain
parameters strongly, while flat directions are weakly resolved or blind
\cite{Fisher1922,Rao1945,Cramer1946,Kay1993,VanDerVaart1998}.

For a tangent direction \(v\in T_\theta\Theta\), the directional Fisher
information is
\begin{empheq}[box=\fbox]{equation}
	\mathcal I_E(\theta;v)
	=
	v^{\mathsf T}\mathcal I_E(\theta)v
	=
	\mathbb E_{\theta,E}
	\left[
	\left(
	v^{\mathsf T}
	\nabla_\theta
	\log p_E^{\varnothing}(Y\mid\theta)
	\right)^2
	\right].
	\label{eq:mechanism_directional_fisher_information}
\end{empheq}
The direction \(v\) may represent a change in stiffness, polarizability,
adhesion strength, mucus-binding rate, field-response coefficient,
plaque-forming probability, cell-entry competence, neutralization sensitivity,
or a hidden orientation-sector parameter. If
\(v^{\mathsf T}\mathcal I_Ev\) is large, the protocol is locally sensitive to
that perturbation. If it is small, the perturbation is weakly resolved. If it is
zero, the protocol is locally blind.

The local blind subspace is therefore
\begin{empheq}[box=\fbox]{equation}
	\mathcal B_E(\theta)
	=
	\ker\mathcal I_E(\theta)
	=
	\left\{
	v\in T_\theta\Theta:
	v^{\mathsf T}\mathcal I_E(\theta)v=0
	\right\}.
	\label{eq:mechanism_local_blind_subspace}
\end{empheq}

\begin{remark}[More data versus a different protocol]
	\label{rem:mechanism_more_data_versus_different_protocol}
	For \(N\) independent observations from the same protocol,
	\[
	\mathcal I_{E,N}(\theta)
	=
	N\mathcal I_E(\theta).
	\]
	Thus more data improve precision in directions the protocol already sees.
	However, if \(v^{\mathsf T}\mathcal I_E(\theta)v=0\), then
	\(v^{\mathsf T}\mathcal I_{E,N}(\theta)v=0\) for every \(N\). More data from
	the same protocol do not make a structurally blind direction visible.
	Resolving that direction requires a different observation kernel, controlled
	protocol variation, additional prior structure, or a complementary
	measurement.
\end{remark}

\subsubsection{Practical identifiability rank}
\label{subsubsec:mechanism_practical_identifiability_rank}

Because Fisher matrices carry units and can mix parameters with different
scales, the matrix should be nondimensionalized before comparing eigenvalues.
Let \(D_\theta\) be a diagonal matrix of characteristic parameter scales. The
scaled Fisher-information matrix is
\begin{equation}
	\widehat{\mathcal I}_E(\theta)
	=
	D_\theta
	\mathcal I_E(\theta)
	D_\theta.
	\label{eq:mechanism_scaled_fisher_information}
\end{equation}
Equivalently, \(\widehat{\mathcal I}_E\) is the Fisher matrix expressed in
dimensionless local coordinates.

For a tolerance \(\delta_{\mathrm{id}}>0\), define the practical identifiability
rank by
\begin{empheq}[box=\fbox]{equation}
	r_E(\delta_{\mathrm{id}})
	=
	\#\left\{
	\lambda_j(\widehat{\mathcal I}_E):
	\lambda_j(\widehat{\mathcal I}_E)>\delta_{\mathrm{id}}
	\right\}.
	\label{eq:mechanism_identifiability_rank_protocol}
\end{empheq}
This rank estimates the number of independent parameter combinations that are
practically visible under protocol \(E\) at the chosen tolerance.

\begin{remark}[Interpretation of identifiability rank]
	\label{rem:mechanism_interpretation_identifiability_rank}
	The rank \(r_E(\delta_{\mathrm{id}})\) is not an absolute property of the
	virion. It depends on the protocol, parameter scaling, noise level, sample
	size, model class, and chosen threshold. A protocol with small rank may still
	be scientifically valuable if it measures one important sector cleanly. For
	example, a plaque assay may have limited structural identifiability while
	strongly constraining assay-conditioned infectious activity; AFM may have
	limited information about free dynamics while strongly constraining
	surface-conditioned mechanics.
\end{remark}

\subsubsection{Nuisance parameters and effective information}
\label{subsubsec:mechanism_nuisance_parameters_effective_information}

Many protocol-resolved inverse problems contain nuisance parameters that can
absorb or mimic changes in latent virion parameters. Partition the parameter
vector as
\[
\theta
=
(\theta_a,\lambda),
\]
where \(\theta_a\) is the target parameter block and \(\lambda\) contains
nuisance parameters. Write the Fisher matrix in block form:
\begin{equation}
	\mathcal I_E(\theta)
	=
	\begin{pmatrix}
		\mathcal I_{aa} & \mathcal I_{a\lambda} \\
		\mathcal I_{\lambda a} & \mathcal I_{\lambda\lambda}
	\end{pmatrix}.
	\label{eq:mechanism_fisher_block_partition}
\end{equation}

When \(\mathcal I_{\lambda\lambda}\) is invertible, the effective Fisher
information for \(\theta_a\), after accounting for nuisance directions, is the
Schur complement
\begin{empheq}[box=\fbox]{equation}
	\mathcal I_{a\mid\lambda}
	=
	\mathcal I_{aa}
	-
	\mathcal I_{a\lambda}
	\mathcal I_{\lambda\lambda}^{-1}
	\mathcal I_{\lambda a}.
	\label{eq:mechanism_effective_fisher_information_schur}
\end{empheq}

\begin{remark}[Interpretation of nuisance-corrected information]
	\label{rem:mechanism_interpretation_nuisance_corrected_information}
	The matrix \(\mathcal I_{a\mid\lambda}\) measures how much information remains
	about the target parameter after nuisance parameters are allowed to adjust. If
	\(\mathcal I_{a\mid\lambda}\) is small, then the protocol may appear sensitive
	to \(\theta_a\) when nuisance parameters are fixed, but weakly informative
	once realistic uncertainty in protocol or environmental parameters is
	included. This is a common route by which experimental collapse becomes
	practical non-identifiability.
\end{remark}

\begin{remark}[If the nuisance block is singular]
	\label{rem:mechanism_if_nuisance_block_singular}
	If \(\mathcal I_{\lambda\lambda}\) is singular or ill-conditioned, the
	nuisance parameters are themselves weakly identifiable. In that case, the
	Schur complement may be replaced by a regularized or generalized-inverse
	version, or the model may require stronger priors, additional calibration data,
	controlled protocol variation, or a reduced parameterization. A singular
	nuisance block is not just a technical inconvenience; it is evidence that the
	protocol does not separate target and nuisance directions well.
\end{remark}

\begin{remark}[Mechanism-level design implication]
	\label{rem:mechanism_level_design_implication}
	The identifiability tools in this subsection show how mechanism-resolved
	collapse can guide experimental design. If a preparation mechanism collapses
	orientation, vary preparation or acquisition geometry. If a medium-filtering
	mechanism confounds virion adhesion with mucus structure, vary particle
	surface chemistry or medium composition. If a biological amplification kernel
	confounds virion competence with cell-line susceptibility, vary cell line,
	adsorption time, overlay, or incubation. If a Fisher eigenvalue remains small,
	the remedy is not simply more data from the same protocol, but a protocol
	change that rotates the sensitivity direction.
\end{remark}

\subsection{Multi-Protocol Inference after Mechanism-Resolved Collapse}
\label{subsec:multi_protocol_inference_after_mechanism_collapse}

The mechanism-resolved view of experimental collapse changes how protocol
disagreement should be interpreted. Disagreement across protocols need not be a
direct contradiction, because different protocols may report different
conditioned projections of the same latent virion--environment system. A
cryo-EM reconstruction, an AFM force curve, a dielectrophoretic trajectory, a
mucus-tracking ensemble, and a plaque count are not competing measurements of
the same mathematical object. They are different protocol-conditioned outputs.

The appropriate multi-protocol question is therefore not whether all protocols
produce the same observed distribution. They generally should not. The question
is whether a common latent model can explain all protocol-conditioned datasets
after each protocol kernel, nuisance structure, readout map, and null channel has
been applied:
\begin{empheq}[box=\fbox]{equation}
	\text{one latent virion--environment model}
	\xrightarrow{\;\text{distinct protocol kernels}\;}
	\text{multiple observed ensembles}.
	\label{eq:multi_protocol_inference_mechanism_at_a_glance}
\end{empheq}
This is the constructive counterpart of protocol blindness. A single protocol
may collapse, select, or project away one latent sector, while another protocol
may probe, perturb, or amplify that same sector. Multi-protocol inference uses
these differences as information rather than treating them only as disagreement.

\begin{remark}[Why raw agreement is the wrong criterion]
	\label{rem:mechanism_multi_protocol_raw_agreement_wrong}
	Raw agreement is usually the wrong criterion for comparing protocols. Cryo-EM
	may report selected structural classes; AFM may report surface-conditioned
	mechanical response; DEP or electrorotation may report field-conditioned
	dielectric response; mucus tracking may report medium-filtered transport; and
	plaque assays may report assay-conditioned infectious activity. Agreement
	should be tested after each protocol kernel has acted, not before.
\end{remark}

\subsubsection{Shared latent parameters and protocol-local parameters}
\label{subsubsec:mechanism_shared_latent_protocol_local_parameters}

Multi-protocol inference begins by separating parameters that are intended to be
common across protocols from parameters that belong to individual protocol
kernels. This separation is essential because mechanism-resolved collapse often
enters through protocol-local quantities.

\begin{definition}[Multi-protocol parameter structure]
	\label{def:mechanism_multi_protocol_parameter_structure}
	Let
	\[
	\theta
	=
	\left(
	\theta_{\mathrm{vir}},
	\theta_{\mathrm{env}}
	\right)
	\]
	denote the shared latent parameter vector, containing virion and environmental
	parameters intended to describe the same underlying system across protocols.
	For protocols \(E_1,\ldots,E_M\), let
	\[
	\lambda_j
	=
	\lambda_{E_j},
	\qquad
	j=1,\ldots,M,
	\]
	denote protocol-local nuisance or calibration parameters. The full
	multi-protocol parameter set is
	\begin{equation}
		\Theta_{\mathrm{multi}}
		=
		\left(
		\theta,
		\lambda_1,\ldots,\lambda_M
		\right).
		\label{eq:mechanism_multi_protocol_parameter_set}
	\end{equation}
\end{definition}

The shared parameter vector \(\theta\) contains quantities intended to persist
across protocols: radius distribution, spike-length scale, effective charge
range, polarizability anisotropy, compliance, orientational stiffness, damping
scale, adhesion scale, or latent infectivity structure. The protocol-local
parameters \(\lambda_j\) contain quantities such as field amplitude, field
frequency, medium conductivity, surface chemistry, tip geometry, imaging noise,
frame rate, grid-preparation variables, particle-selection thresholds, overlay
composition, incubation time, staining threshold, or reconstruction filters.

\begin{remark}[Why this separation is essential]
	\label{rem:mechanism_shared_protocol_separation_essential}
	Without the shared/protocol-local separation, multi-protocol comparison can
	become misleading. If an AFM-derived stiffness differs from a stiffness-like
	quantity inferred from another assay, the discrepancy may reflect surface
	adhesion, indentation depth, hydration, or tip geometry rather than a true
	change in the virion. If plaque titers differ across cell lines, the
	discrepancy may reflect cell susceptibility or overlay conditions rather than
	a different physical particle concentration. If tracking statistics differ
	across mucus samples, the difference may reflect the medium rather than the
	virion alone. The decomposition
	\((\theta,\lambda_1,\ldots,\lambda_M)\) is the statistical expression of this
	distinction.
\end{remark}

\subsubsection{Protocol-specific forward models and likelihoods}
\label{subsubsec:mechanism_protocol_specific_forward_models_likelihoods}

Let
\[
\mathcal D
=
\{
\mathcal D_{E_1},\ldots,\mathcal D_{E_M}
\}
\]
be datasets collected under protocols \(E_1,\ldots,E_M\). Each protocol has its
own augmented observation space
\[
\mathcal Y_{E_j}^{\varnothing}
=
\mathcal Y_{E_j}\cup\{\varnothing\},
\]
its own null-inclusive observation operator
\[
\mathcal M_{E_j}^{\varnothing}(\lambda_j),
\]
and its own protocol-local nuisance parameters \(\lambda_j\).

For each protocol,
\begin{empheq}[box=\fbox]{equation}
	P_{\mathrm{obs},t}^{(j),\varnothing}
	(\cdot\mid \theta,\lambda_j)
	=
	\mathcal M_{E_j}^{\varnothing}(\lambda_j)
	P_{\mathrm{ref},t}(\cdot\mid\theta).
	\label{eq:mechanism_protocol_specific_forward_model_multi}
\end{empheq}
Thus each dataset is interpreted as a distinct protocol-conditioned image of the
same shared latent model.

\begin{definition}[Multi-protocol likelihood]
	\label{def:mechanism_multi_protocol_likelihood}
	Suppose that the datasets are conditionally independent given the shared
	latent parameter vector \(\theta\) and protocol-local nuisance parameters
	\(\lambda_1,\ldots,\lambda_M\). The multi-protocol likelihood is
	\begin{empheq}[box=\fbox]{equation}
		\mathcal L_{\mathrm{multi}}
		(\theta,\lambda_1,\ldots,\lambda_M)
		=
		\prod_{j=1}^{M}
		\mathcal L_{E_j}^{\varnothing}
		(\theta,\lambda_j;\mathcal D_{E_j}).
		\label{eq:mechanism_multi_protocol_likelihood}
	\end{empheq}
	Equivalently,
	\begin{equation}
		\log \mathcal L_{\mathrm{multi}}
		=
		\sum_{j=1}^{M}
		\log
		\mathcal L_{E_j}^{\varnothing}
		(\theta,\lambda_j;\mathcal D_{E_j}).
		\label{eq:mechanism_multi_protocol_log_likelihood}
	\end{equation}
\end{definition}

\begin{remark}[Conditional independence is a modeling assumption]
	\label{rem:mechanism_conditional_independence_multi_protocol}
	The factorized likelihood assumes conditional independence of the datasets
	given the shared and protocol-local parameters. This is a useful first model,
	but it is not automatic. Shared batches, common sample preparation, common
	calibration errors, correlated cell-culture conditions, repeated measurements
	of the same stock, or shared field-of-view effects can introduce dependence
	between datasets. In such cases, the joint likelihood should include shared
	nuisance variables or an explicit dependence structure.
\end{remark}

\begin{definition}[Bayesian multi-protocol posterior]
	\label{def:mechanism_bayesian_multi_protocol_posterior}
	Given a prior
	\[
	q_0(\theta,\lambda_1,\ldots,\lambda_M),
	\]
	the multi-protocol posterior is
	\begin{empheq}[box=\fbox]{equation}
		q(\theta,\lambda_1,\ldots,\lambda_M\mid\mathcal D)
		=
		\frac{
			\mathcal L_{\mathrm{multi}}
			(\theta,\lambda_1,\ldots,\lambda_M)
			q_0(\theta,\lambda_1,\ldots,\lambda_M)
		}{
			\displaystyle
			\int
			\mathcal L_{\mathrm{multi}}
			(\vartheta,\ell_1,\ldots,\ell_M)
			q_0(\vartheta,\ell_1,\ldots,\ell_M)
			\,
			d\vartheta\,d\ell_1\cdots d\ell_M
		}.
		\label{eq:mechanism_bayesian_multi_protocol_posterior}
	\end{empheq}
	The marginal posterior for the shared latent parameters is obtained by
	integrating over protocol-local nuisance parameters:
	\begin{equation}
		q(\theta\mid\mathcal D)
		=
		\int
		q(\theta,\lambda_1,\ldots,\lambda_M\mid\mathcal D)
		\,
		d\lambda_1\cdots d\lambda_M.
		\label{eq:mechanism_multi_protocol_marginal_shared_posterior}
	\end{equation}
\end{definition}

\begin{remark}[Why marginalization matters]
	\label{rem:mechanism_marginalization_matters_multi_protocol}
	The marginal posterior \(q(\theta\mid\mathcal D)\) represents what all
	protocols jointly imply about the shared latent model after protocol-level
	uncertainty has been accounted for. Point estimates that ignore the
	\(\lambda_j\) can overstate agreement or create artificial disagreement by
	treating protocol-specific effects as intrinsic latent properties.
\end{remark}

\subsubsection{Multi-protocol consistency as shared latent explanation}
\label{subsubsec:mechanism_multi_protocol_consistency}

Multi-protocol consistency is not equality of raw datasets. It is the existence
of a shared latent explanation after the appropriate protocol kernels have acted.

\begin{definition}[Multi-protocol consistency]
	\label{def:mechanism_multi_protocol_consistency}
	A latent model
	\[
	P_{\mathrm{ref},t}(dx\mid\theta)
	\]
	is \emph{multi-protocol consistent} with observed datasets
	\(\mathcal D_{E_1},\ldots,\mathcal D_{E_M}\) if there exists a shared
	parameter \(\theta\) and admissible protocol-local nuisance parameters
	\(\lambda_1,\ldots,\lambda_M\) such that
	\begin{empheq}[box=\fbox]{equation}
		P_{\mathrm{obs},t}^{(j),\varnothing}(\cdot)
		\approx
		\mathcal M_{E_j}^{\varnothing}(\lambda_j)
		P_{\mathrm{ref},t}(\cdot\mid\theta),
		\qquad
		j=1,\ldots,M.
		\label{eq:mechanism_multi_protocol_consistency_condition}
	\end{empheq}
	The approximation symbol represents statistical agreement under a specified
	likelihood, discrepancy, posterior predictive check, calibration criterion, or
	model-comparison rule.
\end{definition}

\begin{remark}[What consistency means]
	\label{rem:mechanism_multi_protocol_consistency_meaning}
	Consistency means that the same latent model can generate each observed
	dataset after the corresponding protocol kernel has acted. In this sense,
	apparent disagreement can be diagnostic. It may reveal which states survive
	preparation, which orientations are pinned by surfaces, which particles
	respond electrically, which trajectories are filtered by mucus
	microstructure, or which states are infectious under a given biological assay.
\end{remark}

\begin{definition}[Protocol inconsistency]
	\label{def:mechanism_protocol_inconsistency_multi}
	A collection of protocols is \emph{protocol-inconsistent} relative to a model
	class \(\mathcal P\) if no admissible latent model
	\(P_{\mathrm{ref},t}(\cdot\mid\theta)\in\mathcal P\), together with
	admissible protocol-local nuisance parameters, can explain all observed
	datasets under the specified protocol kernels.
\end{definition}

\begin{remark}[Interpreting protocol inconsistency]
	\label{rem:mechanism_interpreting_protocol_inconsistency_multi}
	Protocol inconsistency is not automatically experimental failure. It may mean
	that the latent model is missing a relevant state variable, that a protocol
	kernel is misspecified, that a nuisance parameter was uncontrolled, that the
	sample changed between protocols, that a biological assay selected a
	functional subpopulation absent from the mechanical model, or that a hidden
	environmental variable is driving the discrepancy. Inconsistency is therefore
	a guide to model revision.
\end{remark}

\subsubsection{Complementary Fisher information}
\label{subsubsec:mechanism_complementary_fisher_information}

Multi-protocol inference improves identifiability when protocols provide
information in different parameter directions. The local version of this
statement is Fisher-information additivity.

\begin{proposition}[Complementary protocols improve local identifiability]
	\label{prop:mechanism_complementary_protocols_improve_identifiability}
	Let \(\mathcal I_{E_j}(\theta)\) be the Fisher-information matrix associated
	with protocol \(E_j\). If the datasets are conditionally independent given
	\(\theta\) and nuisance parameters are fixed, calibrated, marginalized, or
	appropriately profiled, then the Fisher information for the shared parameter
	has the additive form
	\begin{empheq}[box=\fbox]{equation}
		\mathcal I_{\mathrm{multi}}(\theta)
		=
		\sum_{j=1}^{M}
		\mathcal I_{E_j}(\theta).
		\label{eq:mechanism_multi_protocol_fisher_additivity}
	\end{empheq}
	Consequently, protocols with different sensitivity directions can jointly
	identify parameter combinations that none of them identify alone.
\end{proposition}

\begin{proof}
	For conditionally independent datasets, the joint likelihood factorizes:
	\[
	\mathcal L_{\mathrm{multi}}(\theta)
	=
	\prod_{j=1}^{M}
	\mathcal L_{E_j}(\theta).
	\]
	Therefore,
	\[
	\log\mathcal L_{\mathrm{multi}}(\theta)
	=
	\sum_{j=1}^{M}
	\log\mathcal L_{E_j}(\theta).
	\]
	The score is the gradient of the log-likelihood, so the total score is the sum
	of the protocol-specific scores. Under the usual regularity conditions, Fisher
	information is the expected outer product of the score, or equivalently the
	negative expected Hessian of the log-likelihood. Conditional independence
	implies that the protocol-specific score cross terms vanish in expectation,
	so the Fisher-information matrices add.
\end{proof}

\begin{remark}[How complementary information works]
	\label{rem:mechanism_complementary_information_works}
	A geometry-rich protocol may identify size and morphology but not dielectric
	response. A field-driven protocol may identify polarizability but not native
	morphology. An AFM protocol may identify surface-conditioned stiffness but not
	free rotational mobility. A mucus-tracking protocol may identify transport and
	binding parameters but not internal branch structure. A plaque assay may
	identify infectious activity but not total physical particle number. Together,
	these protocols can constrain a latent model more tightly than any single
	protocol can.
\end{remark}

\begin{definition}[Incremental directional information gain]
	\label{def:mechanism_incremental_information_gain_multi_protocol}
	Let \(\mathcal E_{\mathrm{old}}\) denote an existing protocol set and
	\(E_{\mathrm{new}}\) a proposed additional protocol. For a target direction
	\(v\in T_\theta\Theta\), define the incremental directional information gain
	by
	\begin{equation}
		\Delta\mathcal I(E_{\mathrm{new}};v)
		=
		v^{\mathsf T}
		\mathcal I_{E_{\mathrm{new}}}(\theta)
		v.
		\label{eq:mechanism_incremental_information_gain}
	\end{equation}
	If \(v\) is weakly observed by the existing protocol set but strongly observed
	by \(E_{\mathrm{new}}\), then \(E_{\mathrm{new}}\) is complementary for that
	parameter direction.
\end{definition}

\begin{remark}[Design principle]
	\label{rem:mechanism_multi_protocol_design_principle}
	The best complementary protocol is not necessarily the one that resembles the
	existing protocol most closely. It is the protocol whose kernel exposes a
	previously blind or nuisance-confounded direction. In experimental-design
	terms, one should choose protocol variations that rotate, enlarge, or sharpen
	the Fisher-information ellipsoid in the target parameter space.
\end{remark}

\subsubsection{Cross-Protocol Prediction}
\label{subsubsec:mechanism_cross_protocol_prediction}

A useful test of a shared latent model is cross-protocol prediction. One fits,
constrains, or updates the latent model using one protocol and then asks whether
the same latent model predicts the outcome of a second protocol after the second
protocol kernel is applied. This is stronger than asking whether the first
protocol is internally well fit. It asks whether the inferred latent
virion--environment model transfers across a different observation mechanism.

\begin{definition}[Cross-protocol prediction]
	\label{def:mechanism_cross_protocol_prediction}
	Suppose protocol \(E_a\) has produced data \(\mathcal D_{E_a}\). Let
	\(q(\theta\mid\mathcal D_{E_a})\) denote the posterior or fitted uncertainty
	distribution for the shared latent parameters after accounting for
	\(E_a\)-specific nuisance parameters. The predicted null-inclusive observed
	law under a second protocol \(E_b\) is
	\begin{empheq}[box=\fbox]{equation}
		\begin{aligned}
			P_{\mathrm{pred}}^{(b),\varnothing}
			(S\mid \mathcal D_{E_a})
			&=
			\int
			P_{\mathrm{obs},t}^{(b),\varnothing}
			(S\mid E_b,\theta,\lambda_b)
			\\
			&\quad \times
			q(\theta,\lambda_b\mid\mathcal D_{E_a})
			\,d\theta\,d\lambda_b,
			\qquad
			S\in\Sigma_{Y,E_b}^{\varnothing}.
		\end{aligned}
		\label{eq:mechanism_cross_protocol_prediction}
	\end{empheq}
	Here \(\lambda_b\) denotes nuisance or calibration parameters for the
	predicted protocol \(E_b\).
\end{definition}

\begin{remark}[Why cross-protocol prediction is useful]
	\label{rem:mechanism_cross_protocol_prediction_useful}
	Cross-protocol prediction tests whether a latent model inferred from one
	projection predicts another projection. A structural model inferred from
	cryo-EM may be asked to predict an AFM deformation response, an
	electrorotation spectrum, a mucus-transport distribution, or an infectivity
	readout only after the corresponding protocol kernel is applied. Failure of
	this prediction may indicate a missing latent variable, a misspecified
	protocol kernel, an uncontrolled nuisance parameter, or a functional
	subpopulation not captured by the original model.
\end{remark}

\begin{remark}[Prediction should include protocol uncertainty]
	\label{rem:mechanism_prediction_should_include_protocol_uncertainty}
	Cross-protocol prediction should propagate uncertainty in the predicted
	protocol's nuisance parameters \(\lambda_b\). A prediction that conditions on
	a single fixed overlay composition, surface adhesion value, field calibration,
	reconstruction threshold, or staining criterion may appear overly precise.
	Protocol uncertainty is part of the predictive distribution, not an external
	correction.
\end{remark}

\begin{remark}[Predictive failure as information]
	\label{rem:mechanism_predictive_failure_as_information}
	A failed cross-protocol prediction is not automatically a negative result. It
	can identify where the latent model or protocol kernel is incomplete. For
	example, a cryo-EM-derived structural model that fails to predict AFM response
	may be missing hydration, surface adhesion, or deformation variables. A
	tracking-derived transport model that fails to predict plaque formation may be
	missing infectivity, entry, or replication competence. A plaque-derived
	infectivity model that fails to predict particle-count or genome-count data may
	be collapsing defective, damaged, neutralized, or non-plaque-forming sectors
	into a single functional readout.
\end{remark}

\subsubsection{Practical Workflow for Multi-Protocol Inference}
\label{subsubsec:mechanism_practical_workflow_multi_protocol}

A practical multi-protocol analysis can be organized as a sequence of modeling
decisions. The purpose is not to force all assays into the same observed space,
but to specify how each protocol samples, transforms, selects, amplifies, and
reports the shared latent system.

\begin{enumerate}[label=(\roman*),leftmargin=2.2em]
	
	\item \textbf{Define the shared latent target.}
	Specify which parameters are intended to be common across protocols:
	morphology, charge, compliance, polarizability, adhesion, infectivity
	competence, environmental transport parameters, or collective branch
	structure.
	
	\item \textbf{Define protocol-specific kernels.}
	For each protocol, specify
	\[
	\Pi_{E_j}^{\mathrm{lat}},
	\qquad
	s_{E_j},
	\qquad
	R_{E_j},
	\qquad
	\varnothing.
	\]
	State which components are known, calibrated, estimated, approximated, or
	left as nuisance structure.
	
	\item \textbf{Separate shared parameters from protocol-local parameters.}
	Decide which parameters belong to the shared latent vector \(\theta\) and
	which belong to the protocol-local nuisance vectors \(\lambda_j\). This
	prevents preparation, readout, surface, medium, field, cell-line, or
	threshold effects from being absorbed into intrinsic virion parameters.
	
	\item \textbf{Write protocol-specific likelihoods.}
	Use null-inclusive likelihoods when null outcomes, rejection rates, failed
	tracks, missing plaques, endpoint-negative wells, below-threshold signals, or
	detection yields are available.
	
	\item \textbf{Fit jointly or test cross-protocol prediction.}
	Either estimate all protocols jointly through
	\(\mathcal L_{\mathrm{multi}}\), or fit one protocol and test the predicted
	output of another protocol through
	\(P_{\mathrm{pred}}^{(b),\varnothing}\).
	
	\item \textbf{Inspect identifiability.}
	Use sensitivity matrices, Fisher information, profile likelihoods,
	posterior correlations, posterior predictive checks, or discrepancy
	functionals to identify blind, weakly constrained, or nuisance-confounded
	directions.
	
	\item \textbf{Revise the latent model or protocol kernels.}
	If consistency fails, determine whether the failure is best explained by a
	missing latent sector, a misspecified kernel, uncontrolled nuisance variation,
	sample heterogeneity, batch effects, or a genuine biological difference.
\end{enumerate}

\begin{table}[H]
	\centering
	\caption{
		Examples of protocol-specific inferential strengths. The entries are
		schematic: real studies may combine several sectors, and each entry
		depends on the detailed protocol kernel.
	}
	\label{tab:mechanism_protocol_inferential_strengths}
	\renewcommand{\arraystretch}{1.18}
	\begin{tabularx}{0.98\linewidth}{@{}p{0.19\linewidth}p{0.33\linewidth}X@{}}
		\toprule
		\textbf{Protocol}
		&
		\textbf{Strongly visible sectors}
		&
		\textbf{Typical latent quantities inferred}
		\\
		\midrule
		
		Cryo-EM / cryo-ET
		&
		Geometry, conformation, spike distribution, particle classes, selected
		orientations
		&
		Shape distributions, preserved conformational classes,
		orientation-selection profiles, structural heterogeneity.
		\\[0.45em]
		
		AFM
		&
		Surface-conditioned morphology, indentation response, rupture or
		deformation pathways
		&
		Effective stiffness, compliance, adhesion, surface-conditioned height,
		mechanical fragility.
		\\[0.45em]
		
		DEP / electrorotation
		&
		Field response, trapping, crossover behavior, signed rotation,
		polarization-dependent motion
		&
		Effective polarizability, dielectric relaxation scale, charge response,
		rotational drag under forcing.
		\\[0.45em]
		
		Mucus or gel tracking
		&
		Transport, confinement, residence time, anomalous diffusion, adhesive
		trapping
		&
		Effective diffusivity, binding strength, mesh interaction, viscoelastic
		or heterogeneous transport parameters.
		\\[0.45em]
		
		Plaque / focus / endpoint assay
		&
		Infectious amplification, local spread, cytopathic effect, staining or
		reporter visibility
		&
		Infectious-unit concentration, cell-entry competence, assay-conditioned
		replication or spread efficiency.
		\\
		\bottomrule
	\end{tabularx}
\end{table}

\subsubsection{Branch-Level Inverse Inference}
\label{subsubsec:mechanism_branch_level_inverse_inference}

If the latent theory contains collective branches, then experimental collapse
also determines which branches are visible. A branch may exist mechanically but
remain experimentally hidden if the protocol does not couple to the sector in
which that branch primarily lives. Conversely, an active protocol can reveal a
branch by coupling to a variable that passive imaging, center tracking, or
scalar counting would not resolve.

This point is the dynamical analogue of protocol-conditioned observation. Many
mechanically interesting degrees of freedom are not reported as complete normal
modes. They may appear indirectly through relaxation rates, force--indentation
response, field-driven rotation, fluctuation spectra, orientation statistics,
transport-state switching, or biological response. Branch-level inference is
therefore an inverse problem:
\begin{empheq}[box=\fbox]{equation}
	\text{branch-sensitive data}
	\quad
	\Longrightarrow
	\quad
	\text{latent branch parameters}
	\quad
	\text{through a protocol-specific visibility map}.
	\label{eq:mechanism_branch_level_inverse_inference_at_a_glance}
\end{empheq}

\begin{definition}[Latent branch parameter vector]
	\label{def:mechanism_latent_branch_parameter_vector}
	For a collective branch \(n\), define a reduced latent branch parameter vector
	\begin{equation}
		\theta_{\mathsf B_n}
		=
		\left(
		\omega_n(\mathbf k),
		\Gamma_n(\mathbf k),
		P_{\mathrm{tilt}}^{(n)}(\mathbf k),
		\Pi_{\mathrm{hyb}}^{(n)}(\mathbf k),
		\Pi_{\mathrm{dir}}^{(n)}(\mathbf k),
		\Pi_{\mathrm{stab}}^{(n)}(\mathbf k)
		\right)_{\mathbf k\in\mathsf K}.
		\label{eq:mechanism_latent_branch_parameter_vector}
	\end{equation}
	Here \(\omega_n\) is a branch frequency, \(\Gamma_n\) is a damping or decay
	rate, \(P_{\mathrm{tilt}}^{(n)}\) is a tilt or presentation participation
	weight, \(\Pi_{\mathrm{hyb}}^{(n)}\) measures hybridization between sectors,
	\(\Pi_{\mathrm{dir}}^{(n)}\) measures directional anisotropy, and
	\(\Pi_{\mathrm{stab}}^{(n)}\) measures stability or persistence. The set
	\(\mathsf K\) denotes the wavevectors, spatial scales, finite-size modes, or
	experimentally accessible fluctuation modes probed by the protocol or
	simulation.
\end{definition}

\begin{remark}[Interpretation of branch parameters]
	\label{rem:mechanism_interpretation_branch_parameters}
	The vector \(\theta_{\mathsf B_n}\) is a reduced summary of a branch, not a
	claim that all branch information is directly measurable. In simulation, these
	quantities may be extracted from a dynamical matrix, covariance spectrum,
	normal-mode decomposition, or response calculation. In experiments, they may
	be inferred only through protocol-sensitive observables. This is why branch
	visibility must be tied to a specific readout and protocol kernel
	\cite{Hsieh2023Dynamics,PerillaSchulten2017,Bruinsma2021}.
\end{remark}

\begin{definition}[Branch-readout kernel]
	\label{def:mechanism_branch_readout_kernel}
	A \emph{branch-readout kernel} is a protocol-dependent map
	\begin{equation}
		R_E^{\mathrm{br}}
		(dy_{\mathrm{br}}\mid\theta_{\mathsf B_n})
		\label{eq:mechanism_branch_readout_kernel}
	\end{equation}
	that describes how latent branch information is projected into an observed
	branch-sensitive datum \(y_{\mathrm{br}}\). Examples include a relaxation
	rate, fluctuation amplitude, deformation pattern, alignment statistic,
	finite-frequency response peak, field-driven rotation signal, or
	orientation-sensitive response amplitude.
\end{definition}

\begin{remark}[Why \(R_E^{\mathrm{br}}\) is reduced notation]
	\label{rem:mechanism_branch_readout_kernel_reduced_notation}
	The notation \(R_E^{\mathrm{br}}\) denotes a reduced branch-sensitive readout
	model, not the full null-inclusive protocol kernel
	\(K_E^{\varnothing}\). The full kernel still contains latent transformation,
	survival or selection, readout, and null observation. The branch-readout
	kernel isolates the component of the readout that is informative about
	branch-level structure.
\end{remark}

\begin{definition}[Branch-sensitive observable]
	\label{def:mechanism_branch_sensitive_observable}
	Let
	\[
	h_{\mathrm{br}}:\mathcal Y_E\rightarrow\mathbb R^m
	\]
	be a branch-sensitive observable under protocol \(E\). Examples include a
	spectral peak, a relaxation time, a response amplitude, an indentation-mode
	coefficient, a field-driven rotation rate, an orientation-class statistic, or
	a projection of a trajectory onto a mode-like basis. Its protocol-conditioned
	expectation is
	\begin{equation}
		\mathbf H_{\mathrm{br},E}
		(\theta_{\mathsf B_n})
		=
		\mathbb E_E
		\left[
		h_{\mathrm{br}}(Y)
		\mid
		\theta_{\mathsf B_n},
		\mathrm{det}
		\right].
		\label{eq:mechanism_branch_sensitive_expectation}
	\end{equation}
\end{definition}

\begin{definition}[Branch observability score]
	\label{def:mechanism_branch_observability_score}
	The branch observability score for branch \(n\) at wavevector or scale
	\(\mathbf k\) is
	\begin{empheq}[box=\fbox]{equation}
		\mathcal V_n(E,\mathbf k)
		=
		\frac{
			\left\|
			D_{\theta_{\mathsf B_n}(\mathbf k)}
			\mathbf H_{\mathrm{br},E}
			\right\|
		}{
			V_0+
			\left\|
			D_{\theta_{\mathsf B_n}(\mathbf k)}
			\mathbf H_{\mathrm{br},E}
			\right\|
		},
		\qquad
		V_0>0.
		\label{eq:mechanism_branch_observability_score}
	\end{empheq}
	Here \(D_{\theta_{\mathsf B_n}(\mathbf k)}\mathbf H_{\mathrm{br},E}\)
	denotes the local sensitivity of the branch-sensitive observable to the
	branch parameters at \(\mathbf k\). The constant \(V_0\) sets the scale at
	which the score transitions from weak to strong visibility.
\end{definition}

\begin{remark}[Interpretation of branch observability]
	\label{rem:mechanism_interpretation_branch_observability}
	Values near zero indicate that the branch is effectively hidden under the
	protocol and chosen branch readout. Values near one indicate that the protocol
	is sensitive to that branch. The score is not an intrinsic property of the
	branch alone. It depends on the protocol, readout, noise level, scale
	normalization, and nuisance parameters.
\end{remark}

\begin{definition}[Fisher branch observability]
	\label{def:mechanism_fisher_branch_observability}
	If the branch-sensitive observed law admits a density or mass function
	\[
	p_E^{\mathrm{br}}
	(y_{\mathrm{br}}\mid\theta_{\mathsf B_n}),
	\]
	then the branch-level Fisher information is
	\begin{equation}
		\mathcal I_{E}^{\mathrm{br}}
		(\theta_{\mathsf B_n})
		=
		\mathbb E
		\left[
		\nabla_{\theta_{\mathsf B_n}}
		\log p_E^{\mathrm{br}}(Y_{\mathrm{br}}\mid\theta_{\mathsf B_n})
		\nabla_{\theta_{\mathsf B_n}}
		\log p_E^{\mathrm{br}}(Y_{\mathrm{br}}\mid\theta_{\mathsf B_n})^{\mathsf T}
		\right].
		\label{eq:mechanism_branch_level_fisher_information}
	\end{equation}
	A branch direction \(v_{\mathrm{br}}\) is locally invisible under protocol
	\(E\) if
	\begin{equation}
		v_{\mathrm{br}}^{\mathsf T}
		\mathcal I_{E}^{\mathrm{br}}
		(\theta_{\mathsf B_n})
		v_{\mathrm{br}}
		=
		0.
		\label{eq:mechanism_branch_fisher_blind_direction}
	\end{equation}
\end{definition}

\begin{proposition}[Orientation-sensitive protocols can reveal tilt branches]
	\label{prop:mechanism_orientation_sensitive_protocols_reveal_tilt_branches}
	Suppose branch \(n\) has nonzero tilt or presentation participation,
	\[
	P_{\mathrm{tilt}}^{(n)}(\mathbf k)>0.
	\]
	If protocol \(E\) includes a branch-sensitive observable whose expectation
	depends nontrivially on the tilt or presentation sector through field
	coupling, surface alignment, orientation-sensitive contrast, or
	torque-sensitive response, then
	\[
	\mathcal V_n(E,\mathbf k)>0
	\]
	for the corresponding branch readout, even if passive center-tracking
	observability is negligible.
\end{proposition}

\begin{proof}
	By assumption, the branch readout depends nontrivially on the tilt or
	presentation sector. Thus, for at least one component of the branch parameter
	vector,
	\[
	\frac{\partial
		\mathbf H_{\mathrm{br},E}}
	{\partial P_{\mathrm{tilt}}^{(n)}(\mathbf k)}
	\neq
	0.
	\]
	Therefore
	\[
	\left\|
	D_{\theta_{\mathsf B_n}(\mathbf k)}
	\mathbf H_{\mathrm{br},E}
	\right\|
	>
	0,
	\]
	and Eq.~\eqref{eq:mechanism_branch_observability_score} gives
	\(\mathcal V_n(E,\mathbf k)>0\). In a passive center-tracking protocol, the
	corresponding derivative may vanish or be small because the readout projects
	primarily onto translational coordinates rather than orientation or
	presentation variables.
\end{proof}

\begin{remark}[Biological interpretation of branch-level inference]
	\label{rem:mechanism_biological_interpretation_branch_inference}
	For spike-bearing virions, branch-level inference asks whether collective
	motion involves only center displacement or also organized changes in
	spike-bearing presentation. This is relevant to surface-bound layers,
	field-conditioned clusters, dense aggregates, mucus-constrained assemblies,
	and orientation-biased preparation. A branch that appears as simple
	translational relaxation in one protocol may contain hidden presentation
	dynamics that become visible when orientation-sensitive forcing or readout is
	added.
\end{remark}

\begin{remark}[Branch-level inference as experimental design]
	\label{rem:mechanism_branch_level_inference_experimental_design}
	The practical design question is: which protocol should be added to make a
	branch visible? If the branch is predicted to live mostly in translation,
	tracking or fluctuation analysis may be sufficient. If it lives in
	orientation, presentation, or torque response, field coupling,
	polarization-sensitive readout, surface alignment, or orientation-sensitive
	imaging may be required. If it lives in deformation, mechanical loading may be
	the useful perturbation. Branch-level inverse inference therefore turns latent
	branch structure into an experimental design problem.
\end{remark}

\subsubsection{Design Principles for Protocol-Resolved Inference}
\label{subsubsec:mechanism_design_principles_protocol_resolved_inference}

The formalism above suggests practical design principles. These principles are
not additional assumptions. They follow from treating experiments as protocol
kernels: each experimental setting transforms, selects, amplifies, detects, and
reads out the latent virion--environment ensemble in a specific way.

The guiding design question is:
\begin{empheq}[box=\fbox]{equation}
	\text{Which latent sectors does this protocol make visible,}
	\qquad
	\text{and Which sectors does it collapse?}
	\label{eq:mechanism_design_question_protocol_resolved_inference}
\end{empheq}

\begin{enumerate}[label=(\roman*),leftmargin=2.2em]
	
	\item \textbf{Report protocol variables as model parameters.}
	
	Field frequency, ionic strength, conductivity, surface chemistry, frame
	rate, exposure time, tip geometry, force ramp rate, particle-selection
	thresholds, cell line, overlay composition, incubation time, staining rule,
	counting criterion, grid chemistry, vitrification route, and preparation
	timing are not peripheral metadata. They are parameters of
	\(\mathcal M_E^{\varnothing}\).
	
	\item \textbf{Preserve the null channel when possible.}
	
	Rejected particles, nonlocalized trajectories, failed reconstructions,
	particles lost from the field of view, absent plaques, missing foci,
	below-threshold reporter signals, and excluded tracks carry information
	about \(s_E\), \(\eta_E\), and the null channel \(\varnothing\). Removing
	these outcomes too early can erase the denominator of the experiment.
	
	\item \textbf{Use controlled protocol variation.}
	
	A single protocol may leave important parameter directions unidentifiable.
	Controlled variation changes the observation kernel and can expose different
	latent sectors. The purpose of protocol variation is not merely to check
	robustness; it is to change the kernel in a known direction.
	
	\item \textbf{Separate intrinsic parameters from protocol-conditioned
		parameters.}
	
	A stiffness measured by AFM is surface- and loading-conditioned unless
	modeled otherwise. A diffusion coefficient in mucus is medium-conditioned. A
	DEP crossover frequency is field- and medium-conditioned. A cryo-EM
	orientation distribution is preparation- and reconstruction-conditioned. A
	plaque count is cell-line-, overlay-, incubation-, staining-, and
	threshold-conditioned.
	
	\item \textbf{Compare protocols through a shared latent model.}
	
	Cryo-EM, AFM, DEP/electrorotation, plaque assays, and tracking need not agree
	at the observation level. The appropriate consistency test is whether one
	latent model can reproduce each dataset after the corresponding protocol
	kernel is applied:
	\begin{equation}
		P_{\mathrm{obs},t}^{(j),\varnothing}
		(\cdot)
		\approx
		\mathcal M_{E_j}^{\varnothing}(\lambda_j)
		P_{\mathrm{ref},t}(\cdot\mid\theta).
		\label{eq:mechanism_design_principle_shared_latent_model}
	\end{equation}
	
	\item \textbf{Design complementary protocols around blind directions.}
	
	A protocol should be paired with another protocol when its Fisher information
	has weak or zero directions that matter for the scientific question. In
	Fisher-information language, the aim is to choose \(E_{\mathrm{new}}\) such
	that
	\begin{equation}
		v^{\mathsf T}
		\mathcal I_{E_{\mathrm{new}}}(\theta)
		v
		>
		0
		\label{eq:mechanism_design_principle_complementary_fisher}
	\end{equation}
	for a direction \(v\) that is weakly resolved or blind under the existing
	protocol set.
	
	\item \textbf{Do not confuse invisibility with absence.}
	
	A hidden branch, orientation sector, force--torque coupling, adhesive state,
	defective-particle sector, or infectious subpopulation may be real but
	unobservable under a particular readout. The stronger claim that such a
	sector is absent requires a protocol sensitive to that sector, a
	multi-protocol argument that rules it out, or a mechanistic model showing
	that the sector cannot be populated under the relevant conditions.
	
	\item \textbf{Treat protocol design as part of model design.}
	
	Choosing a protocol is choosing which projection of the latent system will be
	measured. Choosing protocol variations is choosing which degeneracies may be
	broken. Choosing what to record as null, rejected, or below threshold is
	choosing whether the detection denominator will be available for inference.
\end{enumerate}

\begin{table}[H]
	\centering
	\caption{
		Design principles for protocol-resolved inference. Each principle
		corresponds to a specific part of the observation model.
	}
	\label{tab:mechanism_design_principles_protocol_resolved_inference}
	\renewcommand{\arraystretch}{1.18}
	\begin{tabularx}{0.98\linewidth}{@{}p{0.25\linewidth}p{0.30\linewidth}X@{}}
		\toprule
		\textbf{Design principle}
		&
		\textbf{Mathematical object}
		&
		\textbf{Practical implication}
		\\
		\midrule
		
		Report protocol variables
		&
		\(\theta_E\), \(\lambda_E\), \(\mathcal M_E^{\varnothing}\)
		&
		Treat field settings, grids, surfaces, overlays, timing, thresholds, and
		instrument settings as model variables.
		\\[0.45em]
		
		Preserve null outcomes
		&
		\(s_E\), \(\eta_E\), \(\varnothing\)
		&
		Record rejected, failed, missing, nonlocalized, below-threshold, or
		non-plaque-forming outcomes when possible.
		\\[0.45em]
		
		Vary protocols deliberately
		&
		\(\Pi_E^{\mathrm{lat}}\), \(s_E\), \(R_E\)
		&
		Change the kernel to expose sectors that are weakly visible under one
		protocol.
		\\[0.45em]
		
		Separate intrinsic and effective quantities
		&
		\(\theta_{\mathrm{vir}}\), \(\theta_{\mathrm{env}}\), \(\theta_E\)
		&
		Do not interpret protocol-conditioned stiffness, diffusivity, dielectric
		response, or PFU as protocol-free properties.
		\\[0.45em]
		
		Compare through a shared latent model
		&
		\(P_{\mathrm{ref},t}(\cdot\mid\theta)\),
		\(\mathcal M_{E_j}^{\varnothing}\)
		&
		Test whether one latent model can generate all observed datasets after
		protocol-specific kernels act.
		\\[0.45em]
		
		Design around blind directions
		&
		\(\mathcal I_E(\theta)\), \(\ker \mathcal I_E(\theta)\)
		&
		Add protocols that contribute information in weakly resolved parameter
		directions.
		\\
		\bottomrule
	\end{tabularx}
\end{table}

\begin{remark}[Section-level summary]
	\label{rem:mechanism_inverse_inference_section_summary}
	Experimental collapse converts the interpretation of experiments from a
	direct readout problem into a protocol-resolved inverse problem. The observed
	data are not simply the latent mechanics. They are the latent mechanics after
	transformation, selection, amplification, detection, and readout by a
	specified protocol. Once this is accepted, the same framework becomes
	constructive: controlled protocol variation can reveal mechanical,
	environmental, or biological parameters that are otherwise hidden.
	
	In this sense, experimental collapse is not only a warning. It is an
	inference principle. A protocol asks a question of the
	virion--environment system, and the observed ensemble is the answer to that
	question.
\end{remark}

\subsubsection{Observation Equivalence and Quotient Latent Spaces}
\label{subsubsec:mechanism_observation_equivalence_quotient_latent_spaces}

Protocol blindness can also be expressed globally. Fisher-information blindness
describes local statistical indistinguishability: it identifies parameter
directions in which the likelihood fails to change to first order. A stronger
question is whether two distinct latent states produce exactly the same observed
law under the entire protocol. This leads naturally to an
observation-equivalence relation.

\begin{definition}[Protocol observation equivalence]
	\label{def:mechanism_protocol_observation_equivalence}
	Two latent states \(x_1,x_2\in\Psi\) are
	\emph{observation-equivalent} under protocol \(E\), written
	\[
	x_1\sim_E x_2,
	\]
	if they induce the same null-inclusive observed law:
	\begin{empheq}[box=\fbox]{equation}
		K_E^{\varnothing}(\cdot\mid x_1)
		=
		K_E^{\varnothing}(\cdot\mid x_2)
		\quad
		\text{as probability measures on }
		\mathcal Y_E^{\varnothing}.
		\label{eq:mechanism_protocol_observation_equivalence}
	\end{empheq}
\end{definition}

\begin{proposition}[Observation equivalence is an equivalence relation]
	\label{prop:mechanism_observation_equivalence_is_equivalence_relation}
	The relation \(\sim_E\) is an equivalence relation on \(\Psi\).
\end{proposition}

\begin{proof}
	For any \(x\in\Psi\),
	\[
	K_E^{\varnothing}(\cdot\mid x)
	=
	K_E^{\varnothing}(\cdot\mid x),
	\]
	so the relation is reflexive. If \(x_1\sim_E x_2\), then equality of
	probability measures gives \(x_2\sim_E x_1\), so the relation is symmetric.
	Finally, if \(x_1\sim_E x_2\) and \(x_2\sim_E x_3\), then
	\[
	K_E^{\varnothing}(\cdot\mid x_1)
	=
	K_E^{\varnothing}(\cdot\mid x_2)
	=
	K_E^{\varnothing}(\cdot\mid x_3),
	\]
	so \(x_1\sim_E x_3\). Thus the relation is transitive.
\end{proof}

\begin{remark}[Interpretation of observation equivalence]
	\label{rem:mechanism_interpretation_observation_equivalence}
	If \(x_1\sim_E x_2\), then no amount of repeated observation under protocol
	\(E\) alone can distinguish \(x_1\) from \(x_2\), unless additional
	assumptions or additional protocols are introduced. The two states may be
	mechanically or biologically different, but they are identical from the
	perspective of the protocol. Protocol blindness can therefore be structural,
	not merely a consequence of noise or finite sample size.
\end{remark}

\begin{definition}[Protocol quotient space]
	\label{def:mechanism_protocol_quotient_space}
	The \emph{protocol quotient space} is the set of observation-equivalence
	classes
	\begin{equation}
		\Psi/{\sim_E}
		=
		\left\{
		[x]_E:x\in\Psi
		\right\},
		\label{eq:mechanism_protocol_quotient_space}
	\end{equation}
	where
	\begin{equation}
		[x]_E
		=
		\left\{
		x'\in\Psi:
		x'\sim_E x
		\right\}.
		\label{eq:mechanism_protocol_equivalence_class}
	\end{equation}
\end{definition}

\begin{remark}[Observed data as a quotient of latent mechanics]
	\label{rem:mechanism_observed_data_as_quotient}
	The quotient space \(\Psi/{\sim_E}\) is the part of latent state space that
	remains distinguishable after protocol collapse. Latent distinctions that lie
	within the same equivalence class are mechanically or biologically present in
	the model but observationally compressed. Thus a protocol does not merely add
	noise to the latent state. It may identify distinct latent states as
	experimentally equivalent.
\end{remark}

\begin{remark}[Measurability of the quotient]
	\label{rem:mechanism_measurability_of_protocol_quotient}
	The quotient notation is used here primarily to express the partition of
	latent state space induced by the protocol. In applications requiring
	integration over \(\Psi/{\sim_E}\), one must specify an appropriate quotient
	\(\sigma\)-algebra or work with finite-dimensional summaries of the
	equivalence classes. For the present theory, the essential point is the
	induced observational partition: different latent states may belong to the
	same protocol-defined observational class.
\end{remark}

\begin{definition}[Approximate protocol equivalence]
	\label{def:mechanism_approximate_protocol_equivalence}
	Exact equality of observed laws is often too strict for finite-resolution
	experiments. Let \(D_E\) be a distance or divergence between probability
	measures on \(\mathcal Y_E^{\varnothing}\). For a tolerance
	\(\varepsilon>0\), two latent states are
	\emph{\(\varepsilon\)-equivalent} under protocol \(E\) if
	\begin{equation}
		D_E
		\left(
		K_E^{\varnothing}(\cdot\mid x_1),
		K_E^{\varnothing}(\cdot\mid x_2)
		\right)
		\leq
		\varepsilon.
		\label{eq:mechanism_approximate_protocol_equivalence}
	\end{equation}
\end{definition}

\begin{remark}[Approximate equivalence and practical blindness]
	\label{rem:mechanism_approximate_equivalence_practical_blindness}
	Approximate equivalence is the global counterpart of practical
	under-resolution. Two latent states may not induce exactly the same observed
	law, but if their observed laws are closer than the experimental noise,
	finite-sample uncertainty, nuisance-parameter uncertainty, or chosen
	resolution threshold, then the protocol cannot reliably distinguish them in
	practice.
\end{remark}

\begin{proposition}[Deterministic blindness as the local tangent form of equivalence]
	\label{prop:mechanism_deterministic_blindness_local_equivalence}
	Suppose the protocol is summarized locally by a differentiable deterministic
	forward map
	\[
	\mathcal F_E:\Psi\rightarrow\mathbb R^m.
	\]
	Let \(x(\epsilon)\) be a smooth curve in \(\Psi\) with \(x(0)=x_0\). If
	\begin{equation}
		\mathcal F_E(x(\epsilon))
		=
		\mathcal F_E(x_0)
		+
		O(\epsilon^2),
		\label{eq:mechanism_deterministic_local_equivalence_curve}
	\end{equation}
	then its tangent vector
	\[
	v
	=
	\left.
	\frac{dx}{d\epsilon}
	\right|_{\epsilon=0}
	\]
	lies in the deterministic protocol-blind subspace:
	\begin{equation}
		v\in\ker D\mathcal F_E(x_0).
		\label{eq:mechanism_tangent_blind_equivalence}
	\end{equation}
\end{proposition}

\begin{proof}
	Differentiating
	Eq.~\eqref{eq:mechanism_deterministic_local_equivalence_curve} at
	\(\epsilon=0\) gives
	\[
	D\mathcal F_E(x_0)
	\left.
	\frac{dx}{d\epsilon}
	\right|_{\epsilon=0}
	=
	0.
	\]
	Therefore \(v\in\ker D\mathcal F_E(x_0)\).
\end{proof}

\begin{remark}[Local versus global blindness]
	\label{rem:mechanism_local_versus_global_blindness_quotient}
	The quotient relation \(\sim_E\) is global: it compares the full observed
	laws generated by two latent states. The deterministic blind subspace
	\(\ker D\mathcal F_E\) is local: it describes tangent directions along which
	a chosen observable summary fails to change to first order. Fisher blindness
	is also local, but likelihood-based rather than deterministic. These three
	notions are related but not identical. Together, they describe protocol
	blindness at three levels: global equivalence, deterministic sensitivity, and
	statistical distinguishability.
\end{remark}

\begin{proposition}[Multi-protocol quotient refinement]
	\label{prop:mechanism_multiprotocol_quotient_refinement}
	Let \(E_1,\ldots,E_M\) be protocols. Define
	\[
	x_1\sim_{\mathrm{multi}} x_2
	\]
	if
	\begin{equation}
		K_{E_j}^{\varnothing}(\cdot\mid x_1)
		=
		K_{E_j}^{\varnothing}(\cdot\mid x_2)
		\qquad
		\text{for every }j=1,\ldots,M.
		\label{eq:mechanism_multiprotocol_observation_equivalence}
	\end{equation}
	Then
	\begin{equation}
		x_1\sim_{\mathrm{multi}}x_2
		\quad
		\Longleftrightarrow
		\quad
		x_1\sim_{E_j}x_2
		\text{ for all }j=1,\ldots,M.
		\label{eq:mechanism_multiprotocol_equivalence_intersection}
	\end{equation}
	Thus multi-protocol observation refines the latent quotient by intersecting
	the equivalence relations induced by the individual protocols.
\end{proposition}

\begin{proof}
	By definition, \(x_1\sim_{\mathrm{multi}}x_2\) holds exactly when the two
	latent states induce the same observed law under every protocol \(E_j\). This
	is precisely the condition \(x_1\sim_{E_j}x_2\) for all \(j=1,\ldots,M\).
\end{proof}

\begin{remark}[Interpretation of quotient refinement]
	\label{rem:mechanism_interpretation_quotient_refinement}
	Each protocol partitions latent state space into observational equivalence
	classes. Adding a complementary protocol can split some of those classes into
	smaller classes, thereby making previously collapsed distinctions visible.
	This is the global analogue of the local statement that multi-protocol
	sensitivity reduces the blind subspace.
\end{remark}

\begin{remark}[Why the quotient viewpoint is useful]
	\label{rem:mechanism_why_quotient_viewpoint_useful}
	The quotient viewpoint separates mechanical existence from experimental
	distinguishability. A sector may be present in the latent mechanics but
	collapsed into the same observational equivalence class as another sector. In
	that case, failure to observe the sector is not evidence of mechanical
	absence. It is evidence that the chosen protocol quotient has removed the
	distinction.
\end{remark}

The next section applies this logic to a concrete virological protocol. The
plaque assay is especially instructive because its observed state space is not a
microscopic mechanical state space. It is a count, morphology, or endpoint
pattern generated by visible infectious lesions. Mechanically and structurally
distinct virion states may therefore be collapsed into the same observational
categories: plaque-forming, focus-forming, endpoint-positive,
non-plaque-forming, or unobserved. This makes the plaque assay a useful worked
example of protocol blindness, biological amplification, null observation, and
protocol-resolved inverse inference.
\newpage
\subsection{Section Reference: Mechanisms, Inference, and Protocol Design}
\label{subsec:section_reference_mechanisms_inference_protocol_design}
\addcontentsline{toc}{subsection}{Section Reference: Mechanisms, Inference, and Protocol Design}

\begingroup
\footnotesize
\setlength{\tabcolsep}{4pt}
\renewcommand{\arraystretch}{1.18}

\begin{xltabular}{0.98\linewidth}{
		@{}
		B{0.21\linewidth}
		L{0.43\linewidth}
		Y
		@{}
	}
	\caption{
		Core mechanisms, equations, and diagnostic concepts introduced in the
		mechanism-resolved experimental-collapse framework. The table summarizes
		how laboratory protocols transform, select, amplify, project, and report
		latent virion--environment states.
	}
	\label{tab:mechanisms_inference_protocol_design_reference}
	\\
	\toprule
	\textbf{Concept or object}
	&
	\CenteredTableHead{Mathematical form}
	&
	\textbf{Interpretation}
	\\
	\midrule
	\endfirsthead
	
	\caption[]{
		Core mechanisms, equations, and diagnostic concepts introduced in the
		mechanism-resolved experimental-collapse framework.
		\emph{Continued from previous page.}
	}
	\\
	\toprule
	\textbf{Concept or object}
	&
	\CenteredTableHead{Mathematical form}
	&
	\textbf{Interpretation}
	\\
	\midrule
	\endhead
	
	\midrule
	\multicolumn{3}{r}{\emph{Continued on next page.}}
	\\
	\endfoot
	
	\bottomrule
	\endlastfoot
	
	Mechanism-resolved protocol pipeline
	&
	\EqCell{
		X
		\rightarrow
		X_{\mathrm{prep}}
		\rightarrow
		X_{\mathrm{constr}}
		\rightarrow
		X_{\mathrm{det}}
		\rightarrow
		Y_E
		\;\text{or}\;
		\varnothing
	}
	&
	Schematic laboratory sequence: preparation, experimental conditioning,
	selection or detection, readout, and null observation.
	\\[0.55em]
	
	Mechanism-resolved kernel factorization
	&
	\EqCell{
		\Pi_E^{\mathrm{lat}}
		=
		\Pi_{E,L}
		\circ
		\cdots
		\circ
		\Pi_{E,1}
	}
	&
	Represents the latent-stage protocol kernel as a sequence of physically,
	chemically, mechanically, biologically, or procedurally interpretable
	stages.
	\\[0.55em]
	
	Mechanism-specific kernels
	&
	\EqCell{
		\Pi_E^{\mathrm{prep}},
		\quad
		\Pi_E^{\mathrm{interface}},
		\quad
		\Pi_E^{\mathrm{surf}},
		\quad
		\Pi_E^{\mathrm{field}},
		\quad
		\Pi_E^{\mathrm{medium}},
		\quad
		\Pi_E^{\mathrm{bio}}
	}
	&
	Representative factors isolating preparation, interface exposure, surface
	interaction, field forcing, medium filtering, and biological amplification.
	\\[0.55em]
	
	Preparation/interface collapse
	&
	\EqCell{
		X
		\xrightarrow{\;\Pi_E^{\mathrm{prep}}\;}
		X_{\mathrm{prep}}
	}
	&
	Collapse induced by sample handling, staining, buffer exchange,
	vitrification, fixation, grid interaction, air--water interface exposure, or
	other pre-readout conditioning.
	\\[0.55em]
	
	Cryo-EM preparation factorization
	&
	\EqCell{
		\Pi_E^{\mathrm{prep}}
		=
		\Pi_E^{\mathrm{vit}}
		\circ
		\Pi_E^{\mathrm{AWI}}
		\circ
		\Pi_E^{\mathrm{thinfilm}}
		\circ
		\Pi_E^{\mathrm{grid}}
	}
	&
	Schematic decomposition of cryo-EM preparation into grid or support
	interaction, thin-film behavior, air--water interface exposure, and
	vitrification.
	\\[0.55em]
	
	Orientation-selection model
	&
	\EqCell{
		p_{E,\mathrm{det}}(Q)
		=
		\frac{
			w_E(Q)p_{\mathrm{ref}}(Q)
		}{
			\int_{SO(3)}
			w_E(Q')p_{\mathrm{ref}}(Q')\,dQ'
		}
	}
	&
	Shows how preparation or selection weights transform a reference orientation
	distribution into a detected, protocol-conditioned orientation distribution.
	\\[0.55em]
	
	Surface/contact collapse
	&
	\EqCell{
		X
		\xrightarrow{\;\Pi_E^{\mathrm{surf}}\;}
		X_{\mathrm{surf}}
	}
	&
	Collapse caused by adsorption, immobilization, receptor contact, substrate
	interaction, cell-surface coupling, or surface-induced deformation.
	\\[0.55em]
	
	Mechanical loading collapse
	&
	\EqCell{
		X_{\mathrm{surf}}
		\xrightarrow{\;\Pi_E^{\mathrm{load}}\;}
		X_{\mathrm{load}}
	}
	&
	Protocol-induced deformation, rupture, indentation response, or
	rate-dependent mechanical conditioning caused by applied force.
	\\[0.55em]
	
	Field/electrical collapse
	&
	\EqCell{
		X
		\xrightarrow{\;\Pi_E^{\mathrm{field}}(\theta_{\mathrm{env}},\theta_E)\;}
		X_{\mathrm{field}}
	}
	&
	State transformation or selection induced by electric fields, dielectric
	contrast, conductivity, polarization, trapping, torque, or field-sensitive
	loss.
	\\[0.55em]
	
	Geometric projection collapse
	&
	\EqCell{
		Y_E=\mathcal O_E(X_E)
	}
	&
	Occurs when a high-dimensional latent state is reported as a projection,
	image, reconstruction, trajectory, curve, count, class, spectrum, or summary
	statistic.
	\\[0.55em]
	
	Observation fiber
	&
	\EqCell{
		\mathcal F_E(y_E)
		=
		\{x_E\in\Psi_E:
		\mathcal O_E(x_E)=y_E\}
	}
	&
	Set of protocol-conditioned latent states that map to the same observed
	datum. Large fibers indicate collapsed latent distinctions.
	\\[0.55em]
	
	Posterior fiber
	&
	\EqCell{
		P_E(dx_E\mid y_E,\mathrm{det})
	}
	&
	Probabilistic analogue of the observation fiber for stochastic readout,
	reconstruction uncertainty, thresholding, noise, or model-based inversion.
	\\[0.55em]
	
	Medium-filtering collapse
	&
	\EqCell{
		\Pi_E^{\mathrm{medium}}
		(dx_{\mathrm{medium}}\mid x;\theta_{\mathrm{env}})
	}
	&
	State-dependent slowing, trapping, exclusion, confinement, immobilization, or
	mobility switching caused by mucus, gels, extracellular matrix, droplets,
	overlays, or other structured media.
	\\[0.55em]
	
	Medium-filtered trajectory model
	&
	\EqCell{
		\begin{aligned}
			d\mathbf r_t
			&=
			\mathbf b_E\,dt
			+
			\sqrt{2}\mathbf B_E\,d\mathbf W_t
			+
			d\mathbf J_t,
			\\
			D_E
			&=
			\mathbf B_E\mathbf B_E^{\mathsf T}
		\end{aligned}
	}
	&
	Represents drift, anisotropic or heterogeneous diffusion, and jump, sticking,
	release, or state-switching events in structured media.
	\\[0.55em]
	
	Hidden mobility state
	&
	\EqCell{
		M_t\in
		\{\mathrm{mobile},
		\mathrm{confined},
		\mathrm{adhesive},
		\mathrm{immobile},
		\mathrm{released}\}
	}
	&
	Latent transport mode that may be only partially visible through the
	reported trajectory.
	\\[0.55em]
	
	Joint medium-filtered latent process
	&
	\EqCell{
		X_t
		=
		(\mathbf r_t,Q_t,M_t,X_{\mathrm{env},t})
	}
	&
	Coupled process containing position, orientation, hidden mobility state, and
	local environmental state.
	\\[0.55em]
	
	Time-window collapse
	&
	\EqCell{
		\Pi_E^{\mathrm{time}}
		(dx_{\mathrm{time}}\mid
		x;t_0,t_1,\Delta t_{\mathrm{samp}},\tau_{\mathrm{exp}})
	}
	&
	Collapse caused by finite observation duration, frame interval, exposure
	time, preparation delay, incubation time, or analysis window.
	\\[0.55em]
	
	Finite-exposure observation
	&
	\EqCell{
		Y_m
		=
		\frac{1}{\tau_{\mathrm{exp}}}
		\int_{t_m}^{t_m+\tau_{\mathrm{exp}}}
		\mathcal O_E(X_t)\,dt
		+
		\nu_m
	}
	&
	Makes explicit that fast latent motion, transient binding, or short-lived
	states can be blurred or averaged within one recorded frame.
	\\[0.55em]
	
	Time-window ratios
	&
	\EqCell{
		\Pi_{\mathrm{samp}}
		=
		\frac{\tau_{\mathrm{state}}}{\Delta t_{\mathrm{samp}}},
		\qquad
		\Pi_{\mathrm{win}}
		=
		\frac{\tau_{\mathrm{state}}}{t_1-t_0},
		\qquad
		\Pi_{\mathrm{exp}}
		=
		\frac{\tau_{\mathrm{state}}}{\tau_{\mathrm{exp}}}
	}
	&
	Dimensionless measures of whether a latent state is sampled, persistent,
	averaged, motion-blurred, or missed under the protocol timing.
	\\[0.55em]
	
	Persistence-biased detection
	&
	\EqCell{
		s_E^{\mathrm{time}}(A)
		=
		\Pr(\tau_A\geq\tau_{\min}\mid E)
	}
	&
	States that persist long enough are more likely to enter the observed
	ensemble than short-lived, rare, or rapidly switching states.
	\\[0.55em]
	
	Time-integrated readout
	&
	\EqCell{
		Y_E
		=
		\int_{t_0}^{t_1}
		h_E(X_t)\,dt+\nu_E
	}
	&
	Readout formed by accumulated biological, optical, reporter, cytopathic, or
	mechanical signal over an incubation or acquisition window.
	\\[0.55em]
	
	Biological amplification collapse
	&
	\EqCell{
		X
		\rightarrow
		X_{\mathrm{bio}}
		\rightarrow
		Y_{\mathrm{bio}}
		\;\text{or}\;
		\varnothing
	}
	&
	Assay reports only latent states that complete a biological pathway
	sufficient to generate a plaque, focus, reporter signal, endpoint response,
	or cytopathic effect.
	\\[0.55em]
	
	Biological pathway probability
	&
	\EqCell{
		\pi_{\mathrm{bio}}(x;E)
		=
		\Pr(A_0\mid x,E)
		\prod_{k=1}^{K}
		\Pr(A_k\mid A_0,\ldots,A_{k-1},x,E)
	}
	&
	Stage-resolved probability that a latent state passes through the required
	biological assay events.
	\\[0.55em]
	
	Binary biological observation kernel
	&
	\EqCell{
		\begin{aligned}
			K_E^{\mathrm{bio},\varnothing}
			(\{\mathrm{signal}\}\mid x)
			&=
			\pi_{\mathrm{bio}}(x;E),
			\\
			K_E^{\mathrm{bio},\varnothing}
			(\{\varnothing\}\mid x)
			&=
			1-\pi_{\mathrm{bio}}(x;E)
		\end{aligned}
	}
	&
	Maps latent states into a biological signal channel or the biological null
	channel.
	\\[0.55em]
	
	Plaque-forming probability
	&
	\EqCell{
		\pi_{\mathrm{PFU}}(x;E_{\mathrm{PFU}})
		=
		\Pr(
		x\text{gen}E_{\mathrm{PFU}}
		)
	}
	&
	Protocol-conditioned probability that a latent virion, aggregate, or
	infectious unit generates a countable plaque.
	\\[0.55em]
	
	Biological readout spaces
	&
	\EqCell{
		\mathcal Y_{\mathrm{PFU}}=\{0,1,2,\ldots\},
		\qquad
		\mathcal Y_{\mathrm{ED}}=\{0,1\}^{J\times R}
	}
	&
	Plaque and focus assays report counts; endpoint dilution assays report binary
	or categorical infection patterns across dilutions and replicate wells.
	\\[0.55em]
	
	Biological visibility threshold
	&
	\EqCell{
		s_E^{\mathrm{bio}}(x_{\mathrm{bio}})
		=
		\mathbf 1\{
		A_E(x_{\mathrm{bio}})\geq a_E^\ast
		\}
	}
	&
	Assay event becomes visible, countable, or scoreable only after crossing a
	protocol-specific threshold.
	\\[0.55em]
	
	Amplification gain
	&
	\EqCell{
		G_E(x)
		=
		\frac{
			A_E(x_{\mathrm{bio}},t_1)+\epsilon_A
		}{
			A_E(x_{\mathrm{bio}},t_0)+\epsilon_A
		}
	}
	&
	Quantifies how a biological process expands a small latent difference into a
	visible assay signal.
	\\[0.55em]
	
	Stage-resolved protocol family
	&
	\EqCell{
		\mathcal E_{\mathrm{bio}}
		=
		\{E^{(1)},E^{(2)},\ldots,E^{(M)}\}
	}
	&
	Collection of related biological assays formed by varying cell line,
	adsorption time, overlay, incubation, reporter threshold, neutralization, or
	another controlled stage parameter.
	\\[0.55em]
	
	Stage-sensitivity index
	&
	\EqCell{
		S_r(E)
		=
		\frac{\partial}{\partial \xi_r}
		\log[
		\Lambda_{\mathrm{bio}}(E;\xi_r)
		+\epsilon_\Lambda]
	}
	&
	Measures how strongly the biological signal changes when a controlled
	assay-stage parameter \(\xi_r\) is varied.
	\\[0.55em]
	
	Biological bottleneck score
	&
	\EqCell{
		\mathcal B_k(E)
		=
		1-
		\int_{\Psi}
		\Pr(A_k\mid A_0,\ldots,A_{k-1},x,E)
		P_{\mathrm{ref},t}(dx)
	}
	&
	Identifies assay stages that strongly filter the latent population.
	\\[0.55em]
	
	Ideal Poisson count model
	&
	\EqCell{
		N_d
		\mid
		\Lambda_{\mathrm{bio}}
		\sim
		\operatorname{Poisson}
		(
		V_{\mathrm{inoc}}f_d
		\Lambda_{\mathrm{bio}}(E)
		)
	}
	&
	Baseline model for independent biological events in the dilute, well-mixed,
	non-overlapping, countable regime.
	\\[0.55em]
	
	Negative-binomial count model
	&
	\EqCell{
		\begin{aligned}
			N_d
			&\sim
			\operatorname{NegBin}(\mu_d,\kappa),
			\\
			\operatorname{Var}(N_d)
			&=
			\mu_d+\frac{\mu_d^2}{\kappa}
		\end{aligned}
	}
	&
	Models overdispersed plaque or focus counts caused by aggregation,
	heterogeneous cell susceptibility, well-to-well variation, clustered input,
	or other replicate-level heterogeneity.
	\\[0.55em]
	
	Zero-inflated count model
	&
	\EqCell{
		\Pr(N_d=0)
		=
		\zeta_E+(1-\zeta_E)e^{-\mu_d}
	}
	&
	Models excess zero counts caused by structural null outcomes, failed
	delivery, failed amplification, nonpermissive wells, or subthreshold signal.
	\\[0.55em]
	
	Censored/countability model
	&
	\EqCell{
		N_{\mathrm{obs}}
		=
		\begin{cases}
			N, & N_{\min}\leq N\leq N_{\max},\\
			\text{censored}, & \text{otherwise}.
		\end{cases}
	}
	&
	Represents wells that are too dense, too sparse, merged, or otherwise outside
	the exact-count regime.
	\\[0.55em]
	
	Spatial plaque process
	&
	\EqCell{
		\Phi_E=\{z_1,\ldots,z_N\}
		\subset\Omega_{\mathrm{well}},
		\qquad
		\mathbb E[N]
		=
		\int_{\Omega_{\mathrm{well}}}
		\lambda_E(z)\,dz
	}
	&
	Represents plaques or foci as spatial events rather than independent scalar
	counts.
	\\[0.55em]
	
	Plaque-merging map
	&
	\EqCell{
		\mathcal S_E:
		\Phi_E^{\mathrm{true}}
		\rightarrow
		\Phi_E^{\mathrm{obs}},
		\qquad
		N_{\mathrm{obs}}
		=
		\#\mathcal S_E(\Phi_E^{\mathrm{true}})
	}
	&
	Models overlap, merging, segmentation failure, or loss of countability at
	high inoculum, weak overlay restriction, or long incubation.
	\\[0.55em]
	
	Endpoint dilution model
	&
	\EqCell{
		\Pr(Z_{jr}=1\mid\theta,E)
		=
		1-
		\exp[
		-
		V_{\mathrm{inoc}}f_j
		\Lambda_{\mathrm{ED}}(E,\theta)
		]
	}
	&
	Represents TCID\(_{50}\)-type assays as binary or categorical censored
	readouts across dilution and replicate wells.
	\\[0.55em]
	
	Endpoint threshold model
	&
	\EqCell{
		Z_{jr}
		=
		\mathbf 1\{A_{jr}\geq a_E^\ast\}
	}
	&
	Represents endpoint positivity as a thresholded observation of a latent
	biological signal.
	\\[0.55em]
	
	Neutralization as kernel deformation
	&
	\EqCell{
		\begin{aligned}
			\pi_{\mathrm{bio}}(x;E,c)
			&=
			\Pr(A_0\mid x,E,c)
			\\
			&\quad\times
			\prod_{k=1}^{K}
			\Pr(A_k\mid A_0,\ldots,A_{k-1},x,E,c)
		\end{aligned}
	}
	&
	Neutralizing agents deform the biological kernel by altering one or more
	stage probabilities and shifting latent states toward null biological
	channels.
	\\[0.55em]
	
	Neutralization signal loss
	&
	\EqCell{
		F_{\mathrm{neutral}}(c)
		=
		1
		-
		\frac{
			\Lambda_{\mathrm{bio}}(E,c)
		}{
			\Lambda_{\mathrm{bio}}(E,0)+\epsilon_\Lambda
		}
	}
	&
	Summarizes the concentration-dependent loss of biological signal under a
	neutralizing or inhibitory condition.
	\\[0.55em]
	
	Morphology-augmented plaque readout
	&
	\EqCell{
		Y_{\mathrm{PFU}}^{\mathrm{morph}}
		=
		(N,M_1,\ldots,M_N)
	}
	&
	Extends PFU beyond scalar count by including plaque size, shape, opacity,
	time-to-visibility, boundary structure, or growth features.
	\\[0.55em]
	
	Morphology-augmented biological kernel
	&
	\EqCell{
		K_E^{\mathrm{morph},\varnothing}
		(dN,dM_1,\ldots,dM_N\mid x)
	}
	&
	Maps latent states to count and morphology distributions, retaining more
	information than a count-only plaque readout.
	\\
\end{xltabular}

\endgroup
\newpage
\section{Worked Example: The Plaque Assay as Experimental Collapse}
\label{sec:worked_example_plaque_assay_experimental_collapse}

The plaque assay provides a simple and instructive example of experimental
collapse. In a standard plaque assay, a virus-containing sample is serially
diluted, applied to a susceptible cell monolayer, allowed to adsorb, covered
with a semi-solid or otherwise spatially restrictive overlay, incubated until
localized infection events become visible, and then stained or otherwise read
out as a number of plaques. The final observable is a count. The latent process
that produces that count is much richer
\cite{Dulbecco1952,DulbeccoVogt1954,Cooper1961,Baer2014,Mendoza2020}.

This makes the plaque assay an especially useful worked example for
protocol-resolved virophysics. It is familiar, experimentally concrete, and
quantitative, but it is also a strong example of protocol-conditioned
projection. A plaque assay does not directly report virion position,
orientation, spike presentation, angular velocity, contact geometry, local force
history, deformation pathway, or collective branch structure. It also does not
directly report total physical particles, total genomes, or total antigen.
Instead, it reports a biologically defined functional projection of the latent
population: the number of infectious units that successfully generate visible,
countable lesions under a specified cell-line, adsorption, overlay, incubation,
staining, and counting protocol.

At the most compact level, the plaque assay performs the reduction
\begin{empheq}[box=\fbox]{equation}
	\begin{gathered}
		\text{\textbf{Latent Virion or Infectious-Unit Population}}
		\\[0.35em]
		\Downarrow
		\\[0.35em]
		\text{\textbf{Protocol-Conditioned Infectious Events}}
		\\[0.35em]
		\Downarrow
		\\[0.35em]
		\text{\textbf{Visible Plaque Count}}
	\end{gathered}
	\label{eq:plaque_assay_collapse_at_a_glance}
\end{empheq}
Thus, a plaque count is not a direct count of total virions. It is a count of
visible infectious lesions generated under particular assay conditions.

\begin{remark}[Plaque count versus physical particle count]
	\label{rem:plaque_count_versus_physical_particle_count}
	The distinction between physical particles and plaque-forming units is
	essential. A physical particle may be structurally intact but noninfectious,
	damaged, neutralized, genome-defective, aggregation-associated, unable to
	attach to the chosen cells, unable to enter, unable to replicate
	productively, unable to spread under the overlay, or unable to generate a
	visible lesion above the counting threshold. Conversely, in the dilute and
	countable regime, a counted plaque is often interpreted as arising from one
	infectious unit. That infectious unit need not be identical to one isolated
	physical virion in every case, especially when aggregation, co-delivery, or
	collective infectivity are present
	\cite{Klasse2015,Sanjuan2018,Brooke2014}.
\end{remark}

\begin{remark}[Why the plaque assay is a useful worked example]
	\label{rem:why_plaque_assay_useful_worked_example}
	The plaque assay compresses a high-dimensional latent population into a
	one-dimensional count. This makes the collapse structure unusually clear.
	Physical particles that differ in morphology, genome integrity,
	receptor-binding competence, aggregation state, medium history,
	neutralization state, or cell-entry probability may all be mapped to the same
	observed category: plaque-forming or non-plaque-forming under the chosen
	protocol. The assay is therefore valuable precisely because it reports an
	experimentally defined functional projection of the population.
\end{remark}

\begin{remark}[Why this is not a criticism of plaque assays]
	\label{rem:plaque_assay_not_criticism}
	The point is not that plaque assays are unreliable. The point is that they are
	specific. A plaque assay is powerful because it asks a biologically meaningful
	question: how much of this sample can produce visible infectious lesions under
	these conditions? Protocol-resolved notation makes that question explicit
	rather than treating PFU as a protocol-free particle count.
\end{remark}

\subsection{Plaque-Assay Protocol as a Collapse Map}
\label{subsec:plaque_assay_protocol_as_collapse_map}

The plaque assay is not defined only by the act of counting plaques. It is
defined by the full biological and procedural pathway through which a latent
particle population is converted into visible lesions. In the notation of the
general framework, the plaque assay is a biological observation protocol with a
null channel: many latent states enter the assay, but only some generate
countable plaques.

\begin{definition}[Plaque-assay protocol]
	\label{def:plaque_assay_protocol}
	Let \(E_{\mathrm{PFU}}\) denote a plaque-assay protocol specified by
	\begin{empheq}[box=\fbox]{equation}
		E_{\mathrm{PFU}}
		=
		\left(
		\begin{gathered}
			\text{virus preparation},
			\text{cell line},
			\text{cell density or monolayer state},
			\text{dilution series},
			V_{\mathrm{inoc}},
			t_{\mathrm{ads}},
			\\
			\text{wash or medium exchange},
			\text{overlay composition and thickness},
			t_{\mathrm{inc}},
			\text{temperature},
			\\
			\text{fixation or staining method},
			\text{plaque-detection threshold},
			\text{counting rule}
		\end{gathered}
		\right).
		\label{eq:plaque_assay_protocol_tuple}
	\end{empheq}
	Here \(V_{\mathrm{inoc}}\) is the plated inoculum volume,
	\(t_{\mathrm{ads}}\) is the adsorption time, and \(t_{\mathrm{inc}}\) is the
	incubation time before plaque visualization.
\end{definition}

\begin{remark}[What the plaque protocol fixes]
	\label{rem:what_plaque_protocol_fixes}
	The protocol \(E_{\mathrm{PFU}}\) fixes more than the counting procedure. It
	fixes the biological environment in which the virus is tested, including the
	susceptible cell type, receptor and entry context, adsorption window, washing
	or medium exchange, spatial restriction imposed by the overlay, duration
	allowed for local amplification, and criterion by which an infection focus
	becomes a visible plaque. Changing any of these choices can change the
	plaque-forming probability of the same physical virion population
	\cite{Baer2014,Mendoza2020}.
\end{remark}

\begin{remark}[Overlay as a physical and biological constraint]
	\label{rem:overlay_physical_biological_constraint}
	The overlay is a useful example of protocol mechanics. Its purpose is not
	merely to make the assay convenient. It restricts long-range spread so that
	local infection events remain spatially localized and can develop into
	countable plaques. Overlay composition, viscosity, thickness, and
	compatibility with the virus--cell system can therefore affect plaque size,
	shape, visibility, and countability \cite{Baer2014,Mendoza2020}.
\end{remark}

\begin{definition}[Latent virion state for the plaque assay]
	\label{def:latent_state_plaque_assay}
	Let \(X\in\Psi\) denote the latent state of a virion, particle, aggregate,
	or infectious unit immediately before the plaque-assay protocol is applied.
	For this worked example, \(X\) may include
	\begin{equation}
		X
		=
		\left(
		\begin{array}{l}
			\text{Particle integrity},\\
			\text{Genome integrity and replication competence},\\
			\text{Spike or receptor-binding competence},\\
			\text{Surface charge and transport state},\\
			\text{Orientation or presentation variables},\\
			\text{Aggregation or co-delivery state},\\
			\text{Neutralization or damage state},\\
			\text{Local medium and handling history}
		\end{array}
		\right).
		\label{eq:latent_plaque_assay_state_components}
	\end{equation}
	The assay does not report these components separately. It reports whether the
	latent state contributes to a visible plaque-forming event.
\end{definition}

\begin{remark}[Why the latent object may be an infectious unit]
	\label{rem:latent_object_may_be_infectious_unit}
	In the simplest dilute interpretation, the latent object is an individual
	virion. More generally, the plaque-relevant latent object may be an
	infectious unit: an isolated virion, an aggregate, a co-delivered packet, or
	another experimentally inseparable unit that behaves as one plaque-producing
	event under dilution. This wording avoids over-interpreting PFU as a literal
	count of individual physical virions.
\end{remark}

\begin{remark}[Why plaque competence is composite]
	\label{rem:why_plaque_competence_composite}
	A virion does not become plaque-forming merely by existing as a physical
	particle. It must remain structurally and genetically competent, encounter a
	susceptible cell, adsorb productively, enter, initiate productive
	replication, spread locally under the overlay, and generate a visible lesion
	under the readout conditions. For many viruses, particle number and
	infectious-unit number can differ substantially; high particle-to-PFU ratios
	may reflect inert particles, structural defects, incomplete or defective
	genomes, lethal mutations, aggregation, assay incompatibility, neutralization,
	decay, or failure to complete one of the required assay steps
	\cite{Klasse2015,Sanjuan2018,McCormick2021,Brooke2014}.
\end{remark}

\subsection{Plaque-Forming Probability as a Protocol Weight}
\label{subsec:plaque_forming_probability_protocol_weight}

The central protocol-resolved quantity is the probability that a latent state
becomes a visible plaque under the specified plaque-assay protocol. For a
virologist, this is the state-dependent probability that a particle or
infectious unit is plaque-forming in the chosen assay. For the mathematical
framework, it is a protocol-dependent biological amplification and detection
weight.

\begin{definition}[Plaque-forming probability of a latent state]
	\label{def:plaque_forming_probability_latent_state}
	Define
	\begin{equation}
		\pi_{\mathrm{PFU}}(x;E_{\mathrm{PFU}})
		=
		\Pr
		\left(
		\begin{array}{c}
			\text{latent state }x
			\text{ generates a visible, countable plaque}\\
			\text{under the plaque-assay protocol }E_{\mathrm{PFU}}
		\end{array}
		\right).
		\label{eq:plaque_forming_probability_definition}
	\end{equation}
	This probability is the plaque-assay analogue of a protocol survival,
	amplification, and detection weight. It maps a latent state to its probability
	of producing a visible plaque under a specified assay.
\end{definition}

The simplest conceptual statement is
\[
\text{Plaque-Forming Probability}
=
\text{Probability of Passing the Plaque-Assay Pathway}.
\]
A more explicit protocol-chain representation is
\begin{empheq}[box=\fbox]{align}
	\pi_{\mathrm{PFU}}(x;E_{\mathrm{PFU}})
	&=
	p_{\mathrm{surv}}(x;E_{\mathrm{PFU}})
	p_{\mathrm{deliv}}(x;E_{\mathrm{PFU}})
	p_{\mathrm{ads}}(x;E_{\mathrm{PFU}})
	\notag\\
	&\qquad\times
	p_{\mathrm{entry}}(x;E_{\mathrm{PFU}})
	p_{\mathrm{rep}}(x;E_{\mathrm{PFU}})
	p_{\mathrm{spread}}(x;E_{\mathrm{PFU}})
	p_{\mathrm{vis}}(x;E_{\mathrm{PFU}}).
	\label{eq:plaque_probability_factorization}
\end{empheq}
The factors represent survival through handling, delivery to the monolayer,
productive adsorption, entry, productive replication, local spread under the
overlay, and visibility under the final readout criterion.

\begin{remark}[How to interpret the factorization]
	\label{rem:interpretation_plaque_forming_probability}
	Equation~\eqref{eq:plaque_probability_factorization} is a reduced
	protocol-chain model, not a claim that the stages are microscopically
	independent. More formally, the factors should be read as conditional
	probabilities along the assay pathway. For example,
	\[
	p_{\mathrm{entry}}
	=
	\Pr(
	\text{entry}
	\mid
	\text{survival, delivery, adsorption},x,E_{\mathrm{PFU}}
	),
	\]
	and similarly for later factors. The important point is that plaque formation
	requires the virion or infectious unit to pass through several biological and
	physical filters. A particle may be physically present but fail to survive
	handling, fail to reach the cell layer, fail to adsorb, fail to enter, fail to
	replicate productively, fail to spread locally, or fail to generate a visible
	plaque above threshold.
\end{remark}

\begin{remark}[Why this probability is protocol-conditioned]
	\label{rem:why_plaque_probability_protocol_conditioned}
	The probability \(\pi_{\mathrm{PFU}}(x;E_{\mathrm{PFU}})\) belongs to the
	state--protocol pair. It is not a context-free property of the particle alone.
	A particle that fails to form a plaque in one cell line, overlay, or
	incubation condition may be plaque-forming under another protocol. Likewise, a
	particle that is physically intact may be non-plaque-forming under a protocol
	that does not support its entry, replication, local spread, or visibility.
\end{remark}

\begin{definition}[Effective plaque-forming concentration]
	\label{def:effective_plaque_forming_concentration}
	Let \(n_{\mathrm{ref}}(x)\) denote the latent number-density distribution of
	physical particles or infectious units per unit sample volume over the state
	space \(\Psi\). The \emph{effective plaque-forming concentration} under
	protocol \(E_{\mathrm{PFU}}\) is
	\begin{empheq}[box=\fbox]{equation}
		\Lambda_{\mathrm{PFU}}(E_{\mathrm{PFU}})
		=
		\int_{\Psi}
		\pi_{\mathrm{PFU}}(x;E_{\mathrm{PFU}})
		n_{\mathrm{ref}}(x)\,dx.
		\label{eq:effective_plaque_forming_concentration}
	\end{empheq}
	More generally, if the latent population is represented by a measure rather
	than a density, the same expression is written as an integral with respect to
	that measure. The quantity \(\Lambda_{\mathrm{PFU}}\) has units of visible
	plaque-forming events per unit sample volume.
\end{definition}

\begin{remark}[PFU is a protocol-conditioned projection]
	\label{rem:pfu_is_protocol_conditioned}
	The concentration \(\Lambda_{\mathrm{PFU}}\) is a protocol-conditioned
	infectious projection of the latent ensemble. It is not the same as the total
	physical particle concentration
	\begin{equation}
		C_{\mathrm{part}}
		=
		\int_{\Psi}
		n_{\mathrm{ref}}(x)\,dx.
		\label{eq:physical_particle_concentration}
	\end{equation}
	The two quantities are related only through the state-dependent plaque-forming
	probability \(\pi_{\mathrm{PFU}}\). This is why two samples with similar
	particle counts can have different plaque titers, and why two samples with
	different latent state distributions can produce similar plaque counts if
	their protocol-weighted plaque-forming concentrations agree.
\end{remark}

\begin{definition}[Particle-to-PFU ratio]
	\label{def:particle_to_pfu_ratio}
	When an independent particle-counting, genome-counting, or antigen-counting
	measurement is available and \(\Lambda_{\mathrm{PFU}}>0\), define the
	particle-to-PFU ratio by
	\begin{empheq}[box=\fbox]{equation}
		\mathcal R_{\mathrm{part/PFU}}
		=
		\frac{
			C_{\mathrm{part}}
		}{
			\Lambda_{\mathrm{PFU}}
		}.
		\label{eq:particle_to_pfu_ratio}
	\end{empheq}
	This ratio summarizes how strongly the physical particle ensemble is
	compressed by the plaque-forming projection.
\end{definition}

\begin{remark}[Interpretation of the particle-to-PFU ratio]
	\label{rem:interpretation_particle_to_pfu_ratio}
	A large \(\mathcal R_{\mathrm{part/PFU}}\) does not by itself identify a
	single cause. It may reflect noninfectious particles, damaged particles,
	incomplete or defective genomes, lethal mutations, unfavorable adsorption, low
	entry efficiency, aggregation, protocol mismatch, host-cell incompatibility,
	neutralization, or readout limitations. In the present framework, all of these
	possibilities correspond to different mechanisms by which the latent ensemble
	is filtered before becoming a visible plaque count
	\cite{Klasse2015,Sanjuan2018,McCormick2021}.
\end{remark}

\begin{remark}[Why independent particle measurements matter]
	\label{rem:why_independent_particle_measurements_matter}
	A plaque assay by itself estimates a protocol-weighted infectious
	concentration. It does not determine the total particle concentration or the
	fraction of particles that are structurally intact, genome-containing,
	antigen-positive, entry-competent, or replication-competent. Independent
	measurements such as particle counting, genome quantification, antigen
	measurement, or structural imaging can therefore be combined with PFU data to
	separate physical abundance from plaque-forming efficiency. In the language of
	this paper, such measurements add additional protocol kernels that constrain
	different latent sectors.
\end{remark}

\subsection{Plaque-Count Statistics in the Dilute Regime}
\label{subsec:plaque_count_statistics_dilute_regime}

Once the effective plaque-forming concentration
\(\Lambda_{\mathrm{PFU}}\) has been defined, the standard plaque-count
statistics arise as a dilute-regime closure of the biological kernel. This is
the regime in which plaque-forming events are sufficiently rare to be spatially
separable, sufficiently independent to be treated as isolated infectious
events, and sufficiently numerous in the selected dilution range to support
quantitative counting.
\medskip 

\noindent The simplest statistical statement is
\begin{empheq}[box=\fbox]{equation}
	\begin{aligned}
		\text{\bfseries Expected plaque count}
		&=
		\text{\bfseries Plated volume}
		\\
		&\quad {}\times
		\text{\bfseries Dilution fraction}
		\\
		&\quad {}\times
		\parbox[t]{0.42\linewidth}{\centering\bfseries Effective plaque-forming\\ concentration}.
	\end{aligned}
	\label{eq:expected_plaque_count_plain_language}
\end{empheq}
The effective plaque-forming concentration is not the total physical particle
concentration. It is the protocol-weighted concentration
\(\Lambda_{\mathrm{PFU}}\) produced by the state-dependent plaque-forming
probability \(\pi_{\mathrm{PFU}}(x;E_{\mathrm{PFU}})\).

\begin{proposition}[Poisson plaque-count model in the dilute limit]
	\label{prop:poisson_plaque_count_model}
	Let \(f_d\in(0,1]\) denote the dilution fraction plated in a given well, and
	let \(V_{\mathrm{inoc}}\) be the plated inoculum volume. In the dilute,
	independent, non-overlapping-plaque regime, the observed plaque count \(N_d\)
	may be modeled as
	\begin{empheq}[box=\fbox]{equation}
		N_d
		\sim
		\operatorname{Poisson}
		\left(
		V_{\mathrm{inoc}} f_d
		\Lambda_{\mathrm{PFU}}(E_{\mathrm{PFU}})
		\right).
		\label{eq:poisson_plaque_count_model}
	\end{empheq}
	Consequently, a basic single-dilution estimator of the effective
	plaque-forming concentration is
	\begin{equation}
		\widehat{\Lambda}_{\mathrm{PFU},d}
		=
		\frac{N_d}{V_{\mathrm{inoc}}f_d}.
		\label{eq:pfu_concentration_estimator}
	\end{equation}
\end{proposition}

\begin{proof}
	The expected number of plated latent units with state in \(dx\) is
	\[
	V_{\mathrm{inoc}} f_d n_{\mathrm{ref}}(x)\,dx.
	\]
	Each such latent unit contributes a visible plaque with probability
	\(\pi_{\mathrm{PFU}}(x;E_{\mathrm{PFU}})\). Under dilute independent
	thinning, the plaque-forming events form an approximately Poisson process
	with intensity
	\[
	V_{\mathrm{inoc}} f_d
	\pi_{\mathrm{PFU}}(x;E_{\mathrm{PFU}})
	n_{\mathrm{ref}}(x)\,dx.
	\]
	Integrating over the latent state space gives the mean
	\[
	V_{\mathrm{inoc}} f_d
	\int_{\Psi}
	\pi_{\mathrm{PFU}}(x;E_{\mathrm{PFU}})
	n_{\mathrm{ref}}(x)\,dx
	=
	V_{\mathrm{inoc}} f_d
	\Lambda_{\mathrm{PFU}}(E_{\mathrm{PFU}}).
	\]
	Thus \(N_d\) is Poisson with the stated mean in the dilute
	independent-event limit.
\end{proof}

\begin{remark}[What the Poisson model assumes biologically]
	\label{rem:what_poisson_model_assumes_biologically}
	The Poisson approximation assumes that plaque-forming events are rare,
	independent, spatially separable, and countable after dilution. It also
	assumes that each counted plaque corresponds to one effective
	plaque-forming event in the selected dilution range. These assumptions are
	useful for standard plaque-titer estimation, but they are protocol-regime
	assumptions rather than universal features of plaque assays. Aggregation,
	plaque merging, cell-layer heterogeneity, uneven adsorption, localized spread
	effects, and counting thresholds can all produce departures from the Poisson
	ideal.
\end{remark}

\begin{remark}[Connection to the usual titer formula]
	\label{rem:connection_to_usual_titer_formula}
	If the dilution is reported as a dilution factor \(D_d=1/f_d\), then
	Eq.~\eqref{eq:pfu_concentration_estimator} becomes the usual plaque-titer
	form
	\begin{equation}
		\widehat{\Lambda}_{\mathrm{PFU},d}
		=
		\frac{N_dD_d}{V_{\mathrm{inoc}}}.
		\label{eq:usual_pfu_titer_formula}
	\end{equation}
	Thus the standard PFU per unit volume calculation is recovered as the
	simplest estimator of the protocol-weighted concentration
	\(\Lambda_{\mathrm{PFU}}\), usually after selecting a dilution whose plaques
	are countable and sufficiently separated.
\end{remark}

\begin{definition}[Dilution-series likelihood]
	\label{def:dilution_series_likelihood}
	Suppose plaque counts are collected across dilution fractions
	\(f_1,\ldots,f_J\), with replicate wells indexed by
	\(r=1,\ldots,R_j\). Let \(N_{jr}\) be the plaque count in replicate \(r\) at
	dilution \(f_j\). Under the dilute-regime Poisson model,
	\begin{equation}
		N_{jr}
		\sim
		\operatorname{Poisson}
		\left(
		V_{\mathrm{inoc}} f_j \Lambda_{\mathrm{PFU}}
		\right).
		\label{eq:poisson_dilution_series_model}
	\end{equation}
	The corresponding likelihood is
	\begin{empheq}[box=\fbox]{equation}
		\mathcal L_{\mathrm{PFU}}(\Lambda_{\mathrm{PFU}})
		=
		\prod_{j=1}^{J}
		\prod_{r=1}^{R_j}
		\frac{
			\left(
			V_{\mathrm{inoc}}f_j\Lambda_{\mathrm{PFU}}
			\right)^{N_{jr}}
			e^{-V_{\mathrm{inoc}}f_j\Lambda_{\mathrm{PFU}}}
		}{
			N_{jr}!
		}.
		\label{eq:dilution_series_likelihood}
	\end{empheq}
\end{definition}

\begin{proposition}[Maximum-likelihood estimator for a dilution series]
	\label{prop:mle_pfu_dilution_series}
	Under the likelihood in Eq.~\eqref{eq:dilution_series_likelihood}, using
	only wells treated as exact count observations, the maximum-likelihood
	estimator is
	\begin{empheq}[box=\fbox]{equation}
		\widehat{\Lambda}_{\mathrm{PFU}}^{\mathrm{MLE}}
		=
		\frac{
			\displaystyle
			\sum_{j=1}^{J}
			\sum_{r=1}^{R_j}
			N_{jr}
		}{
			\displaystyle
			V_{\mathrm{inoc}}
			\sum_{j=1}^{J}
			R_j f_j
		}.
		\label{eq:mle_pfu_dilution_series}
	\end{empheq}
\end{proposition}

\begin{proof}
	The log-likelihood is
	\[
	\log \mathcal L_{\mathrm{PFU}}
	=
	\sum_{j,r}
	\left[
	N_{jr}\log
	\left(
	V_{\mathrm{inoc}}f_j\Lambda_{\mathrm{PFU}}
	\right)
	-
	V_{\mathrm{inoc}}f_j\Lambda_{\mathrm{PFU}}
	-
	\log(N_{jr}!)
	\right].
	\]
	Differentiating with respect to \(\Lambda_{\mathrm{PFU}}\) gives
	\[
	\frac{\partial}{\partial \Lambda_{\mathrm{PFU}}}
	\log \mathcal L_{\mathrm{PFU}}
	=
	\frac{1}{\Lambda_{\mathrm{PFU}}}
	\sum_{j,r}N_{jr}
	-
	V_{\mathrm{inoc}}\sum_j R_j f_j.
	\]
	Setting this derivative equal to zero gives
	Eq.~\eqref{eq:mle_pfu_dilution_series}.
\end{proof}

\begin{remark}[Why the dilution-series likelihood is useful]
	\label{rem:why_dilution_series_likelihood_useful}
	The usual plaque-titer calculation often uses a countable dilution range and a
	simple dilution correction. The likelihood form makes the statistical
	assumptions explicit and allows replicate wells, multiple dilutions,
	uncertainty intervals, and model extensions to be handled in a unified way. It
	also makes clear that the inferred quantity is
	\(\Lambda_{\mathrm{PFU}}\), the protocol-weighted plaque-forming
	concentration, not the total physical particle concentration.
\end{remark}

\begin{remark}[Countable dilution ranges and censored wells]
	\label{rem:countable_dilution_ranges_and_censoring}
	In laboratory practice, very dense wells may be recorded as too numerous to
	count, while very sparse wells may provide little information beyond a zero or
	near-zero count. A more complete likelihood can treat such wells as censored
	or thresholded observations rather than as exact counts. This does not change
	the definition of \(\Lambda_{\mathrm{PFU}}\). It changes the statistical
	readout model \(R_E\) for how plaque-forming events become recorded data.
\end{remark}

\begin{remark}[Scope of the Poisson approximation]
	\label{rem:scope_poisson_plaque_model}
	The Poisson model is the natural first closure when plaque-forming events are
	rare, independent, and spatially non-overlapping. At high concentrations,
	plaques can overlap, merge, compete for susceptible cells, or become difficult
	to count. At very low concentrations, sampling noise dominates. Across
	replicate wells, additional variability may arise from plating heterogeneity,
	cell-monolayer variation, pipetting error, aggregation, local differences in
	adsorption and spread, overlay thickness, incubation conditions, staining
	contrast, or counting thresholds. These effects can be represented by
	overdispersed, zero-inflated, spatial, or censored count models when needed.
\end{remark}

\begin{definition}[Overdispersed plaque-count model]
	\label{def:overdispersed_plaque_count_model}
	When replicate counts exhibit variance larger than the Poisson mean, an
	overdispersed count model may be used. One convenient representation is
	\begin{equation}
		N_{jr}
		\sim
		\operatorname{NegBin}
		\left(
		\mu_{jr},
		\kappa
		\right),
		\qquad
		\mu_{jr}
		=
		V_{\mathrm{inoc}} f_j \Lambda_{\mathrm{PFU}},
		\label{eq:negative_binomial_plaque_model}
	\end{equation}
	with variance, for one common parameterization,
	\begin{equation}
		\operatorname{Var}(N_{jr})
		=
		\mu_{jr}
		+
		\frac{\mu_{jr}^{2}}{\kappa}.
		\label{eq:negative_binomial_plaque_variance}
	\end{equation}
	The parameter \(\kappa>0\) controls overdispersion; the Poisson limit is
	recovered as \(\kappa\rightarrow\infty\). This changes the statistical model
	for count fluctuations, not the definition of
	\(\Lambda_{\mathrm{PFU}}\).
\end{definition}

\begin{remark}[Aggregation and non-independent infectious units]
	\label{rem:aggregation_nonindependent_infectious_units}
	The dilute Poisson model treats plaque-forming events as independent. Viral
	aggregation, co-delivery, cell-to-cell heterogeneity, local clustering,
	collective infection, or spatially correlated susceptibility can violate this
	assumption. In the experimental-collapse framework, such effects do not
	invalidate the plaque assay. They indicate that the latent object being
	counted may not be a single isolated physical virion, but an
	assay-conditioned infectious unit, aggregate, or co-delivered packet. This is
	one reason \(\Lambda_{\mathrm{PFU}}\) should be interpreted as an effective
	plaque-forming concentration rather than as a direct concentration of
	individual physical virions.
\end{remark}

\subsection{Single-Unit Readout Kernel and Null Channel}
\label{subsec:single_unit_readout_kernel_null_channel}

The preceding count model describes the aggregate number of visible plaques. It
is also useful to write the corresponding single-unit readout, because this
makes the null channel explicit. In this subsection, a ``unit'' means whatever
latent assay unit behaves as one independent plaque-forming opportunity under
the dilution model: often a single virion, but potentially an aggregate,
co-delivered packet, or other experimentally inseparable infectious unit.

\begin{definition}[Single-unit plaque-assay readout kernel]
	\label{def:plaque_assay_observation_kernel}
	
	For a single latent state \(x\), define the binary plaque-event variable
	\[
	B\in\{0,1\},
	\]
	with the interpretation
	\begin{equation}
		B =
		\begin{cases}
			1, & \text{\(x\) produces a visible, countable plaque under }
			E_{\mathrm{PFU}}, \\[0.25em]
			0, & \text{\(x\) does not produce a visible, countable plaque.}
		\end{cases}
		\label{eq:single_unit_plaque_binary_variable_definition}
	\end{equation}
	The protocol-conditioned plaque-forming probability is
	\begin{equation}
		\Pr(B=1\mid x,E_{\mathrm{PFU}})
		=
		\pi_{\mathrm{PFU}}(x;E_{\mathrm{PFU}}).
		\label{eq:single_unit_plaque_binary_variable}
	\end{equation}
	Equivalently, the single-unit readout is described by the Bernoulli kernel
	\begin{empheq}[box=\fbox]{equation}
		\begin{aligned}
			K_{\mathrm{PFU}}^{\varnothing}(B\mid x)
			&=
			\left[
			\pi_{\mathrm{PFU}}(x;E_{\mathrm{PFU}})
			\right]^B
			\left[
			1-\pi_{\mathrm{PFU}}(x;E_{\mathrm{PFU}})
			\right]^{1-B},
			\\[-0.1em]
			&\hspace{8em}
			B\in\{0,1\}.
		\end{aligned}
		\label{eq:single_unit_plaque_readout_kernel}
	\end{empheq}
	Thus, \(K_{\mathrm{PFU}}^{\varnothing}\) assigns probability
	\(\pi_{\mathrm{PFU}}(x;E_{\mathrm{PFU}})\) to the plaque-producing outcome
	and probability \(1-\pi_{\mathrm{PFU}}(x;E_{\mathrm{PFU}})\) to the
	non-plaque outcome. The macroscopic plaque count \(N_d\) is then obtained by
	summing these binary plaque events over the diluted plated population, under
	the additional assumptions of dilution, independence, and non-overlap.
\end{definition}

\begin{remark}[Why ``single-unit'' is used]
	\label{rem:why_single_unit_used_plaque_kernel}
	The phrase ``single-unit'' is used instead of ``single-virion'' because the
	plaque-producing unit may not always be one isolated physical virion. In many
	standard dilute applications, the single-virion approximation is appropriate.
	However, aggregation, co-delivery, or collective infection can make the
	plaque-relevant unit larger than one physical particle. The Bernoulli readout
	kernel applies to whatever latent unit the assay treats as one independent
	plaque-forming opportunity.
\end{remark}

\begin{remark}[Null observation in the plaque assay]
	\label{rem:null_observation_in_plaque_assay}
	For a plaque assay, the null channel includes all latent states that fail to
	produce a visible counted plaque. The null state is therefore not a single
	biological mechanism. It includes units that are absent from the plated
	inoculum, fail to survive handling, do not reach a susceptible cell, fail to
	adsorb or enter, enter but do not replicate productively, spread too weakly to
	remain visible under the overlay, merge with neighboring lesions, or remain
	below the detection or counting threshold.
\end{remark}

\begin{remark}[What collapses in the plaque assay]
	\label{rem:what_collapses_in_plaque_assay}
	The plaque assay collapses many latent distinctions into one binary event:
	plaque or no plaque. A virion that is structurally intact but noninfectious, a
	virion that is infectious in principle but fails to adsorb during
	\(t_{\mathrm{ads}}\), a virion that enters but does not replicate
	productively in the chosen cell line, a virion whose local spread is
	suppressed by the overlay or incubation conditions, and a virion whose plaque
	is obscured, merged, or below threshold may all be absent from the final count.
	The observed plaque number therefore compresses structural, mechanical,
	environmental, and biological variation into a protocol-conditioned
	visible-infection event.
\end{remark}

\begin{table}[H]
	\centering
	\caption{
		Experimental-collapse mechanisms in the plaque assay. A standard
		virological protocol implements the abstract collapse structure through
		transport, contact, replication, spread, visibility, and counting stages.
	}
	\label{tab:plaque_assay_collapse_mechanisms}
	\renewcommand{\arraystretch}{1.18}
	\begin{tabularx}{0.98\linewidth}{@{}p{0.24\linewidth}p{0.36\linewidth}X@{}}
		\toprule
		\textbf{Collapse mechanism}
		&
		\textbf{Plaque-assay realization}
		&
		\textbf{Observed consequence}
		\\
		\midrule
		
		Transport conditioning
		&
		Dilution, mixing, inoculum volume, and local transport toward the cell
		layer
		&
		Only units delivered to susceptible cells within the adsorption window can
		contribute.
		\\[0.45em]
		
		Surface or contact conditioning
		&
		Adsorption to the cell monolayer, receptor engagement, and local membrane
		contact
		&
		States with poor adsorption, receptor binding, or entry competence are
		suppressed.
		\\[0.45em]
		
		Preparation or handling conditioning
		&
		Storage, freeze--thaw history, buffer exchange, dilution, washing, and
		overlay addition
		&
		Units that lose infectious competence before readout do not contribute.
		\\[0.45em]
		
		Replication filtering
		&
		Cell-line compatibility, entry, genome release, replication, and
		production of infectious progeny
		&
		Units that enter but fail to amplify locally are not counted as PFU.
		\\[0.45em]
		
		Spread and overlay filtering
		&
		Semi-solid overlay, local diffusion limits, cell-to-cell spread, and
		incubation time
		&
		Only infection events that grow into visible local lesions are counted.
		\\[0.45em]
		
		Geometry or visibility selection
		&
		Plaque size, plaque separation, staining contrast, thresholding, and
		countability
		&
		Foci that are merged, too small, too diffuse, or below threshold may be
		undercounted.
		\\
		\bottomrule
	\end{tabularx}
\end{table}

\subsection{Plaque-Assay Protocol Blindness}
\label{subsec:plaque_assay_protocol_blindness}

The plaque assay is highly sensitive to one composite direction in latent state
space: the direction that changes visible plaque formation under the chosen
protocol. It is comparatively blind to latent distinctions that do not alter the
state-dependent plaque-forming probability
\[
\pi_{\mathrm{PFU}}(x;E_{\mathrm{PFU}})
\]
under those conditions. Thus, plaque-count data are informative about
assay-conditioned infectious activity, but they are not an injective observation
of the full physical, structural, mechanical, or genomic virion ensemble.

\begin{remark}[Protocol blindness in the plaque assay]
	\label{rem:protocol_blindness_in_plaque_assay}
	Two latent virion subpopulations may differ strongly in orientation
	distribution, spike presentation mechanics, rotational diffusion, collision
	history, aggregation pathway, charge state, genome integrity, morphology, or
	preparation history. If those differences do not change
	\(\pi_{\mathrm{PFU}}\) under \(E_{\mathrm{PFU}}\), then they are invisible to
	the plaque count at leading order. Conversely, a small physical or biochemical
	difference that strongly affects adsorption, entry, replication, local spread,
	or visibility may have a large effect on PFU.
\end{remark}

\begin{proposition}[Many latent ensembles can yield the same expected plaque count]
	\label{prop:many_latent_ensembles_same_plaque_count}
	Let \(n_1(x)\) and \(n_2(x)\) be two latent number-density distributions over
	\(\Psi\). If
	\begin{equation}
		\int_{\Psi}
		\pi_{\mathrm{PFU}}(x;E_{\mathrm{PFU}})
		n_1(x)\,dx
		=
		\int_{\Psi}
		\pi_{\mathrm{PFU}}(x;E_{\mathrm{PFU}})
		n_2(x)\,dx,
		\label{eq:latent_ensembles_same_effective_pfu}
	\end{equation}
	then the two ensembles have the same expected plaque count at every dilution
	in the dilute Poisson regime, even if \(n_1\) and \(n_2\) differ substantially
	as latent mechanical, structural, genomic, or biochemical ensembles.
\end{proposition}

\begin{proof}
	Under the dilute Poisson plaque-count model,
	\[
	\mathbb E[N_d]
	=
	V_{\mathrm{inoc}}f_d
	\int_{\Psi}
	\pi_{\mathrm{PFU}}(x;E_{\mathrm{PFU}})
	n(x)\,dx.
	\]
	If the protocol-weighted integrals agree for \(n_1\) and \(n_2\), then
	\[
	\mathbb E_{n_1}[N_d]
	=
	\mathbb E_{n_2}[N_d]
	\]
	for every dilution fraction \(f_d\). Therefore plaque counts alone cannot
	distinguish these two latent ensembles in the dilute regime.
\end{proof}

\begin{remark}[Interpretation of the non-identifiability result]
	\label{rem:interpretation_plaque_nonidentifiability}
	This proposition is the plaque-assay version of protocol blindness. The assay
	is extremely valuable for measuring infectious titer under specified
	conditions, but it is not an injective map from latent virion mechanics to
	observed data. It reports a biologically important projection of the ensemble,
	not the ensemble itself.
\end{remark}

\begin{definition}[Plaque-assay observation equivalence]
	\label{def:plaque_assay_observation_equivalence}
	Two latent number-density distributions \(n_1\) and \(n_2\) are
	\emph{PFU-equivalent} under protocol \(E_{\mathrm{PFU}}\), written
	\[
	n_1\sim_{\mathrm{PFU}} n_2,
	\]
	if they induce the same effective plaque-forming concentration:
	\begin{empheq}[box=\fbox]{equation}
		\int_{\Psi}
		\pi_{\mathrm{PFU}}(x;E_{\mathrm{PFU}})
		n_1(x)\,dx
		=
		\int_{\Psi}
		\pi_{\mathrm{PFU}}(x;E_{\mathrm{PFU}})
		n_2(x)\,dx.
		\label{eq:pfu_equivalence_of_latent_distributions}
	\end{empheq}
\end{definition}

\begin{remark}[Plaque assay as a quotient of latent ensemble space]
	\label{rem:plaque_assay_as_quotient}
	The relation in Eq.~\eqref{eq:pfu_equivalence_of_latent_distributions} is the
	plaque-assay version of the quotient viewpoint introduced earlier. Many
	distinct latent ensembles are collapsed into the same PFU-equivalence class.
	The observed plaque count distinguishes those classes only through the scalar
	\(\Lambda_{\mathrm{PFU}}\), unless additional measurements, richer plaque
	morphology, endpoint structure, or protocol variations are introduced.
\end{remark}

\begin{remark}[What PFU-equivalence does and does not imply]
	\label{rem:what_pfu_equivalence_does_not_imply}
	PFU-equivalence does not imply that two samples have the same number of
	physical particles, the same genome content, the same structural integrity, or
	the same distribution of latent states. It means only that the plaque protocol
	assigns them the same weighted plaque-forming concentration. This distinction
	is central to interpreting particle-to-PFU ratios and to combining PFU assays
	with particle counting, genome quantification, antigen measurement, structural
	imaging, or single-particle measurements.
\end{remark}

\begin{remark}[Transition to the two-subpopulation reduction]
	\label{rem:transition_to_two_subpopulation_reduction}
	The next step is to reduce the latent ensemble to two sectors: a
	plaque-competent sector and a plaque-incompetent, weakly competent, or
	assay-incompatible sector. This reduction is intentionally simple, but it
	exposes the essential inverse problem. A plaque count estimates a weighted sum
	over latent subpopulations; it does not, by itself, reveal how much of the
	sample belongs to each sector.
\end{remark}

\subsection{Two-Subpopulation Reduction}
\label{subsec:two_subpopulation_plaque_reduction}

The general plaque-assay collapse integral
\[
\Lambda_{\mathrm{PFU}}(E_{\mathrm{PFU}})
=
\int_{\Psi}
\pi_{\mathrm{PFU}}(x;E_{\mathrm{PFU}})
n_{\mathrm{ref}}(x)\,dx
\]
is the essential mathematical statement of the worked example. The plaque assay
does not report the full latent ensemble. It reports a protocol-weighted
infectious projection of that ensemble. To make the resulting inverse problem
explicit, it is useful to reduce the latent population to two subpopulations.
\medskip 

\noindent This reduction should be read as protocol-aware. The two sectors are not
necessarily ``infectious'' and ``noninfectious'' in an absolute biological
sense. They are plaque-competent and plaque-incompetent, weakly competent, or
assay-incompatible relative to the specified protocol \(E_{\mathrm{PFU}}\). A
particle that is non-plaque-forming in one cell line, overlay, incubation time,
or staining condition may be more competent under another protocol.

\begin{definition}[Two-subpopulation plaque reduction]
	\label{def:two_subpopulation_plaque_reduction}
	Suppose the latent number-density distribution decomposes as
	\begin{equation}
		n_{\mathrm{ref}}(x)
		=
		n_{\mathrm C}(x)
		+
		n_{\mathrm I}(x),
		\label{eq:two_subpopulation_latent_ensemble}
	\end{equation}
	where \(n_{\mathrm C}\) denotes plaque-competent states and
	\(n_{\mathrm I}\) denotes plaque-incompetent, weakly competent, damaged,
	defective, neutralized, aggregated, or assay-incompatible states. Define the
	corresponding physical concentrations
	\begin{equation}
		C_{\mathrm C}
		=
		\int_{\Psi}
		n_{\mathrm C}(x)\,dx,
		\qquad
		C_{\mathrm I}
		=
		\int_{\Psi}
		n_{\mathrm I}(x)\,dx,
		\qquad
		C_{\mathrm{part}}
		=
		C_{\mathrm C}+C_{\mathrm I}.
		\label{eq:competent_incompetent_concentrations}
	\end{equation}
	When \(C_{\mathrm C}>0\) and \(C_{\mathrm I}>0\), define the
	subpopulation-averaged plaque-forming probabilities
	\begin{equation}
		\overline p_{\mathrm C}(E_{\mathrm{PFU}})
		=
		\frac{
			\displaystyle
			\int_{\Psi}
			\pi_{\mathrm{PFU}}(x;E_{\mathrm{PFU}})
			n_{\mathrm C}(x)\,dx
		}{
			C_{\mathrm C}
		},
		\qquad
		\overline p_{\mathrm I}(E_{\mathrm{PFU}})
		=
		\frac{
			\displaystyle
			\int_{\Psi}
			\pi_{\mathrm{PFU}}(x;E_{\mathrm{PFU}})
			n_{\mathrm I}(x)\,dx
		}{
			C_{\mathrm I}
		}.
		\label{eq:subpopulation_averaged_plaque_probabilities}
	\end{equation}
	Then the effective plaque-forming concentration is exactly
	\begin{empheq}[box=\fbox]{equation}
		\Lambda_{\mathrm{PFU}}
		=
		\overline p_{\mathrm C}C_{\mathrm C}
		+
		\overline p_{\mathrm I}C_{\mathrm I}.
		\label{eq:two_subpopulation_effective_pfu_exact}
	\end{empheq}
\end{definition}

\begin{proof}
	Using
	\(n_{\mathrm{ref}}=n_{\mathrm C}+n_{\mathrm I}\),
	\[
	\Lambda_{\mathrm{PFU}}
	=
	\int_{\Psi}
	\pi_{\mathrm{PFU}}(x;E_{\mathrm{PFU}})
	n_{\mathrm C}(x)\,dx
	+
	\int_{\Psi}
	\pi_{\mathrm{PFU}}(x;E_{\mathrm{PFU}})
	n_{\mathrm I}(x)\,dx.
	\]
	By the definitions of
	\(\overline p_{\mathrm C}\) and \(\overline p_{\mathrm I}\), these two terms
	are
	\[
	\overline p_{\mathrm C}C_{\mathrm C}
	\qquad\text{and}\qquad
	\overline p_{\mathrm I}C_{\mathrm I},
	\]
	respectively. Hence
	\[
	\Lambda_{\mathrm{PFU}}
	=
	\overline p_{\mathrm C}C_{\mathrm C}
	+
	\overline p_{\mathrm I}C_{\mathrm I}.
	\]
\end{proof}

\begin{remark}[Why the averaged probabilities are preferable]
	\label{rem:why_averaged_probabilities_preferable}
	The averaged quantities \(\overline p_{\mathrm C}\) and
	\(\overline p_{\mathrm I}\) are more rigorous than assigning a single
	microscopic probability to every member of a subpopulation. They allow each
	sector to remain internally heterogeneous. A plaque-competent sector may
	contain particles with different adsorption probabilities, entry
	probabilities, replication efficiencies, and local-spread efficiencies. A
	plaque-incompetent sector may contain particles that are completely
	non-plaque-forming together with particles that are weakly competent under
	some assay conditions. Thus the two-subpopulation model reduces the latent
	ensemble without assuming that either sector is microscopically uniform.
\end{remark}

\begin{remark}[Protocol dependence of the two sectors]
	\label{rem:protocol_dependence_two_subpopulation_sectors}
	The labels \(C\) and \(I\) should be interpreted relative to the protocol. A
	state may belong effectively to the plaque-competent sector for one cell line
	or overlay condition and to a weakly competent or assay-incompatible sector for
	another. This is not a contradiction. It reflects the fact that plaque
	competence is a property of the latent state together with the assay
	environment.
\end{remark}

\begin{corollary}[Constant-probability approximation]
	\label{cor:constant_probability_two_subpopulation_approximation}
	If the plaque-forming probability is approximately constant within each
	subpopulation,
\begin{equation}
	\begin{aligned}
		\pi_{\mathrm{PFU}}(x;E_{\mathrm{PFU}})
		&\approx
		p_{\mathrm C},
		&&
		x\in\Psi_{\mathrm C},
		\\[0.25em]
		\pi_{\mathrm{PFU}}(x;E_{\mathrm{PFU}})
		&\approx
		p_{\mathrm I},
		&&
		x\in\Psi_{\mathrm I}.
	\end{aligned}
	\label{eq:two_subpopulation_plaque_probabilities}
\end{equation}
	where \(\Psi_{\mathrm C}\) denotes the plaque-competent sector and
	\(\Psi_{\mathrm I}\) denotes the plaque-incompetent or weakly competent sector. Then,
	\begin{empheq}[box=\fbox]{equation}
		\Lambda_{\mathrm{PFU}}
		\approx
		p_{\mathrm C}C_{\mathrm C}
		+
		p_{\mathrm I}C_{\mathrm I}.
		\label{eq:two_subpopulation_effective_pfu}
	\end{empheq}
\end{corollary}

\begin{remark}[What the two-subpopulation reduction demonstrates]
	\label{rem:what_two_subpopulation_reduction_demonstrates}
	If \(p_{\mathrm I}\approx0\), the plaque assay is effectively blind to the
	plaque-incompetent subpopulation, even if that subpopulation dominates the
	physical particle count. Conversely, changes in \(p_{\mathrm C}\) caused by
	cell line, adsorption time, buffer conditions, overlay composition,
	incubation time, temperature, or readout threshold can change
	\(\Lambda_{\mathrm{PFU}}\) without changing the total number of physical
	particles. This is the simplest mathematical demonstration of why PFU is a
	protocol-conditioned infectious projection rather than a direct physical
	particle count.
\end{remark}
\subsubsection{Competent Fraction and Particle-to-PFU Compression}
\label{subsubsec:competent_fraction_particle_to_pfu_compression}

The two-subpopulation model becomes especially transparent when written in terms
of the plaque-competent fraction of the physical particle population. This form
separates three quantities that are often compressed into a single PFU
measurement: total physical abundance, the fraction of particles in a
plaque-competent sector, and the protocol-dependent efficiency with which each
sector produces visible plaques.

\begin{definition}[Plaque-competent fraction]
	\label{def:plaque_competent_fraction}
	Define the plaque-competent fraction by
	\begin{equation}
		\varphi_{\mathrm C}
		=
		\frac{C_{\mathrm C}}{C_{\mathrm{part}}},
		\qquad
		1-\varphi_{\mathrm C}
		=
		\frac{C_{\mathrm I}}{C_{\mathrm{part}}},
		\label{eq:plaque_competent_fraction}
	\end{equation}
	where
	\[
	C_{\mathrm{part}}
	=
	C_{\mathrm C}+C_{\mathrm I}.
	\]
	Under the constant-probability approximation, the effective plaque-forming
	concentration can be written as
	\begin{empheq}[box=\fbox]{equation}
		\Lambda_{\mathrm{PFU}}
		=
		C_{\mathrm{part}}
		\left[
		p_{\mathrm I}
		+
		(p_{\mathrm C}-p_{\mathrm I})\varphi_{\mathrm C}
		\right].
		\label{eq:lambda_pfu_competent_fraction_form}
	\end{empheq}
\end{definition}

\begin{proof}
	From the two-subpopulation approximation,
	\[
	\Lambda_{\mathrm{PFU}}
	=
	p_{\mathrm C}C_{\mathrm C}
	+
	p_{\mathrm I}C_{\mathrm I}.
	\]
	Using
	\[
	C_{\mathrm C}
	=
	\varphi_{\mathrm C}C_{\mathrm{part}},
	\qquad
	C_{\mathrm I}
	=
	(1-\varphi_{\mathrm C})C_{\mathrm{part}},
	\]
	gives
	\[
	\Lambda_{\mathrm{PFU}}
	=
	C_{\mathrm{part}}
	\left[
	p_{\mathrm C}\varphi_{\mathrm C}
	+
	p_{\mathrm I}(1-\varphi_{\mathrm C})
	\right]
	=
	C_{\mathrm{part}}
	\left[
	p_{\mathrm I}
	+
	(p_{\mathrm C}-p_{\mathrm I})\varphi_{\mathrm C}
	\right].
	\]
\end{proof}

\begin{remark}[What a plaque assay can and cannot infer]
	\label{rem:what_plaque_assay_can_infer_two_population}
	Equation~\eqref{eq:lambda_pfu_competent_fraction_form} shows that a plaque
	assay alone identifies the combination
	\[
	C_{\mathrm{part}}
	\left[
	p_{\mathrm I}
	+
	(p_{\mathrm C}-p_{\mathrm I})\varphi_{\mathrm C}
	\right],
	\]
	not \(C_{\mathrm{part}}\), \(p_{\mathrm C}\), \(p_{\mathrm I}\), and
	\(\varphi_{\mathrm C}\) separately. Additional information is required to
	separate physical particle concentration from plaque competence and
	protocol-dependent plaque-forming efficiency. This is the two-subpopulation
	version of protocol non-identifiability.
\end{remark}

\begin{definition}[Two-subpopulation particle-to-PFU ratio]
	\label{def:two_subpopulation_particle_to_pfu_ratio}
	Assume \(\Lambda_{\mathrm{PFU}}>0\). Under the constant-probability
	two-subpopulation approximation, the particle-to-PFU ratio is
	\begin{empheq}[box=\fbox]{equation}
		\mathcal R_{\mathrm{part/PFU}}
		=
		\frac{C_{\mathrm{part}}}{\Lambda_{\mathrm{PFU}}}
		=
		\frac{1}{
			p_{\mathrm I}
			+
			(p_{\mathrm C}-p_{\mathrm I})\varphi_{\mathrm C}
		}.
		\label{eq:two_subpopulation_particle_to_pfu_ratio}
	\end{empheq}
\end{definition}

\begin{remark}[Interpretation of the two-subpopulation ratio]
	\label{rem:interpretation_two_subpopulation_ratio}
	A large particle-to-PFU ratio can arise for several distinct reasons:
	\(\varphi_{\mathrm C}\) may be small; \(p_{\mathrm C}\) may be small under
	the chosen protocol; \(p_{\mathrm I}\) may be effectively zero; or the assay
	conditions may suppress one or more stages of plaque formation. The ratio is
	therefore not a single mechanistic explanation. It is a compressed summary of
	latent composition and protocol efficiency
	\cite{Klasse2015,Sanjuan2018,McCormick2021,Brooke2014}.
\end{remark}

\begin{remark}[Compression interpretation]
	\label{rem:particle_to_pfu_compression_interpretation}
	The factor
	\[
	p_{\mathrm I}
	+
	(p_{\mathrm C}-p_{\mathrm I})\varphi_{\mathrm C}
	\]
	is the average plaque-forming probability of a randomly selected physical
	particle or assay unit under the protocol. The particle-to-PFU ratio is the
	reciprocal of this average probability. Thus particle-to-PFU compression
	quantifies how strongly the physical particle ensemble is reduced when viewed
	through the plaque-forming projection.
\end{remark}

\subsubsection{Special Cases}
\label{subsubsec:special_cases_two_subpopulation_plaque_model}

The two-subpopulation formula is useful because several familiar virological
situations appear as simple limiting cases.

\begin{remark}[Nearly ideal competent sector]
	\label{rem:nearly_ideal_competent_sector}
	If \(p_{\mathrm C}\approx1\) and \(p_{\mathrm I}\approx0\), then
	\[
	\Lambda_{\mathrm{PFU}}
	\approx
	C_{\mathrm C},
	\qquad
	\mathcal R_{\mathrm{part/PFU}}
	\approx
	\frac{1}{\varphi_{\mathrm C}}.
	\]
	In this idealized limit, the particle-to-PFU ratio approximately reports the
	inverse of the competent fraction. This interpretation fails if competent
	particles have \(p_{\mathrm C}<1\), if weakly competent particles have
	\(p_{\mathrm I}>0\), or if aggregation, co-delivery, or collective infection
	changes the effective plaque-forming unit.
\end{remark}

\begin{remark}[Assay suppression of otherwise competent particles]
	\label{rem:assay_suppression_competent_particles}
	If \(p_{\mathrm C}\ll1\) under a given protocol, then even a sample with a
	large competent physical subpopulation may produce a low PFU titer.
	Biologically, this may occur because the chosen cell line is poorly
	susceptible, adsorption time is short, the overlay suppresses local spread,
	incubation time is insufficient, temperature or medium conditions are
	unfavorable, or the readout threshold is high. In this case, low PFU does not
	necessarily imply absence of physically competent particles; it may indicate
	poor assay coupling to the competent sector.
\end{remark}

\begin{remark}[Weakly competent or conditionally competent sector]
	\label{rem:weakly_competent_sector}
	If \(p_{\mathrm I}>0\), the nominally incompetent sector is not truly
	invisible. It may contain weakly competent, partially damaged,
	semi-infectious, cell-line-dependent, or conditionally competent states. Such
	states contribute weakly to \(\Lambda_{\mathrm{PFU}}\) and may become more
	visible under protocol variation. This is one reason that the labels
	\(C\) and \(I\) should be treated as reduced modeling sectors rather than
	absolute biological categories.
\end{remark}

\begin{remark}[Protocol change can move states between effective sectors]
	\label{rem:protocol_change_moves_effective_sectors}
	The same physical state may be effectively plaque-incompetent under one
	protocol and weakly or strongly plaque-competent under another. Changing the
	cell line, adsorption time, overlay composition, incubation time, temperature,
	neutralization condition, or staining threshold changes the biological kernel
	and can therefore change the effective values of \(p_{\mathrm C}\),
	\(p_{\mathrm I}\), and even the sector assignment implicit in
	\(\varphi_{\mathrm C}\). This prepares the protocol-variation argument in the
	next subsection.
\end{remark}

\subsubsection{Non-Identifiability from PFU Alone}
\label{subsubsec:nonidentifiability_from_pfu_alone}

\begin{proposition}[Non-identifiability from PFU alone]
	\label{prop:two_subpopulation_nonidentifiability_from_pfu_alone}
	Suppose the plaque count identifies only
	\[
	\Lambda_{\mathrm{PFU}}
	=
	p_{\mathrm C}C_{\mathrm C}
	+
	p_{\mathrm I}C_{\mathrm I}.
	\]
	Then \(C_{\mathrm C}\), \(C_{\mathrm I}\), \(p_{\mathrm C}\), and
	\(p_{\mathrm I}\) are not jointly identifiable from plaque counts alone.
\end{proposition}

\begin{proof}
	A plaque count in the dilute Poisson regime depends on the subpopulation
	parameters only through the scalar
	\[
	\Lambda_{\mathrm{PFU}}
	=
	p_{\mathrm C}C_{\mathrm C}
	+
	p_{\mathrm I}C_{\mathrm I}.
	\]
	There are infinitely many parameter quadruples
	\[
	(C_{\mathrm C},C_{\mathrm I},p_{\mathrm C},p_{\mathrm I})
	\]
	that yield the same scalar value. For example, if \(p_{\mathrm I}=0\), then
	\[
	\Lambda_{\mathrm{PFU}}
	=
	p_{\mathrm C}C_{\mathrm C}.
	\]
	Multiplying \(p_{\mathrm C}\) by a factor \(a>0\) and dividing
	\(C_{\mathrm C}\) by the same factor leaves \(\Lambda_{\mathrm{PFU}}\)
	unchanged whenever the transformed probability remains in the admissible
	range \(0\leq ap_{\mathrm C}\leq1\). Therefore the individual subpopulation
	sizes and plaque-forming probabilities are not identifiable from PFU data
	alone.
\end{proof}

\begin{remark}[Interpretation of the non-identifiability]
	\label{rem:interpretation_two_subpopulation_nonidentifiability}
	This proposition does not weaken the plaque assay. It clarifies its target.
	The assay is designed to estimate an effective infectious concentration under
	a specified protocol. It is not designed, by itself, to decompose that
	concentration into physical particle abundance, competent fraction,
	adsorption probability, entry probability, replication competence, local
	spread, and visibility threshold. Those decompositions require additional
	measurements, controlled protocol variation, or prior biological assumptions.
\end{remark}

\begin{remark}[A single PFU value is a weighted sum, not a composition]
	\label{rem:pfu_weighted_sum_not_composition}
	A PFU estimate reports the weighted sum
	\[
	p_{\mathrm C}C_{\mathrm C}
	+
	p_{\mathrm I}C_{\mathrm I}.
	\]
	It does not report the mixture composition
	\((C_{\mathrm C},C_{\mathrm I})\), nor does it report the stage-specific
	probabilities that determine \(p_{\mathrm C}\) and \(p_{\mathrm I}\). This is
	why samples with different latent compositions can yield the same PFU titer,
	and why identical physical particle counts can yield different PFU titers.
\end{remark}

\subsubsection{Using Additional Measurements to Resolve the Two Populations}
\label{subsubsec:additional_measurements_resolve_two_populations}

The two-subpopulation model also shows how complementary protocols can reduce
plaque-assay blindness. Suppose an independent particle-counting,
genome-counting, antigen-counting, or structural measurement provides
information about \(C_{\mathrm{part}}\) or about the composition of the physical
particle population. Then the plaque assay and the auxiliary measurement
together can begin to separate physical abundance from plaque-forming
efficiency.

\begin{proposition}[Competent fraction from particle count and PFU count]
	\label{prop:competent_fraction_from_particle_and_pfu}
	Assume the constant-probability two-subpopulation model, with
	\(p_{\mathrm C}\neq p_{\mathrm I}\), and suppose both
	\(C_{\mathrm{part}}\) and \(\Lambda_{\mathrm{PFU}}\) are known. Then
	\begin{empheq}[box=\fbox]{equation}
		\varphi_{\mathrm C}
		=
		\frac{
			\Lambda_{\mathrm{PFU}}/C_{\mathrm{part}} - p_{\mathrm I}
		}{
			p_{\mathrm C}-p_{\mathrm I}
		}.
		\label{eq:competent_fraction_from_particle_and_pfu}
	\end{empheq}
\end{proposition}

\begin{proof}
	Starting from Eq.~\eqref{eq:lambda_pfu_competent_fraction_form},
	\[
	\frac{\Lambda_{\mathrm{PFU}}}{C_{\mathrm{part}}}
	=
	p_{\mathrm I}
	+
	(p_{\mathrm C}-p_{\mathrm I})\varphi_{\mathrm C}.
	\]
	Solving for \(\varphi_{\mathrm C}\) gives
	Eq.~\eqref{eq:competent_fraction_from_particle_and_pfu}.
\end{proof}

\begin{remark}[Why calibration matters]
	\label{rem:why_calibration_matters_two_population}
	Equation~\eqref{eq:competent_fraction_from_particle_and_pfu} is useful only
	if \(p_{\mathrm C}\) and \(p_{\mathrm I}\) are known, estimated, or constrained
	by additional experiments. If these probabilities are unknown, then adding a
	particle count separates total physical abundance from the infectious
	projection, but it still does not uniquely determine which biological stage
	limits plaque formation.
\end{remark}

\begin{remark}[What auxiliary measurements contribute]
	\label{rem:what_auxiliary_measurements_contribute_two_population}
	Different auxiliary measurements constrain different latent sectors. A
	particle count constrains physical abundance. A genome count constrains
	genome-containing material. An antigen measurement constrains protein-bearing
	material. Structural imaging constrains morphology and particle integrity. A
	neutralization or receptor-binding assay constrains entry-related competence.
	None of these measurements is identical to PFU, but that is precisely why they
	are useful: each adds a different protocol-conditioned projection of the
	latent population.
\end{remark}

\begin{remark}[Transition to protocol variation]
	\label{rem:transition_to_protocol_variation_two_population}
	Auxiliary measurements constrain the latent composition from outside the
	plaque assay. Protocol variation provides a complementary strategy from within
	the biological assay itself. By changing cell line, adsorption time, overlay,
	incubation time, neutralization condition, or readout threshold, one changes
	the effective values of \(p_{\mathrm C}\) and \(p_{\mathrm I}\). The next
	subsection uses this idea to treat protocol variation as a two-subpopulation
	design problem.
\end{remark}
\subsubsection{Protocol Variation as a Two-Subpopulation Design}
\label{subsubsec:protocol_variation_two_subpopulation_design}

Controlled plaque-assay variation can reduce the non-identifiability exposed by
the two-subpopulation model. The goal is not to vary assay conditions
arbitrarily, but to vary them in ways that change the plaque-forming
probabilities of the two latent sectors differently. In this sense, protocol
variation is a design tool: it changes the biological observation kernel so that
previously confounded subpopulations may become distinguishable.

\begin{definition}[Protocol variation as a two-subpopulation design]
	\label{def:protocol_variation_two_subpopulation_design}
	Let \(E_1,\ldots,E_M\) be plaque-assay protocols that differ in cell line,
	adsorption time, overlay composition, incubation time, temperature, readout
	threshold, neutralization condition, receptor availability, or another
	controlled protocol variable. Under the constant-probability
	two-subpopulation approximation, each protocol produces an effective
	plaque-forming concentration
	\begin{equation}
		\Lambda_m
		=
		p_{\mathrm C}^{(m)}C_{\mathrm C}
		+
		p_{\mathrm I}^{(m)}C_{\mathrm I},
		\qquad
		m=1,\ldots,M,
		\label{eq:multi_protocol_two_subpopulation_linear_system}
	\end{equation}
	where \(p_{\mathrm C}^{(m)}\) and \(p_{\mathrm I}^{(m)}\) are the
	protocol-conditioned plaque-forming probabilities of the competent and
	incompetent or weakly competent sectors under \(E_m\). In matrix form,
	\begin{empheq}[box=\fbox]{equation}
		\boldsymbol{\Lambda}
		=
		\mathbf P\mathbf C,
		\qquad
		\mathbf P
		=
		\begin{pmatrix}
			p_{\mathrm C}^{(1)} & p_{\mathrm I}^{(1)}\\
			\vdots & \vdots\\
			p_{\mathrm C}^{(M)} & p_{\mathrm I}^{(M)}
		\end{pmatrix},
		\qquad
		\mathbf C
		=
		\begin{pmatrix}
			C_{\mathrm C}\\
			C_{\mathrm I}
		\end{pmatrix}.
		\label{eq:two_subpopulation_linear_inverse_problem}
	\end{empheq}
\end{definition}

\begin{proposition}[Rank condition for resolving two subpopulations]
	\label{prop:rank_condition_two_subpopulation_resolution}
	In the ideal noiseless setting, the concentrations
	\((C_{\mathrm C},C_{\mathrm I})\) are identifiable from
	\(\boldsymbol{\Lambda}\) only if
	\begin{empheq}[box=\fbox]{equation}
		\operatorname{rank}(\mathbf P)=2.
		\label{eq:rank_condition_two_subpopulation_resolution}
	\end{empheq}
\end{proposition}

\begin{proof}
	The system
	\[
	\boldsymbol{\Lambda}
	=
	\mathbf P\mathbf C
	\]
	has a unique solution for the two-dimensional vector \(\mathbf C\) only if
	the two columns of \(\mathbf P\) are linearly independent. This is exactly the
	condition
	\[
	\operatorname{rank}(\mathbf P)=2.
	\]
	If the rank is one or zero, then at least one nonzero direction in
	\((C_{\mathrm C},C_{\mathrm I})\)-space lies in the null space of
	\(\mathbf P\), so different subpopulation concentrations produce the same
	vector of effective plaque-forming concentrations.
\end{proof}

\begin{remark}[Interpretation of the rank condition]
	\label{rem:interpretation_rank_condition_two_subpopulation}
	Varying a protocol is useful only if it changes the plaque-forming
	probabilities of the two subpopulations differently. If every protocol merely
	scales \(p_{\mathrm C}\) and \(p_{\mathrm I}\) by the same factor, then the
	columns of \(\mathbf P\) remain effectively collinear and the two sectors
	remain confounded. A useful protocol variation is one that changes the
	observation kernel in a direction that separates latent sectors.
\end{remark}

\begin{remark}[Examples of useful plaque-assay variations]
	\label{rem:examples_useful_plaque_assay_variations}
	Changing cell line can alter receptor usage, entry, replication, or local
	spread. Changing adsorption time can alter the probability that particles
	reach and bind susceptible cells. Changing overlay composition can alter
	local spread, plaque morphology, and countability. Changing incubation time
	can alter whether weak, slow, or delayed infection events become visible.
	Changing staining or reporter threshold can alter final visibility. Adding a
	neutralization or receptor-blocking condition can preferentially suppress
	some latent sectors. Each variation changes the biological protocol kernel
	and can, in principle, expose a different mixture of latent subpopulations.
\end{remark}

\begin{remark}[Calibration versus identification]
	\label{rem:calibration_versus_identification_two_subpopulation}
	The rank condition assumes that the protocol probability matrix
	\(\mathbf P\) is known, calibrated, estimated, or otherwise constrained. If
	the entries \(p_{\mathrm C}^{(m)}\) and \(p_{\mathrm I}^{(m)}\) are unknown,
	then protocol variation alone does not automatically identify
	\((C_{\mathrm C},C_{\mathrm I})\). In that case, the experiment identifies a
	combined inverse problem involving both subpopulation concentrations and
	protocol-specific plaque-forming probabilities. Additional calibration,
	prior biological information, auxiliary measurements, or stronger model
	structure is then required.
\end{remark}

\subsubsection{Fisher Information in the Two-Subpopulation Plaque Model}
\label{subsubsec:fisher_information_two_subpopulation_plaque_model}

The same identifiability issue can be expressed locally through Fisher
information. For a single dilution fraction \(f_d\), the Poisson mean is
\[
\mu_d
=
V_{\mathrm{inoc}} f_d
\left(
p_{\mathrm C}C_{\mathrm C}
+
p_{\mathrm I}C_{\mathrm I}
\right).
\]
Let
\[
\theta
=
(C_{\mathrm C},C_{\mathrm I})^{\mathsf T},
\qquad
\mathbf p
=
(p_{\mathrm C},p_{\mathrm I})^{\mathsf T}.
\]
Then
\[
\mu_d
=
V_{\mathrm{inoc}}f_d\,\mathbf p^{\mathsf T}\theta.
\]

\begin{proposition}[Rank-one Fisher information for a single plaque protocol]
	\label{prop:rank_one_fisher_single_plaque_protocol}
	For a single plaque-assay protocol in the two-subpopulation Poisson model,
	the Fisher information for
	\[
	\theta=(C_{\mathrm C},C_{\mathrm I})^{\mathsf T}
	\]
	is
	\begin{empheq}[box=\fbox]{equation}
		\mathcal I_E(\theta)
		=
		\frac{
			\left(V_{\mathrm{inoc}}f_d\right)^2
		}{
			\mu_d
		}
		\mathbf p\mathbf p^{\mathsf T}.
		\label{eq:rank_one_fisher_two_subpopulation_plaque}
	\end{empheq}
	Therefore, if \(\mathbf p\neq 0\), the Fisher matrix has rank one and only
	the concentration combination
	\[
	p_{\mathrm C}C_{\mathrm C}
	+
	p_{\mathrm I}C_{\mathrm I}
	\]
	is locally identifiable from that protocol alone.
\end{proposition}

\begin{proof}
	For a Poisson random variable with mean \(\mu_d(\theta)\), the Fisher
	information is
	\[
	\mathcal I_E(\theta)
	=
	\frac{1}{\mu_d}
	\nabla_\theta\mu_d
	\nabla_\theta\mu_d^{\mathsf T}.
	\]
	Since
	\[
	\nabla_\theta\mu_d
	=
	V_{\mathrm{inoc}}f_d\,\mathbf p,
	\]
	we obtain Eq.~\eqref{eq:rank_one_fisher_two_subpopulation_plaque}. The matrix
	\(\mathbf p\mathbf p^{\mathsf T}\) has rank one when
	\(\mathbf p\neq0\), so the protocol identifies only one linear combination
	of the two concentrations.
\end{proof}

\begin{remark}[Multiple dilutions under one protocol]
	\label{rem:multiple_dilutions_same_protocol_rank}
	Multiple dilutions and replicate wells improve precision for the scalar
	\(\Lambda_{\mathrm{PFU}}\), but they do not by themselves separate
	\(C_{\mathrm C}\) from \(C_{\mathrm I}\) if all wells share the same
	probability vector
	\[
	\mathbf p=(p_{\mathrm C},p_{\mathrm I})^{\mathsf T}.
	\]
	Dilution changes the scale of the Poisson mean; it does not rotate the
	sensitivity direction in subpopulation space.
\end{remark}

\begin{remark}[Fisher form of plaque-assay blindness]
	\label{rem:fisher_form_plaque_assay_blindness}
	The rank-one Fisher matrix is the local information-theoretic version of
	plaque-assay protocol blindness. A single plaque protocol is sensitive to the
	direction \((p_{\mathrm C},p_{\mathrm I})\) in subpopulation space and blind
	to directions orthogonal to it. Multi-protocol variation can increase the
	Fisher rank only when different protocols have non-collinear probability
	vectors
	\[
	(p_{\mathrm C}^{(m)},p_{\mathrm I}^{(m)}).
	\]
\end{remark}

\begin{proposition}[Fisher information under protocol variation]
	\label{prop:fisher_information_protocol_variation_two_subpopulation}
	For independent plaque-count data from protocols \(E_1,\ldots,E_M\), with
	Poisson means
	\[
	\mu_m
	=
	V_m f_m
	\left(
	p_{\mathrm C}^{(m)}C_{\mathrm C}
	+
	p_{\mathrm I}^{(m)}C_{\mathrm I}
	\right),
	\]
	the Fisher information for
	\[
	\theta=(C_{\mathrm C},C_{\mathrm I})^{\mathsf T}
	\]
	is
	\begin{empheq}[box=\fbox]{equation}
		\mathcal I_{\mathrm{multi}}(\theta)
		=
		\sum_{m=1}^{M}
		\frac{
			(V_m f_m)^2
		}{
			\mu_m
		}
		\mathbf p_m\mathbf p_m^{\mathsf T},
		\qquad
		\mathbf p_m
		=
		\begin{pmatrix}
			p_{\mathrm C}^{(m)}\\
			p_{\mathrm I}^{(m)}
		\end{pmatrix}.
		\label{eq:multi_protocol_fisher_two_subpopulation}
	\end{empheq}
	The Fisher rank can be two only if at least two protocol probability vectors
	are not collinear.
\end{proposition}

\begin{proof}
	Each independent Poisson count contributes Fisher information
	\[
	\frac{1}{\mu_m}
	\nabla_\theta\mu_m
	\nabla_\theta\mu_m^{\mathsf T}.
	\]
	Since
	\[
	\nabla_\theta\mu_m
	=
	V_m f_m\mathbf p_m,
	\]
	summing independent protocol contributions gives
	Eq.~\eqref{eq:multi_protocol_fisher_two_subpopulation}. The resulting Fisher
	matrix can span the two-dimensional parameter space only if the vectors
	\(\mathbf p_m\) span that space.
\end{proof}

\begin{remark}[Design interpretation of the Fisher sum]
	\label{rem:design_interpretation_fisher_sum_two_subpopulation}
	Equation~\eqref{eq:multi_protocol_fisher_two_subpopulation} gives the local
	design principle for plaque-assay variation. Additional wells, replicates,
	and dilutions increase information along the sensitivity directions already
	present. Additional protocol conditions can add new sensitivity directions.
	Thus a second protocol is most useful when its probability vector
	\(\mathbf p_m\) is not merely a rescaled version of the first.
\end{remark}

\begin{remark}[Uncertainty in protocol probabilities]
	\label{rem:uncertainty_protocol_probabilities_two_subpopulation}
	In real applications, the protocol probabilities
	\(p_{\mathrm C}^{(m)}\) and \(p_{\mathrm I}^{(m)}\) may themselves be
	estimated or uncertain. This adds nuisance parameters to the inverse problem
	and can reduce the effective information about
	\((C_{\mathrm C},C_{\mathrm I})\). A full analysis may therefore require
	profiling, marginalizing, or experimentally calibrating the probability
	vectors \(\mathbf p_m\). This is the plaque-assay version of the general
	nuisance-parameter problem described earlier.
\end{remark}
\begin{table}[H]
	\centering
	\caption{
		Protocol variations in the two-subpopulation plaque model. A useful
		variation changes the probability vector
		\(\mathbf p_m=(p_{\mathrm C}^{(m)},p_{\mathrm I}^{(m)})^{\mathsf T}\)
		in a direction that helps separate latent sectors.
	}
	\label{tab:two_subpopulation_protocol_variation_design}
	\renewcommand{\arraystretch}{1.18}
	\begin{tabularx}{0.98\linewidth}{@{}p{0.24\linewidth}p{0.34\linewidth}X@{}}
		\toprule
		\textbf{Protocol variation}
		&
		\textbf{Likely effect on probability vector}
		&
		\textbf{Interpretive use}
		\\
		\midrule
		
		Cell line or receptor condition
		&
		Changes entry, intracellular permissiveness, or local spread
		&
		Helps distinguish physically present particles from cell-line-compatible
		plaque-forming units.
		\\[0.45em]
		
		Adsorption time
		&
		Changes delivery, attachment, and early cell-contact probability
		&
		Tests whether low PFU reflects poor adsorption rather than absence of
		competent particles.
		\\[0.45em]
		
		Overlay composition or viscosity
		&
		Changes local spread, lesion expansion, and plaque countability
		&
		Separates entry/replication competence from spread-limited visibility.
		\\[0.45em]
		
		Incubation time
		&
		Changes whether slow or weak events become visible
		&
		Tests delayed amplification, slow replication, or threshold-limited
		plaque detection.
		\\[0.45em]
		
		Neutralization or receptor-blocking condition
		&
		Changes attachment, entry, aggregation, or null-channel probability
		&
		Identifies sectors sensitive to antibody, inhibitor, or receptor-mediated
		blocking.
		\\[0.45em]
		
		Staining or reporter threshold
		&
		Changes final visibility and counting probability
		&
		Separates biological amplification from readout-threshold collapse.
		\\
		\bottomrule
	\end{tabularx}
\end{table}
\subsubsection{Illustrative Worked Numerical Reading}
\label{subsubsec:worked_numerical_reading_plaque_assay}

The two-subpopulation model can be read numerically. The purpose of the example
is not to assign universal values to plaque competence, but to show how the same
equations distinguish physical particle abundance from protocol-conditioned
plaque-forming concentration.

Suppose a plaque assay is described by the constant-probability approximation
\[
\Lambda_{\mathrm{PFU}}
=
C_{\mathrm{part}}
\left[
p_{\mathrm I}
+
(p_{\mathrm C}-p_{\mathrm I})\varphi_{\mathrm C}
\right],
\]
where \(C_{\mathrm{part}}\) is the total physical particle concentration,
\(\varphi_{\mathrm C}\) is the plaque-competent fraction,
\(p_{\mathrm C}\) is the plaque-forming probability of the competent sector, and
\(p_{\mathrm I}\) is the plaque-forming probability of the weakly competent or
assay-incompatible sector.

\begin{remark}[Same particle concentration, different PFU projection]
	\label{rem:same_particle_concentration_different_pfu_projection}
	Let two samples have the same physical particle concentration
	\[
	C_{\mathrm{part}}
	=
	10^{8}
	\;\text{particles per mL},
	\]
	and suppose the plaque-assay protocol has
	\[
	p_{\mathrm C}=0.80,
	\qquad
	p_{\mathrm I}=10^{-3}.
	\]
	If sample A has
	\[
	\varphi_{\mathrm C}=0.10,
	\]
	then
	\begin{equation}
		\Lambda_{\mathrm{PFU}}^{(A)}
		=
		10^{8}
		\left[
		10^{-3}
		+
		(0.80-10^{-3})(0.10)
		\right]
		\approx
		8.09\times 10^{6}
		\;\mathrm{PFU/mL}.
		\label{eq:numerical_sample_A_pfu}
	\end{equation}
	If sample B has
	\[
	\varphi_{\mathrm C}=0.01,
	\]
	then
	\begin{equation}
		\Lambda_{\mathrm{PFU}}^{(B)}
		=
		10^{8}
		\left[
		10^{-3}
		+
		(0.80-10^{-3})(0.01)
		\right]
		\approx
		8.99\times 10^{5}
		\;\mathrm{PFU/mL}.
		\label{eq:numerical_sample_B_pfu}
	\end{equation}
	The two samples have the same total physical particle concentration, but the
	PFU projections differ by almost an order of magnitude because the
	protocol-weighted plaque-competent fraction differs.
\end{remark}

For a plated inoculum volume
\[
V_{\mathrm{inoc}}=0.1\;\mathrm{mL}
\]
and dilution fraction
\[
f_d=10^{-4},
\]
the expected plaque counts are
\begin{equation}
	\mathbb E[N_d^{(A)}]
	=
	0.1\times10^{-4}\times8.09\times10^{6}
	\approx
	80.9,
	\label{eq:expected_count_sample_A}
\end{equation}
and
\begin{equation}
	\mathbb E[N_d^{(B)}]
	=
	0.1\times10^{-4}\times8.99\times10^{5}
	\approx
	9.0.
	\label{eq:expected_count_sample_B}
\end{equation}
Thus, under the same plating and dilution conditions, two samples with identical
particle concentration may yield very different count regimes.

\begin{remark}[Particle-to-PFU compression in the numerical example]
	\label{rem:particle_to_pfu_compression_numerical_example}
	The particle-to-PFU ratios in the two samples are
	\begin{equation}
		\mathcal R_{\mathrm{part/PFU}}^{(A)}
		=
		\frac{10^{8}}{8.09\times10^{6}}
		\approx
		12.4,
		\qquad
		\mathcal R_{\mathrm{part/PFU}}^{(B)}
		=
		\frac{10^{8}}{8.99\times10^{5}}
		\approx
		111.2.
		\label{eq:numerical_particle_to_pfu_ratios}
	\end{equation}
	The ratio is therefore not an intrinsic constant of the virus preparation
	alone. It reflects the latent composition of the sample and the
	plaque-forming probabilities imposed by the assay protocol.
\end{remark}

\begin{remark}[Same PFU projection, different latent composition]
	\label{rem:same_pfu_different_latent_composition}
	The inverse ambiguity can also be seen directly. Suppose one plaque protocol
	has
	\[
	p_{\mathrm C}=0.80,
	\qquad
	p_{\mathrm I}=0.05,
	\]
	and suppose the observed effective plaque-forming concentration is
	\[
	\Lambda_{\mathrm{PFU}}
	=
	8.0\times10^{6}
	\;\mathrm{PFU/mL}.
	\]
	Then all pairs \((C_{\mathrm C},C_{\mathrm I})\) satisfying
	\begin{equation}
		0.80\,C_{\mathrm C}
		+
		0.05\,C_{\mathrm I}
		=
		8.0\times10^{6}
		\label{eq:numerical_same_pfu_line}
	\end{equation}
	have the same expected plaque count under this protocol. For example,
	\[
	(C_{\mathrm C},C_{\mathrm I})
	=
	(1.0\times10^{7},0)
	\]
	and
	\[
	(C_{\mathrm C},C_{\mathrm I})
	=
	(5.0\times10^{6},8.0\times10^{7})
	\]
	both yield
	\[
	\Lambda_{\mathrm{PFU}}
	=
	8.0\times10^{6}
	\;\mathrm{PFU/mL}.
	\]
	The first sample is dominated by plaque-competent units, while the second has
	a much larger weakly competent or assay-incompatible sector. A single plaque
	protocol cannot distinguish these latent compositions if they lie on the same
	PFU-equivalence line.
\end{remark}

\begin{remark}[Numerical protocol variation]
	\label{rem:numerical_protocol_variation}
	Now suppose two plaque protocols produce different probability vectors:
	\[
	\mathbf p_1
	=
	\begin{pmatrix}
		0.80\\
		0.05
	\end{pmatrix},
	\qquad
	\mathbf p_2
	=
	\begin{pmatrix}
		0.20\\
		0.15
	\end{pmatrix}.
	\]
	These vectors are not collinear. If the true subpopulation concentrations are
	\[
	\mathbf C
	=
	\begin{pmatrix}
		C_{\mathrm C}\\
		C_{\mathrm I}
	\end{pmatrix}
	=
	\begin{pmatrix}
		1.0\times10^{7}\\
		8.0\times10^{7}
	\end{pmatrix}
	\;\mathrm{mL}^{-1},
	\]
	then the two effective plaque-forming concentrations are
	\begin{equation}
		\Lambda_1
		=
		0.80(1.0\times10^{7})
		+
		0.05(8.0\times10^{7})
		=
		1.2\times10^{7}
		\;\mathrm{PFU/mL},
		\label{eq:numerical_protocol_one_lambda}
	\end{equation}
	and
	\begin{equation}
		\Lambda_2
		=
		0.20(1.0\times10^{7})
		+
		0.15(8.0\times10^{7})
		=
		1.4\times10^{7}
		\;\mathrm{PFU/mL}.
		\label{eq:numerical_protocol_two_lambda}
	\end{equation}
	The corresponding design matrix is
	\[
	\mathbf P
	=
	\begin{pmatrix}
		0.80 & 0.05\\
		0.20 & 0.15
	\end{pmatrix},
	\]
	with determinant
	\begin{equation}
		\det \mathbf P
		=
		0.80(0.15)-0.05(0.20)
		=
		0.11
		\neq
		0.
		\label{eq:numerical_protocol_design_determinant}
	\end{equation}
	In the ideal noiseless and calibrated setting, the two protocols therefore
	provide enough independent directions to resolve the two concentrations.
\end{remark}

\begin{table}[H]
	\centering
	\caption{
		Numerical illustration of particle-to-PFU compression and protocol
		variation in the two-subpopulation plaque model. The numbers are
		illustrative and are chosen only to show the structure of the inverse
		problem.
	}
	\label{tab:numerical_two_subpopulation_plaque_reading}
	\renewcommand{\arraystretch}{1.18}
	\begin{tabularx}{0.98\linewidth}{@{}p{0.22\linewidth}p{0.24\linewidth}p{0.24\linewidth}X@{}}
		\toprule
		\textbf{Scenario}
		&
		\textbf{Latent composition}
		&
		\textbf{Protocol weights}
		&
		\textbf{Observed implication}
		\\
		\midrule
		
		Same particle count, different competence
		&
		\(C_{\mathrm{part}}=10^8\), with
		\(\varphi_{\mathrm C}=0.10\) versus \(0.01\)
		&
		\(p_{\mathrm C}=0.80\),
		\(p_{\mathrm I}=10^{-3}\)
		&
		Same particle concentration gives
		\(\Lambda_{\mathrm{PFU}}\approx8.09\times10^6\) versus
		\(8.99\times10^5\;\mathrm{PFU/mL}\).
		\\[0.45em]
		
		Same PFU, different composition
		&
		\((C_{\mathrm C},C_{\mathrm I})=(10^7,0)\) or
		\((5\times10^6,8\times10^7)\)
		&
		\(p_{\mathrm C}=0.80\),
		\(p_{\mathrm I}=0.05\)
		&
		Both lie on the same PFU-equivalence line and give
		\(\Lambda_{\mathrm{PFU}}=8.0\times10^6\;\mathrm{PFU/mL}\).
		\\[0.45em]
		
		Two non-collinear protocols
		&
		\((C_{\mathrm C},C_{\mathrm I})=(10^7,8\times10^7)\)
		&
		\(\mathbf p_1=(0.80,0.05)^{\mathsf T}\),
		\(\mathbf p_2=(0.20,0.15)^{\mathsf T}\)
		&
		The design matrix has nonzero determinant, so the two protocols add
		independent sensitivity directions in the ideal calibrated case.
		\\
		\bottomrule
	\end{tabularx}
\end{table}

\subsubsection{Closing Interpretation of the Plaque-Assay Worked Example}
\label{subsubsec:closing_interpretation_plaque_assay_worked_example}

\begin{remark}[Why the plaque assay closes the worked example]
	\label{rem:why_plaque_assay_closes_worked_example}
	The plaque assay closes the worked example well because it makes the abstract
	framework immediately concrete. Experimental collapse is not an exotic
	phenomenon reserved for advanced imaging, electrorotation, or nanomechanical
	manipulation. It is already present in one of the most basic measurements in
	virology. A plaque count is powerful precisely because it compresses a complex
	biological process into a simple observable. The theory developed in this paper
	does not weaken that observable. It clarifies what the observable means, what
	it measures, and what latent sectors it necessarily leaves unresolved.
\end{remark}

\begin{remark}[Main lesson of the worked example]
	\label{rem:main_lesson_plaque_worked_example}
	The plaque assay estimates a protocol-conditioned infectious concentration,
	\[
	\Lambda_{\mathrm{PFU}},
	\]
	not the full physical virion population. In the dilute regime, this
	concentration enters the familiar Poisson counting model and gives the
	standard PFU titer formula. In the latent-state view, however,
	\[
	\Lambda_{\mathrm{PFU}}
	=
	\int_{\Psi}
	\pi_{\mathrm{PFU}}(x;E_{\mathrm{PFU}})
	n_{\mathrm{ref}}(x)\,dx,
	\]
	so the titer is a weighted projection of particle abundance,
	plaque-forming competence, assay environment, and readout threshold. This is
	exactly what makes PFU biologically meaningful, but it is also what makes it
	protocol-conditioned. 
\end{remark}

\newpage 
\subsection*{Section Reference: Plaque Assay as Experimental Collapse}
\label{subsec:section_reference_plaque_assay_experimental_collapse}
\addcontentsline{toc}{subsection}{Section Reference: Plaque Assay as Experimental Collapse}

\begingroup
\footnotesize
\setlength{\tabcolsep}{3.5pt}
\renewcommand{\arraystretch}{1.22}

\begin{xltabular}{0.98\linewidth}{
		@{}
		B{0.20\linewidth}
		L{0.40\linewidth}
		Y
		@{}
	}
	\caption{
		Core objects, equations, and interpretations introduced in the
		plaque-assay worked example. The table summarizes how a plaque assay maps
		a latent virion or infectious-unit ensemble into a protocol-conditioned
		infectious readout.
	}
	\label{tab:plaque_assay_worked_example_reference}
	\\
	\toprule
	\textbf{Concept or object}
	&
	\CenteredTableHead{Mathematical form}
	&
	\textbf{Interpretation}
	\\
	\midrule
	\endfirsthead
	
	\caption[]{
		Core objects, equations, and interpretations introduced in the
		plaque-assay worked example.
		\emph{Continued from previous page.}
	}
	\\
	\toprule
	\textbf{Concept or object}
	&
	\CenteredTableHead{Mathematical form}
	&
	\textbf{Interpretation}
	\\
	\midrule
	\endhead
	
	\midrule
	\multicolumn{3}{r}{\emph{Continued on next page.}}
	\\
	\endfoot
	
	\bottomrule
	\endlastfoot
	
	Plaque-assay collapse map
	&
	\EqCell{
		\begin{gathered}
			\text{latent virion or infectious-unit population}
			\\
			\Downarrow
			\\
			\text{protocol-conditioned infectious events}
			\\
			\Downarrow
			\\
			\text{visible plaque count}
		\end{gathered}
	}
	&
	Compact statement that a plaque assay reports a visible infectious-lesion
	count, not a direct census of total virions, genomes, antigens, or latent
	mechanical states.
	\\[0.6em]
	
	Plaque-assay protocol
	&
	\EqCell{
		E_{\mathrm{PFU}}
		=
		\left(
		\begin{gathered}
			\text{virus preparation},\ \text{cell line},\
			\text{monolayer state}
			\\
			\text{dilution series},\ V_{\mathrm{inoc}},\
			t_{\mathrm{ads}}
			\\
			\text{overlay},\ t_{\mathrm{inc}},\
			\text{temperature}
			\\
			\text{staining},\ \text{threshold},\
			\text{counting rule}
		\end{gathered}
		\right)
	}
	&
	Specifies the biological and procedural conditions that define the plaque
	assay, beyond the final act of counting plaques.
	\\[0.6em]
	
	Latent plaque-assay state
	&
	\EqCell{
		X\in\Psi
	}
	&
	Latent virion, particle, aggregate, or infectious-unit state before the
	plaque protocol acts. It may include integrity, genome competence, receptor
	binding, aggregation, neutralization, orientation, charge, and handling
	history.
	\\[0.6em]
	
	Plaque-forming probability
	&
	\EqCell{
		\begin{aligned}
			\pi_{\mathrm{PFU}}(x;E_{\mathrm{PFU}})
			&=
			\Pr\!\left(
			\begin{gathered}
				x\text{ generates a visible,}
				\\
				\text{countable plaque under }E_{\mathrm{PFU}}
			\end{gathered}
			\right)
		\end{aligned}
	}
	&
	State- and protocol-dependent probability that a latent unit passes the
	plaque-assay pathway and produces a visible countable plaque.
	\\[0.6em]
	
	Staged plaque-pathway factorization
	&
	\EqCell{
		\begin{aligned}
			\pi_{\mathrm{PFU}}(x;E_{\mathrm{PFU}})
			&=
			p_{\mathrm{surv}}\,
			p_{\mathrm{deliv}}\,
			p_{\mathrm{ads}}\,
			p_{\mathrm{entry}}
			\\
			&\quad{}\times
			p_{\mathrm{rep}}\,
			p_{\mathrm{spread}}\,
			p_{\mathrm{vis}}
		\end{aligned}
	}
	&
	Reduced representation of plaque formation as a staged biological pathway:
	survival, delivery, adsorption, entry, replication, local spread, and
	visibility. The factors should be interpreted as conditional pathway
	probabilities.
	\\[0.6em]
	
	Effective plaque-forming concentration
	&
	\EqCell{
		\Lambda_{\mathrm{PFU}}(E_{\mathrm{PFU}})
		=
		\int_{\Psi}
		\pi_{\mathrm{PFU}}(x;E_{\mathrm{PFU}})
		n_{\mathrm{ref}}(x)\,dx
	}
	&
	Central collapse equation for the plaque assay: the protocol-weighted
	concentration of latent units that generate visible plaque-forming events.
	\\[0.6em]
	
	Physical particle concentration
	&
	\EqCell{
		C_{\mathrm{part}}
		=
		\int_{\Psi}
		n_{\mathrm{ref}}(x)\,dx
	}
	&
	Total physical particle or assay-unit concentration before plaque-forming
	weighting. In general, this is not equal to \(\Lambda_{\mathrm{PFU}}\).
	\\[0.6em]
	
	Particle-to-PFU ratio
	&
	\EqCell{
		\mathcal R_{\mathrm{part/PFU}}
		=
		\frac{C_{\mathrm{part}}}{\Lambda_{\mathrm{PFU}}}
	}
	&
	Summarizes how strongly the physical particle ensemble is compressed by the
	plaque-forming projection.
	\\[0.6em]
	
	Dilute-regime expected count
	&
	\EqCell{
		\mathbb E[N_d]
		=
		V_{\mathrm{inoc}}\,f_d\,\Lambda_{\mathrm{PFU}}
	}
	&
	Expected plaque count equals plated volume times dilution fraction times the
	effective plaque-forming concentration.
	\\[0.6em]
	
	Dilute Poisson plaque-count model
	&
	\EqCell{
		\begin{aligned}
			N_d
			&\sim
			\operatorname{Poisson}(\mu_d),
			\\
			\mu_d
			&=
			V_{\mathrm{inoc}}\,f_d\,
			\Lambda_{\mathrm{PFU}}(E_{\mathrm{PFU}})
		\end{aligned}
	}
	&
	Baseline count model when plaque-forming events are dilute, independent,
	spatially separated, and countable.
	\\[0.6em]
	
	Single-dilution PFU estimator
	&
	\EqCell{
		\widehat{\Lambda}_{\mathrm{PFU},d}
		=
		\frac{N_d}{V_{\mathrm{inoc}}\,f_d}
	}
	&
	Basic estimator of the effective plaque-forming concentration from one
	countable dilution.
	\\[0.6em]
	
	Usual PFU titer formula
	&
	\EqCell{
		\widehat{\Lambda}_{\mathrm{PFU},d}
		=
		\frac{N_dD_d}{V_{\mathrm{inoc}}},
		\qquad
		D_d=\frac{1}{f_d}
	}
	&
	Standard titer formula expressed as an estimator of
	\(\Lambda_{\mathrm{PFU}}\), the protocol-conditioned infectious
	concentration.
	\\[0.6em]
	
	Dilution-series likelihood
	&
	\EqCell{
		\begin{aligned}
			\mathcal L_{\mathrm{PFU}}(\Lambda_{\mathrm{PFU}})
			&=
			\prod_{j=1}^{J}
			\prod_{r=1}^{R_j}
			\frac{
				\mu_{jr}^{N_{jr}}
				e^{-\mu_{jr}}
			}{
				N_{jr}!
			},
			\\
			\mu_{jr}
			&=
			V_{\mathrm{inoc}}f_j
			\Lambda_{\mathrm{PFU}}
		\end{aligned}
	}
	&
	Likelihood for replicate plaque counts across dilution fractions under the
	dilute Poisson model.
	\\[0.6em]
	
	Dilution-series MLE
	&
	\EqCell{
		\widehat{\Lambda}_{\mathrm{PFU}}^{\mathrm{MLE}}
		=
		\frac{
			\sum_{j=1}^{J}\sum_{r=1}^{R_j}N_{jr}
		}{
			V_{\mathrm{inoc}}
			\sum_{j=1}^{J}R_j f_j
		}
	}
	&
	Maximum-likelihood estimator for \(\Lambda_{\mathrm{PFU}}\) using exact
	count observations in a dilution series.
	\\[0.6em]
	
	Overdispersed plaque-count model
	&
	\EqCell{
		\begin{aligned}
			N_{jr}
			&\sim
			\operatorname{NegBin}(\mu_{jr},\kappa),
			\\
			\mu_{jr}
			&=
			V_{\mathrm{inoc}}f_j\Lambda_{\mathrm{PFU}},
			\\
			\operatorname{Var}(N_{jr})
			&=
			\mu_{jr}
			+
			\frac{\mu_{jr}^{2}}{\kappa}
		\end{aligned}
	}
	&
	Optional extension when replicate counts vary more than expected under the
	Poisson model, due to aggregation, cell-layer heterogeneity, plating
	variation, or other nonidealities.
	\\[0.6em]
	
	Single-unit plaque event
	&
	\EqCell{
		\begin{gathered}
			B\in\{0,1\},
			\\
			\Pr(B=1\mid x,E_{\mathrm{PFU}})
			=
			\pi_{\mathrm{PFU}}(x;E_{\mathrm{PFU}})
		\end{gathered}
	}
	&
	Binary event describing whether one latent assay unit produces a visible
	countable plaque.
	\\[0.6em]
	
	Single-unit readout kernel
	&
	\EqCell{
		\begin{aligned}
			K_{\mathrm{PFU}}^{\varnothing}(B\mid x)
			&=
			[
			\pi_{\mathrm{PFU}}(x;E_{\mathrm{PFU}})
			]^B
			\\
			&\quad{}\times
			[
			1-\pi_{\mathrm{PFU}}(x;E_{\mathrm{PFU}})
			]^{1-B}
		\end{aligned}
	}
	&
	Bernoulli null-inclusive readout kernel for an individual latent
	plaque-forming opportunity.
	\\[0.6em]
	
	Plaque-assay null channel
	&
	\EqCell{
		B=0
	}
	&
	Null outcome containing all latent states that fail to produce a visible
	counted plaque because of loss, failed adsorption, failed entry, failed
	replication, weak spread, merging, or subthreshold visibility.
	\\[0.6em]
	
	PFU-equivalence of latent distributions
	&
	\EqCell{
		\begin{aligned}
			n_1\sim_{\mathrm{PFU}} n_2
			\Longleftrightarrow\quad
			&\int_{\Psi}
			\pi_{\mathrm{PFU}}(x;E_{\mathrm{PFU}})
			n_1(x)\,dx
			\\
			&=
			\int_{\Psi}
			\pi_{\mathrm{PFU}}(x;E_{\mathrm{PFU}})
			n_2(x)\,dx
		\end{aligned}
	}
	&
	Two latent ensembles are indistinguishable by plaque counts under the same
	protocol when they have the same effective plaque-forming concentration.
	\\[0.6em]
	
	Two-subpopulation decomposition
	&
	\EqCell{
		n_{\mathrm{ref}}(x)
		=
		n_{\mathrm C}(x)
		+
		n_{\mathrm I}(x)
	}
	&
	Reduces the latent ensemble into plaque-competent and plaque-incompetent,
	weakly competent, damaged, neutralized, or assay-incompatible sectors.
	\\[0.6em]
	
	Subpopulation concentrations
	&
	\EqCell{
		\begin{aligned}
			C_{\mathrm C}
			&=
			\int_{\Psi}n_{\mathrm C}(x)\,dx,
			\\
			C_{\mathrm I}
			&=
			\int_{\Psi}n_{\mathrm I}(x)\,dx,
			\\
			C_{\mathrm{part}}
			&=
			C_{\mathrm C}+C_{\mathrm I}
		\end{aligned}
	}
	&
	Physical concentrations of the two reduced latent sectors and their total.
	\\[0.6em]
	
	Subpopulation-averaged plaque probabilities
	&
	\EqCell{
		\begin{aligned}
			\overline p_{\mathrm C}
			&=
			\frac{
				\int_{\Psi}
				\pi_{\mathrm{PFU}}(x;E_{\mathrm{PFU}})
				n_{\mathrm C}(x)\,dx
			}{
				C_{\mathrm C}
			},
			\\[0.25em]
			\overline p_{\mathrm I}
			&=
			\frac{
				\int_{\Psi}
				\pi_{\mathrm{PFU}}(x;E_{\mathrm{PFU}})
				n_{\mathrm I}(x)\,dx
			}{
				C_{\mathrm I}
			}
		\end{aligned}
	}
	&
	Averaged plaque-forming probabilities that allow each sector to remain
	internally heterogeneous.
	\\[0.6em]
	
	Exact two-subpopulation PFU projection
	&
	\EqCell{
		\Lambda_{\mathrm{PFU}}
		=
		\overline p_{\mathrm C}C_{\mathrm C}
		+
		\overline p_{\mathrm I}C_{\mathrm I}
	}
	&
	Exact weighted-sum representation after decomposing the latent ensemble into
	two sectors.
	\\[0.6em]
	
	Constant-probability approximation
	&
	\EqCell{
		\Lambda_{\mathrm{PFU}}
		\approx
		p_{\mathrm C}C_{\mathrm C}
		+
		p_{\mathrm I}C_{\mathrm I}
	}
	&
	Simplified two-sector model when the plaque-forming probability is
	approximately constant within each sector.
	\\[0.6em]
	
	Plaque-competent fraction
	&
	\EqCell{
		\begin{gathered}
			\varphi_{\mathrm C}
			=
			\frac{C_{\mathrm C}}{C_{\mathrm{part}}},
			\\
			1-\varphi_{\mathrm C}
			=
			\frac{C_{\mathrm I}}{C_{\mathrm{part}}}
		\end{gathered}
	}
	&
	Fraction of the physical particle or assay-unit population belonging to the
	plaque-competent sector.
	\\[0.6em]
	
	Competent-fraction form
	&
	\EqCell{
		\begin{aligned}
			\Lambda_{\mathrm{PFU}}
			&=
			C_{\mathrm{part}}
			\left[
			p_{\mathrm I}
			+
			(p_{\mathrm C}-p_{\mathrm I})
			\varphi_{\mathrm C}
			\right]
		\end{aligned}
	}
	&
	Separates physical abundance, latent competent fraction, and
	protocol-dependent plaque-forming probabilities.
	\\[0.6em]
	
	Two-subpopulation particle-to-PFU ratio
	&
	\EqCell{
		\mathcal R_{\mathrm{part/PFU}}
		=
		\frac{1}{
			p_{\mathrm I}
			+
			(p_{\mathrm C}-p_{\mathrm I})
			\varphi_{\mathrm C}
		}
	}
	&
	Particle-to-PFU compression is the reciprocal of the average plaque-forming
	probability of a randomly selected physical particle or assay unit.
	\\[0.6em]
	
	Nearly ideal competent-sector limit
	&
	\EqCell{
		\begin{gathered}
			p_{\mathrm C}\approx1,
			\qquad
			p_{\mathrm I}\approx0,
			\\
			\Lambda_{\mathrm{PFU}}\approx C_{\mathrm C},
			\qquad
			\mathcal R_{\mathrm{part/PFU}}
			\approx
			\frac{1}{\varphi_{\mathrm C}}
		\end{gathered}
	}
	&
	Idealized case in which PFU approximately reports the competent-sector
	concentration and the particle-to-PFU ratio approximates the inverse
	competent fraction.
	\\[0.6em]
	
	Non-identifiability from PFU alone
	&
	\EqCell{
		\Lambda_{\mathrm{PFU}}
		=
		p_{\mathrm C}C_{\mathrm C}
		+
		p_{\mathrm I}C_{\mathrm I}
	}
	&
	A single PFU readout identifies only one weighted sum, not
	\(C_{\mathrm C}\), \(C_{\mathrm I}\), \(p_{\mathrm C}\), and
	\(p_{\mathrm I}\) separately.
	\\[0.6em]
	
	Competent fraction from auxiliary particle count
	&
	\EqCell{
		\varphi_{\mathrm C}
		=
		\frac{
			\Lambda_{\mathrm{PFU}}/C_{\mathrm{part}}
			-
			p_{\mathrm I}
		}{
			p_{\mathrm C}-p_{\mathrm I}
		}
	}
	&
	If \(C_{\mathrm{part}}\), \(\Lambda_{\mathrm{PFU}}\), and the two protocol
	probabilities are known or calibrated, the competent fraction can be
	inferred.
	\\[0.6em]
	
	Protocol-variation linear system
	&
	\EqCell{
		\Lambda_m
		=
		p_{\mathrm C}^{(m)}C_{\mathrm C}
		+
		p_{\mathrm I}^{(m)}C_{\mathrm I},
		\qquad
		m=1,\ldots,M
	}
	&
	Multiple plaque protocols define multiple weighted projections of the same
	two-sector latent composition.
	\\[0.6em]
	
	Two-subpopulation design matrix
	&
	\EqCell{
		\begin{gathered}
			\boldsymbol{\Lambda}
			=
			\mathbf P\mathbf C,
			\\[0.25em]
			\mathbf P
			=
			\begin{pmatrix}
				p_{\mathrm C}^{(1)} & p_{\mathrm I}^{(1)}\\
				\vdots & \vdots\\
				p_{\mathrm C}^{(M)} & p_{\mathrm I}^{(M)}
			\end{pmatrix},
			\qquad
			\mathbf C
			=
			\begin{pmatrix}
				C_{\mathrm C}\\
				C_{\mathrm I}
			\end{pmatrix}
		\end{gathered}
	}
	&
	Matrix form of protocol variation as a two-subpopulation inverse problem.
	\\[0.6em]
	
	Rank condition for two-sector resolution
	&
	\EqCell{
		\operatorname{rank}(\mathbf P)=2
	}
	&
	In the ideal noiseless and calibrated case, the two subpopulation
	concentrations are identifiable only if the protocol probability vectors are
	not collinear.
	\\[0.6em]
	
	Single-protocol Fisher information
	&
	\EqCell{
		\begin{aligned}
			\mathcal I_E(\theta)
			&=
			\frac{
				(V_{\mathrm{inoc}}f_d)^2
			}{
				\mu_d
			}
			\mathbf p\mathbf p^{\mathsf T},
			\\
			\theta
			&=
			(C_{\mathrm C},C_{\mathrm I})^{\mathsf T}
		\end{aligned}
	}
	&
	A single plaque protocol contributes rank-one information and identifies
	only one concentration combination.
	\\[0.6em]
	
	Multi-protocol Fisher information
	&
	\EqCell{
		\begin{aligned}
			\mathcal I_{\mathrm{multi}}(\theta)
			&=
			\sum_{m=1}^{M}
			\frac{(V_m f_m)^2}{\mu_m}
			\mathbf p_m\mathbf p_m^{\mathsf T},
			\\
			\mathbf p_m
			&=
			\begin{pmatrix}
				p_{\mathrm C}^{(m)}\\
				p_{\mathrm I}^{(m)}
			\end{pmatrix}
		\end{aligned}
	}
	&
	Independent protocol contributions add. The Fisher rank can increase only
	when the protocol probability vectors span more than one direction.
	\\[0.6em]
	
	Protocol-variation design principle
	&
	\EqCell{
		\mathbf p_m
		\not\parallel
		\mathbf p_{m'}
	}
	&
	A useful plaque-assay variation changes the biological observation kernel in
	a direction that separates latent sectors rather than merely rescaling the
	same PFU projection.
	\\
\end{xltabular}

\endgroup

\section{Conclusion}
\label{sec:paper_conclusion}

This paper has developed a protocol-resolved formulation of experimental
collapse in virophysics. The central claim is that a virological measurement
does not, in general, report the full latent virion state or the full latent
virion--environment ensemble. It reports a protocol-conditioned observed
ensemble shaped by preparation, forcing, medium interaction, selection, survival,
detection, amplification, and readout. The observed ensemble is therefore not
always the same mathematical object as the reference latent ensemble. It is the
ensemble produced by a specified experimental map.

The paper began by distinguishing reference latent ensembles,
protocol-conditioned latent ensembles, and observed ensembles. This distinction
was expressed through protocol kernels, survival or detection weights, readout
kernels, and null observations. In this language, an experiment is not only a
passive window onto a virion population. It is a map
\[
P_{\mathrm{ref},t}
\longmapsto
P_{\mathrm{obs},t}^{\varnothing}(\cdot\mid E),
\]
whose structure determines which latent states are preserved, transformed,
selected, discarded, amplified, or made visible.
The concept of experimental collapse was then decomposed into concrete
mechanisms. Preparation and interface effects can reshape the ensemble before
measurement. Surfaces can immobilize or deform particles. Fields can steer,
polarize, trap, or rotate virions. Structured media such as mucus can filter
transport by adhesion, confinement, and local rheology. Biological assays can
amplify only those states that complete a required sequence of attachment,
entry, replication, spread, and visibility. Detection thresholds and rejection
criteria can assign large parts of the population to the null channel. These
mechanisms do not make experiments invalid. They define what each experiment
actually measures.
 
The framework also clarifies protocol blindness. A protocol may be insensitive
to mechanically or biologically real latent sectors, either locally through a
deterministic blind subspace, statistically through Fisher-blind directions, or
globally through observational equivalence classes. The quotient-space viewpoint
makes this especially clear: a protocol partitions latent state space into
classes of states that are indistinguishable under that protocol. Multi-protocol
experiments refine this partition. They can make previously collapsed
distinctions visible by probing the latent ensemble through different kernels.

This perspective converts experimental collapse from a warning into an inference
principle. If the protocol kernel is known, estimated, calibrated, or
parametrized, then the observed data can be used to infer latent virion
parameters, environmental parameters, and protocol-specific coupling parameters.
A strongly conditioning protocol can be scientifically powerful precisely because
it generates signal: AFM loading reveals mechanical response, dielectrophoresis
and electrorotation reveal dielectric response, mucus tracking reveals
medium-filtered transport, cryo-EM reveals preparation-conditioned structural
ensembles, and plaque assays reveal assay-conditioned infectious activity.

The plaque-assay worked example illustrated the value of the framework in one of
virology's most familiar measurements. A plaque count is not a count of total
physical virions, total genomes, total particles, or total mechanically
admissible states. It is a count of visible infectious lesions under a specified
cell-line, adsorption, overlay, incubation, staining, and counting protocol. In
the notation of this paper, the assay estimates an effective plaque-forming
concentration
\[
\Lambda_{\mathrm{PFU}}
=
\int_{\Psi}
\pi_{\mathrm{PFU}}(x;E_{\mathrm{PFU}})
n_{\mathrm{ref}}(x)\,dx,
\]
a protocol-weighted projection of the latent ensemble. The two-subpopulation
reduction showed that a single plaque protocol identifies only a weighted
combination of plaque-competent and plaque-incompetent, weakly competent, or
assay-incompatible sectors. Additional measurements or controlled protocol
variation are required to resolve the underlying latent composition.

The broader conclusion is methodological. Virological observables are most
powerful when interpreted together with the protocols that produce them.
Particle counts, plaque counts, force curves, density maps, field-driven
trajectories, and mucus-tracking statistics need not be treated as competing
direct reports of the same object. They can instead be understood as different
protocol-conditioned projections of a shared latent virion--environment system.
Disagreement between protocols is therefore not automatically a problem. It may
be evidence that the protocols are resolving different sectors of the same
latent mechanics.
\medskip 

\noindent The final message of the paper is:
\begin{empheq}[box=\fbox]{equation}
	\text{Latent virion--environment mechanics}
	+
	\text{Protocol mechanics}
	=
	\text{Experimentally visible virophysics}.
	\label{eq:final_paper_summary_equation}
\end{empheq}
A rigorous virophysical interpretation should specify both sides. It should
describe the latent state being modeled, and it should state how the
experimental protocol transforms that state into data. Experimental collapse is
the formal bridge between these two levels.

\end{document}